\documentclass{aa}  

\usepackage{graphicx}
\usepackage{txfonts}
\usepackage{xcolor}
\usepackage{array}
\usepackage{color}
\usepackage{colortbl}
\usepackage{pdflscape}
\begin{document} 

\newcommand{\gmag}{$G$}
\newcommand{\gbp}{$G_{BP}$}
\newcommand{\grp}{$G_{RP}$}
\newcommand{\gaia}{\textit{Gaia~}}
\newcommand{\feh}{${\rm [Fe/H]}$}
   \title{Gaia DR3: Specific processing and validation of all-sky RR Lyrae and Cepheid stars - The RR Lyrae sample}


   \author{G. Clementini$^{1}$,
          V. Ripepi\inst{2},
          A. Garofalo\inst{1},
          R. Molinaro\inst{2},
          T. Muraveva\inst{1},
          S. Leccia\inst{2},
          L. Rimoldini\inst{3},
          B. Holl\inst{3},
          G. Jevardat de Fombelle\inst{3},
          P. Sartoretti\inst{4},
          O. Marchal\inst{5},
          M. Audard\inst{3,6},
          K. Nienartowicz\inst{3,7},
          R. Andrae\inst{8},
          M. Marconi\inst{2},
          L. Szabados\inst{9,10} 
          D.~W. Evans\inst{11},
          I. Lecoeur-Taibi\inst{3},
          N. Mowlavi\inst{3},
          I. Musella\inst{2} and 
          L. Eyer\inst{6}
                        }

  \institute{$^{1}$ INAF,
Osservatorio di Astrofisica e Scienza dello Spazio di Bologna, via Piero Gobetti 93/3,
40129 Bologna, Italy \\
              \email{gisella.clementini@inaf.it}\\
             $^{2}$ INAF-Osservatorio Astronomico di Capodimonte, Via Moiariello 16, I-80131 Naples, Italy\\
             $^{3}$ Department of Astronomy, University of Geneva, Ch. d'Ecogia 16, 1290 Versoix, Switzerland\\
             $^{4}$ GEPI, Observatoire de Paris, Universit\'e PSL, CNRS, 5 Pla Jules Janssen,92190 Meudon, France\\ 
             $^{5}$Observatoire Astronomique de Strasbourg, Universit\'{e} de Strasbourg, CNRS, UMR 7550, 11 rue de l'Universit\'{e}, 67000 Strasbourg, France\\
             $^{6}$Department of Astronomy, University of Geneva, Chemin Pegasi 51, 1290 Versoix, Switzerland\\
             $^{7}$ Sednai S\`arl, Geneva, Switzerland\\
             $^{8}$ Max Planck Institute for Astronomy, K\"{o}nigstuhl 17, 69117 Heidelberg, Germany\\
             $^{9}$ Konkoly Observatory, ELKH Research Centre for Astronomy and Earth Sciences, E\"otv\"os Lor\'and Research Network,  
             Konkoly Thege Mikl\'os \'ut 15-17, 1121 Budapest,  Hungary\\
             $^{10}$ CSFK Lend\"ulet Near-Field Cosmology Research Group, Konkoly Thege Mikl\'os \'ut 15-17, 1121 Budapest, Hungary\\
               $^{11}$ Institute of Astronomy, University of Cambridge, Madingley Road, Cambridge CB3 0HA, UK
             }

   \date{Received Month day, year; accepted Month day, year}

  \abstract
{RR Lyrae stars are  excellent tracers of the oldest stars (ages $\gtrsim$ 9-10 Gyr) 
and standard candles to measure the distance to stellar systems composed prevalently by an old stellar population. 
The {\it Gaia} Third Data Release (DR3) publishes a catalogue  of full-sky RR Lyrae stars observed 
during the initial 34 months of science 
operations, that were processed through the Specific Object Study (SOS) pipeline, developed to validate and  
characterise Cepheids and RR Lyrae stars (SOS Cep\&RRL) observed by {\it Gaia}.}   
{Main steps of the SOS Cep\&RRL pipeline are  described in the documentation and 
papers  accompanying previous {\it Gaia} Data Releases. For DR3 the 
pipeline  was modified as to process, besides  the {\it Gaia} multiband ($G$, $G_{\rm BP}$, $G_{\rm RP}$) time series photometry, 
also epoch 
radial velocities measured for RR Lyrae and Cepheids by the Radial Velocity Spectrometer (RVS)  on board {\it Gaia}.}
{The SOS Cep\&RRL validation of DR3 candidate  RR Lyrae stars relies 
on 
diagnostics tools that include the Period {\it versus} $G$-amplitude diagram and the  Period {\it versus} $\phi_{21}$ and $\phi_{31}$ parameters of the $G$ light curve Fourier decomposition, as defined by a reference sample of bona fide RR Lyrae stars known in the literature (Gold Sample). Great care was devoted to build a large and pure Gold Sample comprising more than 200\,000 RR Lyrae stars.
The SOS processing led to an initial catalogue of 271\,779 RR Lyrae  stars, that are listed in the {\tt vari\_rrlyrae} table of the DR3 archive. 
A thorough cleaning procedure
was then performed,  to produce a final catalogue of  
SOS-confirmed DR3 RR Lyrae stars, by dropping sources that clearly  are  
contaminants, or have an uncertain classification.}
{Multiband time series photometry and characterisation 
are published in {\it Gaia} DR3  for a clean, validated sample of 270\,905 RR Lyrae stars (174\,947 fundamental mode, 93\,952 first overtone and 2\,006 double-mode RR Lyrae) confirmed and fully characterised by the SOS Cep\&RRL pipeline. They  
are  distributed all over the sky, including variables in 95  globular clusters and 25 
Milky Way (MW) companions  (the  Magellanic Clouds, 7 dwarf spheroidal galaxies and 16 ultra-faint dwarf satellites of the MW). RVS time series radial velocities are also published for 1\,096 RR Lyrae and 799 Cepheids of different types (Classical, Anomalous and Type II Cepheids). Of the 270\,905 DR3 RR Lyrae stars, 200\,294 are already known in the literature (Gold Sample) and 70\,611 are, to the best of our knowledge,  new discoveries by {\it Gaia}. 
An estimate of the interstellar absorption is  published for 142\,660 fundamental-mode RR Lyrae stars from a relation based on the $G$-band amplitude and the pulsation period.
Metallicities derived from the Fourier parameters of the light curves are  also released for 133\,559 RR Lyrae stars.}
{The final {\it Gaia} DR3 catalogue of confirmed RR Lyrae stars almost
doubles the DR2 RR Lyrae catalogue. An increase of 
statistical significance, a better characterization of the RR Lyrae pulsational and astrophysical parameters, and the improved astrometry published with {\it Gaia}  EDR3, make the SOS Cep\&RRL DR3 sample, the largest, most homogeneous and parameter-rich catalogue of All-Sky RR Lyrae stars published so far, in the magnitude range from $\langle G \rangle$= 7.64 mag (the magnitude of RR Lyr itsef, the class prototype) to 
$\langle G \rangle$ = 21.14 mag (the faintest RR Lyrae in the 
catalogue).}


\titlerunning{Gaia Data Release 3 - The RR Lyrae sample}
\authorrunning{Clementini et al.}
   \maketitle
%



\section{Introduction}
\label{sec:intro}


RR Lyrae are low mass (typical masses  around 0.6$-$0.8 M$_\sun$),  old (ages larger than  9-10 billion years) stars whose surface  expands and contracts regularly with periods shorter than a day.  The pulsation gives rise to a variation of brightness occurring in two main flavours.
The c-type RR Lyrae (RRc) stars have close to sinusoidal light curves, that repeat in time with periodicities typically between about 0.2 and 0.42 day (d), and an  amplitude of the light variation of  up to about half a magnitude in the visual band. RRc stars 
 oscillate in the radial first overtone pulsation mode. 
The ab-type RR Lyrae (RRab) stars have asymmetrical, sawtooth light curves, periods typically between about 0.42 and 1 d, and visual amplitudes 
 in the {\it Gaia G} band from 
above  0.1 to more than  one magnitude. They pulsate in the fundamental pulsation mode. 
In a small fraction of RR Lyrae stars, fundamental and first overtone pulsation modes are excited at the same time, giving rise to the double-mode (d-type) RR Lyrae (RRd). 
The characteristic light variation makes RR Lyrae stars easily  recognizable even in very crowded region of the sky or high extinction conditions. Their pulsation characteristics (periods, amplitudes, mean magnitudes/colours,  etc) allow an estimate of their metal abundance and reddening, thus  bringing invaluable added values  
 \citep[see, e.g.][for comprehensive reviews on RR Lyrae stars]
{1995CAS....27.....S, Catelan-2015}.

RR Lyrae stars allow to measure the distance to the systems they live in as they follow luminosity – metallicity ($LZ$) relations in the visual bands 
(see, e.g. \citealt[][and references therein]{2003LNP...635..105C, 2003AJ....125.1309C} 
and, e.g. \citealt[][for more recent $LZ$ relations in the Johnson $V$ and $Gaia$ $G$ bands, calibrated on $Gaia$ parallaxes]{2018MNRAS.481.1195M, MNRAS-in-press}) and period – luminosity - (metallicity) ($PLZ$) relations in the infrared \citep[see, e.g.][and references therein, for $PLZ$ relations in the $K_s$ and $W_1$ infrared bands]{2018MNRAS.481.1195M,2021ApJ...909..200B}.  These variable stars are the only  abundant, stellar tracers with a small relative distance error ($\sim$5\%), particularly at large distances, hence allowing to measure the most accurate distances to old, distant stellar populations. But also at 5 - 10 kpc from the Milky Way (MW) disc, they are the stars with the most precise distances. 

RR Lyrae stars are also excellent tracers of the oldest stars which witnessed the first epoch of galaxy formation, thus can provide hints on the ``building blocks'' that contributed to the build up of galactic halos \citep[see, e.g.][and references therein]{Catelan-2015,Martinez-et-al-2019}.
All these properties make the RR Lyrae stars invaluable dynamical tracers of the Galaxy and a crucial tool for a plethora of applications, 
from Galactic archaeology and the search for the fossil records of the Galactic formation, to tidal streams, to dynamics, to validation of distances and extinction maps, and characterization of small scale extinction; therefore explaining the renewed, large interest in these stars within the astronomical community. 

Previous {\it Gaia} data releases 
(\citealt{2016A&A...595A...2G}; \citealt{2018A&A...616A...1G}) already showed the great potential of the mission in the field of stellar variability \citep{2019A&A...623A.110G} 
and for RR Lyrae stars and Cepheids in particular (\citealt{ Clementini-et-al-2017};  \citealt{Clementini-et-al-2016}, hereafter Paper~I;  \citealt{Clementini-et-al-2019}, hereafter Paper~II). 
This is thanks to the multi-epoch nature and almost simultaneous acquisition of the (spectro-)photometry, astrometry and spectroscopy data collected  by {\it Gaia} all over the celestial sphere.

 The small sample of RR Lyrae stars and Cepheids in the Large Magellanic Cloud (LMC; about 3\,200 sources in total) published with {\it Gaia} Data Release 1 (DR1; Paper~I), were only a teaser of the samples and data products published for these pulsating stars 
in  subsequent {\it Gaia} releases. Multi-band time-series photometry and pulsation characteristics were published in 
{\it Gaia} Data Release 2 (DR2; \citealt{2018A&A...616A...1G}) for a catalogue of 
about 140\,000 RR Lyrae stars  and for about 10\,000 Cepheids observed over the whole Galaxy (including  87 globular clusters and 14 MW  companions) during the first 22 months of {\it Gaia} operation 
(Paper II).
The number of RR Lyrae has now almost doubled, and that of Cepheids has increased by 50\%, 
compared to DR2, 
thanks to the larger amount and extended time baseline (34 months) 
spanned by the {\it Gaia} Data Release 3 (DR3). 

On 13 June 2022 with {\it Gaia} DR3 \citep{DR3-DPACP-185} 
multiband (\gmag, \gbp, and \grp) light curves are released for 270\,905 RR Lyrae stars and 15\,006
Cepheids confirmed and fully characterised by the Specific Object Study pipeline for Cepheids and RR Lyrae  (hereafter, SOS Cep\&RRL; Paper I and II)  developed 
within the {\it Gaia} Data Processing and Analysis Consortium (DPAC). 
Radial velocity (RV) time series measurements  obtained with the Radial Velocity Spectrometer (RVS; \citealt{DR2-DPACP-46}) on board \gaia for a sub-sample comprising 1\,096 
RR Lyrae stars and 799 Cepheids \citep{DR3-DPACP-169} 
validated by the SOS Cep\&RRL 
pipeline are also published with DR3. 
The above 
numbers all refer to samples  
cleaned  
from contaminants and sources with an uncertain classification.

In this paper we present changes implemented in the SOS Cep\&RRL pipeline with respect to DR2. We describe in particular a new module activated to process the RVS time series data and explain how the RR Lyrae and Cepheid variables with epoch RVs published in DR3 were selected.
We then focus on the procedures and validations that led to the production of the final, clean sample of 
confirmed RR Lyrae stars released in DR3. The processing 
and validation of the 
confirmed  DR3 Cepheids 
are described instead in a companion paper \citep{DR3-DPACP-169}.

Characteristic parameters derived by the SOS 
pipeline for the final sample of \gaia  
DR3 RR Lyrae stars, that  
are published 
in the \texttt{vari\_rrlyrae} table,   
include: 
pulsation period(s), peak-to-peak amplitudes of the $G$, $G_{BP}$ and $G_{RP}$ [Amp$(G)$, Amp($G_{BP}$), Amp($G_{RP}$)] light curves, classification in type/pulsation mode (RRab, RRc and RRd), mean magnitudes computed as an intensity-average over the full pulsation cycle, along with the $\phi_{21}$, $\phi_{31}$, $R_{21}$ and $R_{31}$ parameters  of the Fourier decomposition of the  $G$-band light curves.  
Metallicities ([Fe/H]) computed 
from a relation found to exist 
with the pulsation period and the $\phi_{31}$ Fourier parameter of the RR Lyrae light curves \citep{1996A&A...312..111J, Nemec-et-al-2013}, and  individual absorption values derived from the relation existing with the pulsation period and the amplitude of the light variation \citep{2002AJ....124.1528P} are also released  
for about a half of the whole SOS  
RR Lyrae sample. 
Amplitudes of the RV curves and mean RV 
values, 
computed 
after modeling the RV curves, are also provided in the DR3 \texttt{vari\_rrlyrae} and \texttt{vari\_cepheid} tables,  
for 
sub-samples of 
RR Lyrae stars and 
Cepheids.

The paper is organised as follows.
In Section~\ref{sec:sos-pip_genover}  we provide a brief overview of the SOS Cep\&RRL pipeline and describe changes and developments implemented to process the {\it Gaia} DR3 photometry and radial velocity time series data.   Section~\ref{sec:app_to_sos} presents the DR3 input data, the selections applied to run the  
pipeline 
on the RR Lyrae candidates and results from the processing.
Section~\ref{sec:s_rvs} focuses on the analysis of the RVS radial velocity (RV) time series data for sub-samples of RR Lyrae stars and Cepheids for which  
RVS RV curves are published in DR3. 
Section~\ref{sec:s_astro-par} 
specifically discusses the astrophysical parameters (metallicity and $G$-band absorption values) we derived from the pulsation characteristics of RR Lyrae stars.
Cleaning from contaminants, re-classifications and final validation 
of the  
RR Lyrae catalogue 
are described in 
Sect.~\ref{sec:s_vali}. 
In Sect.~\ref{sec:results} we discuss the completeness and purity of the clean catalogue of  RR Lyrae stars published in DR3 and present colour-magnitude diagrams and sky maps defined by these  RR Lyrae stars. Finally, conclusions and further developments of the SOS Cep\&RRL pipeline, in preparation for future {\it Gaia} data releases (DR4 and 5), are briefly discussed in Sect.~\ref{sec:conclusions}.

\section{The SOS Cep\&RRL pipeline: general overview and changes for DR3} 
\label{sec:sos-pip_genover}
The variability processing   
\citep[see Chapter~10 in {\it Gaia} DR3 Documentation, ]
[]{DR3-documentation}
\footnote{https://geapre.esac.esa.int/archive/documentation/GDR3. (This link will be updated afterwards).}
is based on the analysis of {\it Gaia} DR3
 calibrated $G$ and integrated $G_{BP}$ and $G_{RP}$ time-series photometry 
 \citep{2021A&A...649A...3R} 
 and, for RR Lyrae stars and Cepheids, also on the 
RVS radial velocity time series
\citep{CU6-DR3-documentation},  through the various steps (modules)  of the general variability analysis approach described in \citet{2017arXiv170203295E} 
and in \citet{DR3-DPACP-162} specifically for DR3. 
This includes identification and classification in different types of the variable 
candidates, that are then fed to the Specific Objects Study (SOS) pipelines,  each of which is  specifically tailored to validate and characterise a certain type of variability. 
Figure~10.1 in  the {\it Gaia} DR3 Documentation \citep{DR3-documentation} 
shows a schematic view of the general variability processing occurring before entering the SOS  
pipelines.

\begin{figure*}
   \centering
   \includegraphics[scale=0.6]{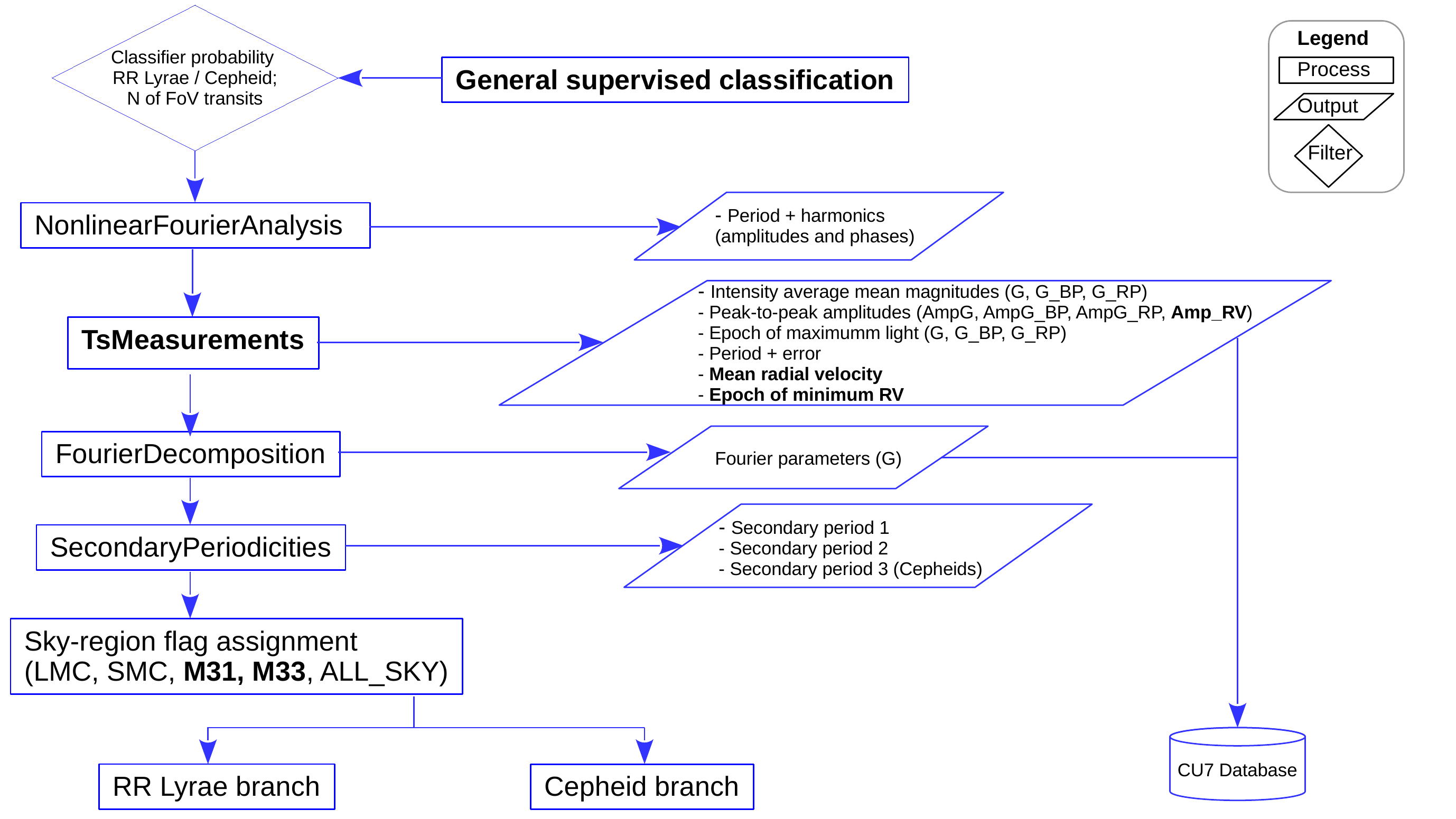}
   \caption{Schematic view of 
   modules (rectangles) and outputs (parallelograms) of the SOS Cep\&RRL pipeline that are in  common to RR Lyrae stars and Cepheids. The figure is an updated version of Fig. 1 in 
   Paper~II. Changes and additions to the pipeline version used in DR2 are highlighted in boldface (see text for details).}
              \label{fig:common-trunk}%
    \end{figure*}

\begin{figure*}
   \centering
    \includegraphics[scale=0.6]{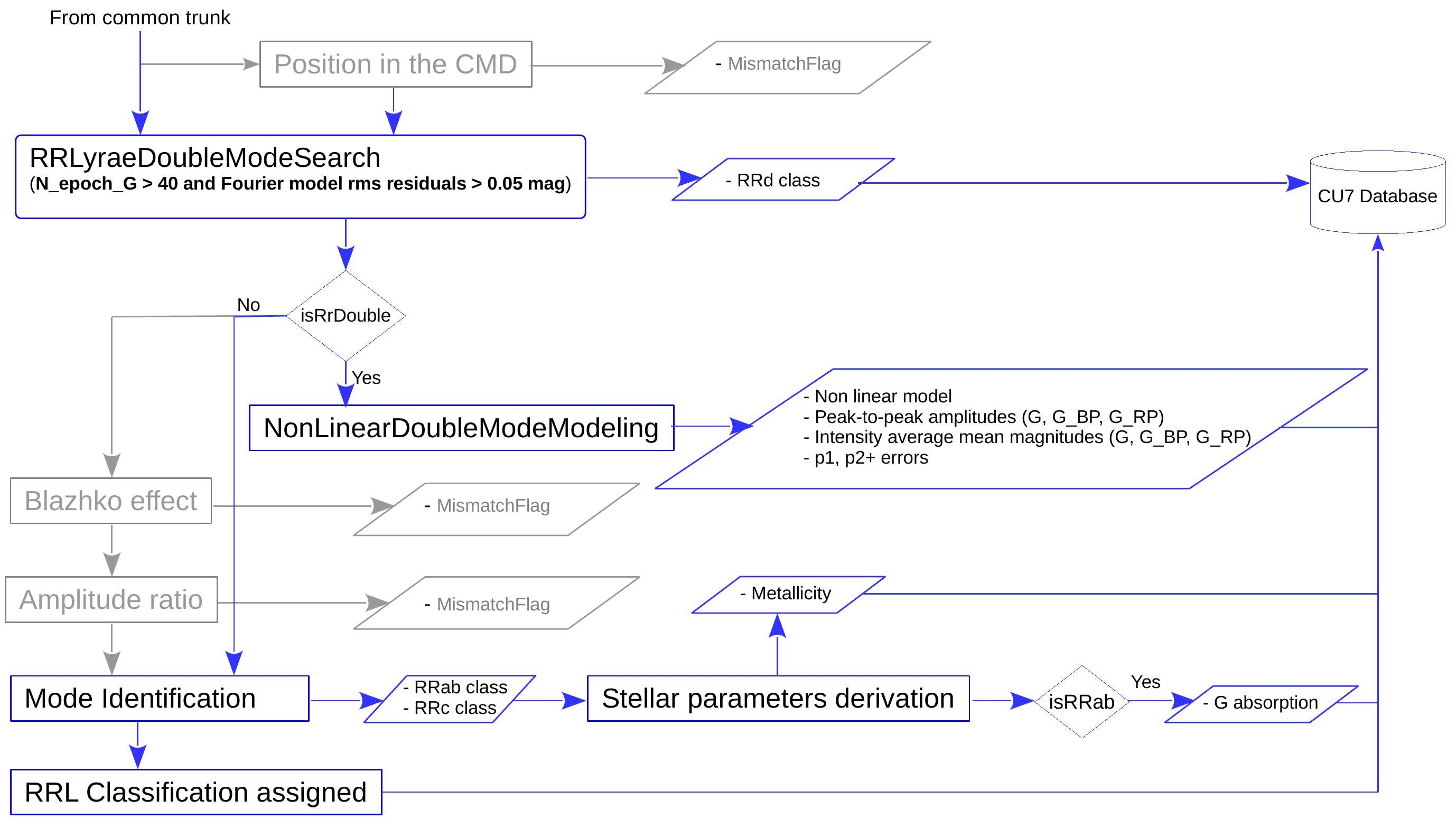}
   \caption{Modules and outputs of the RR Lyrae branch of the SOS Cep\&RRL pipeline. As in Fig.~\ref{fig:common-trunk} changes with respect to DR2 
   are highlighted in boldface. 
  Shown in gray are three modules 
  not activated for DR3 (see text for details).} 
              \label{fig:rrl-branch}%
    \end{figure*}

The SOS Cep\&RRL pipeline is designed to validate and  characterise all-sky candidate  RR Lyrae stars and Cepheids identified in the  \gaia \gmag, \gbp, and \grp ~time series photometry by the General Supervised Classification module (\citealt{DR3-DPACP-165}) in the 
variability analysis pipeline. 
This module 
uses machine learning methods and 
multi-class supervised classifiers  
to classify {\it Gaia} time series of candidate variable objects into categories. A full description of the SOS Cep\&RRL pipeline can be found in Papers~I and II, to which the interested reader is referred for details. In the following, we summarise steps of the pipeline that are in common to RR Lyrae stars
and Cepheids (pipeline main trunk, Fig.~\ref{fig:common-trunk}) and 
the processing taking place in the RR Lyrae branch (Fig.~\ref{fig:rrl-branch}).  The Cepheid branch is discussed in a companion paper \citep{DR3-DPACP-169}.
\par\noindent 
Steps of the SOS Cep\&RRL 
pipeline main trunk include:
\begin{itemize}
\item derivation of the source periodicity ($P$) with the Lomb-Scargle algorithm (\citealt{1976Ap&SS..39..447L}; \citealt{1982ApJ...263..835S}; see Sect. 2.1 in Paper I); 
\item  modeling
of the time series data, folded according to the derived $P$,  using the Levenberg-Marquardt (Levenberg 1944; Marquardt 1963) non-linear fitting
algorithm ({\it NonLinearFourierAnalysis} module in Fig. 1), which refines both the period and model of the light curves;  
\item derivation of the variable star main 
parameters [intensity-averaged  mean magnitudes;
peak-to-peak amplitudes: Amp$(G)$, Amp$(G_{BP})$, Amp$(G_{RP})$; epoch of maximum light; etc.], from the light curves  modelled by the non-linear fitting algorithm; 
\item Fourier decomposition of the $G$-band light curve and derivation of the Fourier parameters:  
$\phi_{21}$, $\phi_{31}$, R$_{21}$ and $R_{31}$;
\item search for secondary periodicities.
\end{itemize}
In the RR Lyrae branch the 
following diagnostic tools and combination of parameters are  used 
to extract bona fide RR Lyrae
stars from the candidates, classify  them in type (RRc, RRab, RRd)/pulsation mode (first-overtone, fundamental mode, double-mode), and derive their astrophysical parameters [metallicity -- [Fe/H], and absorption in the $G$-band -- $A(G)$]:
\begin{itemize}
\item pulsation Period (P) versus 
Amp$(G)$ diagram ($PA$); 
\item $\phi_{21}$ versus $P$ diagram;
\item R$_{21}$ versus $P$ diagram;
\item $\phi_{31}$  versus $P$ diagram; 
\item R$_{31}$ versus $P$
diagram;
\item $[$Fe/H$]$ versus $P$, $\phi_{31}$  relation;
\item $A(G)$ versus Amp$(G)$, $P$, $(G-G_{RP})$ relation.
\end{itemize}
A number 
of changes were implemented and new parts of the pipeline were activated for the processing of the DR3 
RR Lyrae and Cepheid candidates. We have highlighted them in boldface in Figs.~\ref{fig:common-trunk}, ~\ref{fig:rrl-branch} and describe 
below those occurring in (and before) the pipeline main trunk and the RR Lyrae branch. A detailed description of the changes implemented in the Cepheid branch (improved  detection of multimode Cepheids, adoption of new Period-Luminosity and Period-Wesenheit relations directly computed in the {\it Gaia} passbands for the different types of Cepheids, etc.) is presented in \citet{DR3-DPACP-169}. 

\begin{itemize}
    \item Optimization of  outlier-removal operators\\
    Outliers and measurements of poor quality in the {\it Gaia}  photometry and RV 
    time series data are filtered out by operators of the general variability pipeline, whose configuration parameters can be adapted to the specific type of variable sources under investigation.
    In the main chain of operators that are applied sequentially to transform and filter the input photometric time series the
    \texttt{RemoveOutliersFaintAndBrightOperator} (\texttt{ROFABO}),  adopted for DR2, 
    was replaced by the    \texttt{MultibandOutlierRemovalOperator} (\texttt{MORO}), 
     \citep[see Section 10.2.3.2 in {\it Gaia} DR3 Documentation, ][and \citealt{DR3-DPACP-162}]{DR3-documentation}. 
\texttt{MORO} 
identifies outliers 
from the comparison of the measurements in the three {\it Gaia} photometric bands.  However,
a \texttt{ROFABO} step, with  
configuration parameters 
 we specifically tailored to 
RR Lyrae and Cepheid variables, was also added to the operators chain 
to improve the rejection of bad measurements in the time series photometry 
of sources processed through the SOS Cep\&RRL pipeline. 
An \texttt{RvsOutlierRemovalOperator}, that applies a similar approach as the \texttt{ROFABO}, was also applied to clean the RVS time series data \citep[see][for details]{DR3-documentation}.\\

    \item  Processing of RVS radial velocity time series\\
    A specific routine was activated that processes 
    the RVS time series RV measurements.
    The routine constructs the Fourier model of the RV curve obtained by folding the data according to the period inferred from the {\it G}-band time series photometry and returns the Fourier amplitudes and phases, the peak-to-peak amplitude of the RV 
    curve [Amp(RV)], the mean RV and the epoch of minimum RV value.\\
    \item Error estimate via bootstrap technique
    
    Errors for all the parameters of the RR Lyrae stars and Cepheids processed and confirmed by the SOS 
    pipeline for DR3 were estimated via a bootstrap technique, that replaced the Monte Carlo simulations (consisting into randomly varying the data within their errors) adopted for DR1 and DR2 (Paper~I and II).
Specifically, errors on the Fourier fit parameters (period, amplitudes and phases) and the 
quantities characterizing the light and RV curves (mean magnitudes, mean velocity, peak-to-peak amplitudes,  etc.) were estimated by randomly re-sampling the input data (allowing data point repetitions) and 
the parameters were recalculated from the resulting simulated sample. The procedure was repeated 100 times and, for each parameter, the error was estimated from the 
robust standard deviation (1.486 $\times$ MAD,  where MAD is the median absolute deviation) of the  
resulting bootstrap distributions.
A similar procedure was also applied to estimate  random errors of other attributes released for RR Lyrae stars, such as  
 metal abundances  
and absorption values.\\
    \item New and revised sky sub-regions\\
    The last module of the 
    pipeline main trunk subdivides the sources according to different regions on the sky, before sending them to the RR Lyrae and Cepheid branches. This is because different relations are needed to classify the sources, depending on their position on the sky (see Paper~II). For DR2 we divided the sky in three separate regions: Large Magellanic Cloud (LMC), Small Magellanic Cloud (SMC) and, All-Sky comprising the remaining part of the sky. 
    The LMC and SMC regions 
    were expanded for DR3. The new region for the LMC is a box centred on the galaxy at RA$_{\rm LMC}$(J2000)= $82.5^{\circ}$, 
    Dec$_{\rm LMC}$(J2000)=$-68.25^{\circ}$ and extending from $67.50^{\circ}$ to $97.50^{\circ}$ in RA and from $-75.00^{\circ}$ to $-62.00^{\circ}$ in Dec. Similarly, the new region for the SMC is a box centred at RA$_{\rm SMC}$(J2000)= $16^{\circ}$, 
    Dec$_{\rm SMC}$(J2000)=$-73^{\circ}$ and extending from $0.00^{\circ}$ to $30.00^{\circ}$ in RA and from $-76.00^{\circ}$ to $-70.00^{\circ}$ in Dec.
    Two further regions were also defined, one centred on the Andromeda galaxy (M31) and one around the Triangulum galaxy (M33), to allow the processing of bright  Cepheids  within the reach of {\it Gaia} in these galaxies.
    Details on these two additional regions are provided in \cite{DR3-DPACP-169}.\\
    
    In DR2  
    a number of Cepheids with $G$-band amplitude smaller than 0.5 mag and periods shorter than 1 d were missed or misclassified as RR Lyrae stars. A change was implemented in the SOS pipeline for  DR3 by which sources classified as RR Lyrae stars at the end of the common trunk in Fig.~\ref{fig:common-trunk} 
    were sent to both the RR Lyrae and the Cepheid branches if they belonged to the region of the sky and satisfied the conditions described in the following:
    \begin{itemize}
        \item All-Sky sources with a  parallax available were always sent to both branches;
         \item sources in the LMC and in the SMC 
         sub-regions (as defined  above) with amplitude smaller than 0.5 mag and periods shorter than 1 d, and mean $G$ magnitude brigther than 18.5 and 19 mag in the LMC and SMC, respectively;
        \item all sources with $P\le$ 1 d in the M31 and  M33 sub-regions.
    \end{itemize} 
    \end{itemize}
 This procedure 
 caused a number of sources to have a double classification. 
 They were 
 checked and finally classified in just one class (RR Lyrae star or Cepheid; see Sects.~\ref{sec:vali_criteria} and ~\ref{new-RRLs}).
 
  \begin{figure}
   \centering
 \includegraphics[scale=0.18]{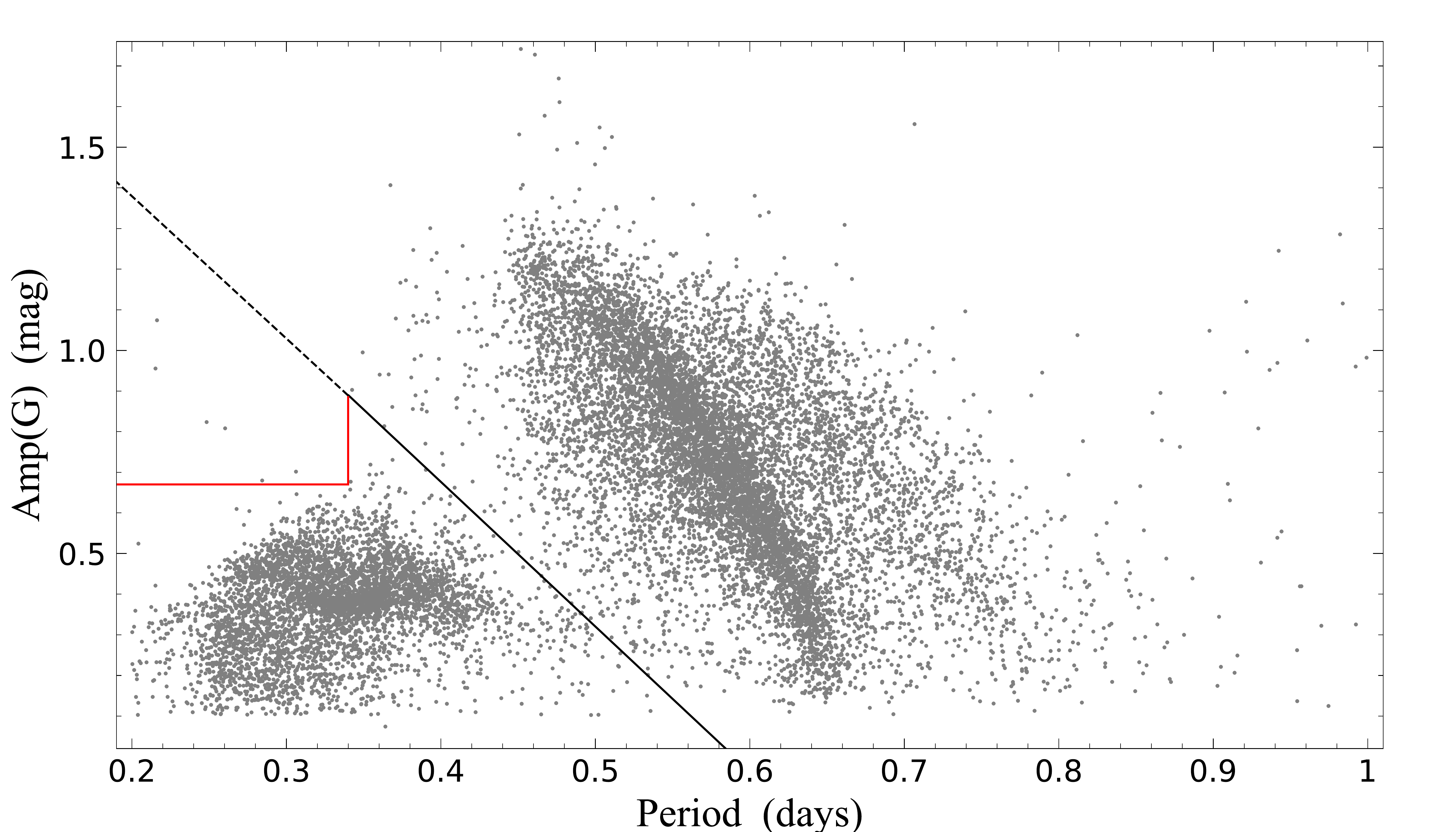}
   \caption{
   New boundaries of the RRab
and RRc regions in the $PA$ diagram defined by a sample of about 15\,000 RR Lyrae stars randomly selected from the final RR Lyrae catalogue  published in the  \texttt{dr3\_vari\_rrlyrae} table of \gaia DR3. 
 The dotted portion of the black line separating  RRab from RRc stars was replaced by the red broken line, in order to properly classify high amplitude
RRab stars with pulsation periods in the same domain of RRc stars (see text for details).
}
\label{TaglioSilvio}
\end{figure}

Two major changes were implemented in  
the RR Lyrae branch (Fig.~\ref{fig:rrl-branch}) of the SOS  
pipeline:
 \begin{itemize}
     \item Improvement of double-mode RR Lyrae detection\\
The module to 
search for a secondary periodicity in the time series photometry of RR Lyrae stars was activated only for sources with more than 40  measurements (FoV transits) in the $G$-band time series data, and when residuals from the Fourier
model best fitting the $G$-band light curve folded with the primary period were larger than 0.05 mag.  
This allowed us to reduce significantly the number of spurious double-mode RR Lyrae detections.\\ 
     \item Implementation of new thresholds in the Period-Amplitude diagnostic\\
The 
 Period-Amplitude ($PA$) diagram, is used by  the SOS pipeline to separate fundamental mode and first overtone RR Lyrae stars (see fig.~ 4 in Paper~II). 
     For DR1 and DR2 we applied a straight line to separate the RRab and RRc regions. For DR3 we have defined new boundaries of the  
     two regions, in order to properly distinguish 
      high amplitude/short period RRab from RRc stars. 
The new boundaries are shown by the red  
segments (described by the condition: 
Amp($G$)$\le 0.64$ mag for $P \le 0.34$ d) that replaced the dotted portion of the black straight line in Fig.~\ref{TaglioSilvio}. All sources below the red segments 
and the black solid line are classified as RRc stars, whereas sources above those lines are RRab stars. 
However, 
the 
new thresholds were not able to capture all such short period RRab stars (see Sect.~\ref{sect:Short-P-RRab}). 
A more efficient procedure will be implemented 
for {\it Gaia} DR4 that,  specifically taking into account the much steeper rise to maximum light of RRab 
compared to RRc stars, will 
help to better  disentangling the two types.
 \end{itemize}
 Finally, shown in gray in Fig.~\ref{fig:rrl-branch} are three modules not activated for DR3. 
 Among them, the Amplitude ratio routine, although 
 not operational during the SOS 
 processing, was used during  validation to compute 
 Amp$G$/Amp$G_{BP}$, Amp$G$/Amp$G_{RP}$, Amp$G_{BP}$/Amp$G_{RP}$ ratios and identify  eclipsing binaries contaminating the RR Lyrae sample (see Sect.~\ref{sec:s_vali}).

\section{Application of the SOS Cep\&RRL pipeline to the
DR3 dataset: input data and source selections}
\label{sec:app_to_sos}
\begin{figure*}
   \centering
    \includegraphics[scale=0.20]{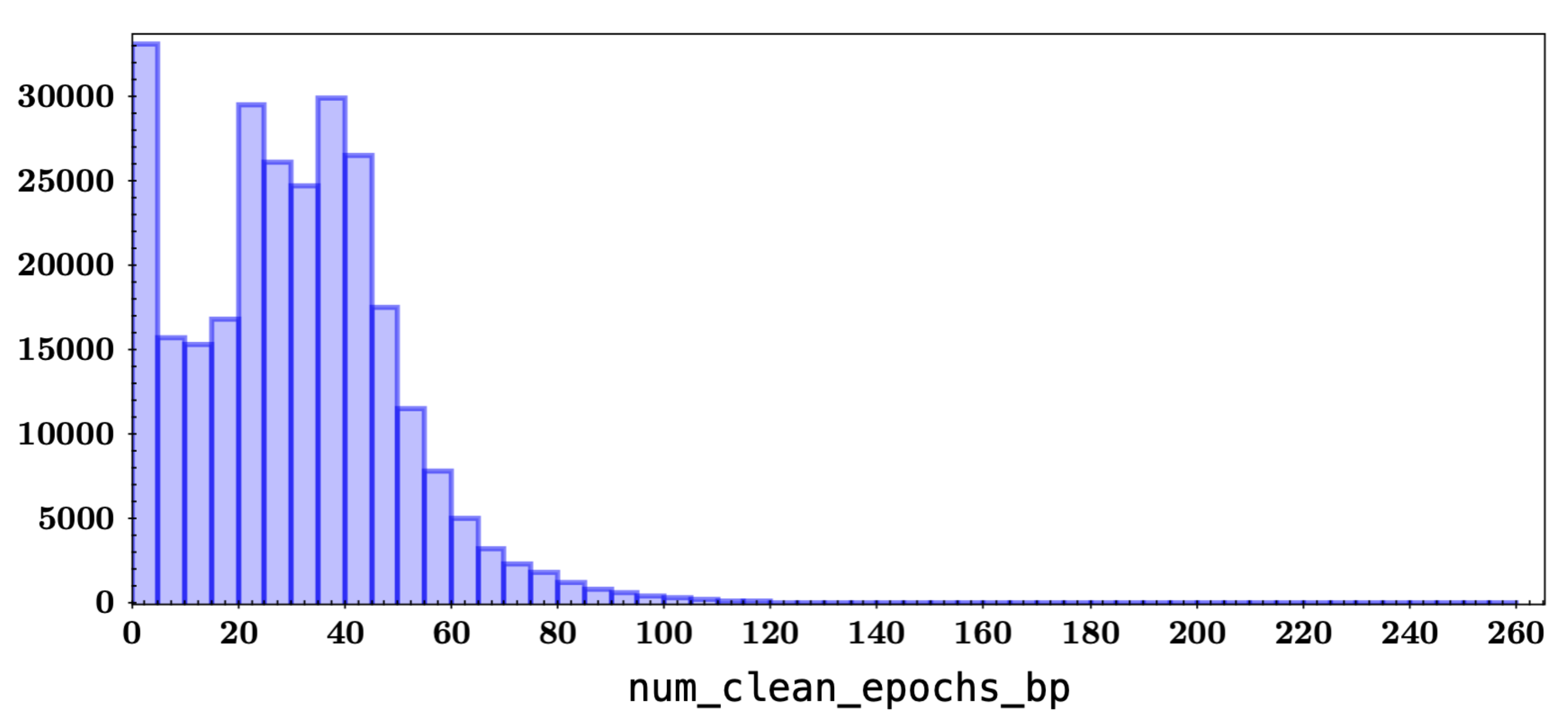}~\includegraphics[scale=0.20]{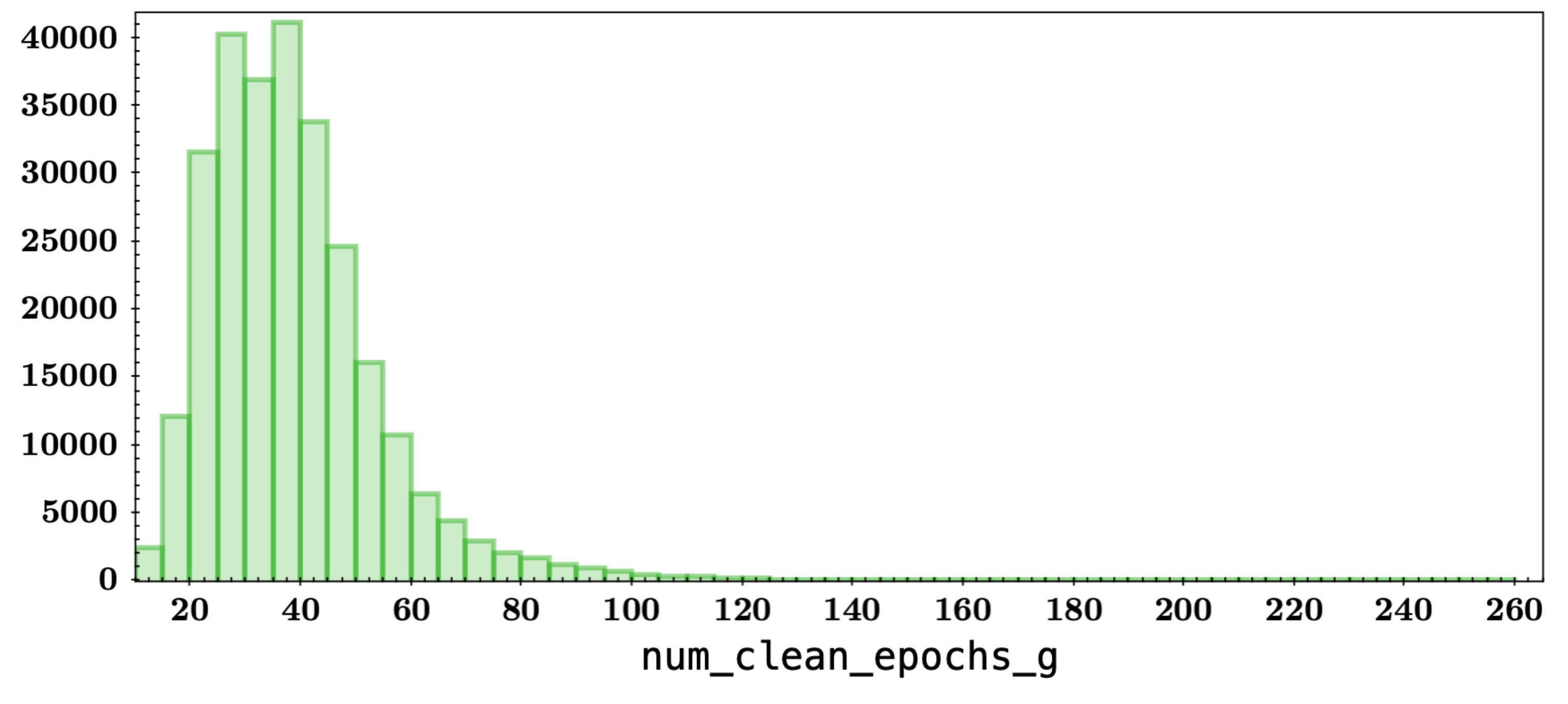}~\includegraphics[scale=0.20]{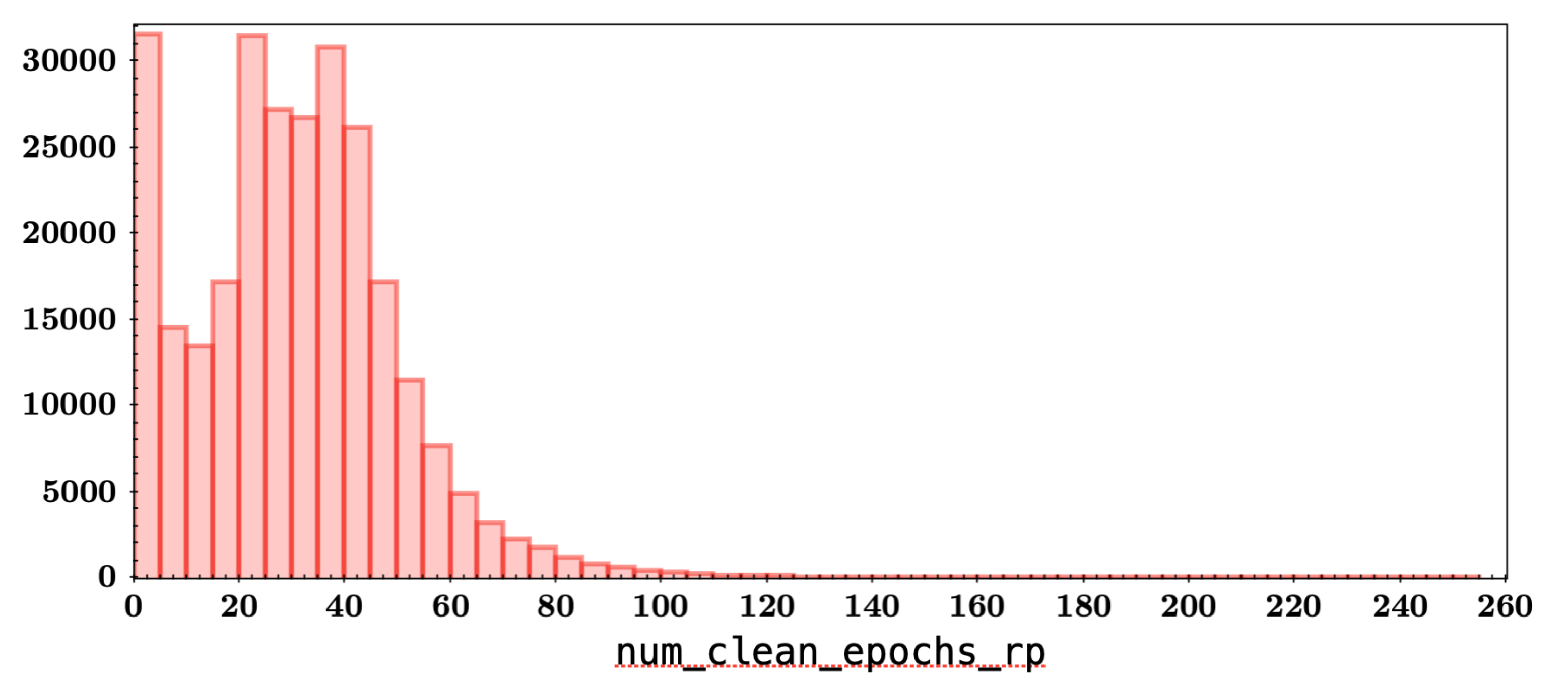}
   \caption{ Distribution in number of epochs  in the {\it Gaia} $G_{BP}$ (left), $G$ (centre) and $G_{RP}$ (right) passbands,  of the 
   271,779 RR Lyrae stars confirmed by the SOS Cep\&RRL pipeline.}
\label{Nepoch-Phot}%
    \end{figure*}

Input data fed into the SOS Cep\&RRL pipeline in DR3 consisted of FoV $G$ and integrated $G_{BP}$ and $G_{RP}$ (when available) time series photometry of all-sky
variable sources classified as RR Lyrae (and Cepheids, see \citealt{DR3-DPACP-169})  candidates   
 with various probability values, by the General Supervised Classification module (\citealt{DR3-DPACP-165}; see Fig.~\ref{fig:common-trunk}) 
of the variability pipeline, as well as RV time series data from the RVS,  for small subsamples of these sources with mean magnitude brighter than  $G_{RVS}\sim$ 14 mag.\footnote{$G_{RVS}$ is the {\it Gaia} magnitude of the flux collected in the RVS passband: 845 - 872 nm. 
For the transformation from  $G_{RVS}$ to $G$ magnitude see Eq. 2 in \citet{2018A&A...616A...1G}. 
In practise, $G - G_{RVS}\sim$ 0.48 and 0.45 mag, for RRc and RRab stars,  
respectively.} 
The photometry and RV datasets 
comprise observations collected by \gaia   over 34 months, from 2014 July 25 to 2017 May 28. In this time frame,  for the confirmed RR Lyrae stars 
discussed in this paper,   \gaia  obtained on average about 38-39 useful photometric measurements in the $G$-band (with min/max values between 12 and 257 measures, depending on the source position on sky, Fig.~\ref{Nepoch-Phot},  centre panel),  to compare with $\sim$30 $G$ measures in DR2, and about 30 measurements in $G_{BP}$ (with  min/max values between 0 and  258, Fig.~\ref{Nepoch-Phot}, left  panel) and $G_{RP}$ (with  min/max values between 0 and  254, Fig.~\ref{Nepoch-Phot}, right panel). 
Similarly, the {\it Gaia} RVS time series data for the sub-samples of RR Lyrae and Cepheids presented in Sect.~\ref{sec:s_rvs}  comprise an average of about 23 RV measurements (min/max values between 5 and 77).

The following criteria were applied to select RR Lyrae candidates to  process with the SOS Cep\&RRL pipeline:
\begin{itemize}
    \item   sources classified as RR Lyrae candidates, by the Classifiers of the General Supervised Classification module, with minimum probability thresholds chosen  as to maximise the recovery of RR Lyrae stars known from different surveys in the literature 
    (see Sect.~\ref{gold-sample});
    \item sources with a minimum number of 12 \gmag-FoV transits, after the application of the {\tt MORO} and {\tt ROFABO} outlier removal operators (see Fig.~\ref{Nepoch-Phot});
    \item  a minimum number of 5 measurements, for the publication of the RV time series and, at least 7 transits for the modeling of the folded RV curve (see Fig.~\ref{Nepoch-RV});
    \item a peak-to-peak amplitude $>$0.1 mag in the \gmag-band;
    \item periods in the range of 0.2 $\leq P < $ 1.0 d.
   \end{itemize}

\subsection{Training sets and literature reference catalogues}\label{gold-sample}
Candidate RR Lyrae stars to be processed through the SOS Cep\&RRL pipeline were selected on the basis of their probability score, as to maximise the recovery of RR Lyrae stars already known in the literature. 
A preliminary step was the creation of 
 a reference database of known RR Lyrae stars,  by collecting 
 catalogues of these variables from major  surveys in the literature,  including 
 those in 152 different stellar systems: the Magellanic Clouds, 107 globular clusters, 9 dwarf spheroidal galaxies (dSphs; Carina, Draco, Fornax, Leo~I, Leo~II, Sagittarius, Sculptor, Sextans and Ursa Minor) and 34 ultra-faint dwarfs (UFDs).  
 This large catalogue, homogenized by eliminating duplicates, contains more than 425,000 unique, all-sky, RR Lyrae entries and 
 spans an about 16 magnitude  range, from RR Lyr itself ($\langle$V$\rangle$=7.76 mag, \citealt{Fernley-et-al-1998}) to V6, 
 the lone  RR Lyrae star identified 
 in Leo~T \citep{Clementini-et-al-2012}, one of the furthest ultra-faint MW satellites, that 
 with a mean magnitude $\langle$V$\rangle$=23.59 mag 
 overcomes {\it Gaia}'s faint limit by about 3 magnitudes. However, we note that 97\% of the sources in our custom RR Lyrae catalogue are within 
 {\it Gaia}'s limiting magnitude and have a counterpart in the  EDR3 {\tt source\_table} (see below). To build our custom catalogue we primarily used as reference the OGLE catalogues for RR Lyrae stars (version IV of the survey, \citealt{Soszynski-et-al-2014,Soszynski-et-al-2019}), which have a  high completeness and purity  for the LMC, SMC and the MW bulge and disc, 
 along with the {\it Gaia} DR2 RR Lyrae catalogue in the {\tt gaiadr2.vari\_rrlyrae} table (Paper~II), 
 CATALINA \citep{Drake-et-al-2013,Drake-et-al-2014,Drake-et-al-2017,Torrealba-et-al-2015}, KEPLER \citep{Molnar-2018}, the Zwicky Transient Facility (ZTF; \citealt{Chen-2020}), the All Sky Automated Survey (ASAS; \citealt{Pojmanski-2000,Pojmanski-2002,Pigulski-2009}), the All-Sky Automated Survey for Supernovae (ASAS-SN; \citealt{Jayasinghe-et-al-2019}), LINEAR \citep{Palaversa-et-al-2013,Sesar-2013}, The Palomar Transient Factory (PTF; \citealt{Sesar-2014}), the VISTA Variables in the Via Lactea survey (VVV; \citealt{Contreras-et-al-2018,Dekany-et-al-2018}), the fifth edition of the General Catalogue  of Variable Stars (GCVS; \citealt{Samus-2017}) and,  the PanSTARRS1 3$\pi$ survey (PS1; \citealt{Sesar-et-al-2017}). 
To integrate our compilation 
with known RR Lyrae stars 
in Galactic globular clusters (GCs), 
we used as main references an updated version of the Variable Stars in Galactic Globular Clusters catalogue \citep{Clement-et-al-2001}, while for 
RR Lyrae stars in the MW UFDs we used the compilation in  \citet{Vivas-et-al-2020}.  
For the MW dSphs we used the RR Lyrae collections by \citet{Muraveva-et-al-2020} for Draco, \citet{Martinez-et-al-2016} for Sculptor, \citet{Bersier-et-al-2002} for Fornax, \citet{Coppola-et-al-2015} for Carina, \citet{Siegel-2000} for Leo~II, \citet{Stetson-et-al-2014} for Leo~I, 
\citet{2019AJ....157...35V} for
Sextans,  \citet{Kunder-et-al-2008} for Sagittarius and, by \citet{Nemec-et-al-1988} and \citet{Bellazzini-et-al-2002} for Ursa Minor. 

We specifically note that in building our  custom RR Lyrae catalogue we privileged  inclusiveness, hence,  we did not make any cut as far as purity is concerned. For the selection procedures  described in the following sections we used a version frozen at December 2020 of our custom RR Lyrae catalogue. However, the catalogue is being regularly  updated as soon  as new RR Lyrae compilations appear in the literature, and we plan to make a most updated version available on-line soon after {\it Gaia} DR3 (Garofalo et al. 2022, in prep.).

We performed a spatial cross-match, within a 2.5$\arcsec$ paring radius, of our custom RR Lyrae catalogue and sources in the {\tt gaia\_source} table of the  \gaia EDR3 archive  (\texttt{gaiaedr3.gaia\_source}) finding a counterpart in the 
EDR3 main catalogue for  414\,082  
(97$\%$)  of the RR Lyrae stars in our collection.\footnote{We refer the reader to \citet{DR3-DPACP-177}, for a comprehensive catalogue of known variable sources cross-matched with {\it Gaia} DR3, extending to also other types of variability the catalogue being discussed here.}
The Classifiers of the General Supervised Classification module  returned these 414,082 sources as RR Lyrae candidates with probabilities ranging from very low values.
Among them, more than 75\% (311\,798)  
were 
 classified as RR Lyrae by the SOS Cep\&RRL pipeline,  
with characteristic parameters ($P$, pulsation mode, etc. ) consistent with the literature recovered  for about 
65\% of them.

\subsection{Gold sample and validation criteria}\label{sec:vali_criteria}

\begin{figure}
 \centering
\includegraphics[scale=0.4]{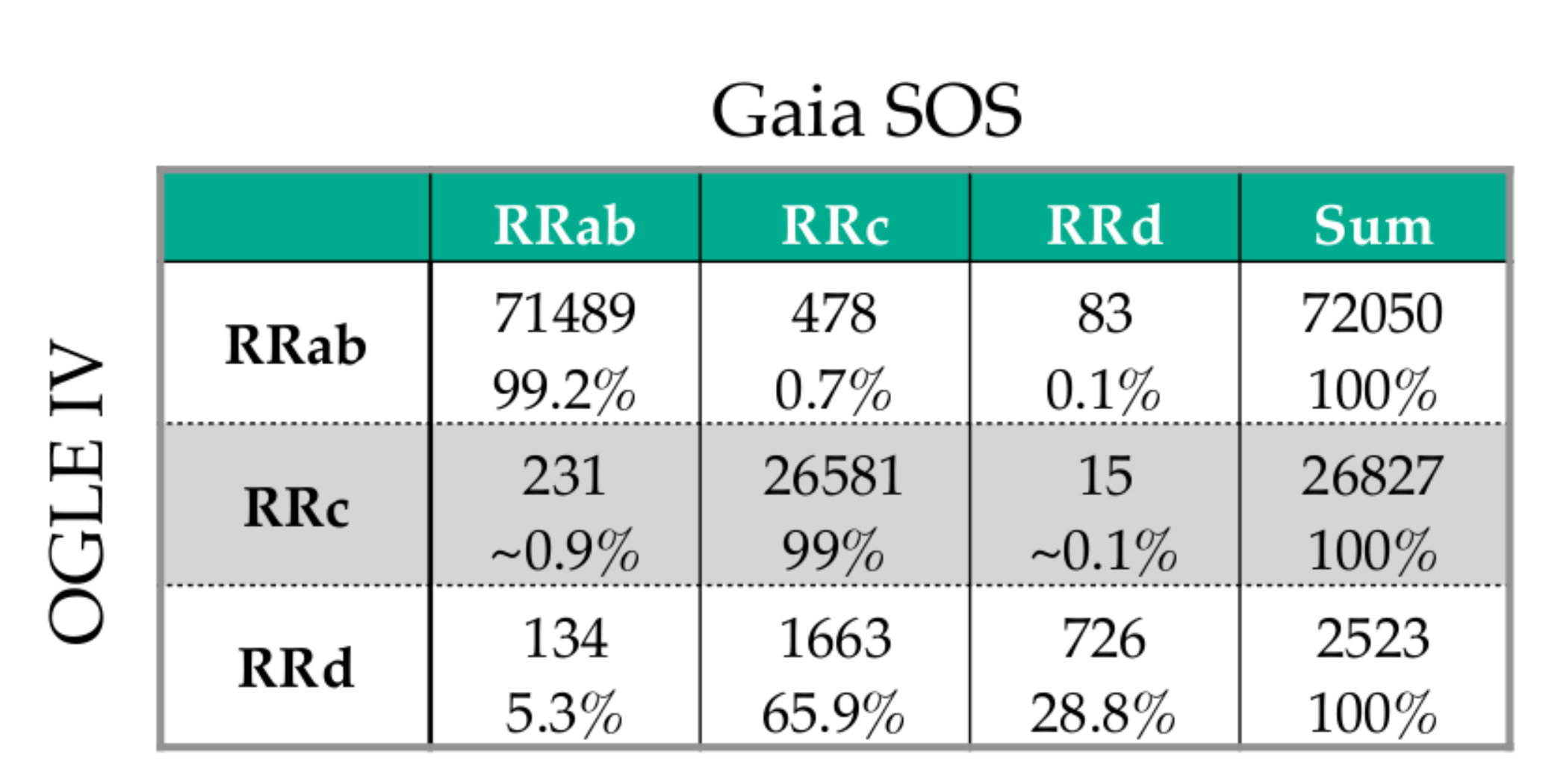}
\caption{Confusion matrix for the RR Lyrae stars. As control sample we used all 
   sources classified as RR Lyrae stars by the OGLE-IV catalogue of variable stars in the LMC, SMC, and MW bulge,  that have a cross-match within a radius of 2.5$\arcsec$ with the SOS-confirmed RR Lyrae stars published in our Gaia DR3 catalogue, 
  for a total of 101\,400 objects.  
  Rows refer to literature results and columns to results of the SOS Cep\&RRL pipeline. The corresponding success percentage is shown in the diagonal cells.}
   \label{matrice-confusione}
    \end{figure}

\begin{figure*}
   \centering
\includegraphics[scale=0.33]{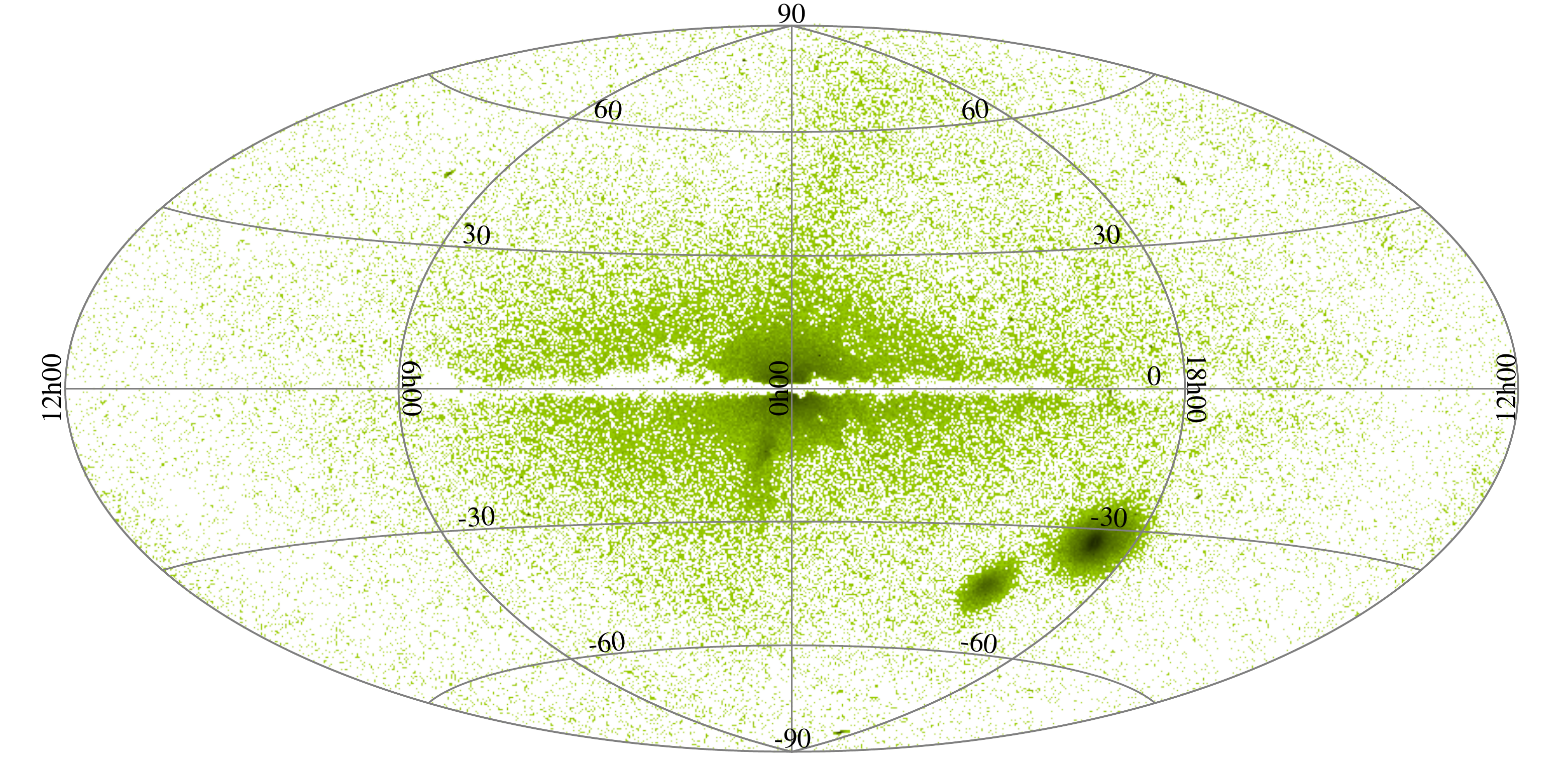}
\includegraphics[scale=0.35]{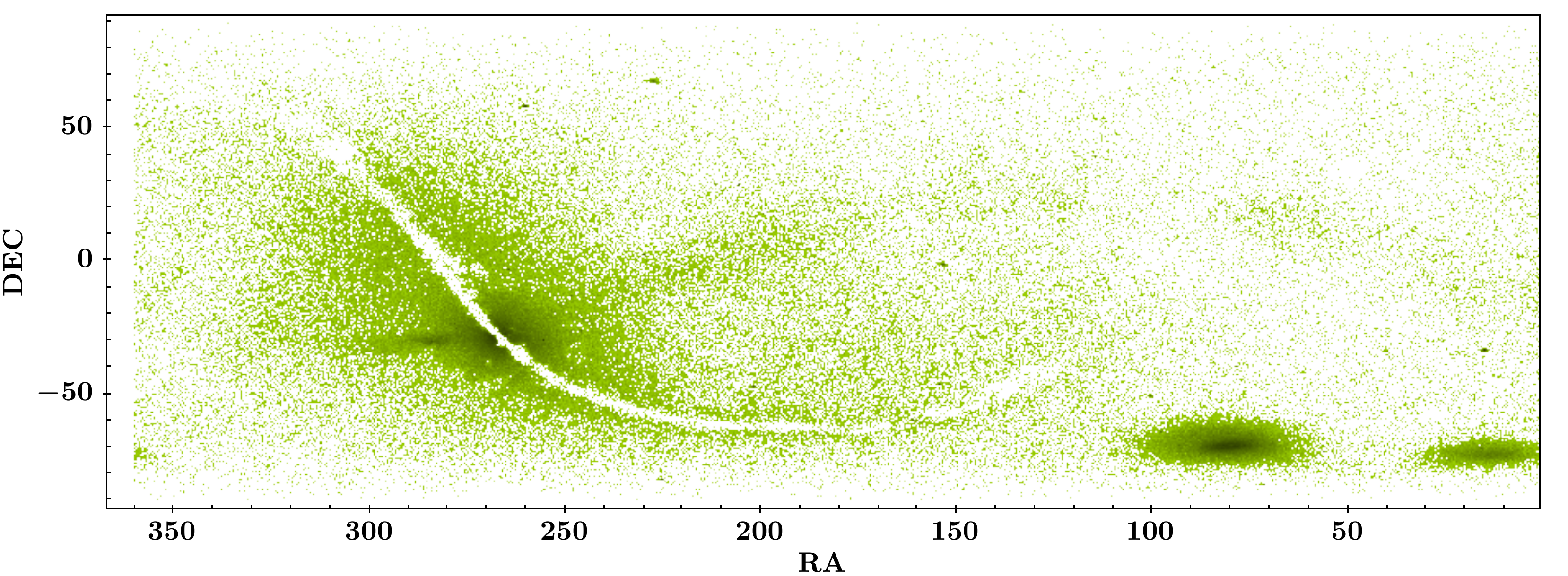}
     \caption{Sky distributions in galactic coordinates (top panel) and equatorial coordinates (bottom panel) of 200\,636  RR Lyrae stars in the Gold Sample.}
     \label{fig:GoldSample}
    \end{figure*}

In order to validate the unexpectedly large number of candidate RR Lyrae produced by the Classifiers and confirmed by the SOS Cep\&RRL pipeline,   and extract a  bona fide sample to  release in DR3, we adopted a strategy  driven by our custom RR Lyrae reference catalogue and  
the RR Lyrae diagnostic tools summarised in Sect~\ref{sec:sos-pip_genover}. 
Our strategy consisted of two main steps: 
\begin{enumerate}
\item we divided the sources between known (311,798 sources) and potentially new/unknown RR Lyrae stars,  based on our custom catalogue;
\item we used the diagnostic tools of known RR Lyrae stars 
correctly characterized by the SOS Cep\&RRL pipeline (Gold Sample RR Lyrae stars, hereafter) as a reference 
to select and validate new/unknown RR Lyrae stars.
\end{enumerate}
A large effort was devoted to build from the initial sample of 311,798 known RR Lyrae stars recovered and characterized by the SOS Cep\&RRL pipeline a Gold Sample as large and pure as possible.
We started by selecting from the 311\,798 known RR Lyrae 
stars those with SOS pulsation properties 
simultaneously satisfying the following conditions:
\begin{itemize}
    \item |$P_{\rm SOS} - P_{\rm Literature}$|  $\leq$ 0.001  d;
    \item the $\phi_{21}$ Fourier parameter within the range appropriate for RR Lyrae stars: 3.0 $\leq \phi_{21} \leq$ 5.8 rad, in the $\phi_{21}$ vs $P$ diagram, 
    for sources with $G$ light curves modeled with at least 2 harmonics.
\end{itemize}  
Then, for sources with $G$ light curves modeled with only 1 harmonic, we used the 
 $PA$ diagram to reject those falling outside the regions appropriate for RR Lyrae stars.
 We also relaxed the period criterion (to |$P_{\rm SOS} - P_{\rm Literature}$|  $\leq$ 0.003 d and to |$P_{\rm SOS} - P_{\rm Panstarrs}$|  $\leq$ 0.01 d) in a  number of cases (5\,617 sources in total), after visual inspection of the light curves. 
 By this procedure the original sample of 311\,798 known RR Lyrae stars reduced to  
 201\,173 sources. Among them, 369 
 were found to be eclipsing binaries,  based on the ratios of the $G$, $G_{BP}$, $G_{RP}$ light curve amplitudes (see end of Sect~\ref{sec:sos-pip_genover}), and then further confirmed by visual inspection of the light curves; and 168 sources with a double classification  were re-classified as Cepheids (see  also  Sect.~\ref{sec:s_vali}).
 
This led us to a final Gold Sample comprising 200\,636 bona fide known RR Lyrae stars. Among them 101\,400 are RR Lyrae stars observed by OGLE in the LMC, SMC and the Galactic bulge and disc,  
of which 71\,489 are classified as RRab, 26\,581 as RRc and 726 as RRd stars for both the SOS Cep\&RRL pipeline and OGLE. 
Figure~\ref{matrice-confusione} shows a confusion matrix for the Gold Sample RR Lyrae stars, drawn using as control sample the OGLE-IV RR Lyrae catalogue. 

Figure ~\ref{fig:GoldSample} shows the sky distribution (using galactic and equatorial coordinates in the top and bottom panels, respectively) 
of the RR Lyrae stars in the Gold Sample. 
The three panels of Fig.~\ref{fig:GoldSample-diagnostics} show their $PA$, $\phi_{21}$  vs $P$ and $\phi_{31}$  vs $P$ diagrams, where the RR Lyrae stars in common with OGLE (101,400 sources) are plotted using different colours for RRab (blue), RRc (magenta) and RRd (black) stars. 

\subsection{Potentially new/unknown RR Lyrae stars}\label{new-RRLs}
\begin{figure}   
\includegraphics[scale=0.32]{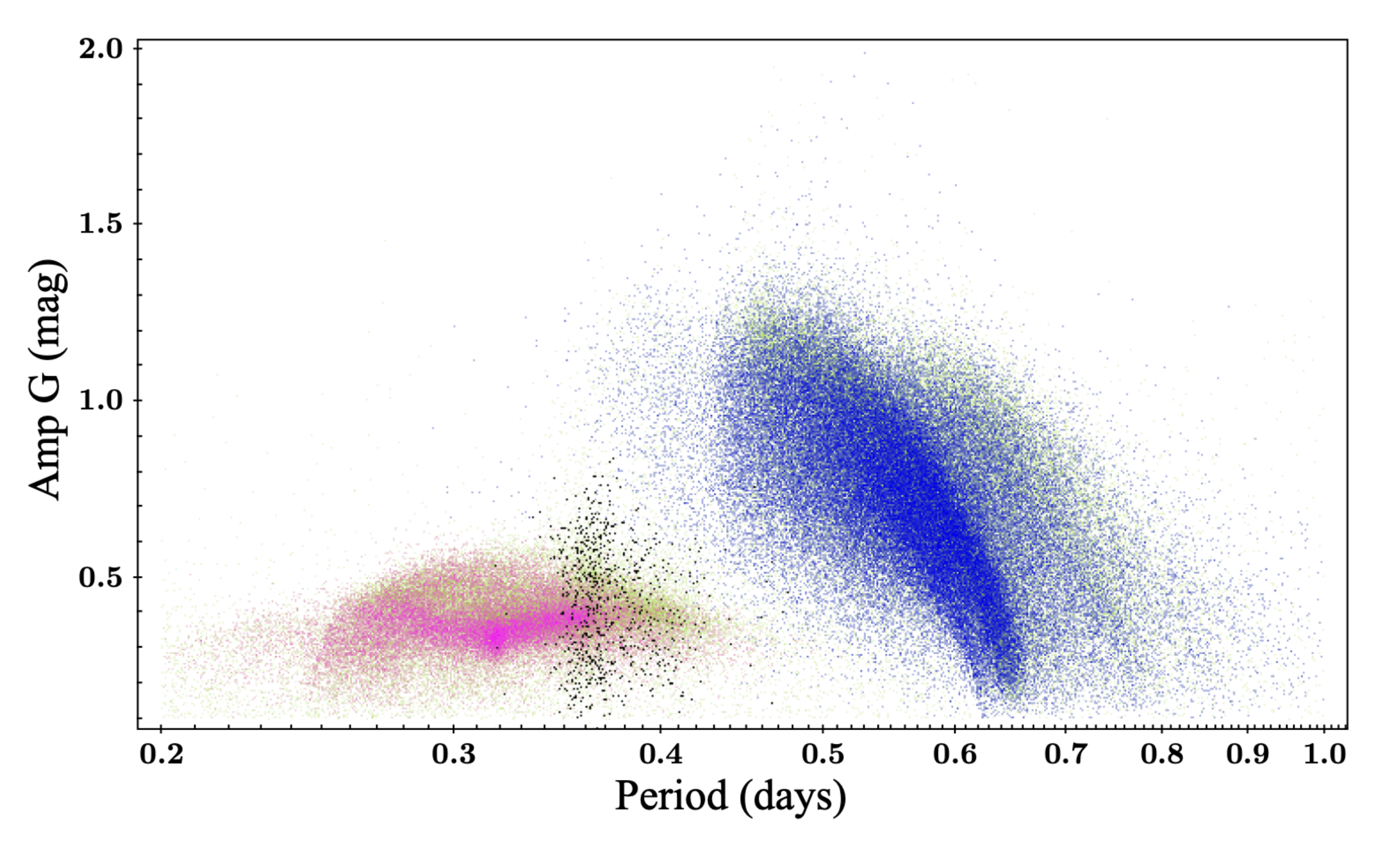}
\includegraphics[scale=0.32]{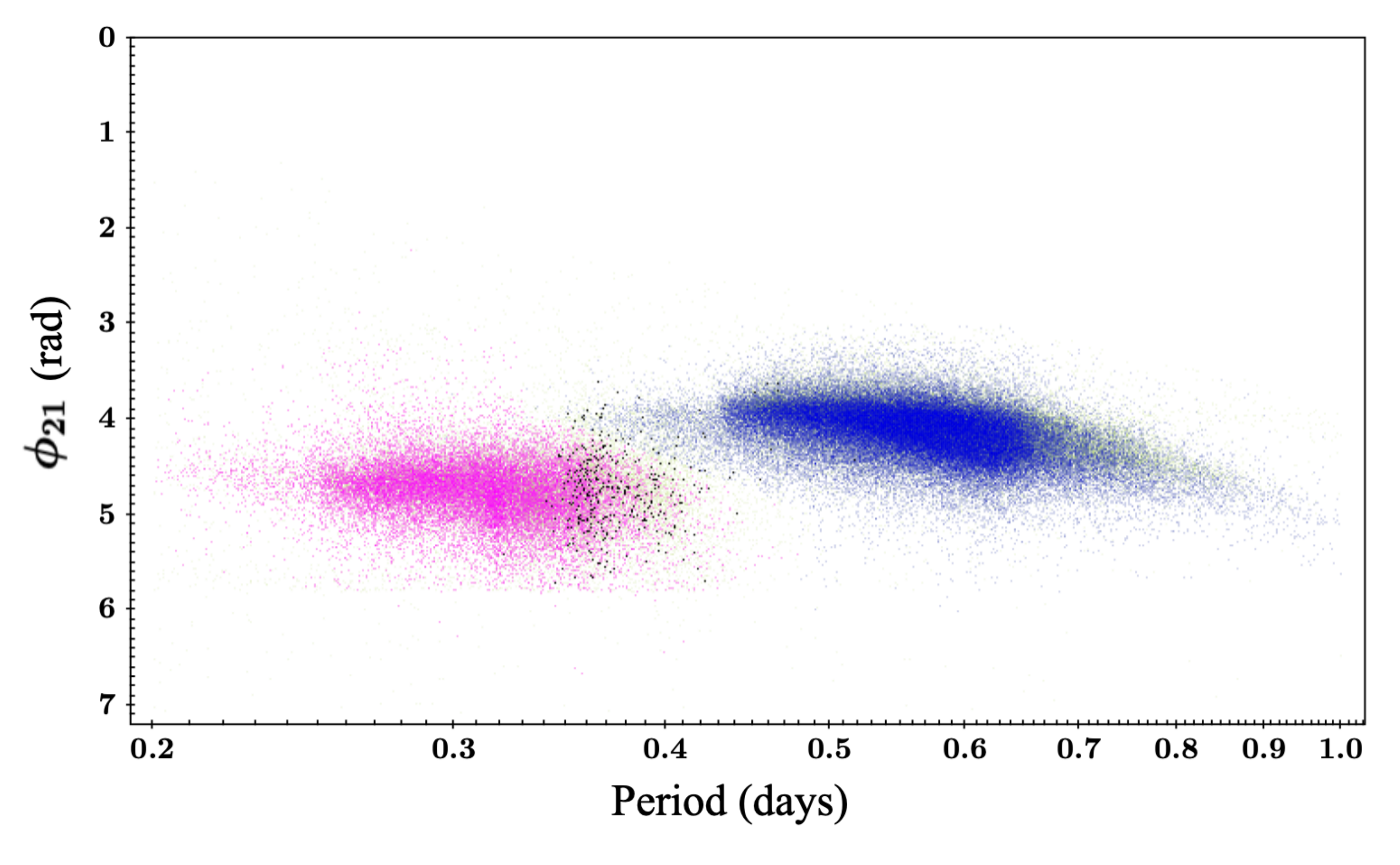}
\includegraphics[scale=0.32]{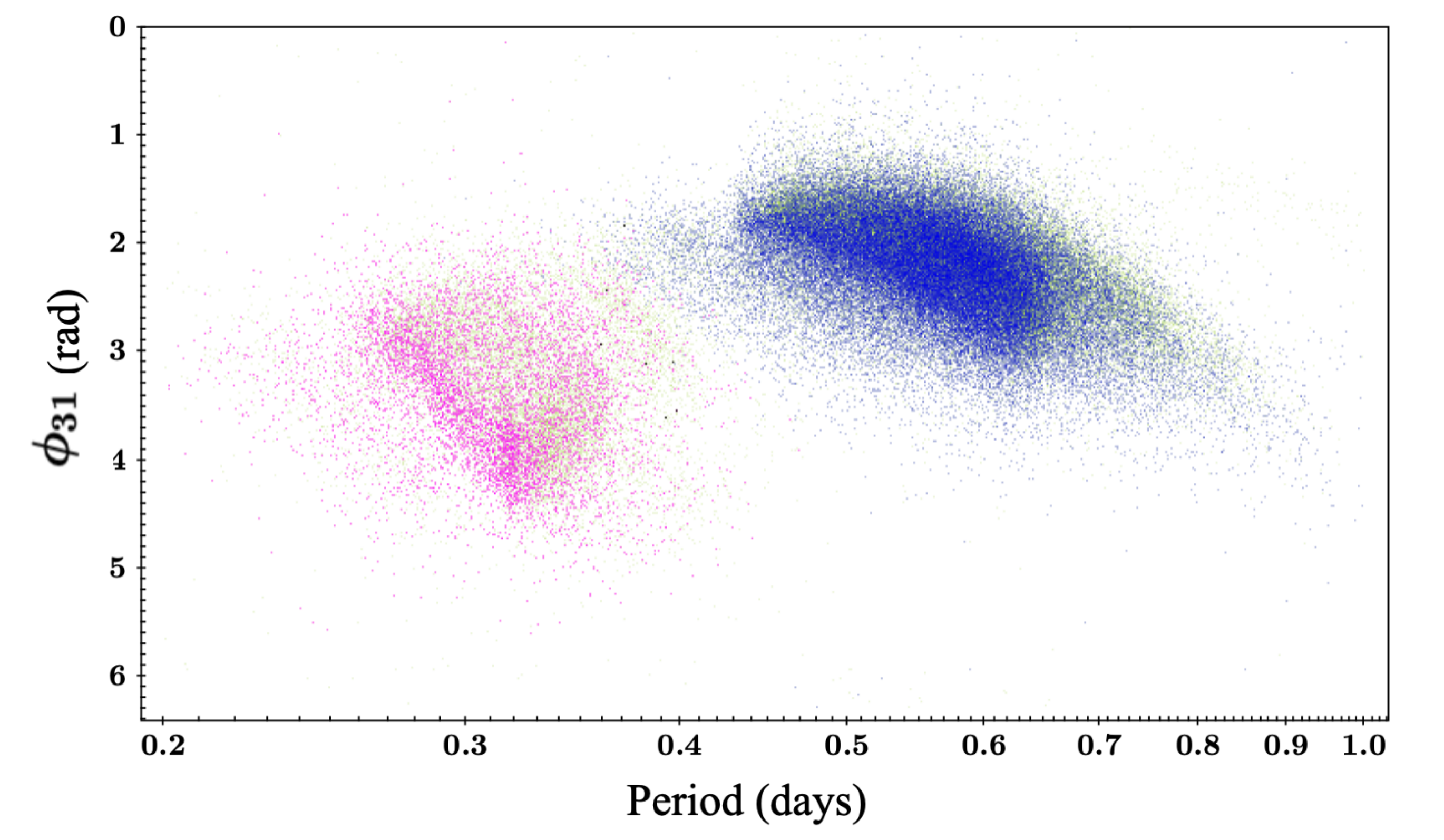}
   \caption{$PA$, $\phi_{21}$  vs $P$ and $\phi_{31}$  vs $P$ diagrams of the 200\,636 RR Lyrae stars in the Gold Sample. The 101\,400 sources in common with OGLE are plotted using different colours for RRab (blue), RRc (magenta) and RRd (black) stars, respectively.}    
   \label{fig:GoldSample-diagnostics}%
    \end{figure}

We have used the Gold Sample 
as a training set, to select bona fide  RR Lyrae stars from the remaining very large sample of 
potentially new/unknown 
RR Lyrae candidates provided by the Classifiers and  processed through the SOS Cep\&RRL pipeline.
Specifically, we used the $PA$, $\phi_{21}$ vs $P$ and $\phi_{31}$ vs $P$ diagrams defined by the Gold Sample RR Lyrae stars (see Fig.~\ref{fig:GoldSample-diagnostics}),  to select, through sequential steps,  
potentially new/unknown, bona fide RR Lyrae stars, as briefly outlined below. 
\begin{itemize}
   \item Among the sources with $G$ light curves modeled with 3 or more harmonics, we retained as bona fide RR Lyrae stars only those with the $\phi_{31}$ parameter in the proper region 
   of the $\phi_{31}$ vs $P$ diagram, as defined by the Gold Sample RR Lyrae stars (0.6 $\leq\phi_{31}\leq$ 5.1 rad,    see, bottom panel of Fig.~\ref{fig:GoldSample-diagnostics}).
   \item Bona fide RR Lyrae  stars were selected from  sources with $G$ light curves modeled with only 2 harmonics,  by making 
a 2-dimensional cartesian cross-match between their 
($P$, $\phi_{21}$) parameters and those of the Gold Sample RR Lyrae stars as reference (see, centre panel of Fig.~\ref{fig:GoldSample-diagnostics}),   and verifying afterwards that they were  properly located also in the Gold Sample $PA$ diagram (see upper panel of Fig.~\ref{fig:GoldSample-diagnostics}).
 \item Finally, bona fide RR Lyrae stars were selected from sources with $G$ light curves modeled with only 1 harmonic,  by making a 2-dimensional cartesian cross-match between their [$P$, Amp$(G)$] parameters and those of the Gold Sample RR Lyrae stars as reference (see, upper panel of Fig.~\ref{fig:GoldSample-diagnostics}). 
\end{itemize}

Some of sources retained as bona fide new RR Lyrae stars,  according to the above selection procedures, turned out to have a double classification. 
This occurred for sources with $P<1$ d and Amp$(G) < 0.5$ mag (see Sect.~\ref{sec:sos-pip_genover}) which could either be  
{\it c-}type RR Lyrae stars or  first-overtone classical Cepheids  (DCEP\_1O). The light curves of these sources were visually inspected and their position on the Period-Luminosity, Period-Wesenheit relations for DCEPs 
\citep[see][]{DR3-DPACP-169} 
checked, to finally classify them in one class or the other. 
At the end of these 
extensive selection and cleaning procedures we were left with a sample of 72\,167 new/unknown (to the best of our knowledge) bona fide RR Lyrae stars. 

\begin{figure*}
   \centering
\includegraphics[scale=0.52]{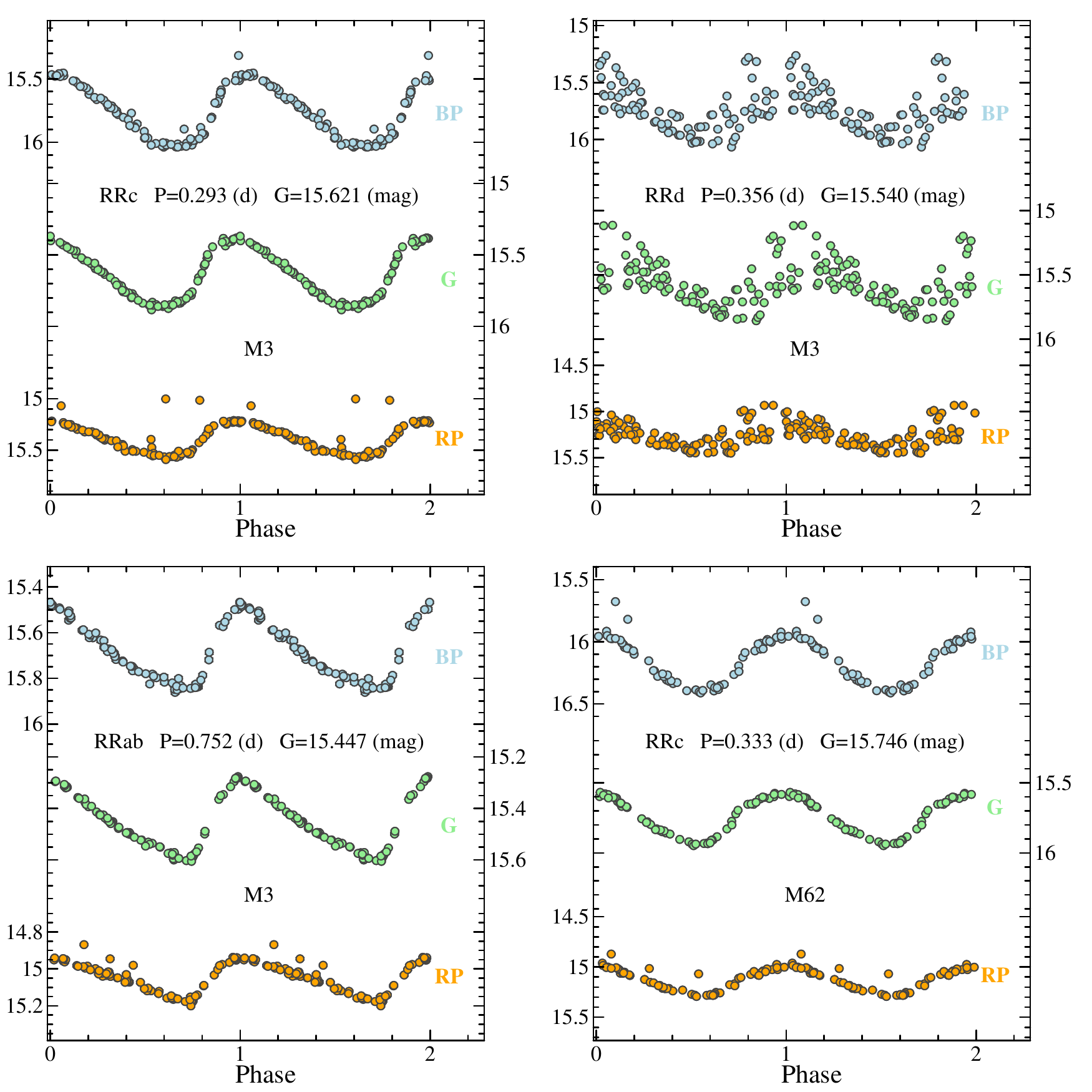}~\includegraphics[scale=0.52]{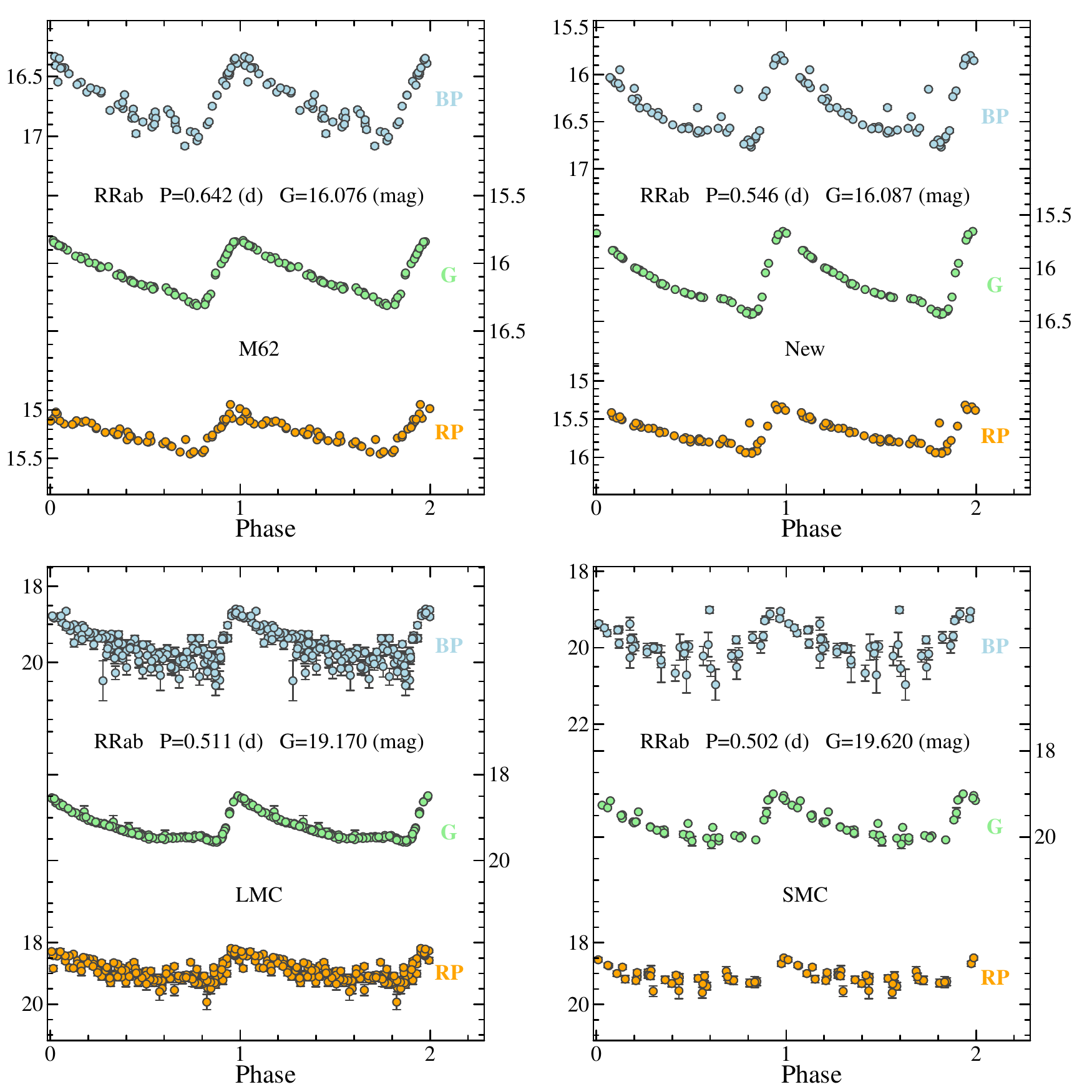}   
         \caption{Examples of light curves for SOS-confirmed DR3 RR Lyrae stars  
      in the 
      globular clusters M3 and  
      M62, 
      in the LMC and the SMC and,  a new All-Sky RR Lyrae star.
      }
         \label{light-curves-DR3-1}
   \end{figure*}

\begin{figure}
   \centering
      \includegraphics[scale=0.45]{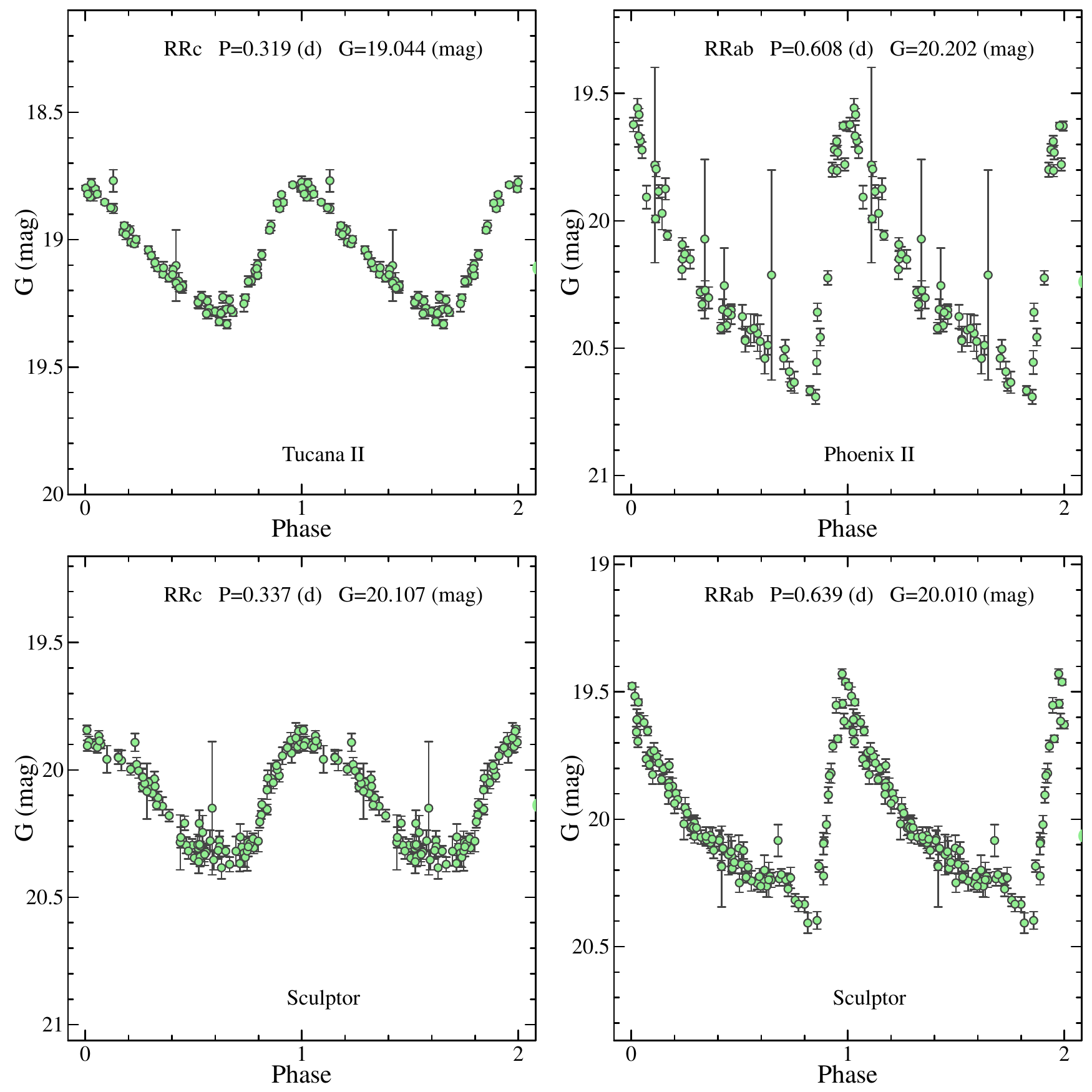}
      \caption{Same as in Fig.~\ref{light-curves-DR3-1}, but for RR Lyrae stars in the Sculptor dSph galaxy and in the Phoenix~II and Tucana~II MW UFDs. Only the $G$-band light curves are shown, because of the poor S/N of the $G_{BP}$ and $G_{RP}$ time series data, at the faint magnitudes of the RR Lyrae stars in these systems.
      }
         \label{light-curves-DR3-2}
   \end{figure}

To summarise, the total number of DR3 RR Lyrae stars  confirmed and fully characterised by the SOS Cep\&RRL pipeline adds to 272\,803 sources (between known RR Lyrae stars in the Gold Sample and 
new discoveries by {\it Gaia}). 
These sources were first cross-matched with catalogues of different types of variable stars produced by other SOS variability pipelines, to check for possible overlaps. A number were found  
with ellipsoidal variables \citep[ELL;][]{DR3-DPACP-174}, eclipsing binaries \citep[ECL;][]{DR3-DPACP-170} and Active Galactic Nuclei (AGN;  \citealt{DR3-DPACP-167}). They were solved by assigning the overlapping sources to a unique variability class after visual inspection of the light curves.
The resulting RR Lyrae catalogue then  went through an iterative  process of cross-validation aimed to ensure consistency and quality of the different data products (astrometry, photometry, variability, etc.) published for the {\it Gaia} DR3 sources. By this whole procedure the RR Lyrae catalogue was trimmed down to 271\,779 
sources, of which 200\,589 
are in the Gold Sample and 71\,190  
are potentially new RR Lyrae stars, that are published in the DR3 {\tt vari\_rrlyrae} table.
Examples of 
the $G$, $G_{BP}$ and $G_{RP}$ light curves 
for DR3 RR Lyrae stars with different pulsation mode (RRab, RRc and RRd) in the MW field and in a number of different stellar systems (the GCs M3 and M62, the LMC and the  SMC,  the Sculptor dSph, and the
Phoenix~II 
 and Tucana~II 
 UFDs) 
are presented in Figs.~\ref{light-curves-DR3-1} and ~\ref{light-curves-DR3-2}.
 
The catalogue of 271\,779 
RR Lyrae  stars 
was further cleaned by contaminants and sources affected by issues which emerged after the final export of the DR3 {\tt vari\_rrlyrae} table. This final cleaning procedure is described  in Sect.~\ref{sec:s_vali}.

\section{RVS radial velocity  time series for RR Lyrae stars and Cepheids}
\label{sec:s_rvs}

With DR3 for the first time epoch RVs 
measured  from the RVS spectra 
are published  for a small sample 
of 
RR Lyrae stars and 
Cepheids. This small dataset provides a first  quality assessment of the RVS transit RVs. 
At the same time, it anticipates RVS data-products that will become available with next {\it Gaia} releases.
In the following subsections we describe how this sample was  selected (Sect.~\ref{sec:RV-selection}), processed by the SOS 
pipeline (Sect.~\ref{sec:RV-SOS-processing}) and validated with the literature (see  Sect.~\ref{sec:RV-validation} for the RR Lyrae stars, and Sect.~6.6  in \citealt{DR3-DPACP-169} for the Cepheids). 

\subsection{Target selection}\label{sec:RV-selection}
The {\tt gaia\_source} table in the 
DR3 archive provides 
RV values\footnote{The RV values of RR Lyrae stars and Cepheids in the DR3 {\tt gaia\_source} table are combined RVs 
obtained in 2 different ways depending on the star $G_{RVS}$ magnitude.
For bright stars ($G_{RVS}\le$12 mag), the combined RV 
is the median of the single-transit RVs. 
For  faint stars,
for which the single-transit RVs 
are less precise, the
single-transit cross-correlation functions are averaged. The mean
cross-correlation function obtained in this way is then used to measure the star RV.}  
for 5\,096 RR Lyrae stars and 3\,190 Cepheids  that were confirmed by the SOS Cep\&RRL pipeline. Among them we selected sources with  5 of more epoch RV measurements and  mean $G$ magnitude brighter than $\sim$15.5 mag. 
A further selection was made by visual inspection of the light and RV curves, ending up with a sample of  
1\,100 RR Lyrae stars and 798 Cepheids.

\begin{figure}[h!]
   \centering
    \includegraphics[scale=0.30]{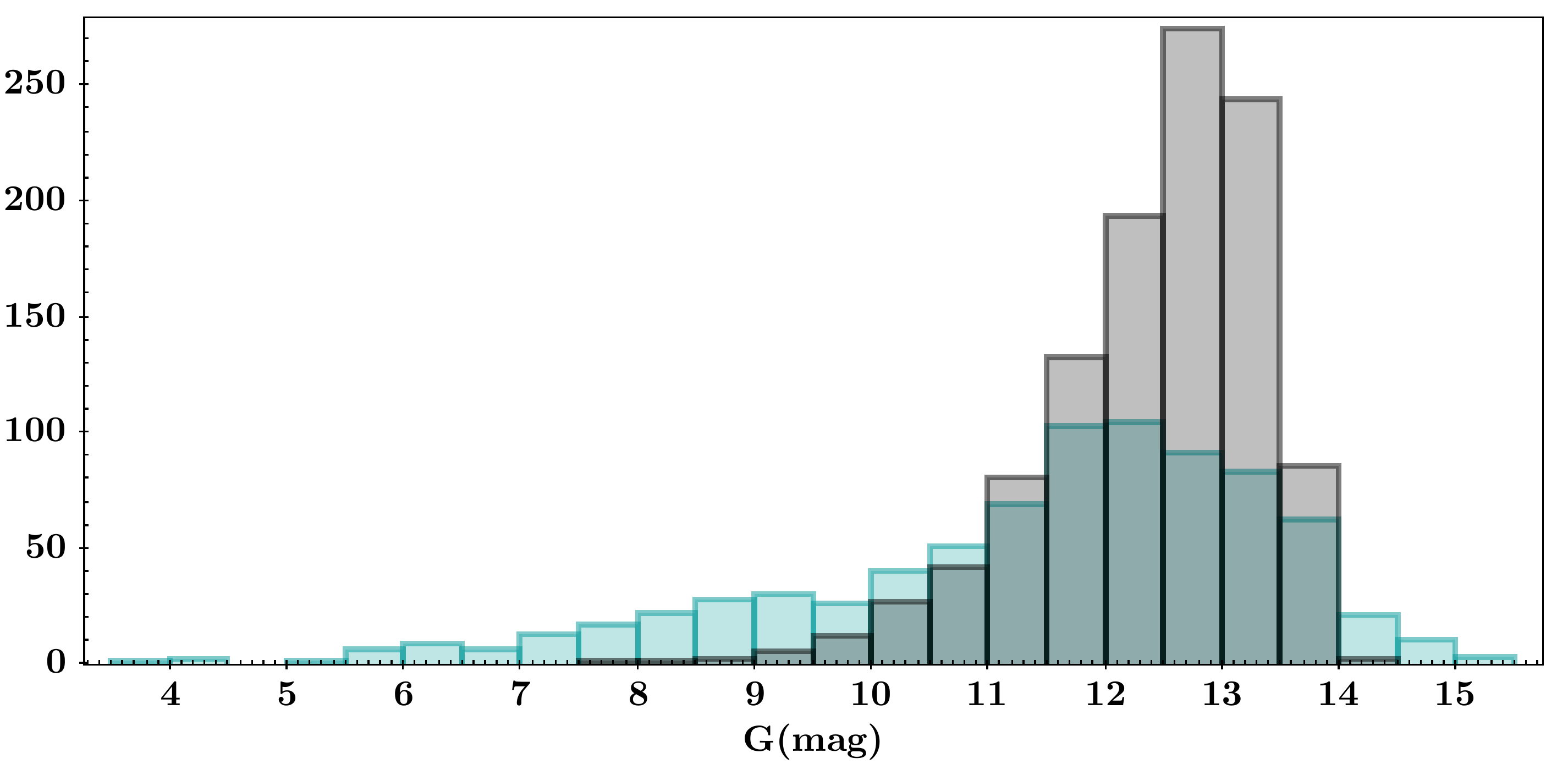}
   \caption{Distribution in $G$ magnitude of the 1\,100 RR Lyrae stars (gray histogram) and 798 Cepheids (cyan histogram) for which RV time series  are published in {\it Gaia} DR3.}
   \label{Gdist-RV}%
    \end{figure}

\begin{figure}[h!]
   \centering
    \includegraphics[scale=0.30]{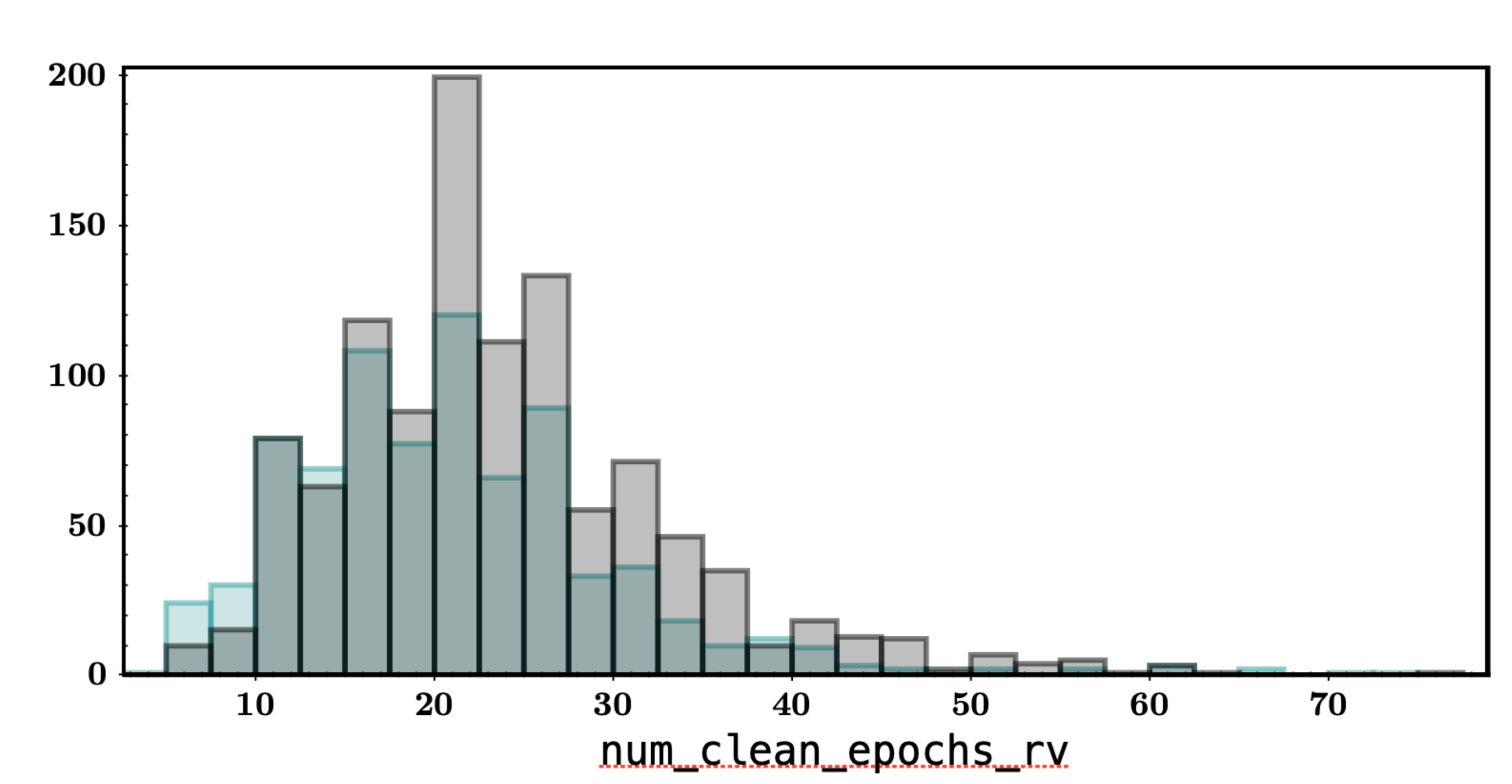}
   \caption{Number of individual measurements in the RV time series  data available for the 1\,100 RR Lyrae stars (gray histogram) and 798 Cepheids (cyan histogram) subsamples.}
   \label{Nepoch-RV}%
    \end{figure}

The 1\,100 RR Lyrae stars  have intensity-averaged $G$ mean magnitudes (as computed by the SOS 
pipeline) in the range  of 7.64 $\leq G \leq$ 14.33 mag 
(Fig.~\ref{Gdist-RV}, gray  histogram)
and a number of  RVS individual measurements ranging from 5 to 77 (Fig.~\ref{Nepoch-RV}, gray histogram).
The 798 Cepheids have intensity-averaged $G$ mean magnitudes in the range  of 3.75$\leq G \leq$ 15.46 mag 
(Fig.~\ref{Gdist-RV}, cyan  histogram) and a number of valid RVS individual measurements  ranging  from  5  to 74 (Fig.~\ref{Nepoch-RV}, cyan  histogram).

The epoch RVs 
are computed using the 
RVS pipeline described in the on-line documentation \citet[Chap. 6]{CU6-DR3-documentation}. 
In practice they are derived by comparing each cleaned and calibrated RVS spectrum with the synthetic reference template of a star with similar atmospheric parameters. See \citet[Section 6.4.8]{CU6-DR3-documentation}
and \citet[Sect. 7]{DR2-DPACP-47} for a detailed description of how epoch RVs 
are calculated, and \citet[Fig. 6.13]{CU6-DR3-documentation} for the median precision of epoch measurements as a function of $G_{RVS}$. 
The quality of the epoch RVs 
depends on several factors: the $G_{RVS}$ magnitude of the star (since the fainter the star, the noisier the spectrum); the appropriate match with a synthetic reference template (see \citet[Section 6.4.8.3]{CU6-DR3-documentation} for how template spectra are associated with RVS spectra); the quality of astrometric data; and potential limitations of the processing software (e.g., for deblending or the identification of peculiar stars). The epoch  
RVs published in DR3 are only those of the 1\,100 RR Lyrae stars and 
798 
Cepheids described in this paper, which are available via the {\it Gaia} archive {\tt vari\_epoch\_radial\_velocity} table. 
Typical uncertainties of the epoch RVs 
for these sources are $\lesssim$ 1 km/s for $G\sim$9-9.5 mag  ($G_{RVS}\sim8.5-9$ mag, for RR Lyrae stars) 
and  $\lesssim$ 6.5 km/s for $G\sim$13 mag ($G_{RVS}\sim12.5$ mag for RR Lyrae stars) 
(see \citet[Fig. 6.13]{CU6-DR3-documentation}. 
The epoch data of all remaining stars will be published in DR4. The other RVS-pipeline products published in DR3 are at source level (i.e., obtained combining data from all epochs). They are listed in Table 6.1 of the on-line documentation and are available via the {\it Gaia} archive {\tt gaia\_source} table.  These products include, for example, information about the $G_{RVS}$ magnitude of the star (called {\tt grvs\_mag} in the {\it Gaia} archive) and the atmospheric parameters of the synthetic reference template used to derive the star epoch RVs
(called {\tt rv\_template\_teff}, {\tt rv\_template\_logg} and 
{\tt rv\_template\_feh}, and 
{\tt rv\_atm\_param\_origin}).
Not all the  
1\,100 RR Lyrae stars and 798  
Cepheids with epoch RVs 
published in DR3 also have source-level RVS pipeline products published. Source-level information is lacking for 51 Cepheids and 45 RR Lyrae stars, for which the all-epoch-combined RV 
was deemed of insufficient quality \citep[see][for the filters applied to the all-epoch-combined RVs]{DR3-DPACP-159}. 
In most cases, the reason was the detection of potential variability, such as double lines in more than 10\% of the epoch spectra, variability in spectral-sample flux from epoch to epoch, or a large dispersion in epoch RV ({\tt radial\_velocity\_error}$>$40 km/s). In other cases, the spectra were deemed too noisy, or the effective temperature $T_\textrm{eff}$ was found to exceed the limit for good-quality RV measurements (i.e., 7\,000~K for $G_{RVS}>$12 mag; and 14\,500~K for $G_{RVS}\le$12 mag). 
However, we note that most of the above issues are normal features, for high amplitude variable sources like RR Lyrae stars and Cepheids, whose RV curves typically exhibit amplitudes of tens of km/s. In fact, we  
 carefully inspected the RV curves of these 51 Cepheids and 45 RR-Lyrae stars and none of them was found to have a peculiar shape. 
 On the contrary, the large majority of these sources have very regular and nice RV curves. Therefore, we kept them for publication and,  
in Appendix~\ref{app:var-nopubb}, provide information about their  $G_{RVS}$ magnitudes 
and 
the atmospheric parameters of the associated templates, along with the reason why they do not have source-level RVS pipeline products
published.

\subsection{Processing 
of the RV data through the SOS Cep\&RRL pipeline and results}\label{sec:RV-SOS-processing}
 \begin{figure*}[h!]
   \centering
   \includegraphics[scale=0.95]{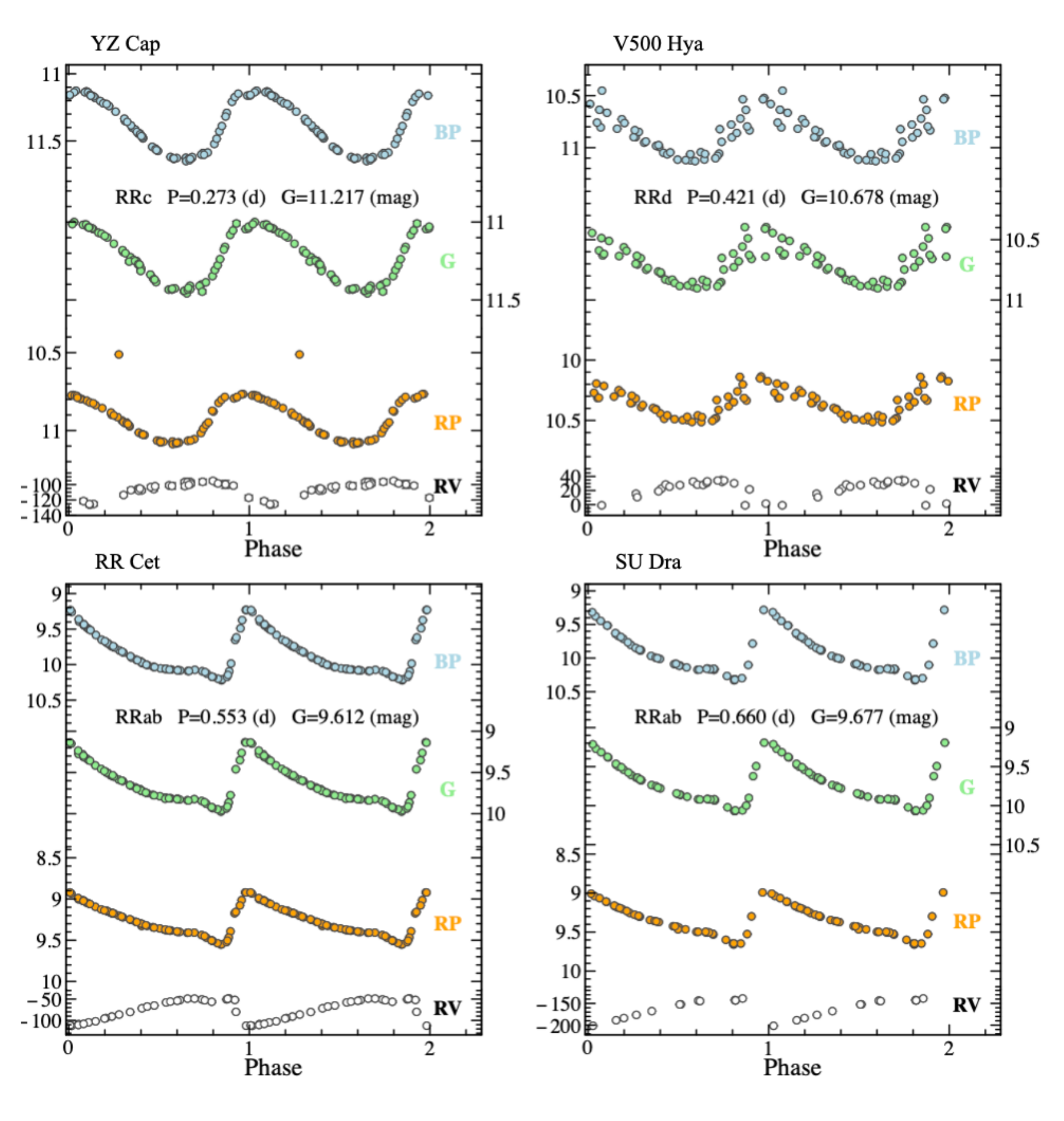}
   \caption{From top to bottom in each panel, $G_{BP}$, $G$, and $G_{RP}$ light curves and RV curve for RR Lyrae stars 
      released in DR3. Error bars of the photometric and RV measurements are smaller than or comparable to symbol size. Sources in the upper row,
from left to right, are a first overtone (RRc) and a double-mode (RRd); in the lower row are two fundamental mode (RRab) RR Lyrae stars. 
 }
         \label{FigVibStab}
   \end{figure*}
The time series RV data of the 1\,100 RR Lyrae stars and 798  
Cepheids were processed by the SOS Cep\&RRL pipeline and for  sources with more than 7 transits (1\,097/1\,100 of  the RR Lyrae stars and 785/798 of the Cepheids) the RV curves obtained by folding the data with the periods inferred  from the $G$-band photometry were 
modelled 
with truncated Fourier functions. 
Mean RV ($\langle$RV$\rangle$), peak-to-peak RV amplitude [Amp(RV)] and epoch of minimum RV were computed from the modelled RV curves, that are published in the DR3 {\tt vari\_rrlyrae} and {\tt vari\_cepheid} tables.
Fig.~\ref{FigVibStab}  shows, from top to bottom, 
examples of $G_{BP}$, $G$,
$G_{RP}$ light and RV curves for RR Lyrae stars 
that are  released in DR3.  
The RV curves have the typical
shape compared to the light curves with the minimum value in RV corresponding to
maximum star brightness.

An atlas of the light and RV  curves for the full sample of 1\,100 RR Lyrae stars with time series RVs published in DR3 is presented in Appendix~\ref{appendix-atlas}. 
Examples of RV curves and an Atlas of the light and RV  curves for the 798 Cepheids is presented in \citet{DR3-DPACP-169}.

\begin{figure}
\includegraphics[scale=0.33]{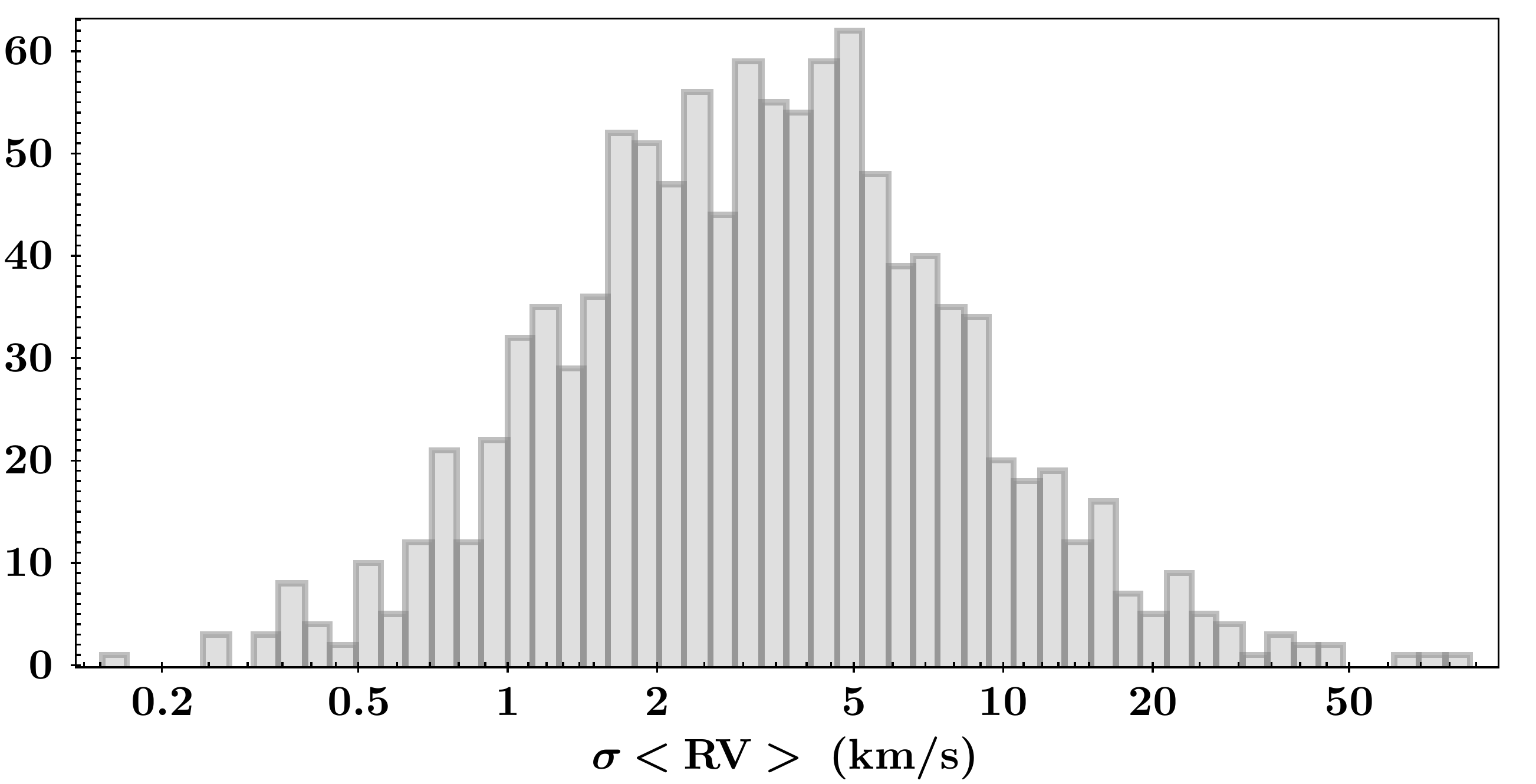}
   \caption{Error distribution
of the mean RV values computed by the SOS pipeline for the 1097/1100 RR Lyrae stars. }
\label{Fig:Isto-sigmarv}  
\end{figure}

\begin{figure}
\includegraphics[scale=0.33]{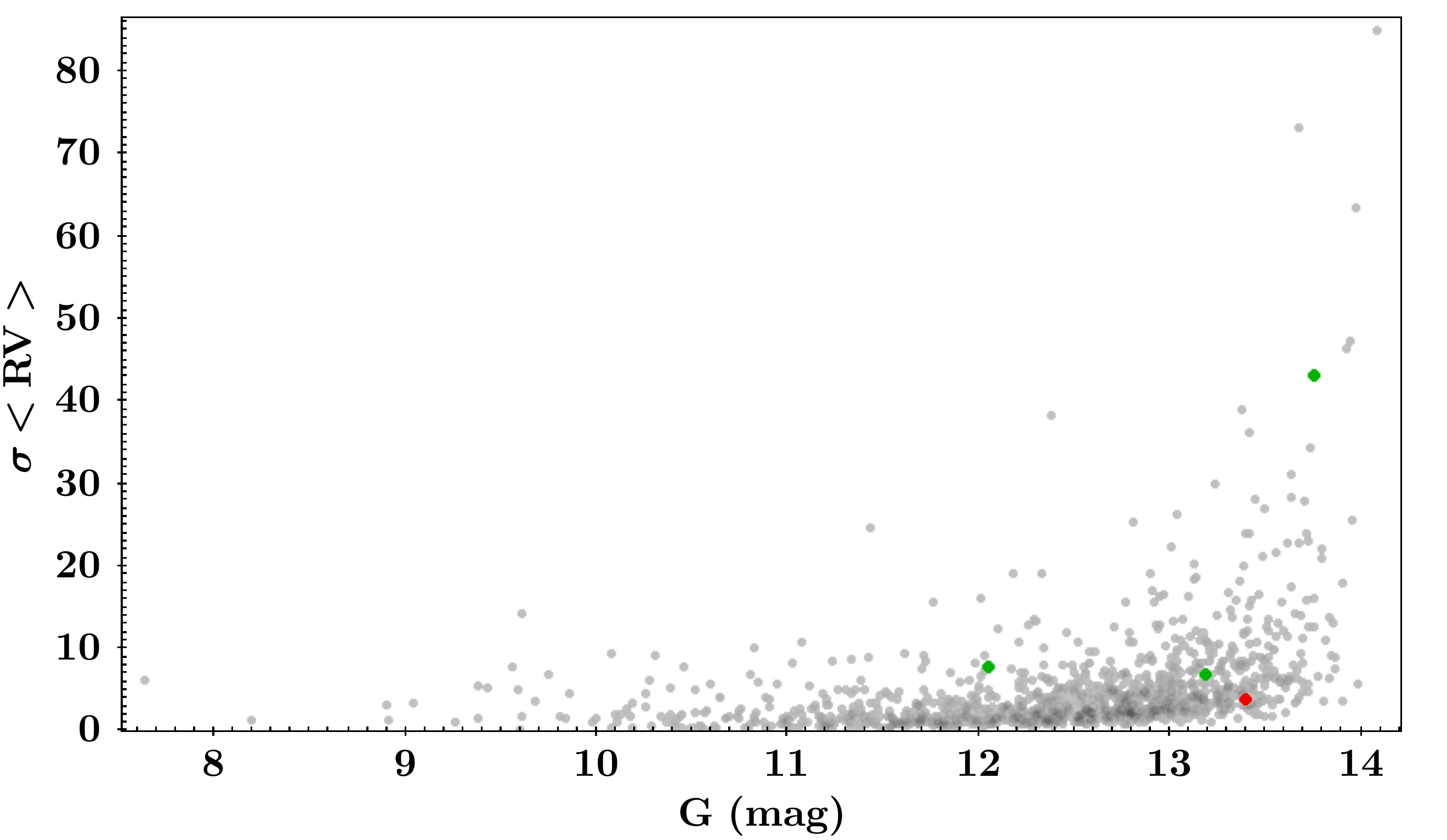}
   \caption{Uncertainty of the mean RV values 
   as a function of the SOS mean <G> magnitude for the 1097/1100 RR Lyrae stars (gray filled points). Plotted with different colours are 1 DCEP (red symbol) and 3 ECLs (green symbols) that contaminate the 1097 RR Lyrae sample (see Sect.~\ref{sec:s_vali} for details).}
\label{Fig:g_sigmarv}  
\end{figure}

Figure~\ref{Fig:Isto-sigmarv} shows the error distribution
of the mean RV values computed by the SOS pipeline 
($\sigma \langle {\rm RV} \rangle$) for the 1097/1100 RR Lyrae stars, whereas Fig.~\ref{Fig:g_sigmarv} shows 
 $\sigma \langle {\rm RV} \rangle$ as a function of 
the mean $G$ magnitude of these sources. The uncertainty ranges from 0.48 km/s to 84.88 km/s for  $\langle G \rangle$ ranging from 7.64 to 14.30 mag, with peak between 3 and 6 km/s,  corresponding to the bulk of RR Lyrae stars with  mean $G$ magnitude between 12 and 13 mag in Fig.~\ref{Fig:g_sigmarv} and consistently with fig. 6.13
in \citet{CU6-DR3-documentation}. Less than 20 sources have $\sigma_{\langle {\rm RV} \rangle}>$ 25 km/s corresponding to sources in the faint magnitude tail of Fig.~\ref{Fig:g_sigmarv}.

\begin{figure}
\includegraphics[scale=0.33]{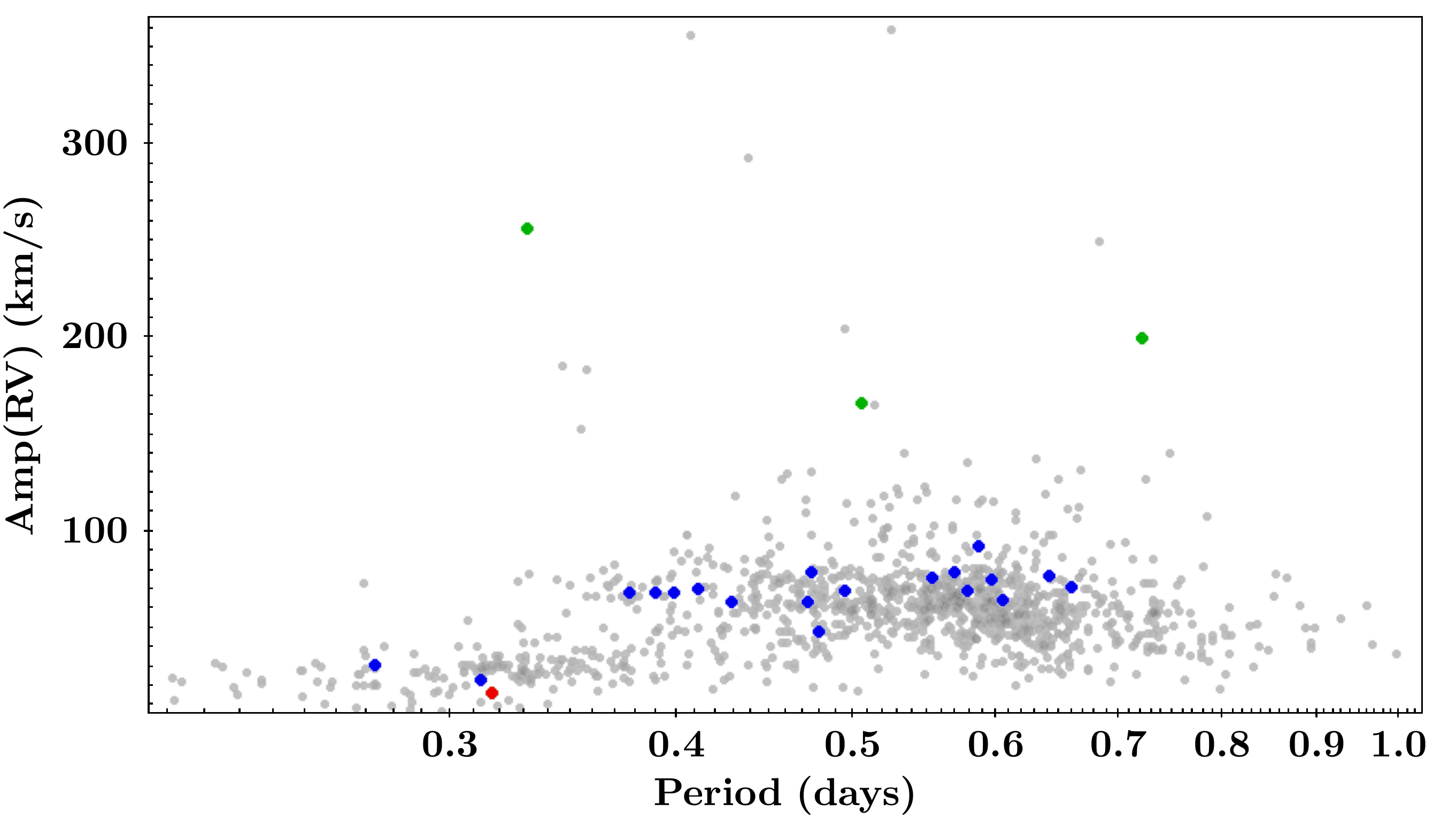}
   \caption{$P$ versus Amp(RV) diagram of the 1097/1100 RR Lyrae stars (gray filled points). Blue filled points show 19 RR Lyrae stars with high accuracy RV curves in the literature used to validate the {\it Gaia} RVs (see Sect.~\ref{sec:RV-validation}).  }
\label{Fig:P-ampRV}%
\end{figure}

Finally, Figure~\ref{Fig:P-ampRV} shows the $P$ vs Amp(RV) diagram of the  1097 RR Lyrae stars. A few sources have very large Amp(RV) values. Three of them are contaminants  (green filled circles), the others have very few RV measurements, hence very uncertain Amp(RV) values.




\begin{figure}
\includegraphics[scale=0.32]{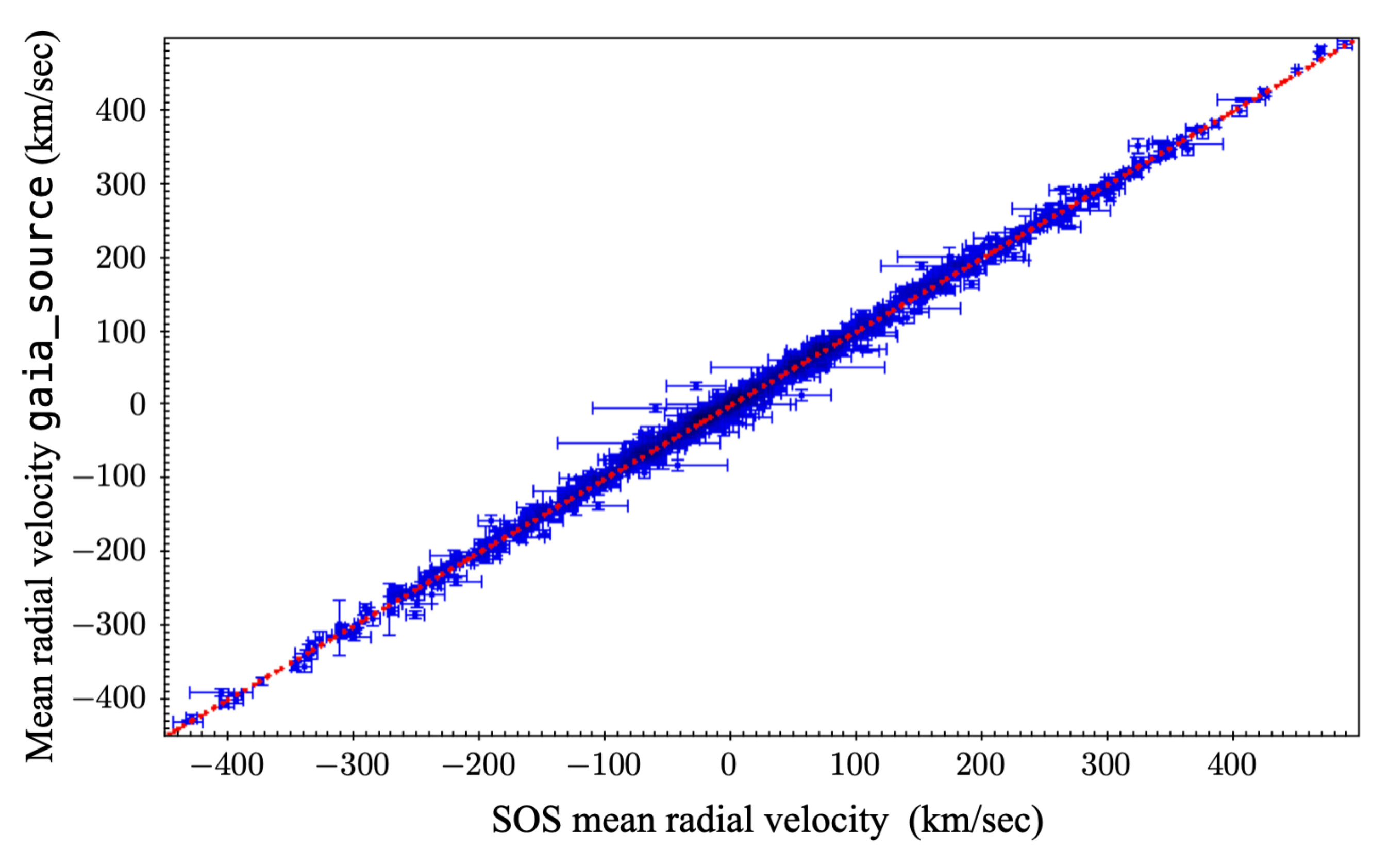}
   \caption{Comparison of the mean RV values 
  in the DR3 {\tt vari\_rrlyrae} table, which  are   computed by the SOS Cep\&RRL pipeline from the modelled RV curves of the RR Lyrae stars, 
    and the combined RV values in the  DR3 {\tt gaia\_source} table, for the 1\,054 RR Lyrae stars for which both values are available. Median and mean difference between the two average values are of 5.21 and  6.88  km/s (with  6.35 km/s  standard deviation), respectively.}
              \label{Fig:RVcompRR}%
\end{figure}

\begin{figure*}
   \centering
     \includegraphics[scale=0.31]{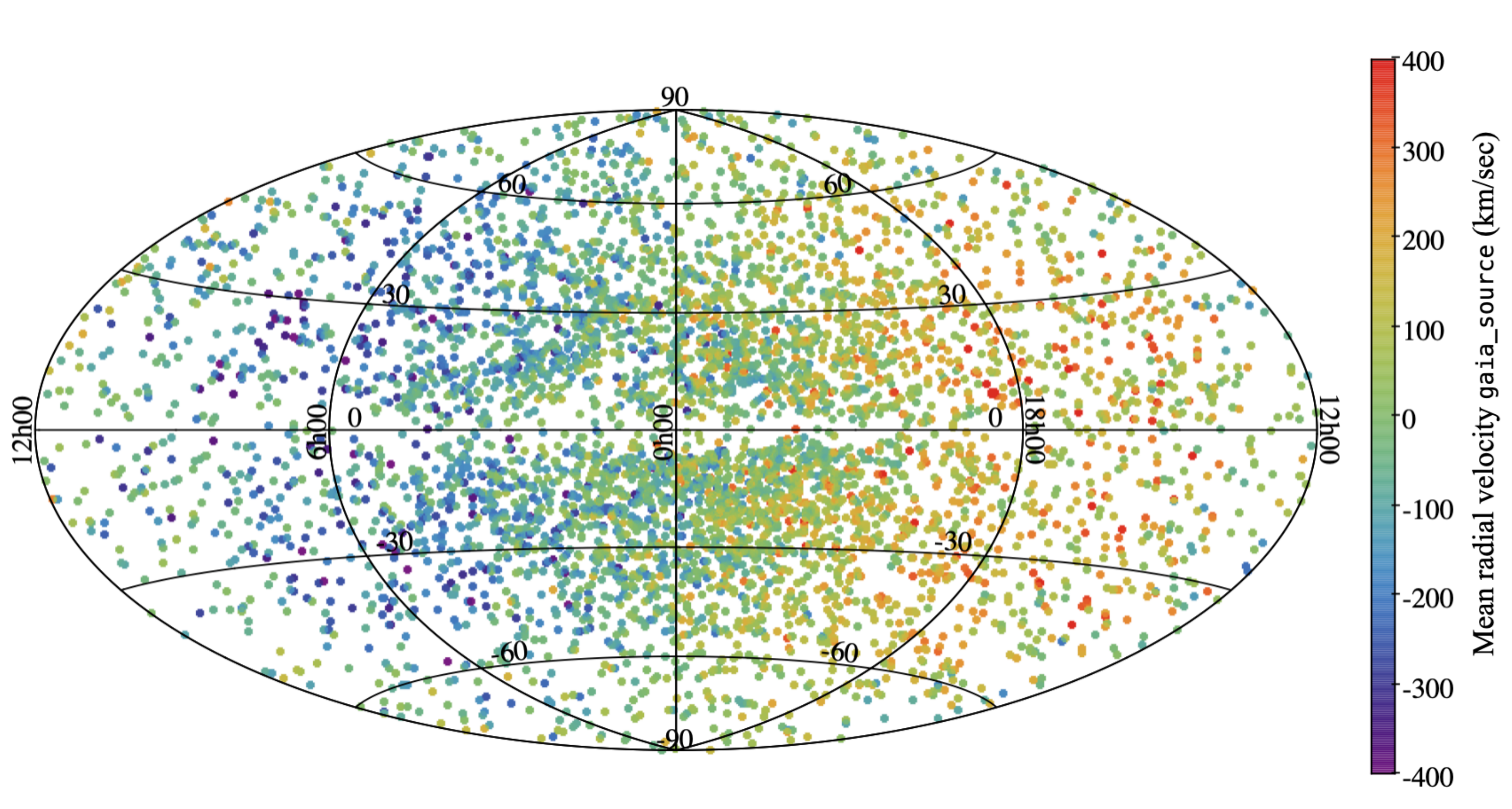}~\includegraphics[scale=0.31]{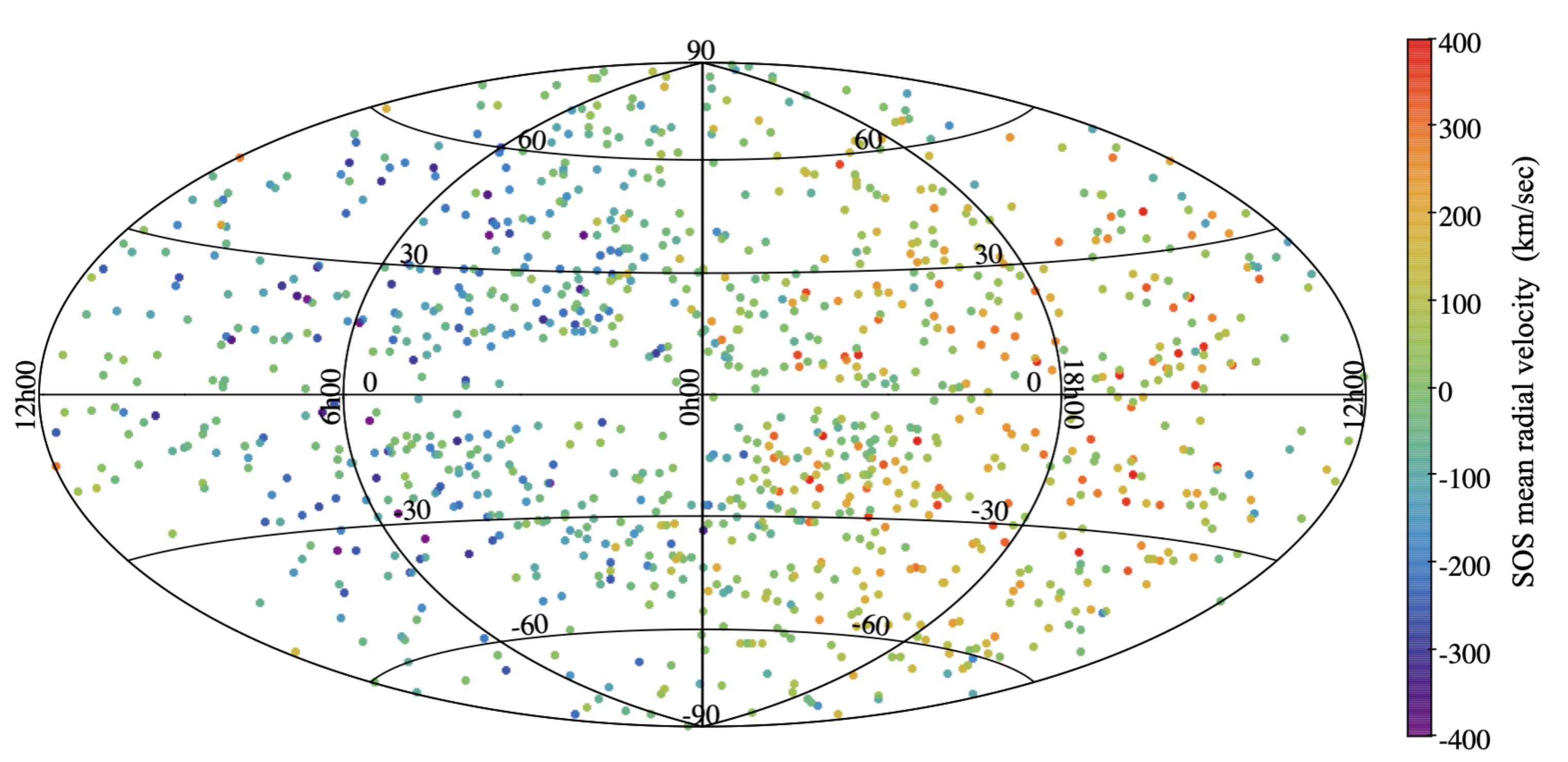}
   \caption{RV maps defined by  1\,097/1\,100 RR Lyrae stars in the DR3 {\tt vari\_rrlyrae} (righ panel) and    5\,096  RR Lyrae stars in the {\tt gaia\_source} table (left panel). The sources have been colour-coded according to their RV values, as encoded in the colour bar on the right.}\label{Fig:RVmapRR}%
    \end{figure*}

A comparison was made between the SOS mean RV values 
of the 1\,097 RR Lyrae stars  
and the combined RV values for  these stars
in the {\it Gaia} DR3 {\tt gaia\_source} table. 
The two  
RV values are available for  1\,054 RR Lyrae stars. 
This comparison is shown in Fig.~\ref{Fig:RVcompRR}. 
There is good agreement, within the errors,  between the two sets of RV measurements.
Median and mean difference between the two average values are of 5.21 and  6.88  km/s (with  6.35 km/s  standard deviation), respectively. 
Finally, 
Fig.~\ref{Fig:RVmapRR} shows  the RV map defined by the 1\,097/1\,100 RR Lyrae stars in the DR3 {\tt vari\_rrlyrae} (righ panel)  and, in the left panel, the map defined by the 5\,096 RR Lyrae stars in the DR3 {\tt gaia\_source} table, 
for comparison. The sources have been colour-coded according their mean RV values. 
The two distributions are very well consistent, as expected from the good agreement between the two sets of RV values (see Fig.~\ref{Fig:RVcompRR}). RR Lyrae stars are distributed at all galactic latitudes and populate the maps with RVs  ranging from large negative to large positive values.

\subsection{Validation with the literature}\label{sec:RV-validation}
  A number of RR Lyrae stars and Cepheids for which 
  epoch RVs are published in DR3 
 were analysed in the past with the Baade-Wesselink (B-W) technique  (e.g. \citealt[and references therein]{1992ApJ...396..219C},  
 for different versions of  the B-W method applied to RR Lyrae stars and,
 \citealt{1989ApJ...342..467G},  \citealt{1997A&A...318..797R},   
 for B-W analyses of Cepheids).  Hence, high accuracy RV measurements  either obtained with the CORAVEL spectrophotometer  \citep{1979VA.....23..279B}, or from high resolution 
 spectra, are available in the literature for these stars. 
We have selected 19 RR Lyrae stars 
with 
10 or more epoch RVS 
measurements  in the RV 
curves 
and have made a detailed comparison with the high accuracy RVs covering the whole star pulsation cycle, available for these variable stars in the literature.
We have mainly used for our comparison data taken from the most recent literature, as datasets collected in/before the early eighties often have very large uncertainties.
A similar comparison for a sample  
of DCEPs is presented in \citet{DR3-DPACP-169}.

The 19 RR Lyrae stars used for this test 
are
listed in Tables~\ref{Tab:Table-rr}. 
In
the table we indicate the literature data used for the comparison, the
star mean astrophysical parameters ($\langle T_{\rm eff}\rangle$, $\langle$log g$\rangle$, and iron abundance: [Fe/H]),  that are, in general, the values adopted for or derived from the B-W analyses.
The sample comprises 
17 RRab and 2 RRc stars. They span a range in [Fe/H] of more than 2 dex. Three of them 
are known or suspected to be affected by Bla$\check {\rm z}$ko effect \citep{1907AN....175..325B}, 
a periodic modulation of the amplitude and shape of the light (and RV) curve, occurring with periodicities from tens to hundreds of days; one (V32) is a member of a globular cluster (NGC~6121 -- M4); whereas UU Vir is an RR Lyrae star suspected to be in a binary system,  according  to anomaly detected in the $Gaia$ DR2 proper motions for the source \citep{2019A&A...623A.116K}.
The 19 RR Lyrae stars have   0.273 $\leq$ P $\leq$ 0.660 d, 9.56 $\leq \langle G \rangle \leq$ 12.90 mag and 10 or more (up to 37) individual RVS measurements, with errors in the range from 0.41 to 16 km/s depending on the magnitude and pulsation phase).

The  RVS measurements of these 19 RR Lyrae  stars are reasonably well distributed over the
pulsation cycle and often there is more than one RV dataset available in the
literature to make the comparison. 
The 
literature RVs and the {\it Gaia} RVS measurements 
of the stars 
in Table~\ref{Tab:Table-rr} span 
time intervals up to a few decades and the 
literature periods 
often do not allow to optimally phase all the  available RV datasets.
We have thus performed  a new period search on each combined (literature plus $Gaia$) RV 
dataset with the 
package Graphical Analyzer of Time Series  (GRATIS;  custom 
software developed at the Bologna Observatory by P. Montegriffo, see e.g. 
\citealt{2000AJ....120.2054C}) 
and defined  
periods allowing to optimally match  the {\it Gaia} and the literature RV curves for all sources used for this comparison\footnote{ This procedure worked quite well as, contrary to Cepheids, RR Lyrae stars generally show slow, if any, period changes. This is consistent with their evolution being much slower than for Cepheids. Furthermore, as far as we know none of the 19 RR Lyrae stars considered for our comparison is known to present  significant period changes. 
Indeed, for some of these RR Lyrae stars, the period is known to be fairly constant.}. {No photometry was involved in the procedure, however, we note that the differences between the periods derived for the 19 RR Lyrae stars from the combined RV datasets  ($P_{\rm RV}$) and the periods derived by the SOS pipeline ($P_{\rm SOS}$) from the DR3 time series photometry are in the range of $2.8 \times 10^{-8}$ d  $<| P_{\rm RV} - P_{\rm SOS}| < 2.0 \times  10^{-5}$ d  for 18 of the 19 stars in our sample.  Only for V440 Sgr, (source\_id= 6771307454464848768) $P_{\rm SOS}$ is 0.048 d shorter than the period derived from the combined RV dataset, and it is clearly wrong, as this star has only sparsely sampled light and RV curves. The correct period for V440 Sgr  is provided in Table~\ref{Tab:Table-rr}.}

\begin{figure*}[h!]
\begin{center}
      \includegraphics[width=9.1cm]{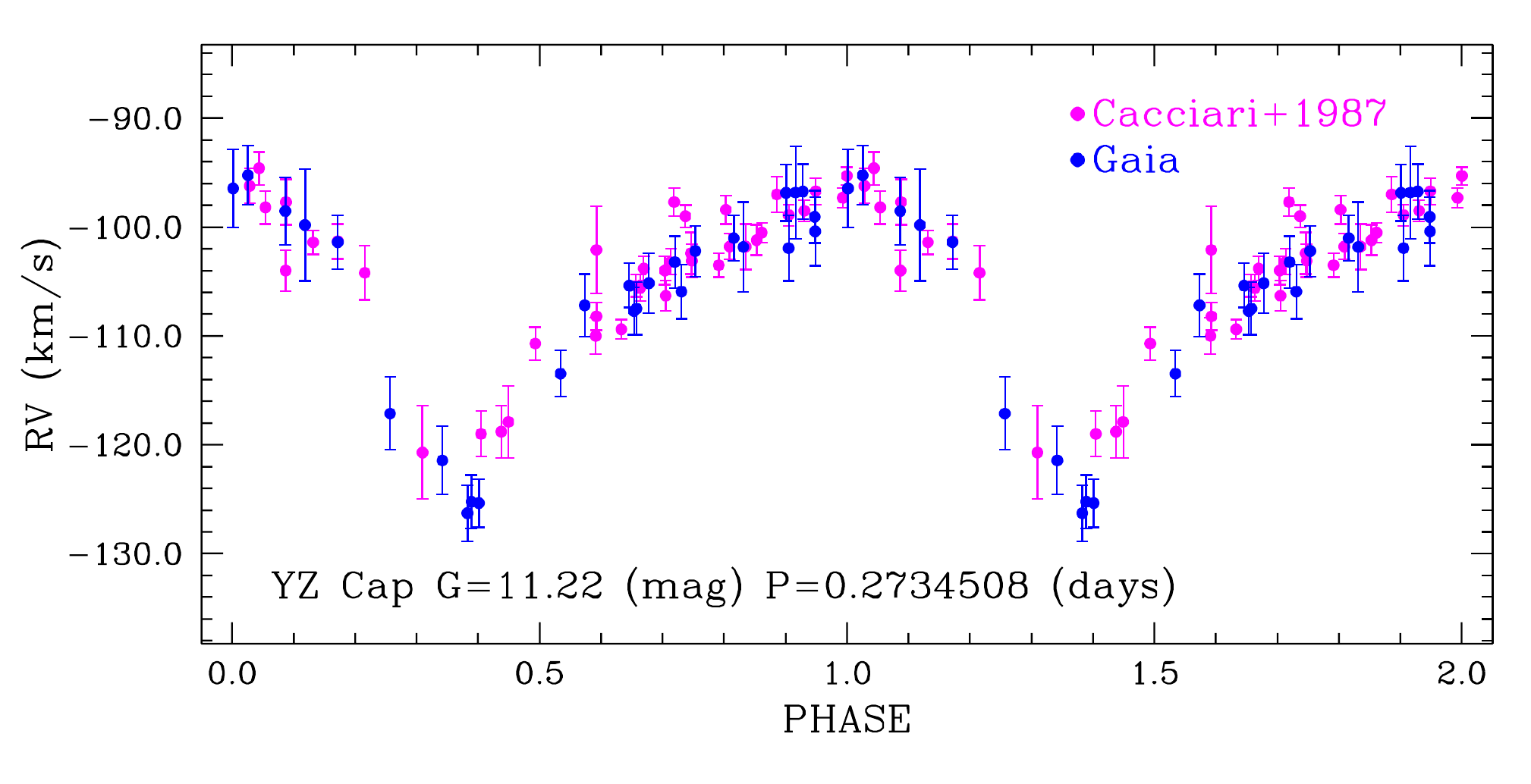}
      ~\includegraphics[width=9.1cm]{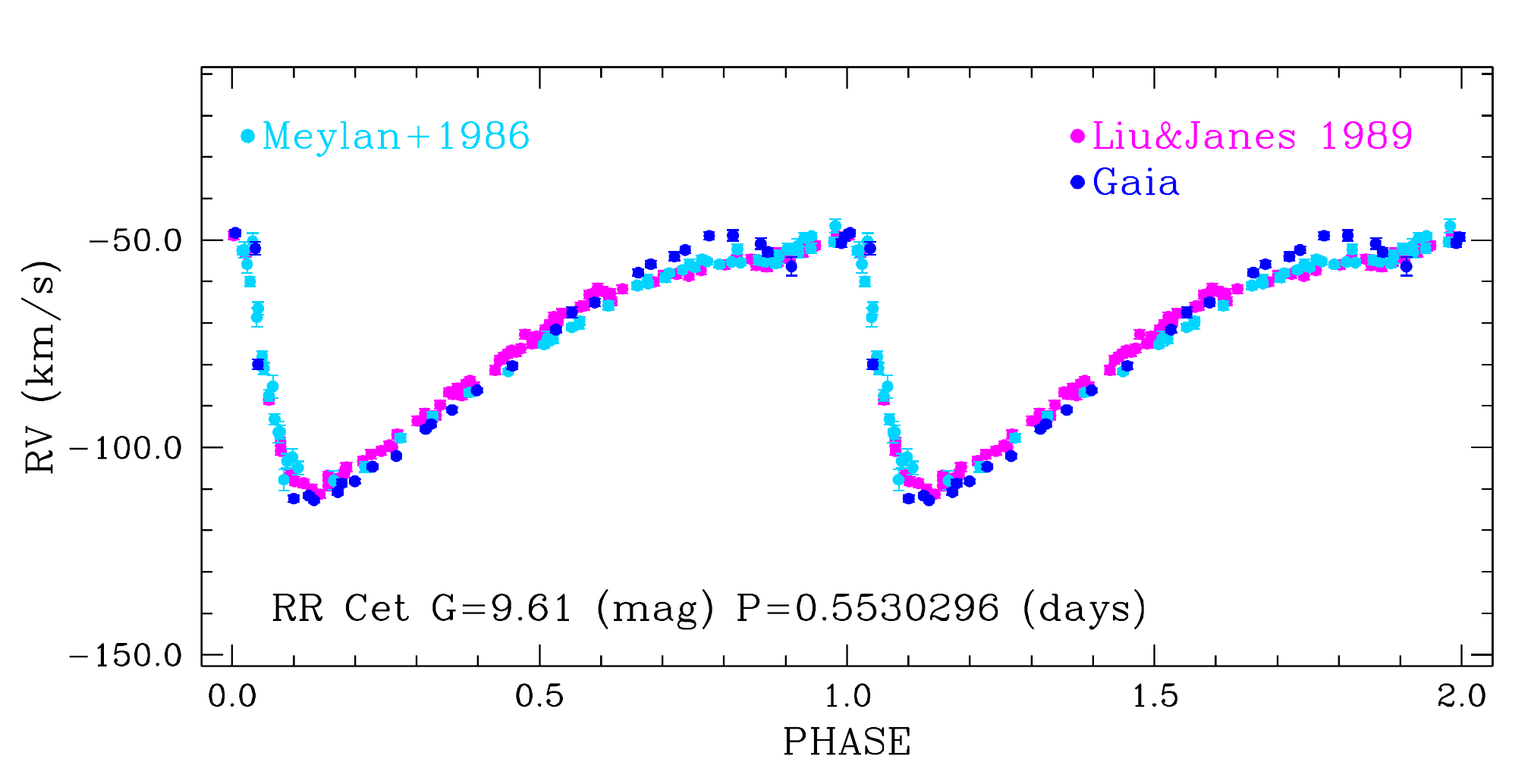}
      \includegraphics[width=9.1cm]{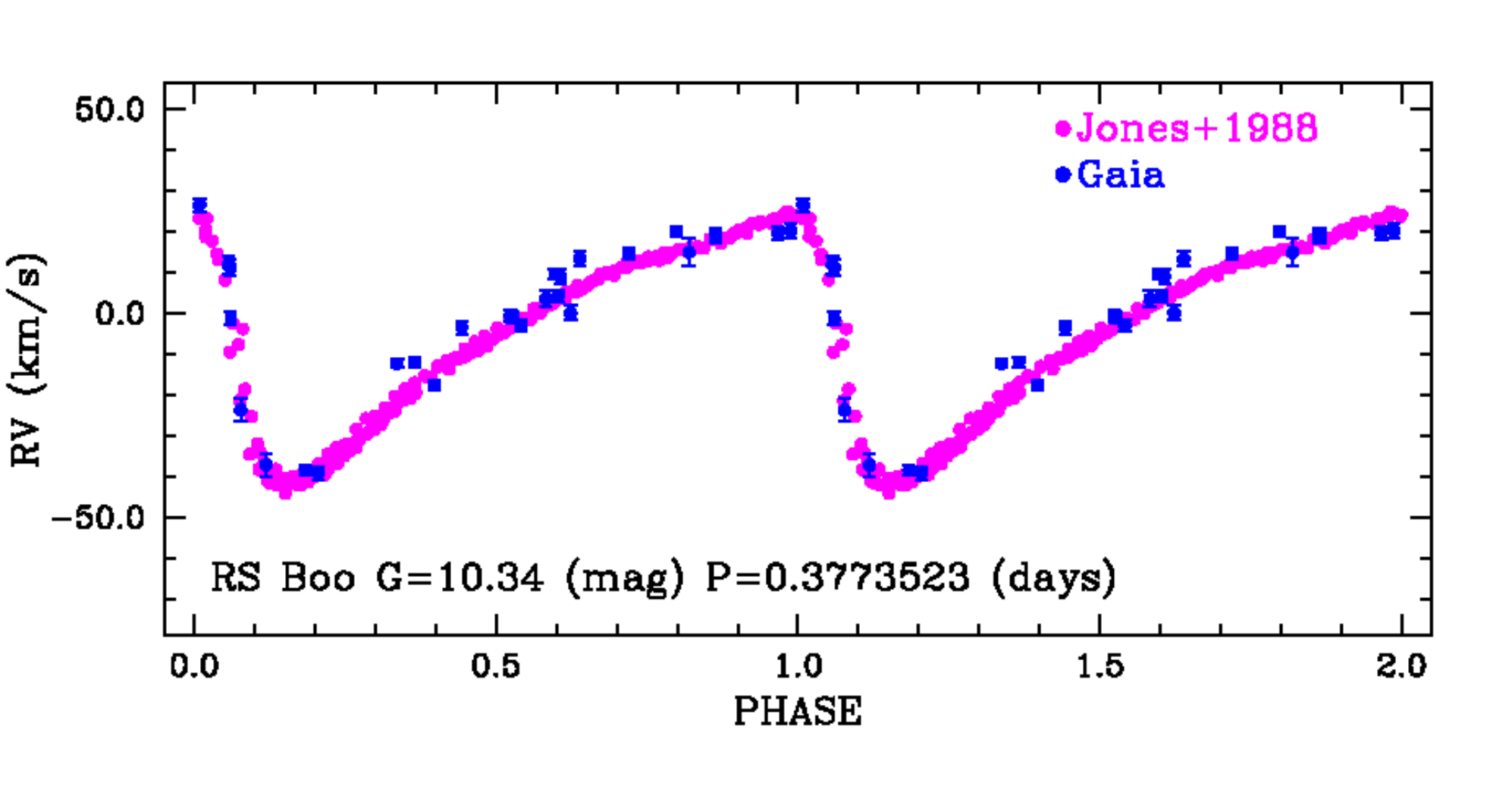}
      ~\includegraphics[width=9.1cm]{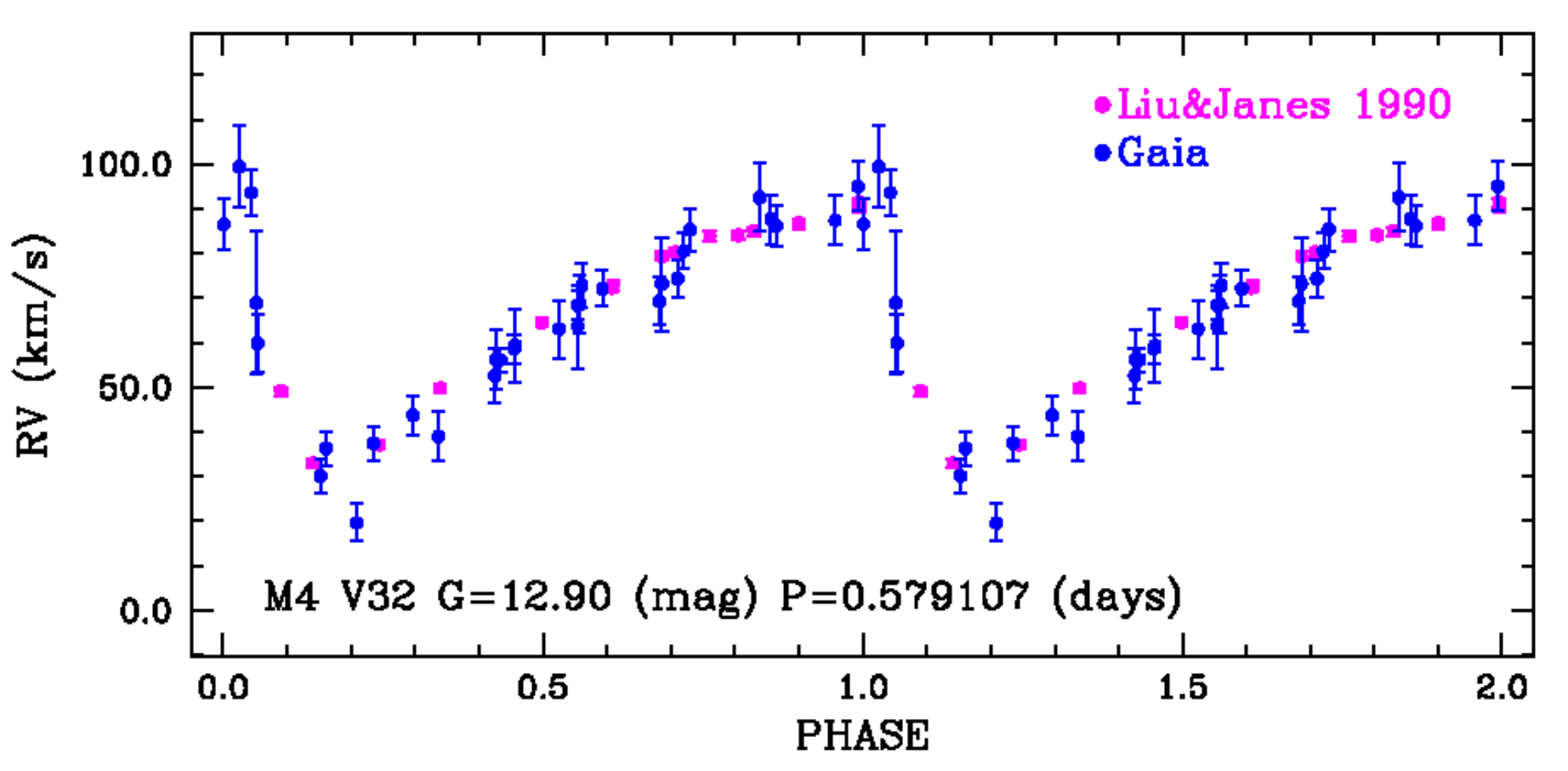} 
      ~
      \includegraphics[width=9.1cm]{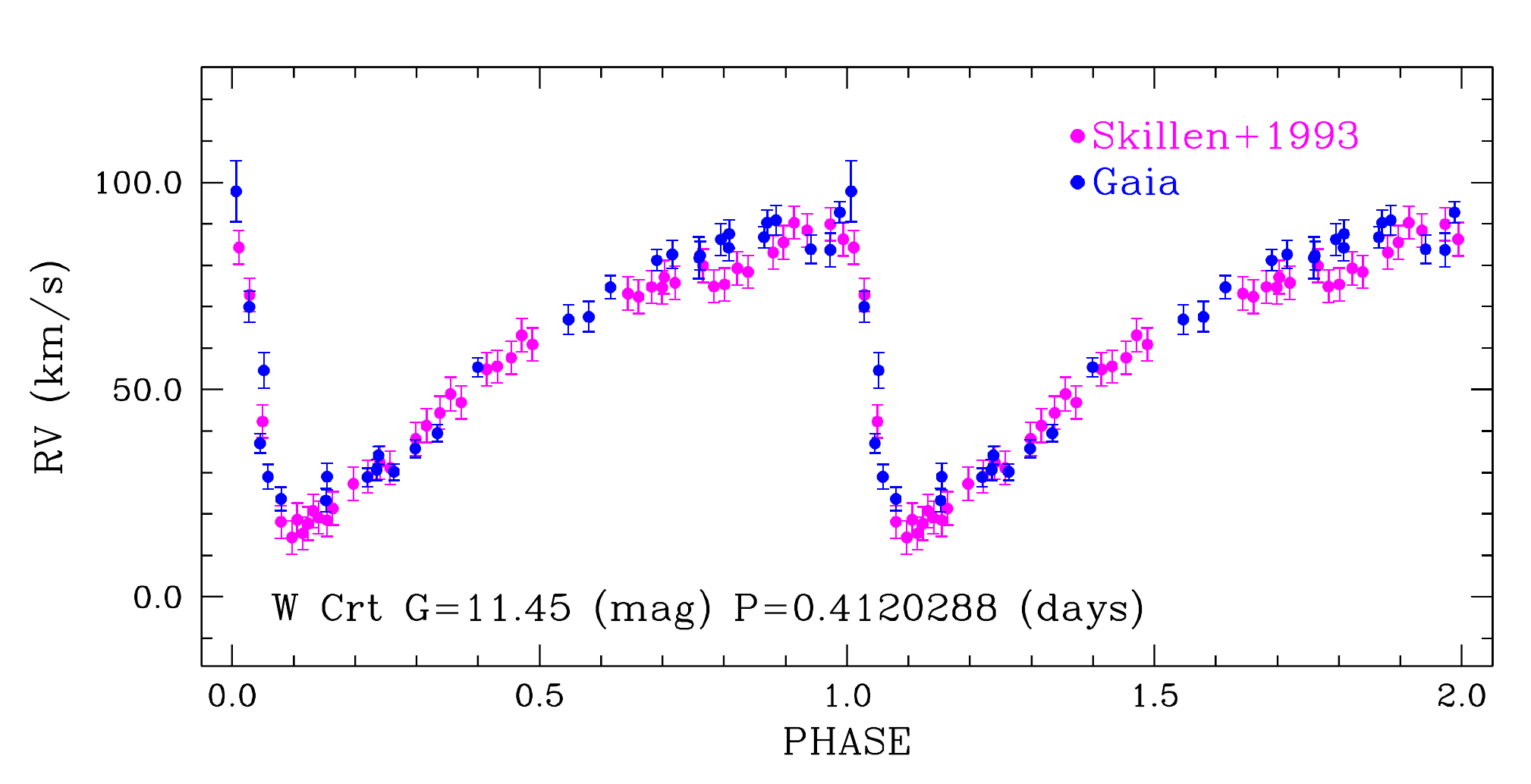}
      ~\includegraphics[width=9.1cm]{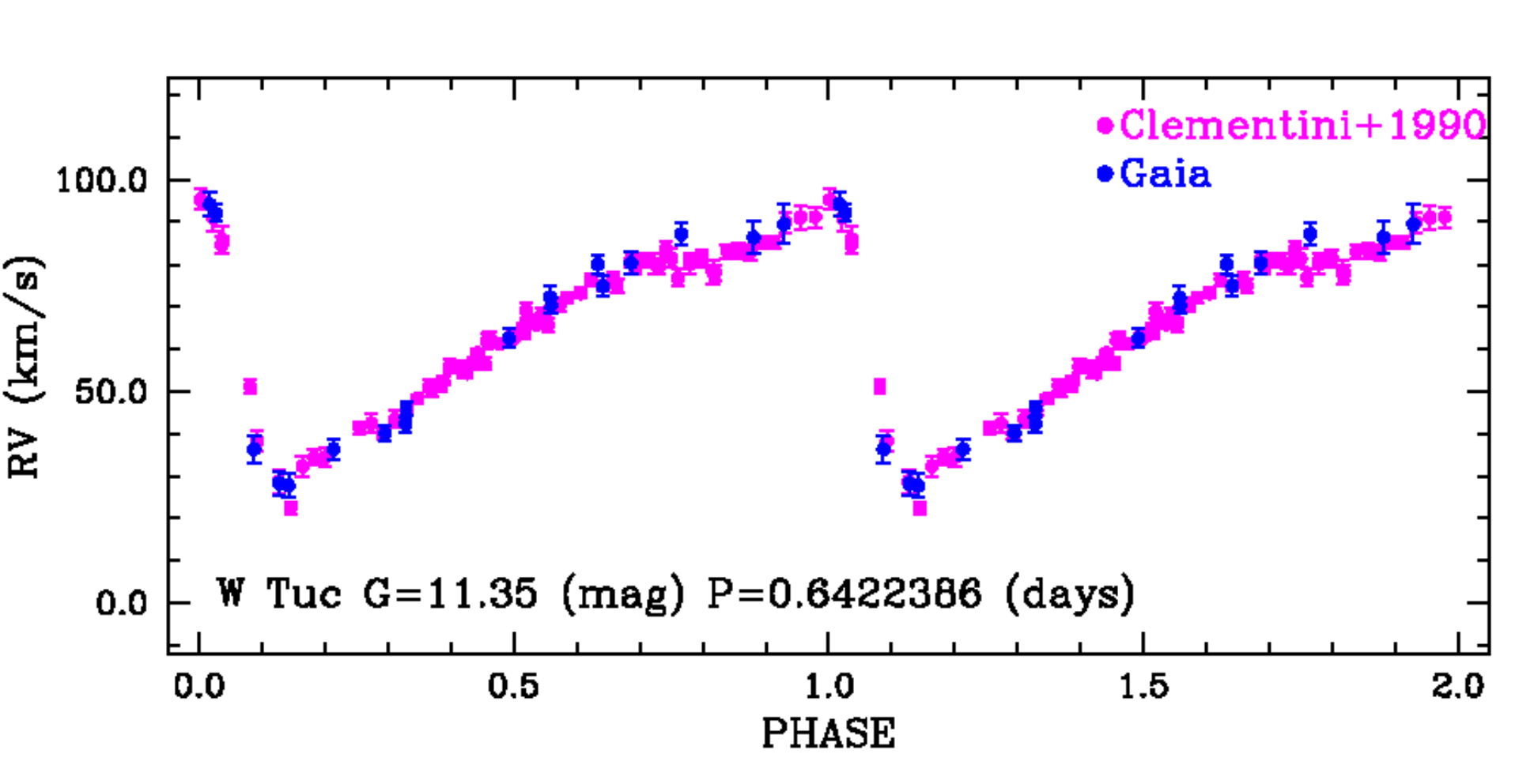}
      \caption{Comparison of the $Gaia$ RVS radial velocities  (blue filled circles) with literature RV datasets (magenta and cyan filled circles), for the 
     RRc star YZ~Cap (top-left), and the RRab stars  RS~Boo (middle-left),  W~Crt (bottom-left), RR Cet (top-right),
     V32 in the GC M4 (middle-right) and W~Tuc (bottom-right). Also shown: pulsation period (as re-derived in the present analysis) and the intensity-averaged $G$-band mean magnitude 
     (as derived by the SOS Cep\&RRL pipeline). The {\it Gaia} RV curve of V32 in M4 was shifted by $-5$ km/s
     to better match the literature RV curve.
     }
     \label{Fig:RR+CC}
      
  \end{center}
\end{figure*}

As an example, Fig.~\ref{Fig:RR+CC} shows 
the radial velocity curves of the RRc star YZ~Cap and the RRab stars RS~Boo, W~Crt, RR~Cet, V32 in  
M4 and W~Tuc,  using different colours for the literature and the $Gaia$ RVS data.  The RV curves of the remaining 13 RR Lyrae stars are presented in Figure~\ref{Fig:13di19}.  
This comparison  
shows that there is a good agreement both in the shape and the amplitude of the {\it Gaia} RVS and the literature RV curves for 14 of the 19 RR Lyrae stars in the sample, thus assessing the good quality of the RVS epoch RV data for these sources. Specifically, the agreement is good for: YZ~Cap, RS~Boo, W~Tuc, V~Ind, DX~Del, UU~Cet V440~Sgr, and also for V32 in M4, TV~Boo, AV~Peg, TW~Her and UU~Vir, upon rigidly shifting their RVS curves by +5 km/s for the first two, $-$3 km/s for the third one and by $-$4 km/s for the latter two, as to better match the literature curves. 
Agreement is also acceptable for RV~Phe and SS~For, in spite of the large scatter 
caused by the Blazhko affecting these two  stars. 
 Conversely, there are noticeable differences between the literature and {\it Gaia} RV curves for RR~Cet, RX~Eri, SU~Dra, SW~Dra and, to a lesser extent, also for W~Crt. These differences are most apparent around the bump just before the RV maxima, around phases 0.7-0.9, where the {\it Gaia} RVS curves reach higher velocities than the literature curves. Investigating the reason causing these differences is beyond the purposes of this paper, however, we note  that shock waves are known to propagate in the atmosphere of RR Lyrae stars in correspondence with the hump (phase$\sim$ 0.9) and  bump (phase$\sim$ 0.7) in the light and RV curves \citep[][and references therein] {1988A&A...199..242G, 1994MNRAS.267...83C, 2014A&A...565A..73G} that cause flux redistribution, H line emission and U-excess. 
 The RVS measurements seem to  better reflect than the 
 literature RVs the 
 flux redistribution caused by shocks 
 propagating in the atmospheres of  these 5 RR Lyrae stars.

\section{Astrophysical parameters}
\label{sec:s_astro-par}

The {\tt StellarParametersDerivation} module of the  SOS Cep\&RRL pipeline (see Fig.~\ref{fig:rrl-branch}) computes two main astrophysical parameters for the confirmed RR Lyrae stars: 
1. a photometric estimate of individual 
metal abundance  
([Fe/H]) for RRab and RRc stars whose light curves have been Fourier-modelled with at least 3 harmonics and, 2. a $G$ band absorption [$A(G)$] for RRab stars for which the ($G - G_{RP}$) colour is available.
The {\tt StellarParametersDerivation} module is fully described in Sect. 2.1 of Paper~II. 
Here, we only note that unlike in DR2, uncertainties in  metallicity and $A(G)$ values (as well as for any other  parameter derived by the SOS pipeline) in DR3 were calculated with the bootstrap. 

In the following we present the [Fe/H] abundances (Sect.~\ref{sec:sos-met}) and $A(G)$ values (Sect.~\ref{sec:sos-A(G)}) computed from 
the pulsation characteristics of the confirmed RR Lyrae stars,  
and compare them with values in the literature and with the astrophysical parameters derived by the {\it Gaia} pipeline that processes the RVS spectra and the $BP$, $RP$  low-resolution prism spectra (Apsis, \citealt{DR3-DPACP-157}; Sects.~\ref{Apsis-metallicity} and ~\ref{Apsis-absorption}). 

\subsection{Metallicity}\label{sec:sos-met}
In the {\it Gaia} DR3 {\tt vari\_rrlyrae} table, individual     
photometric metallicities ([Fe/H]) are published for 133\,557 RR Lyrae stars.  
They are   
computed by the  
{\tt StellarParametersDerivation} module from   
the  pulsation period $P$ and the  $\phi_{31}$ parameter of the $G$ light curve Fourier  decomposition, using the relations for RRab and RRc stars derived in   \citet{Nemec-et-al-2013}.
These relations are calibrated on [Fe/H] values obtained from abundance analysis of high-resolution spectra (R$\sim$36\,000 and 65\,000) of 41 field RR Lyrae stars.

\begin{figure}
     \centering
     \includegraphics[scale=0.30]{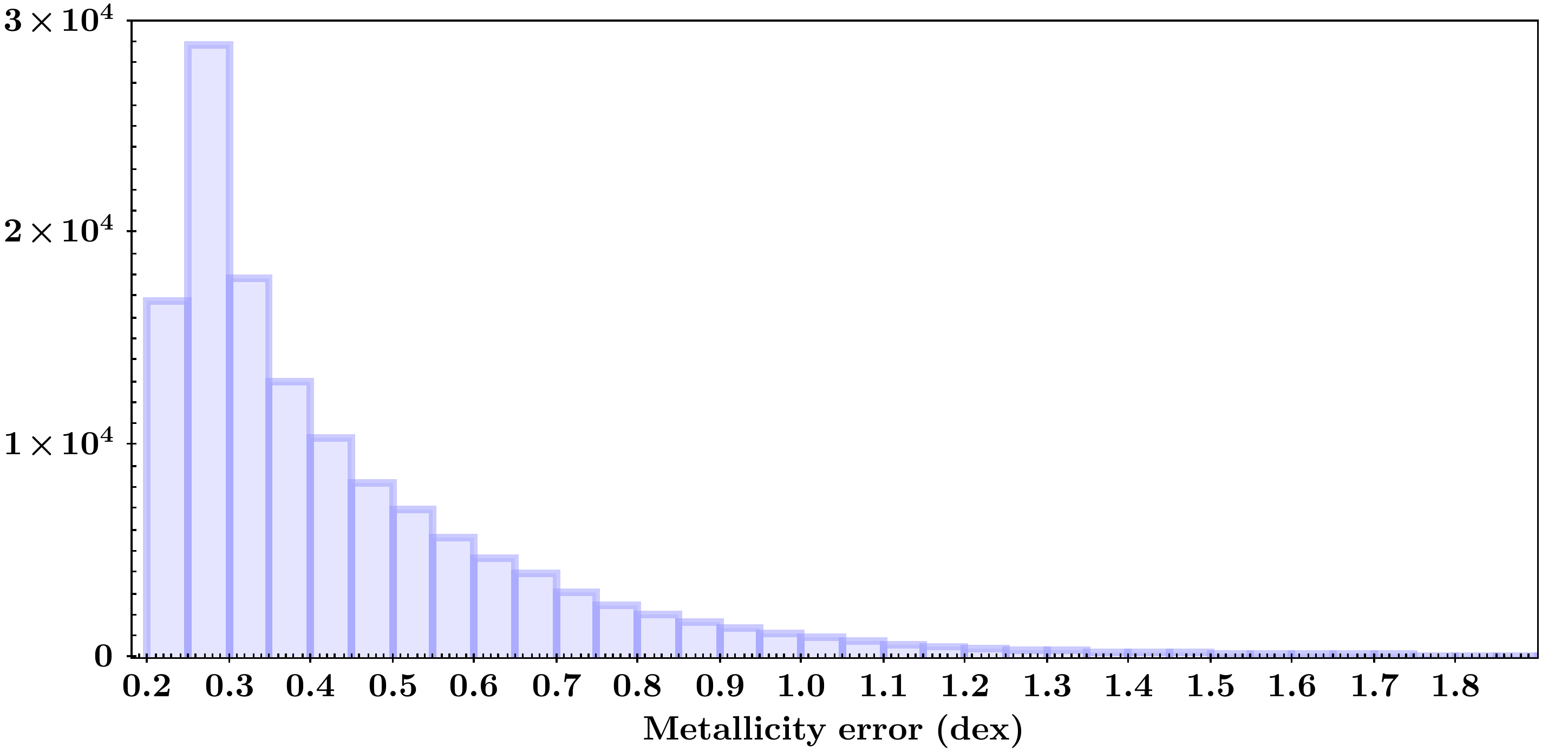}
     \caption{Error distribution of the metallicities derived for 133\,557 DR3 RR Lyrae stars by the SOS  
     pipeline. The mean and median metallicity errors are 0.46 dex (0.33 dex std) and 0.36 dex, respectively. We show in the figure only sources with   $\sigma_{\rm [Fe/H]} \le 2.0$ dex (i.e. the  99.6\% of the sample).
     We also note that for 71\% 
     of the sources in the plot $\sigma_{\rm [Fe/H]} \le 0.5$ dex.
}
     \label{fig:histo_met_err}
 \end{figure}

Figure~\ref{fig:histo_met_err} shows the  error distribution of the  photometric   metallicities for the  133\,557 RR Lyrae in the sample.  
Mean and median metallicity errors are 0.46 dex (with 0.33 dex std) and 0.36 dex, respectively. 
Errors were 
computed 
by adding in quadrature the uncertainty estimated with the bootstrap, the scatter around the calibration relations: 0.084 dex and  0.13 dex for RRab and RRc stars, respectively \citep{Nemec-et-al-2013},  and adopting a rather generous systematic error of 0.2 dex,  
 that likely led to an overestimation of the metallicity uncertainties.

\begin{figure}[h!]
\includegraphics[scale=0.35]{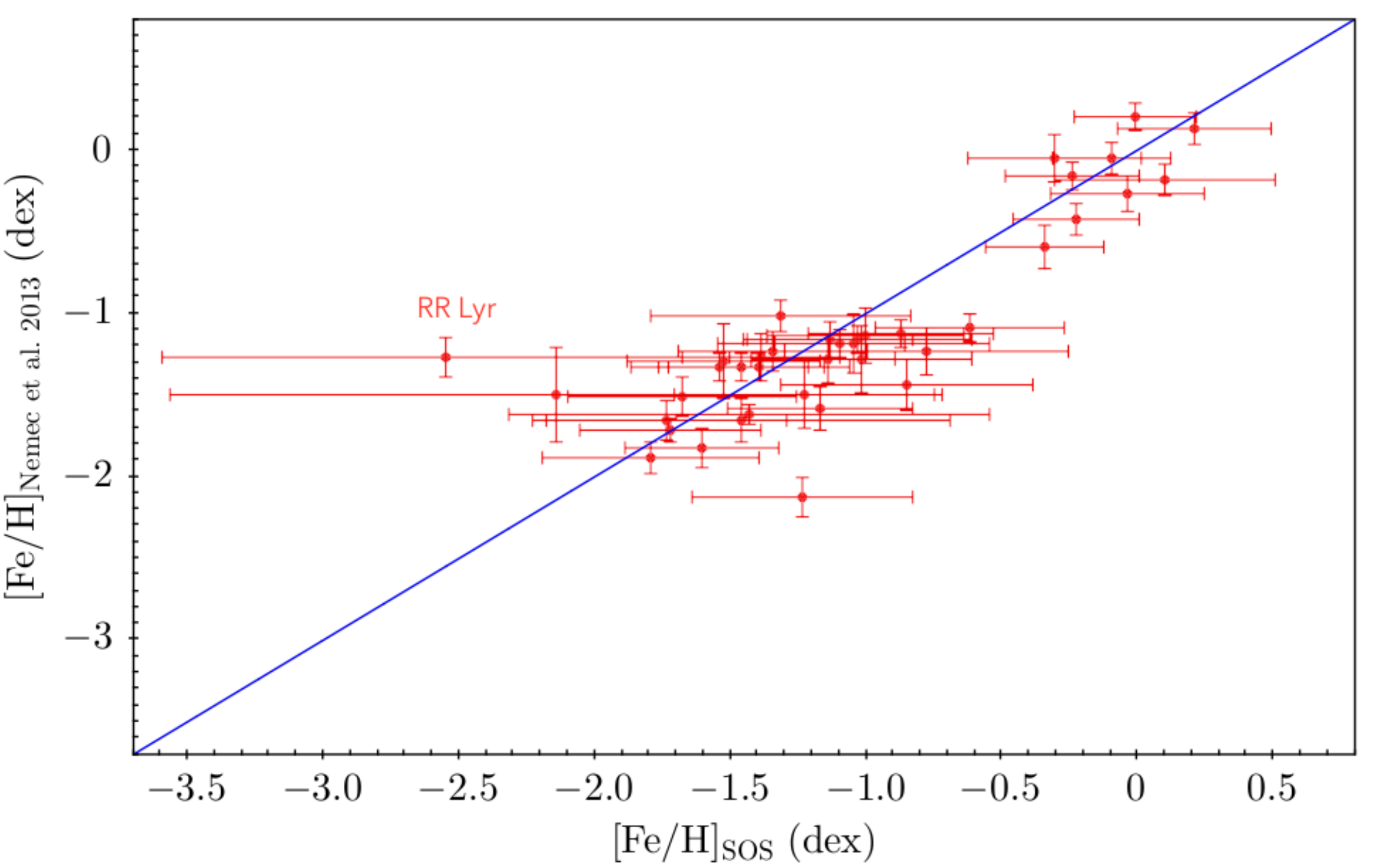}
\caption{Comparison of the photometric metallicities derived by the SOS Cep\&RRL pipeline and the metallicities from high-resolution spectra of 39 RR Lyrae  stars among the \citet{Nemec-et-al-2013}'s  calibrators.
}
\label{fig:Nemec}
\end{figure}

 To validate the photometric metallicities, 
 firstly we checked how well they reproduce the metal abundance of  the \citet{Nemec-et-al-2013}'s calibrators. This is shown in Fig.~\ref{fig:Nemec}, where the [Fe/H] metallicities derived by the SOS Cep\&RRL pipeline are compared with the metal abundances from high resolution spectra of 39 RR Lyrae among the   \citet{Nemec-et-al-2013} calibrators for which both estimates are available. 
The agreement is good within the errors, confirming that the SOS  photometric metallicities of RR Lyrae stars are closely tied to the metallicity scale from abundance analysis of  \citet{Nemec-et-al-2013}. 
We note, however, that the photometric metallicity of RR Lyr, the prototype and  brightest star of the class, is significantly discrepant and poorly determined,  as shown by the very large error bars, likely due to a poor determination of the $\phi_{31}$ parameter for the star. 

As already
noted in Paper~I,  
the photometric metallicities inferred from the [Fe/H], $P$ and $\phi_{31}$ relation 
better represent the mean  metal abundance of a population of RR Lyrae stars than the metallicy  of the individual sources. Furthermore, new calibrations of this 
relation have appeared in the recent  literature 
(e.g. \citealt{2021MNRAS.502.5686I}, \citealt{2021ApJ...912..144M})\footnote{We remind the reader that the pipelines processing the {\it Gaia} data are frozen well in advance of releases. Specifically, the SOS Cep\&RRL pipeline adopted to process the DR3 data was frozen in early 2020.}  Claims have been made by  \citet{2021MNRAS.502.5686I} 
that metallicities based on the 
\citet{Nemec-et-al-2013} calibrations are biased toward high metallicities.
Conversely, 
\citet{2021ApJ...912..144M} found an excellent agreement for higher metallicities
in the \citet{Nemec-et-al-2013} calibration range ($-1.5 \lesssim {\rm [Fe/H]} \lesssim 0.03$ dex) while estimates  diverge for lower metallicities. 

In order to test the reliability of the RR Lyrae  metallicities computed by the SOS pipeline we have compared them with the metal abundances  from high resolution spectroscopy  available in the literature for different samples of RR Lyrae stars.
Since \citet{2021ApJ...912..144M} relations are calibrated on the RR Lyrae metallicities  from high-resolution spectra  (R$\sim$35\,000) 
of \citet{Crestani-et-al-2021}, in the following we discuss the comparison with  \citet{Crestani-et-al-2021}. Comparison with other abundance analyses of RR Lyrae stars in the literature is presented in Sect.~\ref{other-comparisons}. 

\begin{figure*}[h!]
\includegraphics[scale=0.33]{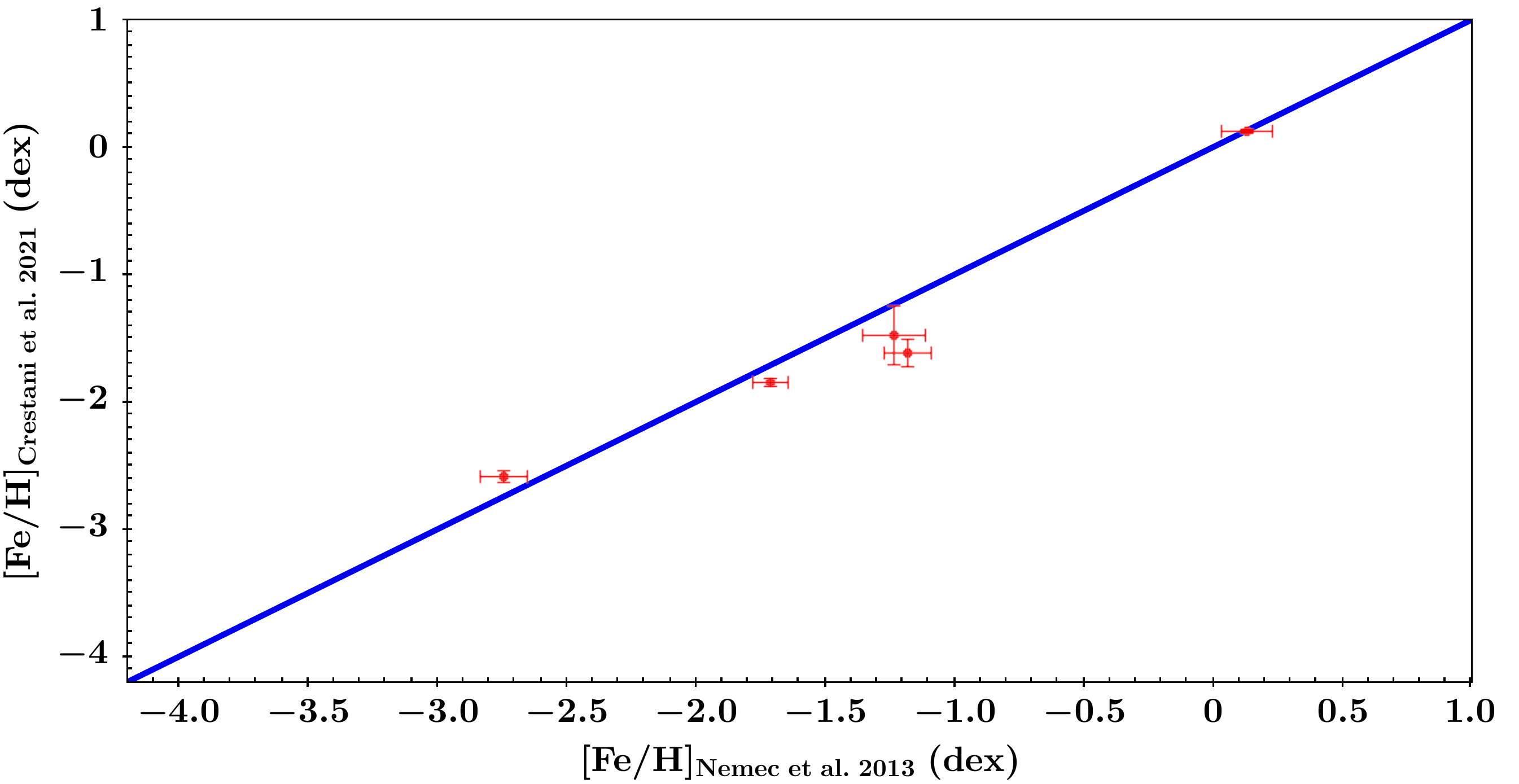}~\includegraphics[scale=0.30]{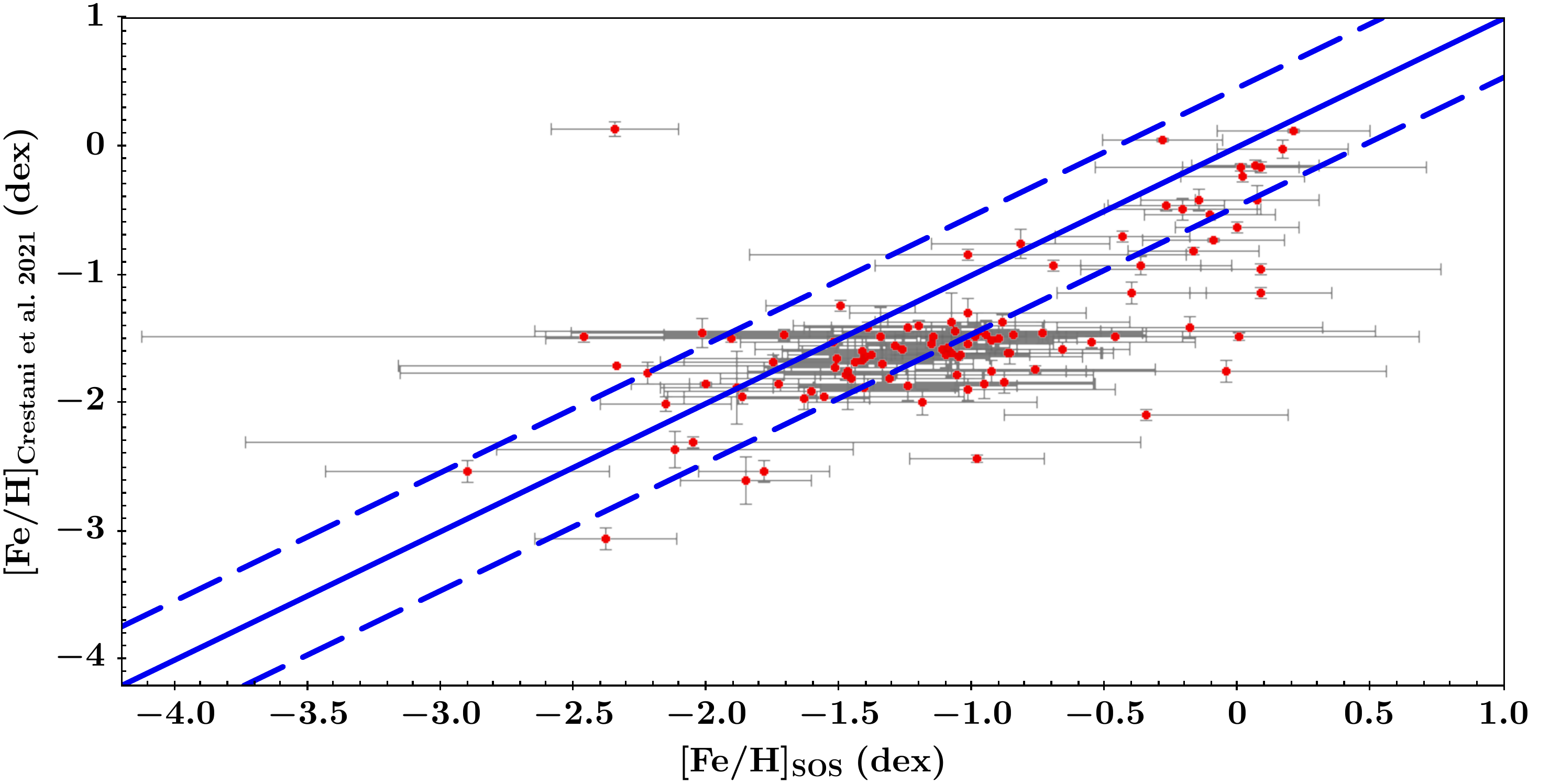}

\caption{{\it Left panel:} comparison of the metallicity from high-resolution spectra for 5 RR Lyrae stars in common between \citet{Nemec-et-al-2013} and 
\citet{Crestani-et-al-2021}; {\it Right panel:} comparison of the photometric metallicities derived by  the SOS pipeline and the spectroscopic metallicities from high resolution spectra  
 derived by 
\citet{Crestani-et-al-2021} for 105 RR Lyrae stars in common between the two samples. The 
two dashed lines are offset by $\pm$
0.46 dex, the mean error of the photometric metallicities (see Fig. 19), from the one-to-one line.}
\label{Nemec-crestani}
\end{figure*}

First we have compared the metal abundances from high resolution spectra 
for RR Lyrae stars in common between \citet{Nemec-et-al-2013} and   
\citet{Crestani-et-al-2021}.  
This comparison is shown in the left panel of  Fig.~\ref{Nemec-crestani}
and seems to confirm that  
\citet{Nemec-et-al-2013}'s metallicities 
may be 0.1-0.2 dex higher than  \citet{Crestani-et-al-2021}'s  for   [Fe/H]$\sim -1.5$ dex.
The right panel of Fig.~\ref{Nemec-crestani} shows instead the comparison between 
 the photometric metallicities derived by the SOS Cep\&RRL pipeline  
  and the metal abundances from 
 high resolution spectra from 
\citet{Crestani-et-al-2021}  
 for 105 RR Lyrae stars in common between the two studies and, as expected, confirms the metallicity offset seen in the left panel of the figure,  nevertheless  the two sets of estimates are still generally consistent,  within the admittedly large errors of the SOS metallicities.

In Table~\ref{tab:metal-abs} we compare metallicities  available in the literature for a sample of globular clusters (GCs) and dwarf spheroidal galaxies (dSphs),  with 
the metal abundances obtained by averaging  individual photometric estimates from the SOS  
pipeline  
for RR Lyrae stars in these systems. The agreement is satisfactory and, except for M62, it is always within the standard deviations of the mean  values estimated from the RR Lyrae stars.

\begin{figure}
   \centering
   \includegraphics[scale=0.31]{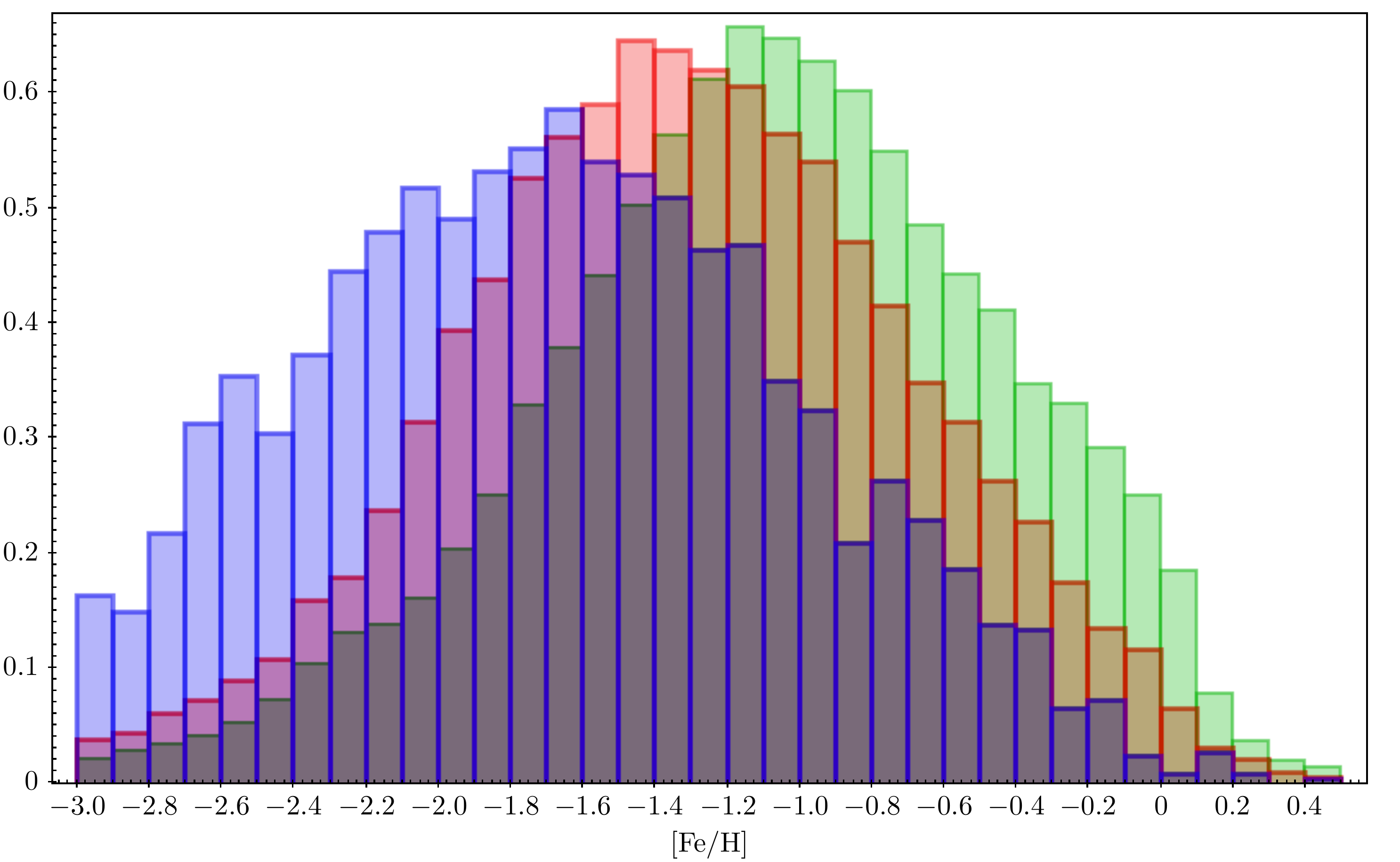}
   \caption{Normalised metallicity distribution of the 133\,577 RR Lyrae stars with photometric metallicities computed by the SOS  
   pipeline. The sources are divided in the three separate regions (All-Sky, LMC, and SMC) defined in Sect.~\ref{sec:sos-pip_genover}. Green, red, and blue histograms represent All-Sky, LMC, and SMC variables, respectively. The three distributions peak at mean values of \feh $=-1.07 \pm$ 0.63, $-1.31 \pm$ 0.62, and $-1.66 \pm  0.66$ dex for the All-Sky, LMC, and SMC variables, respectively.
}
    \label{MET-isto}%
    \end{figure}

Finally, the metallicity distributions of RR Lyrae stars in the All-Sky, LMC, and SMC regions, are shown with different colours in Figure~\ref{MET-isto}. A photometric metallicity estimate is available for 114\,653 
RR Lyrae  in the All-Sky region, for 16\,295 
in the LMC region and, for 2\,634  
in the SMC region. The metallicity distributions peak at mean values of [Fe/H]= $-1.07\pm$ 0.63,
$-1.31\pm$ 0.62, and $-1.66 \pm$ 0.66 dex for the All-Sky, LMC, and SMC
variables, respectively.
The mean value estimated for the LMC variables confirms to be about 0.2 dex higher than 
found by \citet{2004A&A...421..937G} from low resolution (R $\simeq$ 800) spectra of 98 RR Lyrae stars in the bar of the LMC: [Fe/H]$= -1.48 \pm 0.03$ dex (on the Harris 1996 metallicity scale) and, by \citet{2006A&A...460..459B}, [Fe/H]$= -1.53 \pm 0.02$ dex, from 78 RR Lyrae stars,  covering a wide range of distances, out to 2.5 degrees from the LMC centre.

\subsubsection{Comparison with GSP\_Spec   
metallicities}\label{Apsis-metallicity}
The {\tt astrophysical\_parameters} 
table in the {\it Gaia} DR3 archive contains 
metallicities 
derived by specific modules of the Apsis pipeline  \citep{DR3-DPACP-157}, that analyse the {\it Gaia} RVS spectra and the low-resolution $BP$, $RP$  prism
spectra.
In particular, of interest for our comparison are  the metal abundances produced by {\tt GSP\_Spec}, the module that processes the RVS spectra \citep{DR3-DPACP-186}.

\begin{figure*}
   \centering
 \includegraphics[scale=0.36]{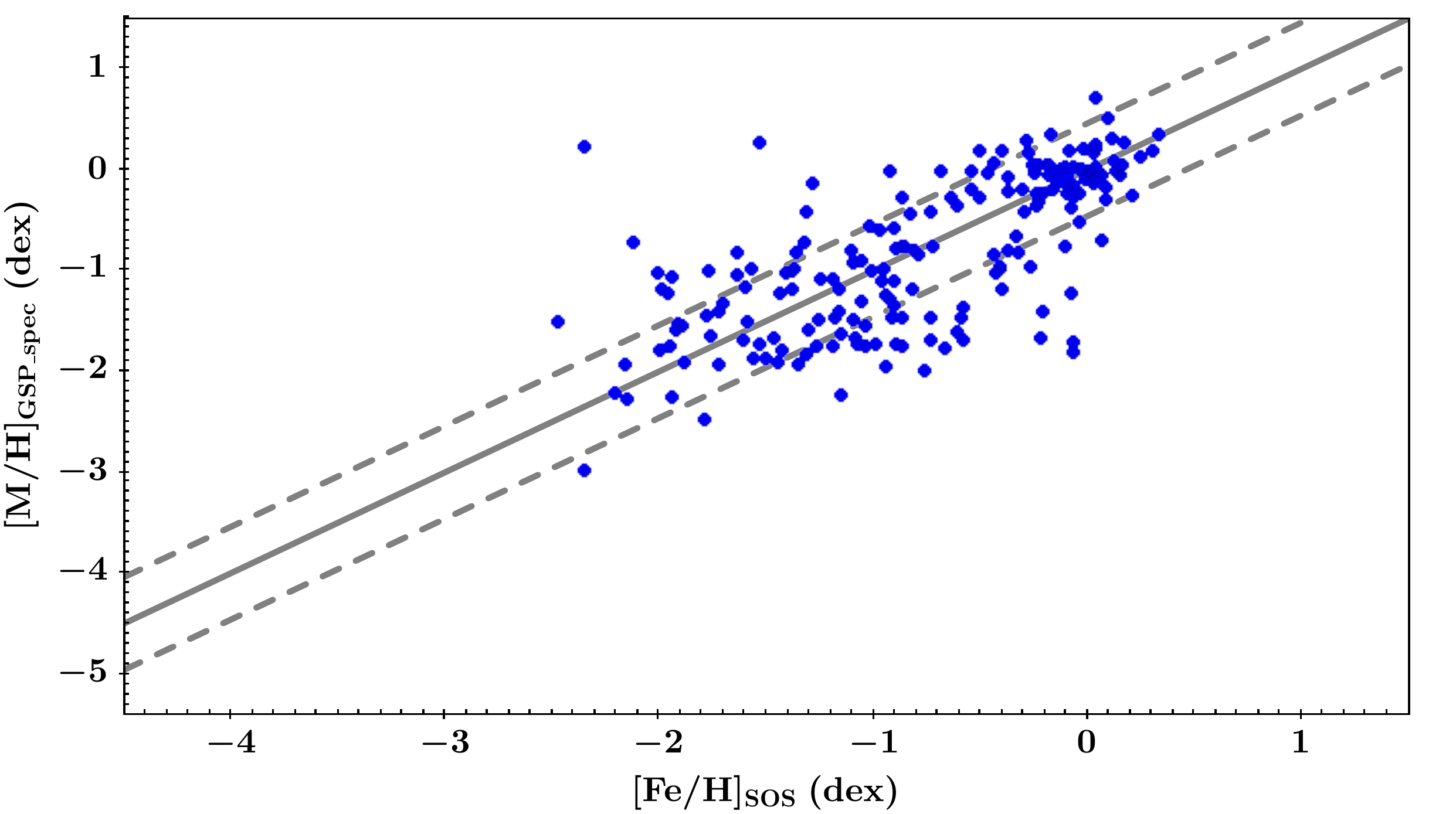}~\includegraphics[scale=0.36]{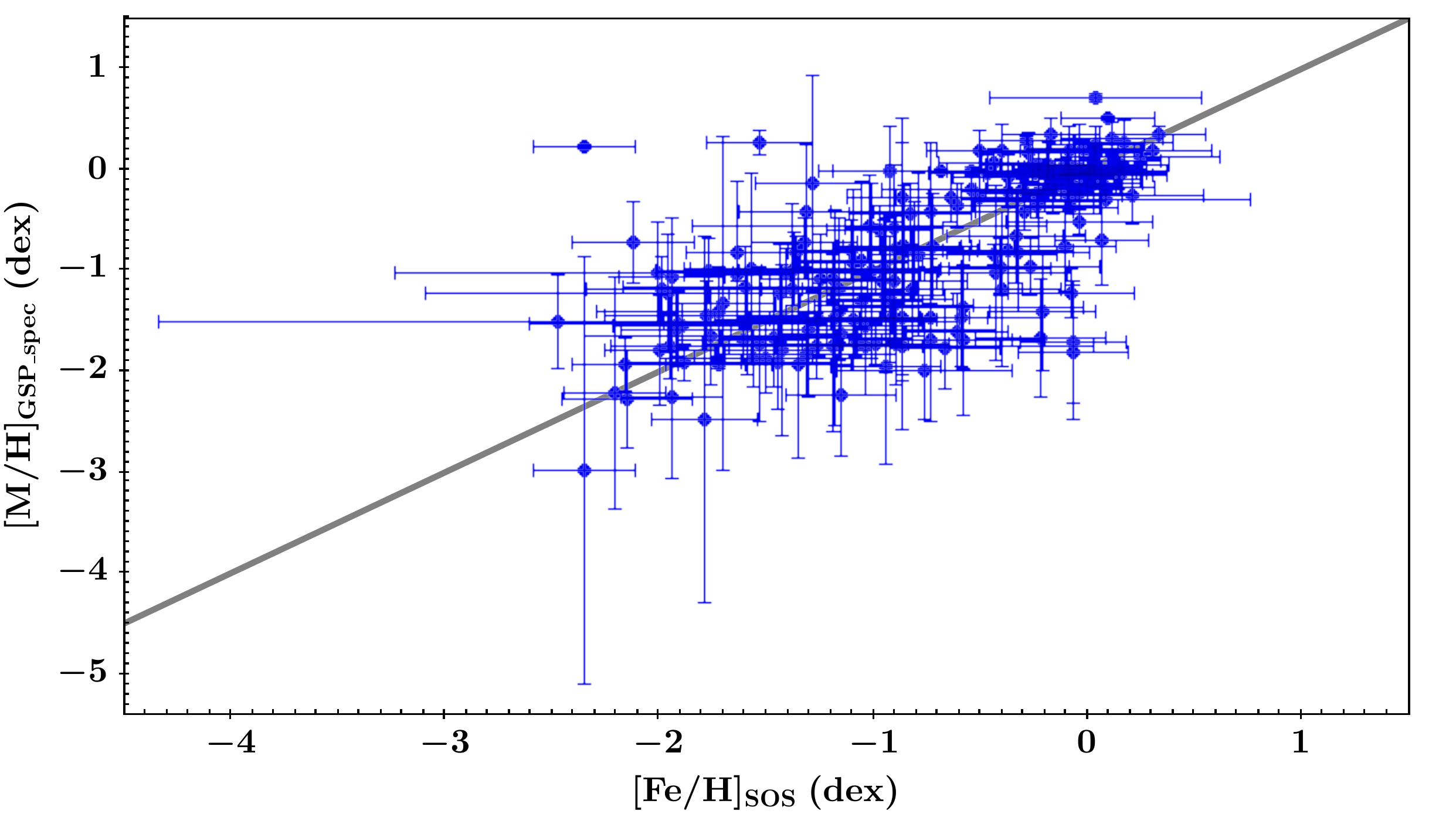}
   \caption{{\it Left panel:} comparison of the [M/H] metal abundances ({\tt mh\_gspspec}) computed by the GSP\_Spec module of Apsis and  [Fe/H] metallicities computed by the SOS Cep\& RRL pipeline for 193 RR Lyrae stars for which both values are available. The two dashed lines are offset by $\pm$ 0.46 dex, the mean error of the photometric  metallicities (see  Fig.~\ref{fig:histo_met_err}), from the one-to-one line; {\it Right panel:} same as in the left panel but showing also the error bars of the  [M/H]$_{\tt GSP\_Spec}$  and 
   [Fe/H]$_{\rm SOS}$ metallicities.
   }
                 \label{cu8-gspspec1}%
    \end{figure*}

[M/H] metal abundances  computed by the GSP\_Spec module ({\tt mh\_gspspec}) are available for 
288 RR Lyrae stars among the sample with photometric metallicity computed by the SOS 
pipeline. Following  recommendations in \citet{DR3-DPACP-186}, we applied 
the GSP\_spec quality flag ({\tt flags\_gspspec}), to filter out sources with lower quality GSP\_Spec abundances. This reduced the sample to 193 RR Lyrae stars. The left panel of Fig.~\ref{cu8-gspspec1} shows the comparison between the [M/H] abundances from  GSP\_Spec 
and the [Fe/H] metallicities computed by the SOS  
pipeline for these 193 RR Lyrae stars. 
The agreement between the two sets of metallicities is 
satisfactory over the whole range from [Fe/H]$\sim -2.5$ to $\sim $+0.3 dex, with about 70\% of the sources laying within $\pm$0.46 dex (the mean metallicity error in Fig.~\ref{fig:histo_met_err})  from the  one-to-one line and 93\% laying  
within $\pm$0.5 dex. This gives us confidence in the reliability  of the photometric metallicities obtained with the SOS Cep\&RRL pipeline. 

[M/H] metal abundances  computed by GSP\_Phot ({\tt mh\_gsphot}), the Apsis module that processes the low resolution  $BP,RP$ prism spectra \citep{DR3-DPACP-156} 
are also available for several thousands RR Lyrae stars confirmed by the SOS pipeline. 
However, the  GSP\_phot metallicities of RR Lyrae stars 
appear to be affected by very large uncertainties and systematic effects.
In particular, for the  
 193 RR Lyrae stars discussed previously,  
 the GSP\_Phot metallicities are systematically offset from both the GSP\_Spec and the SOS  
 estimates by more than 1 dex towards larger abundances. 
 The reason for this  likely residing in the  unusually large flux errors in the time-averaged $BP/RP$ spectra 
  of RR Lyrae stars compared to non-variable sources of same spectral type (see Fig.~\ref{cu8-XP-spectra-RRLs} in Appendix B), 
  as well as in issues of the instrument model of the $BP/RP$ spectra due to a lack of calibrators with emission lines below 400\,nm 
  \citep[see][for details]{DR3-DPACP-93}.
We come back to this issue later when discussing the GSP\_Phot $G$ absorption in  Sect.~\ref{Apsis-absorption}. 

\subsection{$G$-band absorption}\label{sec:sos-A(G)}
 \begin{figure}[h!]
     \centering
     \includegraphics[scale=0.29]{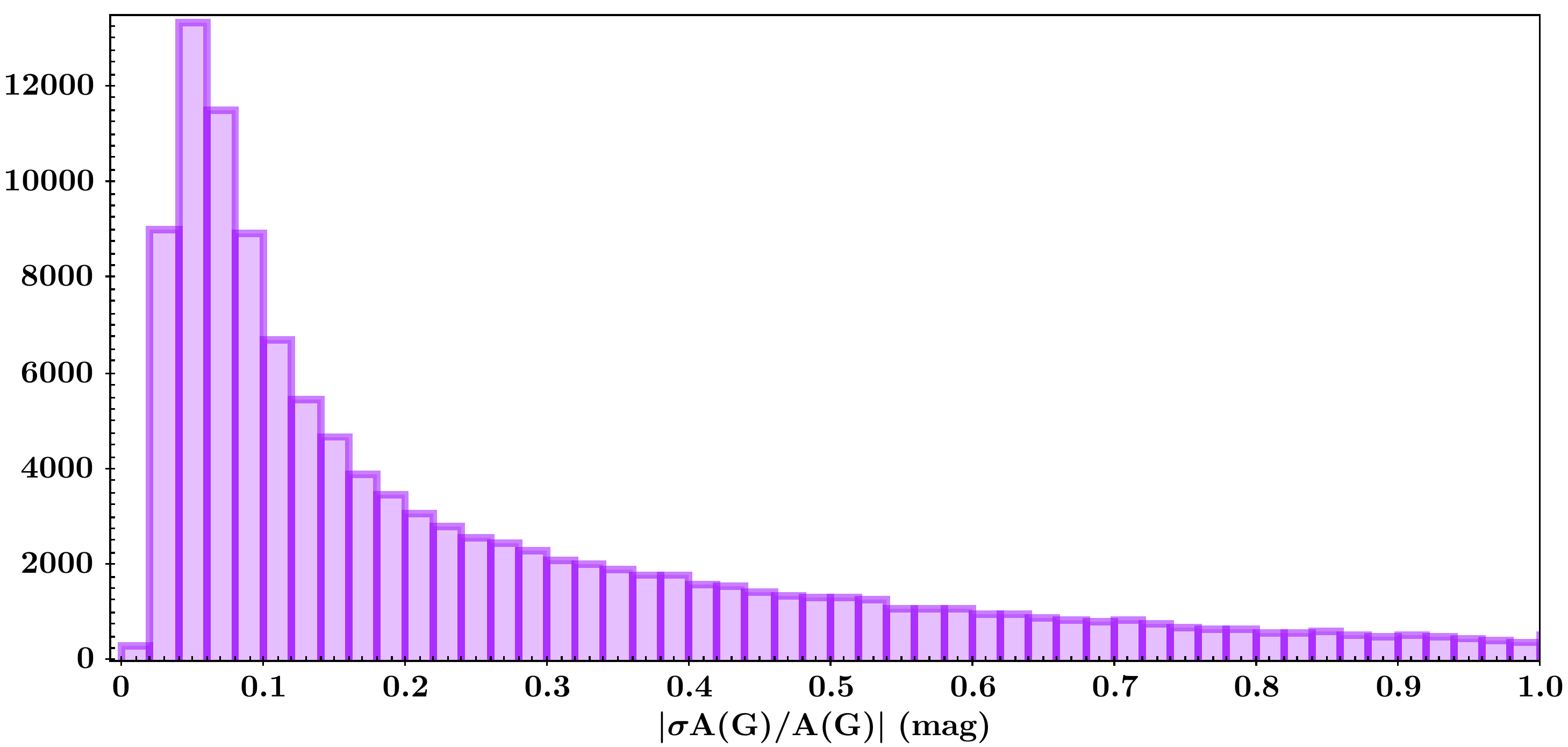}
     \caption{Error distribution, in $\sigma_{A(G)}/A(G)$, of the $A(G)$ values computed  by the  {\tt StellarParametersDerivation} module for 142\,867 fundamental-mode RR Lyrae stars. Only sources for which 
     $\sigma_{A(G)}$ is less than 100\% 
     (81\% of the RR Lyrae stars with an estimate of $A(G)$)   are shown in the plot.}
     \label{fig:histo_absoG}
 \end{figure}
An estimate of individual interstellar absorption in  the $G$-band [$A(G)$] was obtained by the  {\tt StellarParametersDerivation} module for 
142\,867 fundamental-mode RR Lyrae stars from an empirical relation  among the  amplitude of the $G$ light curve [Amp$(G)$], the pulsation period ($P$) and the $(G - G_{RP})$ colour of these RRab stars (see Sect.~2.1.2 in Paper~II, for details). 
Typical uncertainties in  
$A(G)$ are  of about $\pm$ 0.4 mag, corresponding to the mean of the absorption uncertainty (\texttt{g\_absorption\_error}) values in the {\tt vari\_rrlyrae} table for the 142\,867 RR Lyrae stars. 
 Figure~\ref{fig:histo_absoG} shows the error distribution, in $\sigma_{A(G)}/A(G)$, of the $A(G)$ values. 
For 30\% of the 142\,867 RRab stars $\sigma_{A(G)}$ is 
less than 10\%, for 
47\% of the sample  
is less that 20\% and,  for 
68\% of the sample is less than 50\%.
Finally, for 19\% of the sample $\sigma_{A(G)}$ is larger than 100\%.\\
In the right part of Table~\ref{tab:metal-abs} we compare the  $A(G)$ values obtained transforming the $A(V)$ values in the literature for a sample of GCs and dSphs, using the ratio between extinction coefficients in different bands by \citet{Bono-et-al-2019}, (see Notes in Table~\ref{tab:metal-abs})  with  the $A(G)$ values  obtained by averaging  the individual $G$-absorption estimates computed by the  SOS pipeline for RR Lyrae stars in these systems.
 Mean A(G) measurements and relative standard deviations listed in Table~\ref{tab:metal-abs} were computed over the entire sample of RR Lyrae stars contained in each system.  However, the  innermost RR Lyrae stars in the GCs are affected by significant crowding effects that impact the $G_{RP}$ magnitudes in particular, and may thus lead to unreliable (large) $A(G)$ values.
Indeed, if RR Lyrae stars within the half-light radius of each cluster are excluded, the derived mean $A(G)$ values are more compatible with the literature values and their standard deviations are reduced by up to 50\%.
We also recommend 
to use with 
caution the $A(G)$ values obtained by the SOS pipeline for RR Lyrae stars  fainter than $G\geqslant$ 18.5$-$19 mag. These sources may have very uncertain $G_{RP}$ magnitudes, hence, as well leading to 
unreliable $A(G)$ values. This is the case  for some of the dSphs listed in Table~\ref{tab:metal-abs}. Among them, Ursa Major~II, the only dSph among those listed in Table~\ref{tab:metal-abs} containing bright RR Lyrae stars ($G\sim$ 18 mag), has in fact 
a mean $A(G)$ value with small std and in perfect agreement with the literature estimation.
Finally, an absorption  map drawn from the 142\,867 RRab stars is presented in Sect.~\ref{sec:results}.

\subsubsection{Comparison with GSP\_phot absorption values}\label{Apsis-absorption}
\begin{figure*}
   \centering
   \includegraphics[scale=0.37]{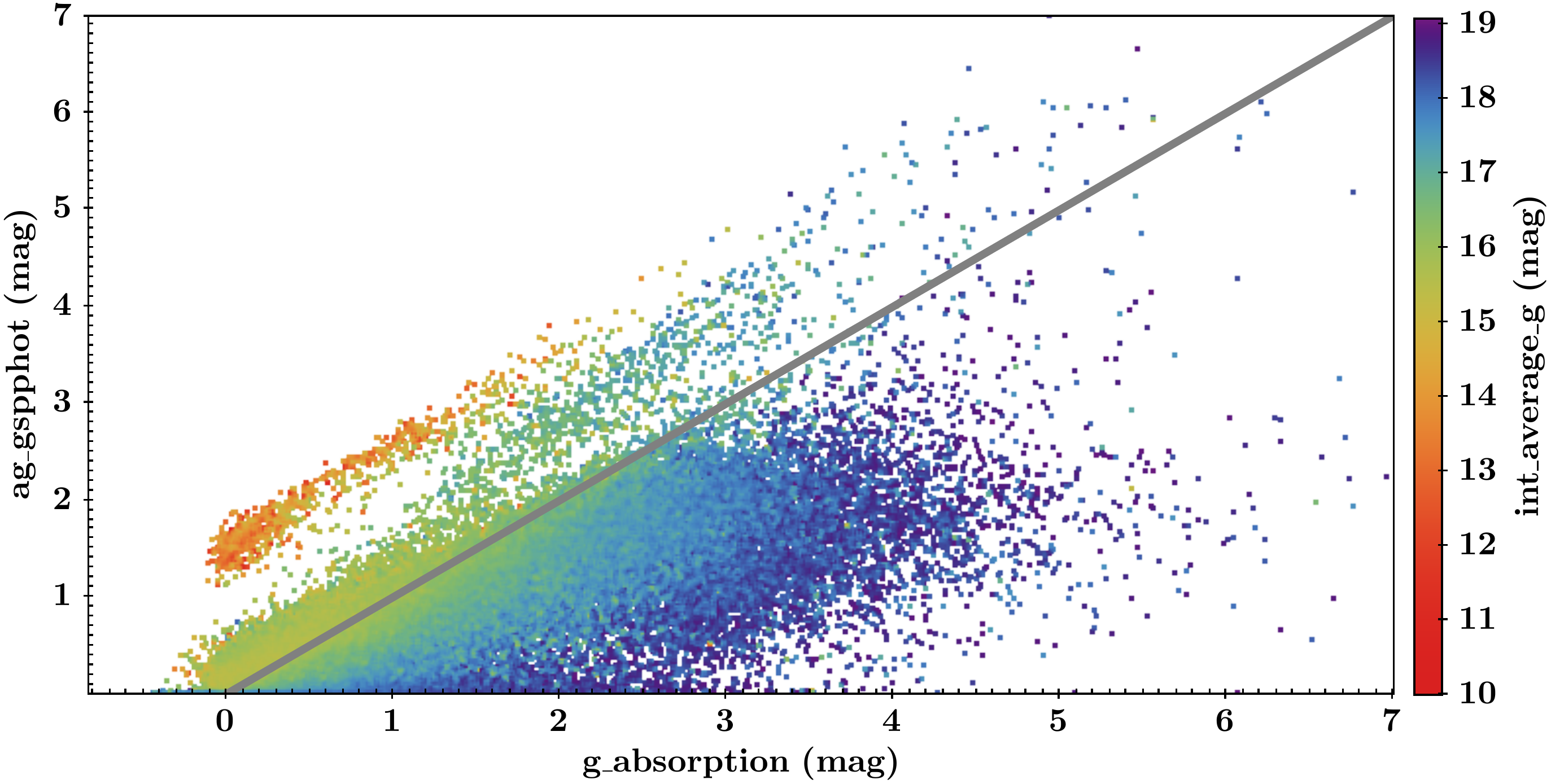}
   \caption{Comparison of $G$ absorption values   derived by the
GSP\_Phot module of the Apsis pipeline ({\tt ag\_gspphot}; \citealt{DR3-DPACP-157}) and the values derived by the SOS pipeline ({\tt g\_absorption}), for 84\,501 
RRab stars for which both estimates are available. 
The $A(G)$ values are colour-coded according to
the intensity-averaged $G$ magnitudes 
of these RR Lyrae stars, as encoded in the colour scale on the right (see text for details).
}
                 \label{cu8-cu7-A(G)}%
    \end{figure*}

The {\tt astrophysical\_parameters} 
table in the {\it Gaia} DR3 archive contains 
 $A(G)$ values ({\tt ag\_gspphot}) 
derived by the GSP\_Phot
 module of the Apsis pipeline  \citep{DR3-DPACP-157}. 
GSP\_Phot $A(G)$ values are available for 84\,501 
of the RRab stars with an $A(G)$ estimate from the SOS pipeline.
In Figure~\ref{cu8-cu7-A(G)} we show the comparison between the two independent $A(G)$ estimates. 
The $A(G)$ values are colour-coded according to the  intensity-averaged $G$ magnitudes 
of the RR Lyrae stars. 
Two sequences can be clearly seen  above the one-to-one line. These bright 
sequences are composed by a mixture of RR Lyrae stars 
for which  absorption values were either computed by the {\tt ESP\_HS} module of Apsis, the routine 
designed to derive astrophysical parameters of bright ($G <$17.65 mag)  O-, B-, and A-type stars,  or 
by the GSP\_Phot module 
(see panel (a) of Fig.~\ref{cu8-plot1} in Appendix B). Misidentification
 as OB stars of the RR Lyrae in the brightest sequence 
in Fig.~\ref{cu8-cu7-A(G)} is the culprit of the overestimated ({\tt ESP\_HS}) absorption values for these sources.
On the other hand, as discussed more in detail in Appendix B, the peak-to-peak variability of about one magnitude in $G$ (Amp$(G)$ parameter in the {\tt vari\_rrlyrae} table)   
of the RRab stars is ``absorbed'' as increased flux uncertainties
in the time-averaged $BP/RP$ spectra used to estimate the GSP\_Phot absorption.  Given the unusually large $BP/RP$ flux uncertainties,
the $\chi^2$ is ``flattened'' thus opening the room for degeneracies. 
In
particular, this appears to be most likely an extreme case of temperature-extinction
degeneracy due to  GSP\_Phot  adopting wrong mean temperatures for these RR Lyrae stars.

To conclude, as already pointed out for the   GSP\_Phot metallicities (Sect.~\ref{Apsis-metallicity}), the comparison of the $A(G)$ values shows the difficulty to  correctly infer astrophysical parameters for high amplitude variables,  such as the RR Lyrae stars, based on time-averaged $BP/RP$ spectra, and 
recommends use for RR Lyrae stars (or for any other type of high amplitude pulsating variable stars) of the epoch $BP/RP$ spectra individually (if their S/N allows for it)  along with epoch $T_{eff}$ and $\log g$ values,  and then average the results, to get more reliable GSP\_Phot metallicities and absorption values. On the other hand, as further discussed in Appendix B,  the SOS Cep\&RRL and GSP\_Phot absorption values agree within $\simeq \pm$ 0.2 mag for RR Lyrae stars for which the GSP\_Phot module uses $T_{\rm eff}$ values in the proper range for fundamental mode RR Lyrae stars (e.g. 6\,300 $\lesssim T_{\rm eff} \lesssim$ 6\,800 K; 
see column 11 in Table~\ref{Tab:Table-rr} and the top panel of Fig.~\ref{cu8-plot2}). This is likely the case for about 14\,000 RR Lyrae stars among the 84\,501  
sources for which both absorption estimates are available. We also find that for these same RR Lyrae strars the SOS Cep\&RRL [Fe/H] values and the GSP\_Phot metallicities agree within $\pm$0.5 dex. 
  
  \tiny
  \begin{table*}
  \caption{Metallicity and absorption in the $G$ band for a sample of GCs and dSphs, obtained by averaging individual values derived by the SOS 
  pipeline for the RR Lyrae stars in these systems.}\label{tab:metal-abs}
\begin{tabular}{l|c c c c |c c c c c}

	\hline
    \hline
   
	Name & [Fe/H]$_{\rm C09}$& RRLs& [Fe/H]$_{Gaia}$ & St. dev.[Fe/H]$_{Gaia}$ & RRLs& $A(G)$& St. dev.[$A(G)$] &  A$(G)^{*}$ & Ref. \\
         &  dex    & N$_{1}$   &   dex      &     dex  &  N$_{2}$ & mag & mag & mag& \\          
	\hline
    NGC~1261 & $-$1.27 &14 & $-$1.13 & $\pm$ 0.49 & 12& 1.32 & $\pm$ 1.47 & 0.02 & (1) \\ 
    NGC~1851 & $-$1.18 &21& $-$1.11 & $\pm$ 0.48 &18 &1.27 & $\pm$ 1.06 & 0.05& (2) \\
    NGC~288 & $-$1.32 &1& $-$1.24 & $\pm$ 0.33 & 1 &0.07 &  $\pm$ 0.08 & 0.08 & (3)\\
    NGC~3201 & $-$1.59 &73 & $-$1.27 & $\pm$ 0.36 & 77 &0.73 & $\pm$ 0.30 & 0.62 & (4) \\
    NGC~5024 & $-$2.10 &25& $-$1.66 & $\pm$ 0.57 &25 & 0.49 & $\pm$ 0.69 & 0.05 & (5) \\ 
    NGC~5139-OCen & $-$1.53 & 36& $-$1.42 & $\pm$ 0.63 & 39&0.43 & $\pm$ 0.36 & 0.31 & (6) \\    
    NGC~5272-M3 &  $-$1.50 &125& $-$1.38 & $\pm$ 0.39 & 105 &0.89 & $\pm$ 1.52 & 0.01 & (7)\\ 
    NGC~6266-M62 & $-$1.18 &105& $-$0.73 & $\pm$ 0.36 & 88& 2.77 & $\pm$ 1.81 & 1.23 & (8)\\ 
    IC~4499 & $-$1.53 &64& $-$1.45 & $\pm$ 0.48 & 58&0.66 & $\pm$ 0.23 & 0.60 & (9)\\ 
    NGC~7078-M15 & $-$2.37 &44& $-$1.95 & $\pm$ 0.54 &30 &0.89 &$\pm$0.83& 0.26 & (10)\\

    \hline
    \hline
    Sculptor & $-$1.68 &289& $-$1.64 & $\pm$ 0.50 &275& 0.53 & $\pm$ 0.63 &0.05& (11)\\
    Draco&  $-$1.93& 258 &$-$1.79 & $\pm$ 0.67 & 216&0.54 & $\pm$ 0.43 & 0.08& (12)\\ 
    Ursa Major I&  $-$2.18& 5 & $-$2.38 & $\pm$ 0.43 & 5& 1.94 & $\pm$ 2.00 & 0.02& (13)\\ 
    Ursa Major II& $-$2.47& 1 &$-$2.24 & $\pm$ 0.45 &3 & 0.26 & $\pm$0.04 & 0.25 & (14) \\ 
    \hline
\end{tabular}
\tablefoot{ $^{*}A(G)$ values obtained transforming the $A(V)$ values in the literature 
using the following ratio between extinction coefficients $A(G)/A(V)$ = 0.840 from \citet{Bono-et-al-2019}}. N$_1$ and N$_2$ are the number of RR Lyrae stars on which the mean [Fe/H]$_{Gaia}$  and $A(G)$ values were computed, with the  corresponding standard deviation of the mean in Columns 5 and 8, respectively.
\tablebib{(1)~\citet{Salinas-et-al-2016};
(2) \citet{Walker-1998}; (3) \citet{Kaluzny-et-al-1997}, \citet{Arellano-et-al-2013}; (4) \citet{Layden-et-al-2003}, \citet{Arellano-et-al-2014};
(5) \citet{Cuffey-1966}, \citet{Goranskij-1976}, \citet{Arellano-et-al-2011}; (6) \citet{Braga-et-al-2018}; (7) \citet{Benko-et-al-2006};
(8) \citet{Contreras-et-al-2010}; (9) \citet{Wlaker-et-al-1996AJ}; (10) \citet{Corwin-et-al-2008};
(11) \citet{Martinez-et-al-2016};(12) \citet{Muraveva-et-al-2020}; (13) \citet{Garofalo-et-al-2013}; (14) \citet{Vivas-et-al-2020}.
}
\end{table*} 

\section{Final cleaning and validation of the DR3 RR Lyrae catalogue}
\label{sec:s_vali}

\normalsize
During the final validation of the catalogue of 271\,779 RR Lyrae stars 
confirmed and characterized  by the SOS pipeline 
a number of issues,  misclassifications and uncertain classifications were detected. In this section we discuss them and provide tables with the identifiers of the affected sources. 
A final statistics of the SOS confirmed DR3 RR Lyrae stars, cleaned for the issues described in this section, is presented  in Sect.~\ref{sec:results}.

\subsection{Mode identification for short period RR Lyrae stars\label{short-P-rrab}}
Short period RRab stars with $P\lesssim 0.4$ d overlap with RRc stars in the $PA$ diagram. In DR3 we have adopted new boundaries of the  RRab and RRc regions in the  $PA$ diagram to separate the fundamental mode from the first overtone RR Lyrae stars (see Sect.~\ref{sec:sos-pip_genover} and Fig.~\ref{TaglioSilvio}). However, some RRab stars were not captured by the new boundaries and were wrongly classified as RRc stars.  This is the case of the RRab stars: 4297554691689496192 ($P$=0.366 d) and 1826975050760883072 ($P$=0.357 d). We warn the reader that there may be a few more cases like these in the DR3 {\tt vari\_rrlyrae} table.
\label{sect:Short-P-RRab}

\subsection{Double-mode RR Lyrae stars with  $Pf\simeq P1O$}
There are 2007 double-mode RR Lyrae stars with parameters published in the DR3 {\tt vari\_rrlyrae} table. For 121 of them the two periodicities, $Pf$ and $P1O$, listed in the table differ by less than 0.01 d.
This was likely caused by 
a re-mapping problem of $p1$ and $p2$, the two periodicities detected in the time series photometry of these sources, into $Pf$ (fundamental pulsation period) and $P1O$ (first overtone pulsation period). 
In fact, for all these 121 sources the $p1$ and $p2$ values differ and are in the proper ratio for double-mode RR Lyrae stars. We provide in Table~\ref{table:121_RRd}  the correct $Pf$ and $P1O$ values computed by the SOS pipeline for these 121 double-mode RR Lyrae stars. 

\begin{table}[h!]
\begin{tiny}
\caption{Correct $Pf$ and $P1O$ periods  for 121 double-mode RR Lyrae stars for which the two periodicites  published in the {\it Gaia} DR3 \texttt {vari\_rrlyrae} table differ by less than 0.01 d. 
This table is available in its entirety in the electronic version of the journal.}
\label{table:121_RRd}      
\centering    
\begin{tabular}{ccc}     
\hline\hline             
\noalign{\smallskip} 
  \gaia sourceid & $Pf$& $P1O$\\
  \noalign{\smallskip} \hline \noalign{\smallskip}
~~408895729987233536  & 0.373330688241618  & 0.512642740772834\\
~~501173342860631808 & 0.356601096360751 & 0.479398897708066\\
~~510602690164934528 & 0.450007942826883 & 0.333771875260333\\
~~711991473282863616 & 0.405938896214094 & 0.544093988121247\\
~~799159586901582464  & ~~0.41546585867929  & 0.556716992103652\\
1009169258447656576 & 0.373435128605576 & 0.501424952889536\\
1019488037636107136 & 0.377019820562923 & 0.505549683973198\\
1300079698489198464 & 0.401148054398303 & 0.537722672814259\\
1323028808262955264 & 0.478288427372862 & ~~0.35013538941965 \\
...&...&...\\
\noalign{\smallskip} \hline \noalign{\smallskip}
\end{tabular}
\end{tiny}
\end{table}

\subsection{RR Lyrae stars with large or null errors for main parameters} 
The {\tt peak\_to\_peak\_bp}, {\tt peak\_to\_peak\_rp}, {\tt int\_average\_bp} and {\tt int\_average\_rp} parameters 
of a number of RR Lyrae stars in the DR3 {\tt vari\_rrlyrae} table  
have very large errors (see e.g. for source: 4660156103061766400), or 
errors set to zero (see e.g. sources:  4660223379454198144, 4685971467924582144 and 
5027218748491264768).
This happens for bona fide RR Lyrae stars with mean $G$ magnitude ({\tt int\_average\_g})  fainter than 18.5-19 mag and sparse measurements with large errors in the $G_{BP}$ and $G_{RP}$ light curves. 
The affected sources are mainly in the Magellanic Clouds, in the Galactic bulge and in a number of MW dSphs and UFDs. 

Additionally, the $R21$ ({\tt r21\_g}) and $R31$ ({\tt r31\_g}) Fourier parameters of the $G$ light curves for $\sim$ 4\,400 
DR3 RR Lyrae stars exceed the expected range of [0,1].
However, these limits are fully respected if the rather large errors associated with the $R21$, $R31$ parameters of these sources are taken into account.

\subsection{RR Lyrae stars with negative or very large $A(G)$ values}

The $G$ absorption ($A(G)$; {\tt g\_absorption} parameter) is  negative (with values ranging from 0 to $-$1.279 mag) for 9\,503 DR3 RR Lyrae stars, 
corresponding to 6.3\%  of the 142\,867 RR Lyrae stars with an $A(G)$ value in the {\tt vari\_rrlyrae} table (see Sect.~\ref{sec:sos-A(G)}), and it is larger than 10 magnitudes (up to 3\,367.32 mag for source 5847675733921242624) for 286 sources (i.e., 0.2\% of the RR Lyrae stars with an $A(G)$ estimate). 

As 
discussed in Sect.~\ref{sec:sos-A(G)}, 19\% of the RR Lyrae stars with an $A(G)$ estimate  
have $|\sigma_{A(G)}/A(G)|>1$. 
This is the case of all the 9\,503 sources with a  negative $A(G)$. Among them, 8\,033   
have $-0.1<A(G)<0$ mag; 9\,297 
 have $-0.2<A(G)<0$ mag, and 9\,487 
 have $-0.5<A(G)<0$ mag. The remaining sources (0.2\%) with $A(G)<-0.5$ mag are sources fainter than 
$G\sim$19 mag, hence, have uncertain $(G-G_{RP})$ colours, that affected the $A(G)$ estimations.  

\begin{figure}[h!]
     \centering
   \includegraphics[scale=0.27]{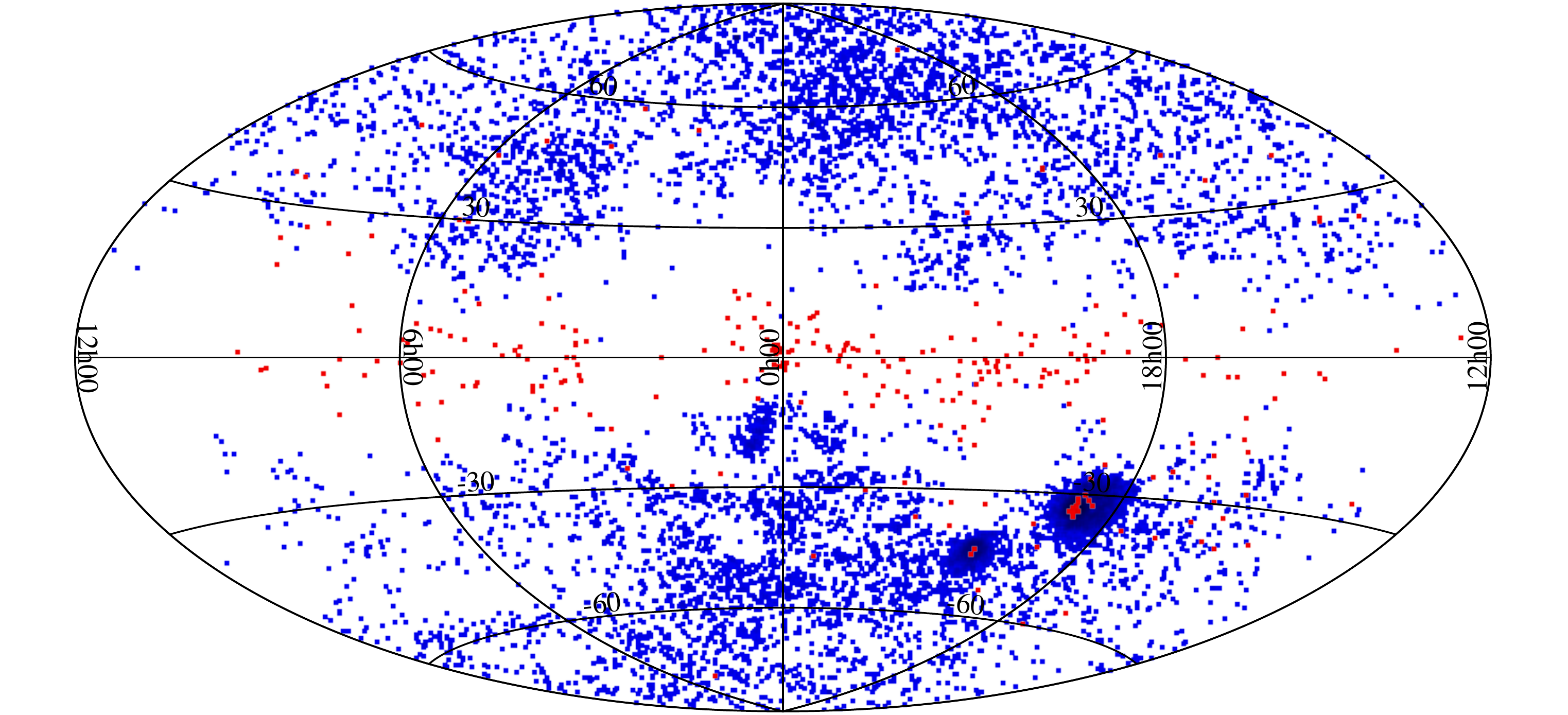}
     \caption{Sky distribution, in galactic coordinates, of 9\,503 RR Lyrae stars with low/negative $A(G)$ values (blue symbols) and  286 RR Lyrae stars with $A(G)$ values larger than 10 magnitudes (red symbols).}
     \label{fig:extreme-A(G)}
 \end{figure}

Regarding the 286 RRLyrae stars with large positive $A(G)$ values, discarding the extreme case of 5847675733921242624,  which clearly has a wrong $A(G)$ value, 
$0.007<|\sigma_{A(G)}/A(G)|<0.2$ for 77\% of them 
 and $0.007<|\sigma_{A(G)}/A(G)|<0.5$ for 91\%.  
 The sky distribution of the RR Lyrae stars with low/negative or very large $A(G)$  values  is shown in Fig.\ref{fig:extreme-A(G)}, using blue symbols for the former and red symbols for the latter. As expected, the RR Lyrae stars with very large $A(G)$  values are mainly confined in the Galactic disc and the regions with highest reddening in the Magellanic Clouds, while RR Lyrae stars with a low/negative $A(G)$ value are mainly located at high latitudes in the MW halo.
 
\subsection{RR Lyrae stars with a different classification in the OGLE catalogue}\label{sec:reclassifications}
In the {\it Gaia} DR3 {\tt vari\_rrlyrae} table there are 31 sources  
that the Classifiers of the General Supervised Classification module \citep{DR3-DPACP-165} classified only as RR Lyrae candidates, hence, 
they were processed only through the RR Lyrae branch of the SOS Cep\&RRL pipeline, which confirmed their classification as RR Lyrae stars.
However, these sources are classified as Cepheids (5 as DCEPs and 26 as ACEPs) in the OGLE-IV catalogue  
of variable stars. 
Using their {\it Gaia} EDR3 parallaxes these sources were plotted on the period-luminosity ($PL$) and period-Wesenheit ($PW$) relations of DR3 Cepheids \citep[see Table~2 in][]{DR3-DPACP-169} 
and for two of them the classification as RRab stars was confirmed, while 22 were reclassified as Cepheids (5 DCEPs and 17 ACEPs) and 7  have an uncertain classification between ACEPs or RRab stars. The sourceids of these 31 sources are provided in Table~\ref{table:5+26_CEPOGLE} along with their final classifications. 
One of the sources re-classified as DCEP (source 5861856101075703552) is included in the list of 1100 RR Lyrae stars with RVS time series RVs  published in DR3. Owing to this re-classification, the number of RR Lyrae stars with epoch RVs drops to 1099 and the Cepheid number increases to 799 
(but see also Sect.~\ref{sec:binaries-et-al}).

\begin{table*}
\caption{List of 31 RR Lyrae stars in the \gaia DR3  \texttt {vari\_rrlyrae} table  that have a different classification in the OGLE catalogue.  
Based on a number of attributes among which their {\it Gaia} EDR3 parallax they 
are either confirmed as fundamental mode RR Lyrae stars (2 sources), or reclassified as DCEPs (5 sources),  as ACEPs (17 sources), or have an uncertain classification between RRab stars and ACEPs (7 sources).}
\label{table:5+26_CEPOGLE}      
\centering 
\begin{tabular}{lcll}     
\hline\hline             
\noalign{\smallskip} 
 ~~~~~~~\gaia sourceid & 
  $\langle G \rangle$ & {\bf Revised} type&~~~~Type\\
  &(mag)&~(this paper)& (OGLE-IV)\\ 
   \noalign{\smallskip} \hline \noalign{\smallskip}
4121189659364789632 & 18.407  $\pm$ 0.040 &RRab&ACEP 1O\\
6724031478150071680 & 14.782  $\pm$ 0.007 &RRab&ACEP 1O\\
3325594070648057344   &13.304 $\pm$ 0.007 & DCEP\_1O& CEP 1O/2O\\
5259472605426820864   & 17.174 $\pm$ 0.002& DCEP\_1O& CEP 1O/2O\\
5337626858335118080   & 17.471 $\pm$ 0.002 & DCEP\_1O& CEP 1O\\
5861856101075703552$^{(*)}$   & 13.398 $\pm$ 0.010 & DCEP\_MULTI& CEP F/1O\\
6025724977547669120   &  15.533 $\pm$ 0.010 & DCEP\_1O& CEP F/1O\\
4086063080350131968 &13.863 $\pm$ 0.025&ACEP\_F  &ACEP F\\
4088429190686315136 & 16.836 $\pm$ 0.021&ACEP\_F  &ACEP F\\
4099638849167163648 & 13.514 $\pm$ 0.024 &ACEP\_1O&ACEP 1O\\
4123275364279164672 & 14.546 $\pm$ 0.004&ACEP\_1O&ACEP 1O\\
5253954057825909504 & 17.249 $\pm$ 0.005&ACEP\_F &ACEP F\\
5333097209687948032 & 19.000 $\pm$ 0.007&ACEP\_F&ACEP 1O\\
5343906650256155392 & 15.834 $\pm$ 0.002&ACEP\_1O&ACEP 1O\\
5355902184686008832 & 14.863 $\pm$ 0.003&ACEP\_1O &ACEP 1O\\
5625331663591871104 & 17.862 $\pm$ 0.002&ACEP\_1O &ACEP 1O\\
5949793765379361152 & 17.784 $\pm$ 0.006&ACEP\_1O &ACEP 1O\\
5956690971798673664 & 15.261 $\pm$ 0.020&ACEP\_F& ACEP F\\
5970892843763951104 & 17.397 $\pm$ 0.003&ACEP\_F& ACEP F\\
5978313726930066048 & 16.533 $\pm$ 0.004&ACEP\_1O&ACEP 1O\\
5979051907235436160 & 17.042 $\pm$ 0.004&ACEP\_F&ACEP F\\
6026530163731660928 &15.445 $\pm$ 0.009 &ACEP\_F& ACEP F\\
6029981290235791232 &15.607  $\pm$ 0.002 &ACEP\_1O&ACEP 1O\\
6724250895118108928 & 15.294  $\pm$ 0.003 &ACEP\_1O & ACEP 1O\\
4048728299707790976 & 15.562 $\pm$ 0.014 &ACEP\_1O/RRab? &ACEP 1O\\
4101192497784043520 & 14.931 $\pm$ 0.010 &ACEP\_1O/RRab?&ACEP 1O\\
4107356012051364480 & 15.654 $\pm$ 0.020 &ACEP\_1O/RRab?&ACEP 1O\\
5956833835251660800 & 15.975 $\pm$ 0.014&RRab/ACEP\_F? &ACEP F\\
5968150764841574912 & 18.733 $\pm$ 0.005 &ACEP\_1O/RRab? &ACEP 1O\\
5980267829660152960 & 17.404 $\pm$ 0.005&ACEP\_1O?/RRab?   &ACEP 1O\\
6061867986376466688 & 16.290  $\pm$ 0.005&ACEP\_F/RRab? &ACEP F\\
\noalign{\smallskip} \hline \noalign{\smallskip}
\end{tabular}
\tablefoot{$^{(*)}$This source is among the  1\,100 RR Lyrae stars with RVS time series RVs published in DR3.}
\end{table*} 

\subsection{RR Lyrae stars missing in the {\tt vari\_rrlyrae} table}\label{sec:15-overlaps}
The catalogue of 271\,779 RR Lyrae stars does not contain 14 bona fide RR Lyrae stars confirmed by the SOS Cep\&RRL pipeline that, by mistake,  were 
removed from the SOS final catalogue, but are listed in the {\tt vari\_classifier\_result} table.
Table~\ref{table:rrl_perdute} gives the sourceids of these 14 RR Lyrae stars and  
in Table~\ref{table:SOS-params-15-RRLs} we provide parameters ($P$, peak-to-peak $G$, $G_{BP}$, $G_{RP}$ amplitudes, mean magnitudes, etc.) computed by the SOS pipeline for these sources.
\begin{table}
\caption{List of 
14 bona fide RR Lyrae stars that, by mistake,  
were removed from the SOS RR Lyrae catalogue, hence, do not appear in the {\tt vari\_rrlyrae} table but are listed in the {\tt vari\_classifier\_result} table. 
The SOS Cep\&RRL parameters for these  
14 RR Lyrae stars are provided in Table~\ref{table:SOS-params-15-RRLs}.}

\label{table:rrl_perdute}      
\centering    
\begin{tabular}{c}
\hline\hline             
\noalign{\smallskip} 
  \gaia DR3 sourceid 
  \\
 \hline
\noalign{\smallskip}
4092009204924599040 
\\
4120414435009794048 
\\
4144246349481643392 
\\
5797652730842515968 
\\
5797917193442176640 
\\
5846086424210395520 
\\
5917239841741208576 
\\
5991733644318583424 
\\ 
6017924835910361344 
\\
6069336998880602240 
\\
6707009423228603904 
\\
5935214760885709440 
\\
4362766825101261952
\\
5967334102579505664
\\
\noalign{\smallskip} \hline \noalign{\smallskip}
\end{tabular}
\end{table}

\subsection{Overlaps with pipelines processing 
galaxies and QSO}\label{sec:galaxies-QSO}
\begin{table}
\caption{List of 521 sources in the {\it Gaia} DR3 \texttt {vari\_rrlyrae} table that  are classified as galaxies or non-single/extended objects by 
other DPAC pipelines. 
They include sources that either are clearly not RR Lyrae stars or  for which the SOS Cep\&RRL  classification as RR Lyrae stars  is 
uncertain (see column 2). This table is available in its entirety in the electronic version of the journal.
}
\label{table:521_RRL_GAL}      
\centering    
\begin{tabular}{rc}     
\hline\hline             
\noalign{\smallskip} 
  \gaia DR3 sourceid~~~~~~~~ &Tentative  class$^{(*)}$\\
  \noalign{\smallskip} \hline \noalign{\smallskip}
252947980027906048&RR Lyrae?\\
367375357718907776&galaxy/QSO\\
392996452549916416& ...\\
406587893736299136& ...\\
419307353484601472& ...\\
3206068021312873728& ...\\  
1056196096618040960& ...\\  
5571533079335954816& ...\\  
4685908211622296576& ... \\ 
4928220400858440704& ... \\ 
...~~~~~~~~~~~~~~~~~~ & ... \\
\noalign{\smallskip} \hline \noalign{\smallskip}
\end{tabular}
\tablefoot{$^{(*)}$ Tentative classification after visual inspection of the light curves.}
\end{table} 

\begin{table}
\caption{List of 282 sources in the \gaia DR3 \texttt {vari\_rrlyrae} table that  are classified as QSO by other DPAC pipelines. They either are clearly not RR Lyrae stars or sources with an uncertain classification.
This table is available in its entirety in the electronic version of the journal.}
\label{table:282_RRL_QSO}      
\centering    
\begin{tabular}{cc}     
\hline\hline             
\noalign{\smallskip} 
  \gaia DR3 sourceid & Class$^{(*)}$\\
   \noalign{\smallskip} \hline \noalign{\smallskip}
  5710645592594769280 & ECL\\
  5246357493290558336 & ECL\\  
4661507982702013312 & ...\\
5116113679481496832 & ...\\
2456881998103809280 & ...\\
1305269977485608192 & ...\\
6242394047412731520 & ...\\
  ... & ... \\
\noalign{\smallskip} \hline \noalign{\smallskip}
\end{tabular}
\tablefoot{$^{(*)}$ Classification after visual inspection of the light curves.}
\end{table} 

A total of 1\,139 sources in the {\tt vari\_rrlyrae} table are classified as galaxies by other DPAC pipelines.
After visual inspection of their  light curves, 618 of them were  confirmed as RR Lyrae stars, of which over 400 are known RR Lyrae stars in the OGLE-IV catalogue or in the catalogue of variable stars in  GCs. 
 The remaining 521 sources are listed in Table~\ref{table:521_RRL_GAL}. The majority of them are rather faint sources ($G>$20 mag), and either clearly not RR Lyrae stars, or have an uncertain classification.
 
About 57\,000 RR Lyrae stars 
in the {\tt vari\_rrlyrae}  table are classified as QSO by other DPAC pipelines. Among them 55\,448 are known RR Lyrae stars ($\sim$30\,000 from the OGLE-IV catalogue and $\sim$25\,400 from other major catalogues of RR Lyrae stars in the literature). 
We visually inspected the light curves of the remaining 1\,635 sources and for 1\,353 of
them  confirmed the classification as RR Lyrae stars (80 of them are classified as  RR Lyrae stars also in the SIMBAD database, \citealt{Wenger-2000}). 
A remaining 282 sources, listed in Table~\ref{table:282_RRL_QSO}, either are clearly not RR Lyrae stars or have an uncertain classification. 

\subsection{Non-single star sources identified by SEAPipe}
\begin{table}[h!]
\caption{Eighty bona fide RRab stars in the \gaia DR3  \texttt {vari\_rrlyrae} table that are identified as non-single star by the {\tt SEAPipe} pipeline.
This table is available in its entirety in the electronic version of the journal.}
\label{table:80_sea}      
\centering 
\begin{tabular}{c}     
\hline\hline             
\noalign{\smallskip} 
  \gaia DR3 sourceid\\
   \noalign{\smallskip} \hline \noalign{\smallskip}
4061010604884154624\\
4097861458597689856\\
4043627562186192768\\
1827951859436521728\\
6045465845526924544\\
...  \\
\noalign{\smallskip} \hline \noalign{\smallskip}
\end{tabular}
\end{table} 


The DPAC-Source
Environment Analysis Pipeline ({\tt SEAPipe}; \citet{2011ExA....31..157H}, Harrison et al., in preparation) was run on the catalogue of 271\,779 RR Lyrae stars  confirmed by the SOS Cep\&RRL pipeline to check whether additional sources are present in the vicinity 
(within a radius of $\sim$2$\arcsec$), which might 
contaminate the light of the RR Lyrae stars.  
SEAPipe returned a list of 80 RR Lyrae stars found to be non-single sources. 
These 80 sources reside mainly in the MW bulge and disk, 79 of them are RR Lyrae stars already known in the literature (67 are in the OGLE-IV catalogue and 12 in the
 ZTF and \gaia DR2 catalogues), only 1 is a new discovery. 
 Visual inspection of the light curves allowed us to
 confirm that they 
are all {\it ab}-type RR Lyrae stars, with different levels of noise  particularly in the $G_{BP}$ and $G_{RP}$  light curves,  often also  accompanied by a reduced amplitude of the $G$-band light curve. These are  clear  signatures of contamination and blending by neighboring stars \citep[see e.g., Fig~16 in][]{2005A&A...430..603D}
that can affect the estimated mean magnitude and peak-to-peak amplitudes 
of these RR Lyrae stars, but 
not the derived  periods. 
We list these 80 RRab stars in Table~\ref{table:80_sea}
and warn the reader that caution should be taken when using amplitudes, mean magnitudes, Fourier parameters  and derived quantities of  these sources. 

\subsection{Contamination by binaries and other types of variable stars}\label{sec:binaries-et-al}
RR Lyrae stars in the {\tt vari\_rrlyrae} with {\it
Gaia} EDR3 parallaxes  accurate to better than 5\% 
were plotted on a 
$PW(G, G_{BP} - G_{RP})$  relation along with 
DR3 Cepheids with same parallax accuracy  
\citep[see][for details]{DR3-DPACP-169}. 
A group of 147 RR Lyrae stars  appear to be  from $\sim$0.5 to $\sim$8.5 mag fainter than expected in this plot. 
These sources are all known RR Lyrae stars classified  by various different surveys: OGLE (44 sources), ASASSN (40), CATALINA (25), PS1(5) and  ZTF,   DR2, GCVS and the catalogue of variable stars in GCs (33 sources altogether). 
The light curves of these 147 
sources were visually inspected and often found to exhibit large scatter, split branches and other issues, despite they are all rather bright sources. 
On the basis of the light curves eyeballing 
the sample contains: 
75 
bona fide RR Lyrae stars 
or 
most likely  
RR Lyrae stars (59 RRab and 16 RRc stars); 55 ECLs \footnote{The classification as ECL was based on the value of the ratio between  peak-to-peak amplitudes of the $G_{BP}$ and  $G_{RP}$  light curves (this ratio for eclipsing binaries typically is Amp$(G_{BP})/$Amp$(G_{RP}) = 1.0 \pm$  0.2, while is $\sim$ 1.6 for RR Lyrae stars) and on the visual inspection of the light curves folded with once and twice the period determined by the pipeline.} 
that contaminate the RR Lyrae sample or  with classification uncertain between ECL and RR Lyrae star,  or 
between ECL and non-RR Lyrae source; 8 sources with a very prominent bump before the minimum in the light curve; and 9 sources with such poor light curves that a classification in type is not possible.
We provide in Table~\ref{table:23di32_subl} 
the identifiers 
of the 72 
sources in this sample that we re-classify as ECLs 
 or possible ECLs (55 in total), and variable stars with poor light curves and uncertain classification in type (17 sources).
 
 Finally, three of the sources re-classified as ECLs (sources: 4130380472726484608, 3062985235999231104 and  5734183009797858176) are included in the list of 1100 RR Lyrae stars with RVS time series RVs  published in DR3. Owing to their re-classification
 and the re-classification as DCEP of the  source discussed in Sect.~\ref{sec:reclassifications},  the number of bona fide RR Lyrae stars with epoch RVs finally drops to 1096.
 
\subsection{Scan-angle-period  correlations}\label{scan-angle-P-connection}
 A clear correlation was found to exist between the 
 scan-angle of the {\it Gaia} observations and the detection of spurious periods for variable stars of different types.
 A detailed analysis of this correlation, and parameters defined to assess the reliability of the periods derived from the time series data analysis of variable sources, are presented in \citep{DR3-DPACP-164}. 
 The SOS 
 sample of 271\,779 RR Lyrae stars was checked and correlation-value plots were produced showing that,  at most, perhaps a few percent of the sources in the sample is  affected by possible scan-angle effects. Hence, a low, but perhaps not fully negligible percentage.

\begin{table}[h!]
\caption{List of sources
  in the \gaia DR3  \texttt {vari\_rrlyrae} table that are re-classified as ECLs or possible ECLs (55 in total), and variable stars with poor light curves and uncertain classification in type (17 sources). This table is available in its entirety in the electronic version of the journal.
}\label{table:23di32_subl}      
\centering 
\begin{tabular}{rcc}     
\hline\hline             
\noalign{\smallskip} 
  \gaia DR3 sourceid~~~~~~ & Class$^{(*)}$&Lit\_Survey\\
   \noalign{\smallskip} \hline \noalign{\smallskip}
87159184382553088  & ECL & CATALINA\\
 568221222620689024 & ECL& ASAS-SN\\
 326383086038685952 & ECL &GCVS\\
 6812645960632807296 & ECL & ASAS\\
 5292689405748796160 & ECL &CATALINA\\
 ...~~~~~~~~~~~~~~~~~ & ... & ...\\
\noalign{\smallskip} \hline \noalign{\smallskip}
\end{tabular}
\tablefoot{$^{(*)}$ Classification after visual inspection of the light curves.}
\end{table}

\section{Results and final accounting}
\label{sec:results}

\normalsize
The 
 validation process described in Sect.~\ref{sec:s_vali} allowed us to  clean the catalogue of 271\,779 RR Lyrae stars that are 
 published in the DR3 
 {\tt vari\_rrlyrae} table, 
 by rejecting  888 sources that are  contaminants (mainly ECLs),  
  and objects with an uncertain classification (Sects.~\ref{sec:reclassifications}, ~\ref{sec:galaxies-QSO},  ~\ref{sec:binaries-et-al}), thus leading  to a 
  clean sample of  270\,891 confirmed RR Lyrae stars. To this number we add the 14 RR Lyrae stars (4 RRab and 10 RRc stars),  discussed in Sect.~\ref{sec:15-overlaps}, thus leading to a final sample of 270\,905 DR3 bona fide RR Lyrae stars, of which 200\,294 are RR Lyrae 
  already known in the literature (Gold Sample) and 70\,611 are new discoveries by {\it Gaia}. 
   For ease of the reader we collect in just one table (Table~\ref{table:888_rejected}) the sourceids of the 888 sources rejected during the final validation process described in Sect.~\ref{sec:s_vali}.  We also note that it turned out that the period of a fraction of the 888 sources we rejected during the final cleaning,  may indeed be  affected by the scan-angle effects mentioned in Sect~\ref{scan-angle-P-connection}. 
  \begin{table}[h!]
\caption{{\it Gaia} DR3 sourceids of the 888 sources rejected during the final validation of the catalogue of 271\,779 RR Lyrae stars that are published in
the DR3 {\tt vari\_rrlyrae} table.
This table is available in its entirety in the electronic version of the journal.}
\label{table:888_rejected}      
\centering 
\begin{tabular}{r}     
\hline\hline             
\noalign{\smallskip} 
  \gaia DR3 sourceid~~~~~~~~\\
   \noalign{\smallskip} \hline \noalign{\smallskip}
72425797289881344\\
87159184382553088\\
252947980027906048\\
271665825465872256\\
279252868010888960\\
...~~~~~~~~~~~~~~~~~\\
\noalign{\smallskip} \hline \noalign{\smallskip}
\end{tabular}
\end{table}

\begin{figure*}
   \centering
   \includegraphics[scale=0.4]{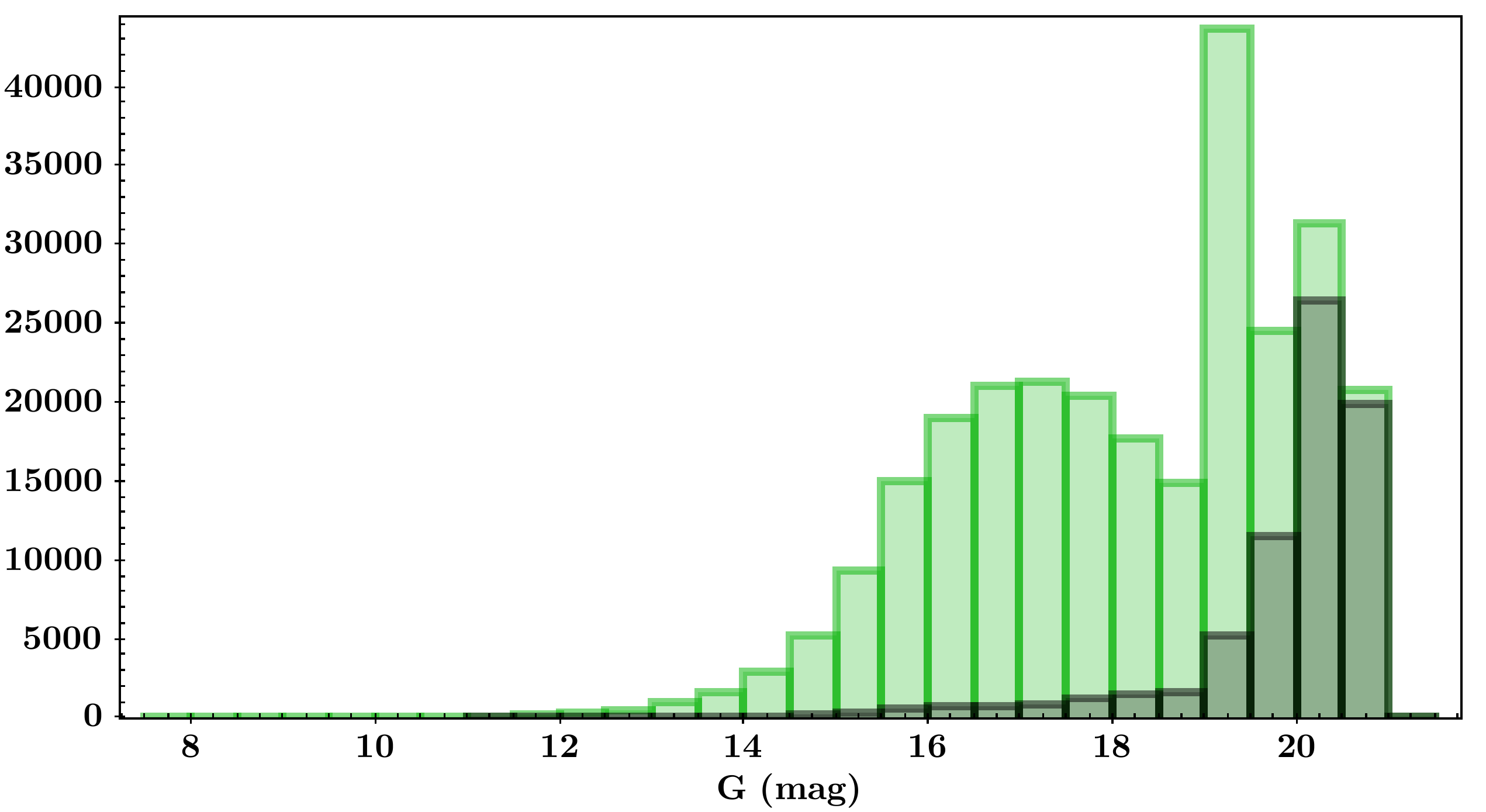}
   \caption{Distribution in $G$ magnitude of the clean sample of 270\,905 
   DR3 RR Lyrae stars confirmed by the SOS Cep\&RRL pipeline. The peaks at $G\sim$ 19.2, 19.7 mag correspond to RR Lyrae stars in the LMC and SMC, respectively.
   The magnitude distribution of the 70\,611  
   new RR Lyrae stars is  highlighted in black.
   }
              \label{G-range}%
    \end{figure*}

  Figure~\ref{G-range} shows the $G$ magnitude distribution of the clean sample of 270\,905 
  DR3 RR Lyrae stars confirmed by the SOS 
  pipeline. The  magnitude distribution of the 
  70\,611 new RR Lyrae stars is  highlighted in black.
  
Table~\ref{table:RRLtypes} summarises some statistic about the SOS 
DR3 RR Lyrae sample, namely, 
the subdivision  
between known (Gold Sample) and new discoveries by {\it Gaia}; the number of RRab, RRc and RRd  types; the number in the LMC, SMC and All-Sky regions; how many RR Lyrae stars have a metallicity or $G$ absorption  estimate and, finally, the number of RR Lyrae stars (and Cepheids) with RVS time series 
RVs published in DR3.
\begin{table*}[h!]
\caption{Statistics of the 
RR Lyrae stars confirmed by the SOS Cep\&RRL pipeline that are published in DR3. Numbers in Columns 2 and 3 correspond to the samples before and after the validation and cleaning described in Sect.~\ref{sec:s_vali}, respectively. Numbers in Column 4 are after adding the 14 RR Lyrae stars discussed in Sect.~\ref{sec:15-overlaps}. 
}             
\label{table:RRLtypes} 
\centering                         
\begin{tabular}{l r r r}       
\hline\hline                 
\noalign{\smallskip}
Sample & N($-$14)~~~~~~~&N$_{\rm clean}$($-$14)&N$_{\rm clean}$($+$14)\\
& \texttt{vari\_rrlyrae}& &\\
\hline                          
\noalign{\smallskip}
Gold Sample& 200\,589~~~~~~~&200\,293~~~~& 200\,294~~~~\\  
New discoveries$^{\rm (a)}$& 71\,190~~~~~~~& 70\,598~~~~&70\,611~~~~\\
\hline                          
\noalign{\smallskip}
Total number of RR Lyrae stars&  271\,779~~~~~~~& 270\,891~~~~&270\,905~~~~\\
\hline                          
LMC region$^{\rm (b)}$&  31\,535~~~~~~~&31\,379~~~~&31\,379~~~~\\
SMC region$^{\rm (b)}$&  4\,828~~~~~~~&4\,788~~~~&4\,788~~~~\\ 
All-Sky region$^{\rm (b)}$& 235\,416~~~~~~~&234\,724~~~~&234\,738~~~~\\
\hline                       
\noalign{\smallskip}
Fundamental mode (RRab)$^{\rm (c)}$ & 175\,350~~~~~~~&174\,943~~~~&174\,947~~~~\\ 
First overtone (RRc)$^{\rm (c)}$ & 94\,422~~~~~~~&93\,942~~~~&93\,952~~~~\\
Double mode (RRd) & 2\,007~~~~~~~& 2\,006~~~~& 2\,006~~~~\\
\hline                        
\noalign{\smallskip}
[Fe/H] values  (RRab and RRc)   & 133\,577~~~~~~~&133\,558~~~~&133\,559~~~~\\
$A(G)$ values (RRab only)     & 142\,867~~~~~~~&142\,656~~~~&142\,660~~~~\\
\hline                        
\noalign{\smallskip}
RV time series (RR Lyrae)$^{\rm (d)}$ & 1\,100~~~~~~~&1\,096~~~~&1\,096~~~~\\
RV time series (Cepheids)$^{\rm (d)}$ & 798~~~~~~~&799~~~~&799~~~~\\
RV time series (other type)
& ...~~~~~~~&3~~~~&3~~~~\\
\hline                          
   
\noalign{\smallskip}
Total number of sources with RV time series data&1\,898~~~~~~~&1\,898~~~~&1\,898~~~~\\
\hline  
\noalign{\smallskip}
\end{tabular}
\tablefoot{$^{(a)}$``New''  means ``new to the best of our knowledge''.\\ 
$^{(b)}$Regions for the LMC, SMC and All-Sky are as defined in Sect.~\ref{sec:sos-pip_genover}.\\
$^{(c)}$Numbers of RRab and RRc stars in column N$_{\rm clean}$($-$14) are corrected for the re-classifications in Sect.~\ref{short-P-rrab}.\\
$^{(d)}$Numbers of RR Lyrae and Cepheids with RV time series data published in DR3 are corrected  in column N$_{\rm clean}$($-$14)  for the re-classifications in Sects.~\ref{sec:reclassifications} and ~\ref{sec:binaries-et-al}.\\
}
\end{table*}
We note that  1\,676 RR Lyrae stars of the  270\,905  
clean sample are distributed among 95 GCs, and 1\,114 are spread among 7 dSph galaxies (of which 505 in Sculptor) and 16 UFD satellites of the MW. 

The  following parameters, computed by the SOS Cep\&RRL  pipeline 
are released for the clean sample of  270\,891  RR Lyrae stars (as well as for the  888 sources filtered out in Sect.~\ref{sec:s_vali}) in the  {\it Gaia} DR3  {\tt vari\_rrlyrae} table, along with the related uncertainties:
\begin{itemize}
\item source pulsation period(s) (main and secondary periodicity, if any);
\item intensity-averaged mean $G$, $G_{\rm BP}$, $G_{\rm RP}$ magnitudes;
\item mean RV (for  1\,096 RR Lyrae stars, one RR Lyrae star re-classified as DCEP\_1O,  and 3 sources re-classified as ECLs);
\item epochs of maximum light in the 3 pass-bands;
\item epoch of minimum RV;
\item $\phi_{21}$ and $R_{21}$ Fourier parameters of the $G$ light curve;
\item $\phi_{31}$ and $R_{31}$ Fourier parameters of the $G$ light curve;
\item peak-to-peak $G$, $G_{\rm BP}$, $G_{\rm RP}$, RV amplitudes [${\rm Amp(}G{\rm )}$, {\rm Amp}($G_{\rm BP}$), {\rm Amp}($G_{\rm RP}$), Amp(RV)];
\item RR Lyrae subclassification into RRab, RRc and RRd types;
\item absorption in the $G$ band, $A(G)$, for RRab stars;
\item photometric metallicity, [Fe/H],  for RRab and RRc stars.
\end{itemize}
The SOS Cep\&RRL parameters for the 14 bona fide RR Lyrae stars that do not appear in the \texttt{vari\_rrlyrae} table (see Sect.~\ref{sec:15-overlaps}) are provided instead in Table~\ref{table:SOS-params-15-RRLs}.

{\it Gaia}'s sourceids, coordinates, $G$, $G_{\rm BP}$ and $G_{\rm RP}$  time series photometry for each of the 270\,905 RR Lyrae stars (as well as for the  888 rejected sources)  and 
RVS time series RVs for 1\,895  between RR Lyrae stars and Cepheids and for 3 RR Lyrae 
stars re-classified as ECLs,  can be retrieved from the  {\it Gaia}  data release archive\footnote{\texttt{http://archives.esac.esa.int/gaia/}}  and its mirror 
nodes. The archives also provide tools for queries and cross-match of {\it Gaia} data with other catalogues available in the literature.
 
 We provide in Table~\ref{table:3aRRL} 
 the specific link to the {\tt vari\_rrlyrae} table of the {\it Gaia} DR3 archive 
 and  summarise the names of the parameters computed by
SOS Cep\&RRL that can be retrieved for RR Lyrae stars from the archive table. Finally, in  Appendix~\ref{app:queries} 
 we give examples of queries to retrieve some of the quantities and parameters listed in Table~\ref{table:3aRRL}.

\subsection{Completeness and purity of the DR3 RR Lyrae sample}
The final, clean catalogue
of 270\,905 RR Lyrae stars confirmed by the SOS 
pipeline released in
\textit{Gaia} DR3 almost doubles the DR2 RR Lyrae catalogue. This is
thanks to the improved sky coverage, the increased statistic, and
the longer time baseline (34 months) covered by the DR3 time series data (see
Sect.~\ref{sec:app_to_sos}), that all contributed to improving the completeness and
purity of the DR3 RR Lyrae catalogue. This is well shown by 
comparing the DR3 and the OGLE-IV catalogues  of RR Lyrae stars,  
that we use in the following 
to estimate the completeness, contamination  and percentage of new RR Lyrae stars in 
the LMC, SMC and Galactic bulge regions.
We also note that  while the completeness of our catalogue can be easily estimated in  these regions, because our reference catalogue,  
OGLE-IV, is complete more or less at the same 
magnitude of the {\it Gaia} data, 
completeness in the All-Sky region is much more difficult to establish,  because we still lack a likewise complete and homogeneous catalogue as OGLE's for All-Sky RR Lyrae stars. In principle, we could use as reference for the All-Sky region our custom catalogue of  
414\,082 literature RR Lyrae stars, with a counterpart in the EDR3 {\tt gaia\_source} table 
and, specifically, the subsample of 311\,798 sources confirmed as RR Lyrae stars  by the SOS Cep\&RRL pipeline. However, this reference catalogue suffers from being  a compilation of many different surveys, each with a rather  different sky coverage, depth/limiting magnitude, completeness and contamination.

With these caveats in  mind 
to estimate  the completeness 
of our clean, final catalogue of 270\,905 DR3 
RR Lyrae stars we have used as reference the following 
catalogues that cover different regions of the sky:\\
i) OGLE IV for the RR Lyrae stars in the 
LMC, SMC and the Galactic Bulge (red areas in Fig.~\ref{OGLE-IV-footprint});\\
ii) the All-Sky catalogue 
in the DR2 {\tt vari\_rrlyrae} table (hereafter, {\it Gaia} DR2 catalogue) cleaned from contaminant galaxies  (139\,801 RR Lyrae stars, see Paper~II);\\ 
iii) the ASAS-SN 
catalogue, that  contains 28\,337 RR Lyrae stars with magnitudes in the range $10.4<V<17.4$ mag, distributed all-sky as shown by Fig.~\ref{ASAS-SN-footprint};\\
iv) the CATALINA 
catalogue, that  contains 42\,775 RR Lyrae  stars with magnitudes in the range $11.0<V<21.1$ mag, distributed 
nearly all-sky as shown in Fig.~\ref{CATALINA-footprint}; and,\\ 
v) the catalogue of 311\,798 all-sky known RR Lyrae stars (extracted from 
our custom catalogue of literature RR Lyrae stars (see Sect.~\ref{gold-sample}), that the SOS pipeline confirms as RR Lyrae stars, along with the 200\,294  
RR Lyrae stars in the clean Gold Sample,  for which periods are  correctly recovered by the SOS  pipeline\footnote {We note that a 
large fraction of the 311\,798 known RR Lyrae stars confirmed by the SOS Cep\&RRL pipeline (see Sect.~\ref{gold-sample}) were finally rejected  because the SOS periods for these sources differ by more than 0.001 d from the literature periods. Part of these rejected sources are listed  in the {\tt  vari\_classifier\_result} table.}. 
Results are summarised in Table~\ref{table:completezza}. 
Completeness values in Col.~2 correspond to the recovery percentages from a cross-match within 2.5$\arcsec$ 
between RR Lyrae stars in the DR3 clean catalogue and the reference catalogues as listed in Col.~1.
Percentages of new discoveries in Col.~3 correspond to the percentage of additional sources in the DR3 catalogue with respect to the number of recovered sources in each reference catalogue.
Contamination in Col.~4 is estimated only for the new sources in the DR3 catalogue, by checking how many of the new sources have a non-RR Lyrae classification in the SIMBAD catalogue, and corresponds to the percentage of contaminants over the number of new sources with respect of  each reference catalogue.
\begin{table}[h!]
\caption{Completeness, contamination and new discoveries of our final clean catalogue of 270\,905 DR3 RR Lyrae stars in different sky regions, from the  comparison with different literature surveys.}
\label{table:completezza}      
\centering 
\begin{tabular}{lrrl}     
\hline\hline             
\noalign{\smallskip} 
 &Completeness& New${^{\rm (a)}}$ & Contamination\\
    \noalign{\smallskip} \hline \noalign{\smallskip}
 OGLE-IV${^{\rm (b)}}$&&&\\ 
     \noalign{\smallskip} \hline \noalign{\smallskip}
LMC & 83\%~~~~~~~~& 4\%~~~~~& ~~~~~~$<$1.8\%\\
SMC & 94\%~~~~~~~~& 6\%~~~~~& ~~~~~~$<$8\%\\
Bulge-up& 79\%~~~~~~~~& 20\%~~~~~& ~~~~~~$<$0.15\%\\
Bulge-down& 82\%~~~~~~~~& 40\%~~~~~& ~~~~~~~~~~...  \\
    \noalign{\smallskip} \hline \noalign{\smallskip} 
    \gaia DR2${^{\rm (c)}}$&90\%~~~~~~~~&103\%~~~~~& ~~~~~~$<$0.12\%\\ 
    \noalign{\smallskip} \hline \noalign{\smallskip}
ASAS-SN${^{\rm (b)}}$& 73.7\%~~~~~~~~&& \\
    \noalign{\smallskip} \hline \noalign{\smallskip}
 CATALINA${^{\rm (b)}}$&85.5\%~~~~~~~~&&\\
    \noalign{\smallskip} \hline \noalign{\smallskip}
All Surveys${^{\rm (d)}}$
&64.2\%~~~~~~~~&&\\ 
\noalign{\smallskip} \hline \noalign{\smallskip}
\end{tabular}
\tablefoot{${^{\rm (a)}}$New means ``new sources to the best of our knowledge''.\\
${^{\rm (b)}}$ Regions used for the comparison with OGLE-IV are shown in red in  Fig.~\ref{OGLE-IV-footprint}, for ASAS-SN are shown in Fig.~\ref{ASAS-SN-footprint} and for 
CATALINA in Fig.~\ref{CATALINA-footprint}.\\
${^{\rm (c)}}$Comparison with the 139\,801 sources in the DR2 RR Lyrae catalogue, after cleaning from contaminant galaxies.\\ 
${^{\rm (d)}}$ Comparison between the 311\,798 all-sky known RR Lyrae stars, and the Gold Sample RR Lyrae stars whose periods are correctly recovered by the SOS pipeline.}
\end{table} 

\subsection{Colour-magnitude diagrams}\label{CMDs}
\begin{figure*}[h!]
\centering
\includegraphics[scale=0.60]{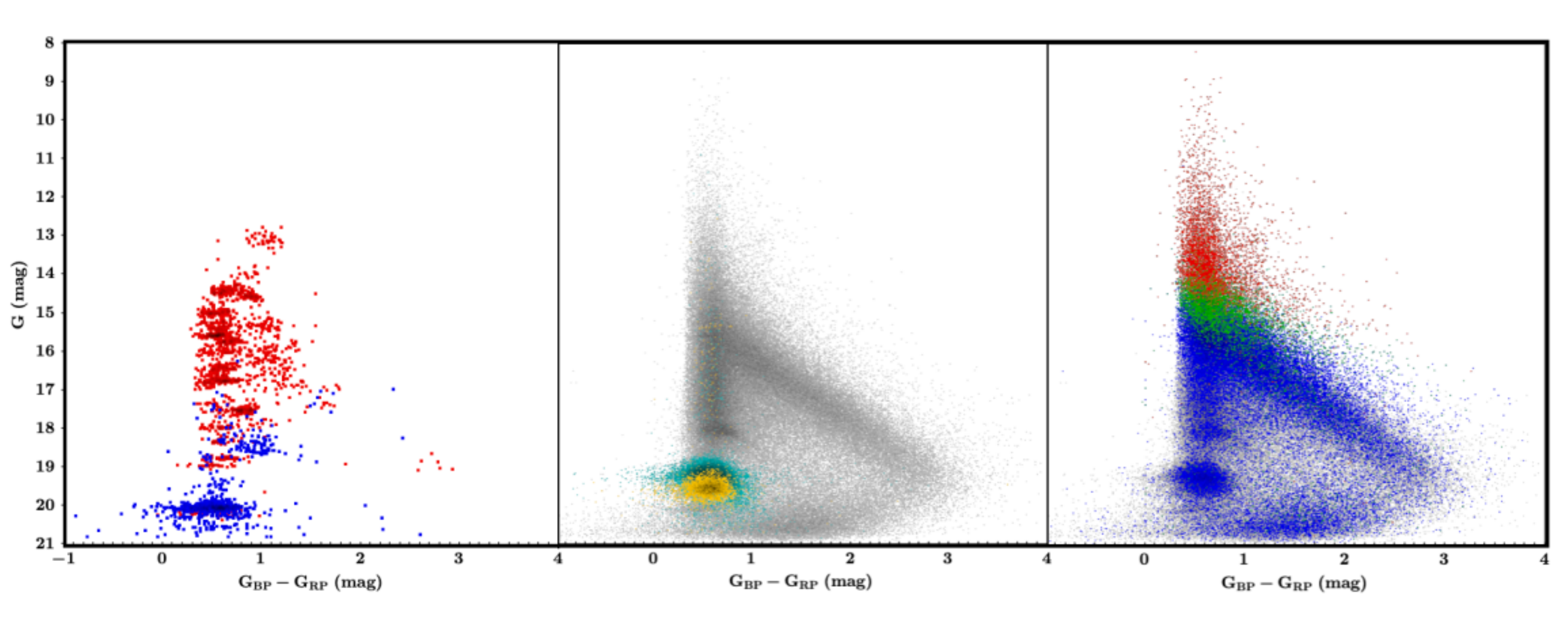}
\caption{{\it Left panel:} \gmag, \gbp$-$\grp~CMD of  
RR Lyrae stars in the DR3 clean catalogue 
that are in GCs (red points) and dSph galaxies (blue points), for which the colour information is available (2\,558 out
of 2\,790 sources). Each concentration of points corresponds to a different system. {\it Middle panel:} \gmag, \gbp$-$\grp~CMD of the whole sample of RR Lyrae stars 
for which the colour information is available, 215\,116  
sources (out of 270\,905) 
and showing with different colours the variable stars in the LMC (26\,468 sources; cyan symbols), SMC (4\,353 sources; yellow symbols) and All-Sky (184\,280 sourcs; gray symbols) regions, as defined in Sect.~\ref{sec:sos-pip_genover}. {\it Right  panel:} Same as in the middle panel but with blue, green and red points marking 
RR Lyrae stars with $\sigma_{\varpi}/\varpi<$0.50 (58\,018 sources), $<$0.2 (13\,377 sources) and $<$0.1 (4\,670 sources), respectively.}
\label{fig:cmd}
\end{figure*}

\begin{figure*}[h!]
\centering
\includegraphics[scale=1.0]{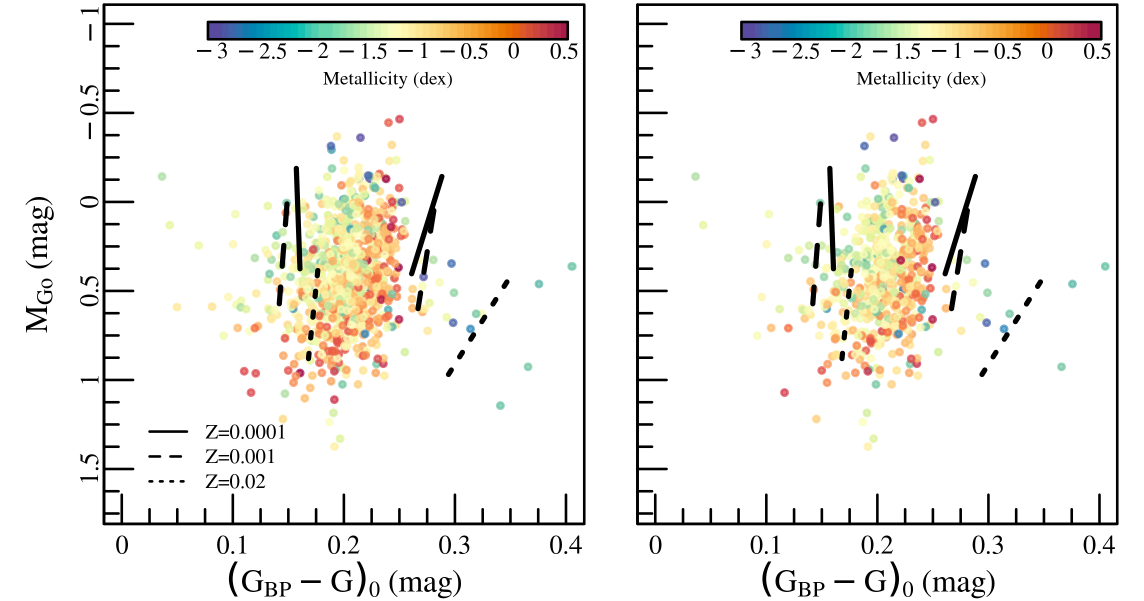}
\caption{{\it Left panel:} M$_{G_{0}}$, (\gbp$-$\gmag)$_{0}$ CMD of  
915 DR3 RRab 
stars 
that have $\sigma_{\varpi}/\varpi$ $<$0.1, RUWE$<$1.4 and $A(G) <$0.2;  
{\it Right panel:} same as in the left panel but for 
620 RR Lyrae of the sample for which $A(G) <$0.1.
Sources are colour-coded according to their metallicity. 
Lines show the boundaries of theoretical instability strips (ISs) for RRab stars of  
three 
different chemical compositions
(Z=0.0001, solid  
lines; Z=0.001, dashed 
lines; and Z=0.02, dotted 
lines), from \citet{2015ApJ...808...50M}.
}
\label{fig:cmd-strips2}
\end{figure*}

Figure~\ref{fig:cmd} shows colour-magnitude diagrams (CMDs) in apparent $G$ vs ($G_{BP}-G_{RP}$) magnitudes defined by RR Lyrae stars in the DR3 clean catalogue. 
In particular, the left panel shows the CMD of RR Lyrae stars in GCs (red symbols) and with blue symbols the RR Lyrae stars in dSphs and UFDs for which the ($G_{BP}-G_{RP}$) colours are available. 
The middle panel shows  instead the CMD of the whole sample of 215\,115 (out of 270\,905) RR Lyrae stars 
for which the ($G_{BP}-G_{RP}$) colours are available,   
using different colours for variables in the LMC, 
SMC 
and All-Sky 
regions, as defined in Sect.~\ref{sec:sos-pip_genover}.
Finally, the right panel shows the CMD of the whole sample (gray symbols), with red, green and blue symbols  showing RR Lyrae stars with parallax better than  10\%, 20\% and 50\%, respectively. In the 
left panel of Fig.~\ref{fig:cmd-strips2} 
we show instead  the HR diagram in 
$G$ absolute magnitude 
(M$_{G_{0}}$)
and dereddend ($G_{BP}-G)_0$ colour of 
915 RRab stars 
that have parallax with $\sigma_{\varpi}/\varpi <$0.1, RUWE$<1.4$\footnote{See Section 14.1.2 of "Gaia Data Release 2 Documentation release 1.2"; https://gea.esac.esa.int/archive/documentation/GDR2/.}, 
and $A(G) <$0.2 mag. Then in the right panel, are shown 620 RRab stars from the same sample, but with $A(G) <$0.1 mag. The observed mean $G$ magnitudes were dereddened using the $A(G)$ values estimated by the SOS pipeline, while the $G_{{BP}_{0}}$ magnitudes were derived from the observed G$_{BP}$ values using the relations:  $A(G)/A(V)$= 0.840 and  $A(G_{BP})/A(V$)= 1.086 from \citet{Bono-et-al-2019}. 
The RR Lyrae stars have been colour-coded according to their metallicity, as  computed by the SOS  
pipeline, and appear to be a mixture of a metal poor and bluer component and a more metallic and redder component. 
We have over-plotted the HR diagrams in Fig.~\ref{fig:cmd-strips2}  with 
the boundaries of 
theoretical  instability strips (ISs) for fundamental mode RR Lyrae stars of three 
different chemical compositions
(Z=0.0001; Z=0.001;
and Z=0.02), from the suite of theoretical models by \citet{2015ApJ...808...50M}. 
The RR Lyrae stars nicely fall within the  boundaries of the theoretical ISs shown in Fig.~\ref{fig:cmd-strips2}.
The slight discrepancy at the red boundary of the IS between observed RR Lyrae position and the predicted edge might be due to the treatment of convection. As extensively discussed in \citep{2004ApJ...612.1092D}
an increase in the mixing length parameter from the adopted 1.5 value to 2.0, makes the  predicted boundary bluer by about 300 K, thus much better matching the observed star distribution.  We also note that the few RR Lyrae stars  with ($G_{BP} - G)$ colours larger than $\sim$0.3 mag, that fall outside the red boundaries of the Z=0.001 and 0.02 ISs, have SOS metallicity estimates with errors larger than  0.46 and up to 1.5  dex. 

\subsection{Sky maps}\label{Skymaps}
\begin{figure*}
   \centering
      \includegraphics[scale=0.5]{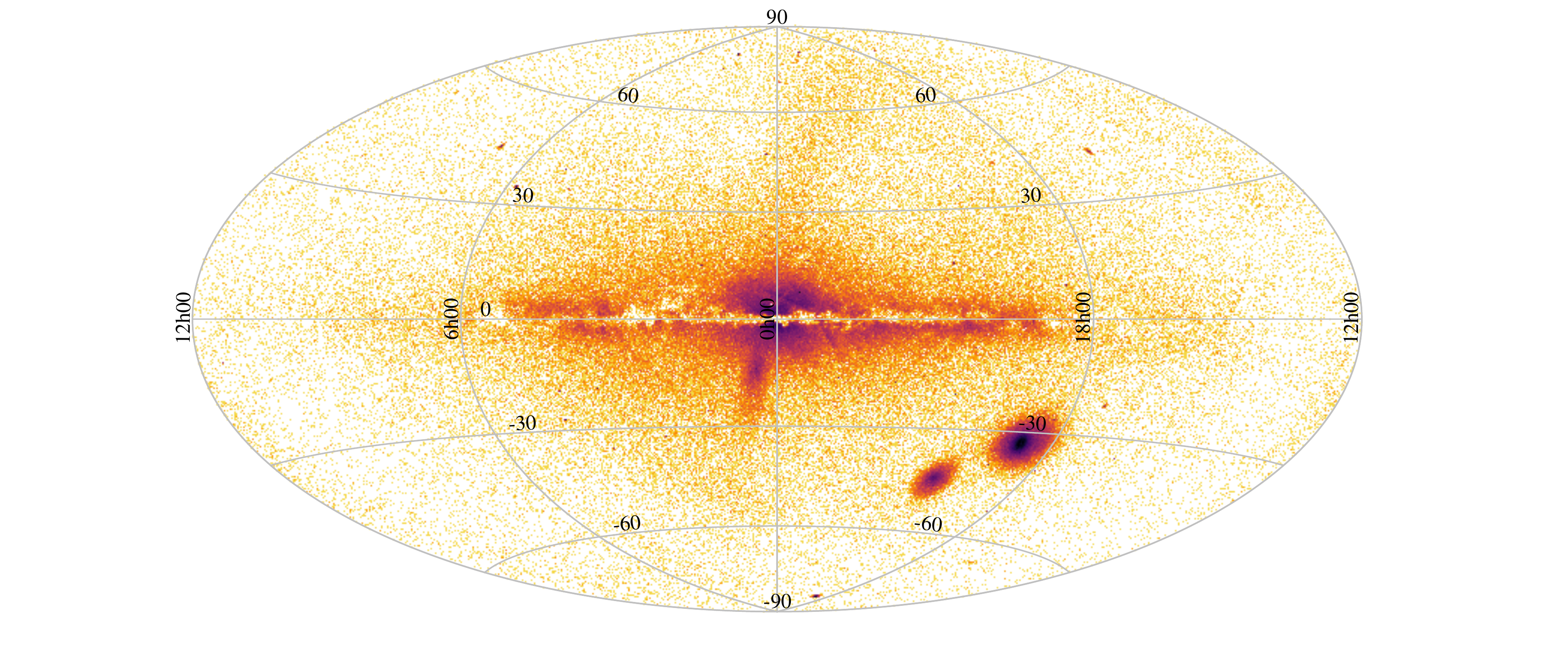}
   \caption{Distribution on sky, in galactic 
   coordinates,  
   of the clean sample of 270\,905  
   DR3 RR Lyrae stars confirmed and characterized by the SOS Cep\&RRL pipeline. 
   The sample comprises 200,294 
   known RR Lyrae stars (Gold Sample) and 
   70,611 new discoveries by {\it Gaia}.}
              \label{skymap}%
    \end{figure*}

\begin{figure}
   \centering
   \includegraphics[scale=0.35]{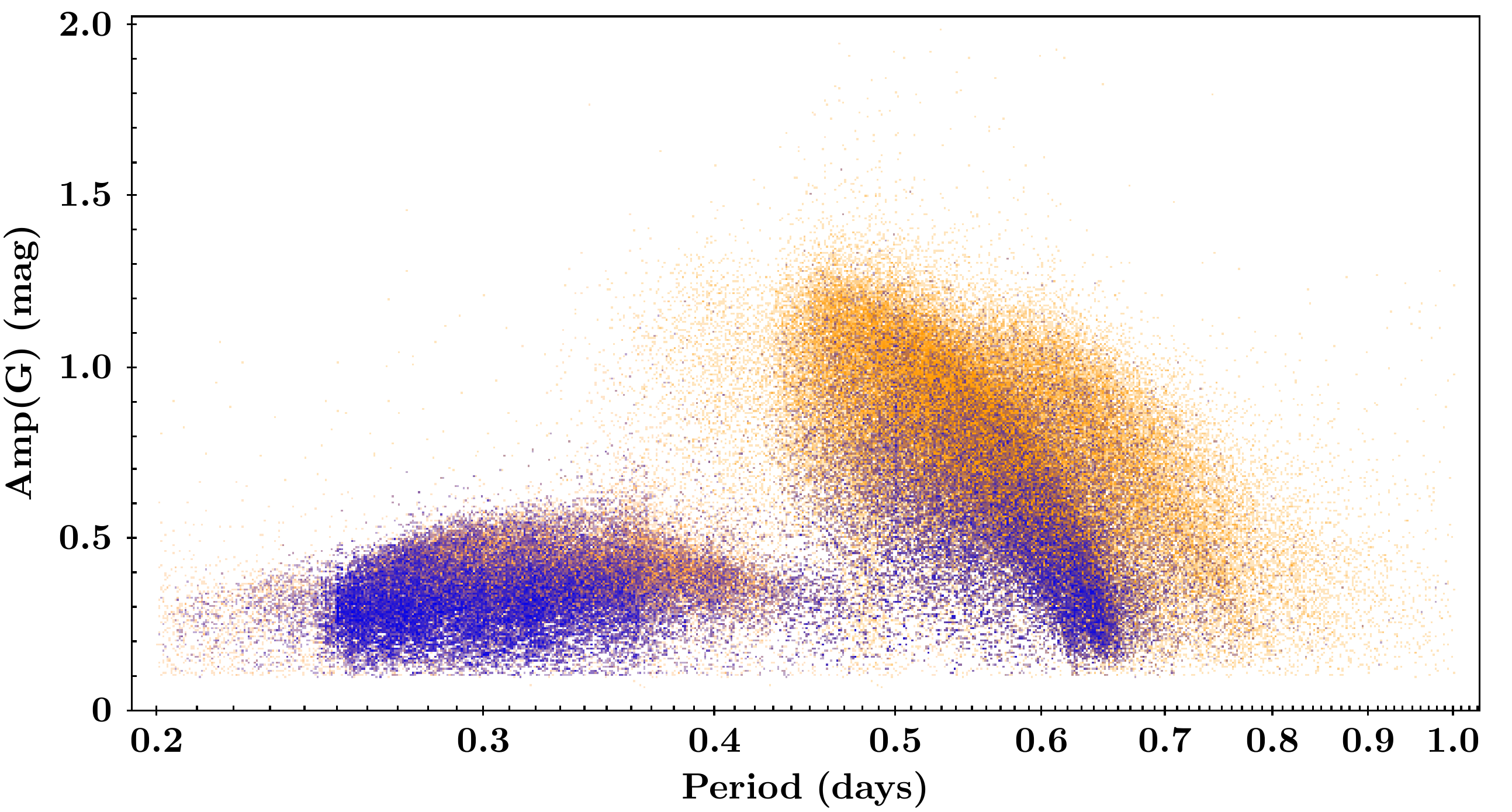}
   \includegraphics[scale=0.35]{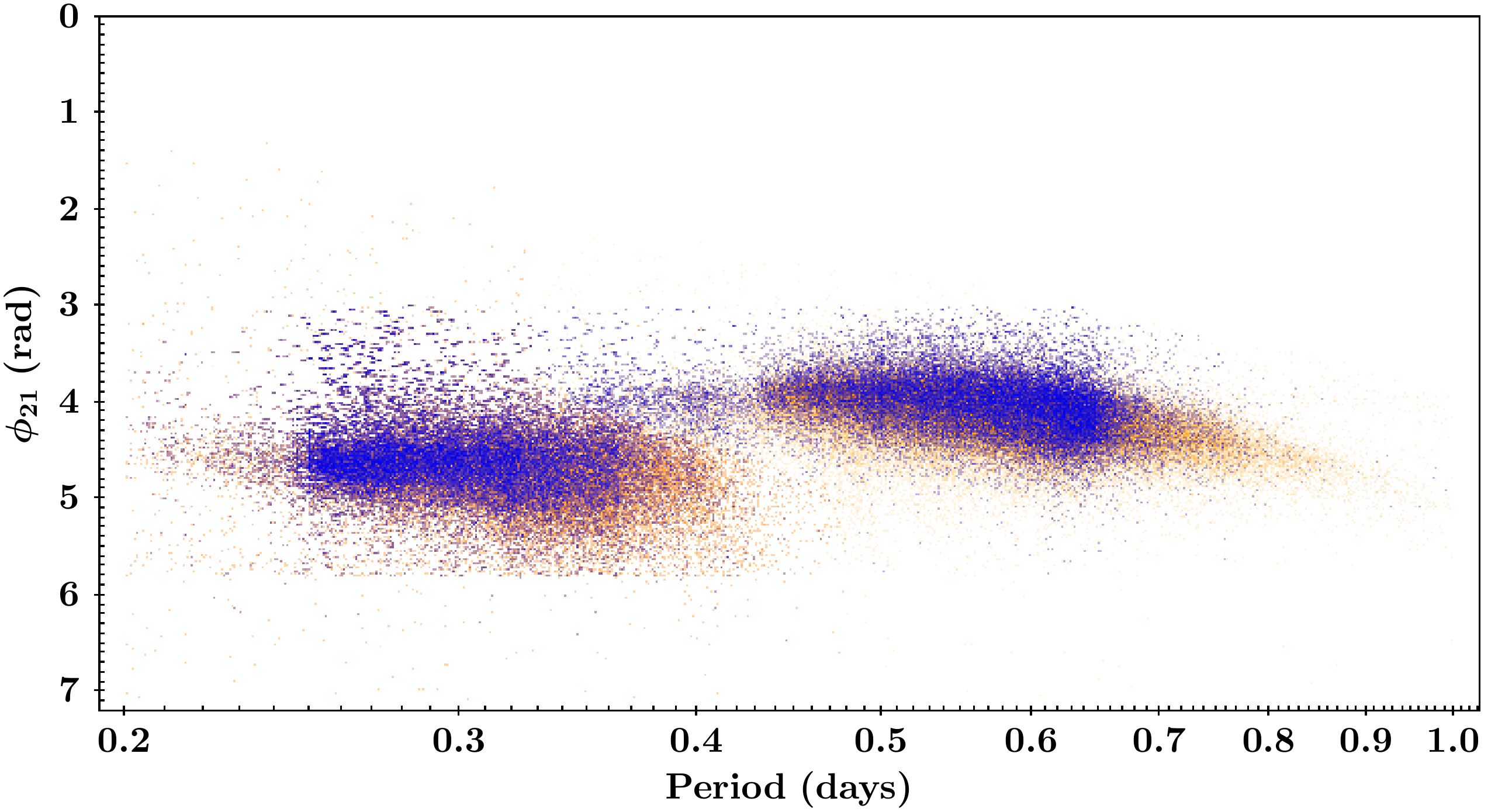}
   \includegraphics[scale=0.35]{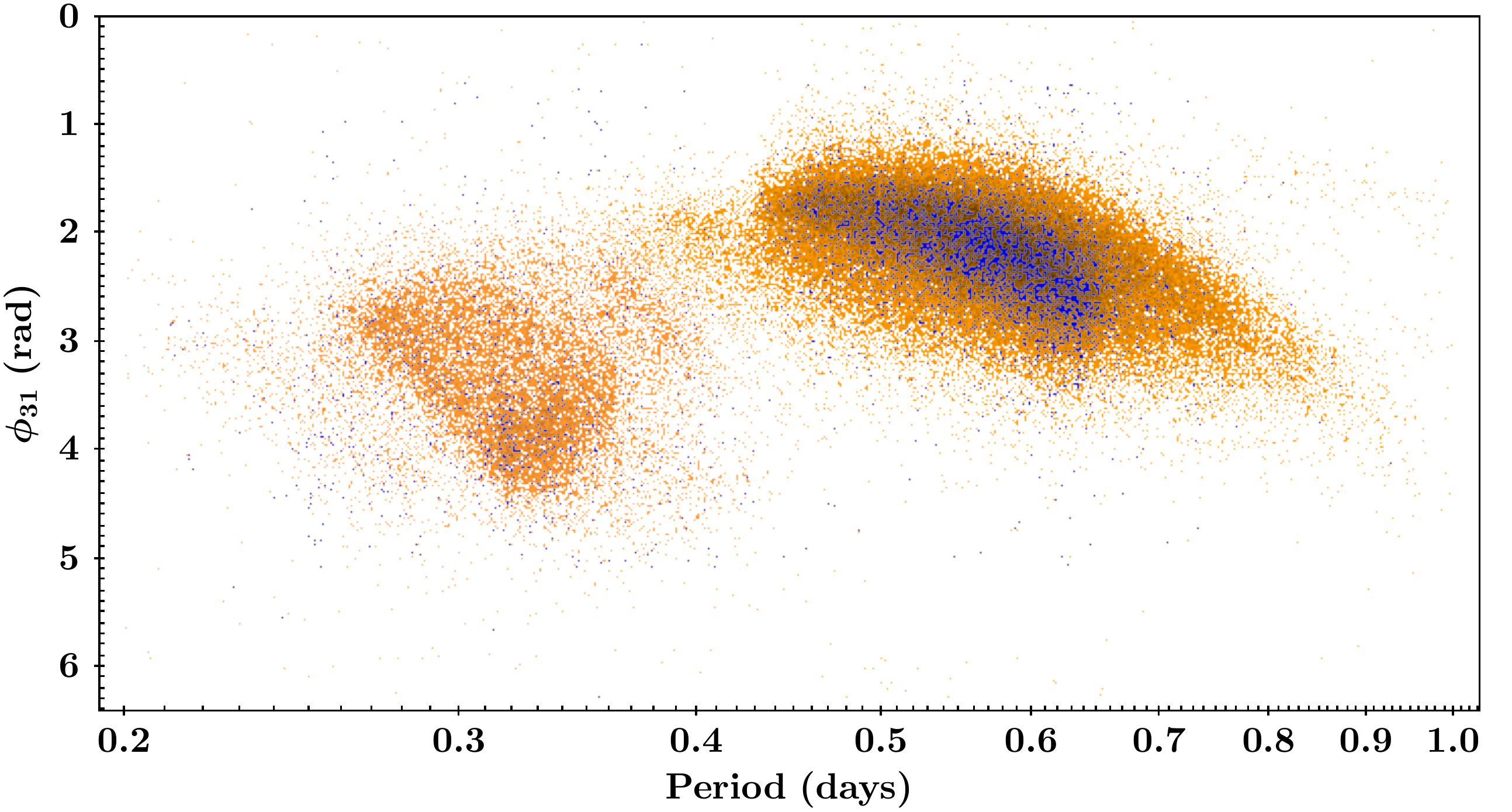}
   \caption{$PA$ (270\,905 
   sources; top panel), $\phi_{21}$   vs $P$ (251\,423 sources; centre panel) and $\phi_{31}$  vs $P$ (136\,015 sources; bottom panel) diagrams of the 
   RR Lyrae stars in the DR3 clean sample. New discoveries by \gaia are marked with blue dots, they are 70\,611 in the top panel, 70\,447 in the central panel and 5\,317 in the bottom panel.}
              \label{PAandDIAGNOSTICS}%
    \end{figure}

Figure~\ref{skymap} shows the distribution on sky, in galactic 
   coordinates, of the clean sample of  270\,905  
   DR3 RR Lyrae stars  
   and Fig.~\ref{PAandDIAGNOSTICS} shows 
   their $PA$, $\phi_{21}$   vs $P$ 
   and   $\phi_{31}$  vs $P$ 
   diagrams. The two extended 
   RR Lyrae overdensities
   in the bottom-right quadrant 
   of Fig.~\ref{skymap}
    trace the halos of the Large and Small Magellanic Clouds. The stream of RR Lyrae stars crossing the Galactic disc is the disrupting Sagittarius dSph galaxy popping out just below to the left from the centre of the map. Other smaller RR Lyrae overdensities can easily be recognised,  like  
   the Sculptor dSph (with more than  500 RR Lyrae) close to the south Galactic pole, the Ursa Minor and Draco dSphs in the upper north-west quadrant, the  Sextans dSph in the upper north-east quadrant and  close to the north Galactic pole the concentration of more than 150 RR Lyrae stars of the M3 GC. 
   Further $\sim$1\,500 RR Lyrae  stars are concentrated  in other 94 different GCs 
   in the map. 
   
\begin{figure*}
   \centering
   \includegraphics[scale=0.38]{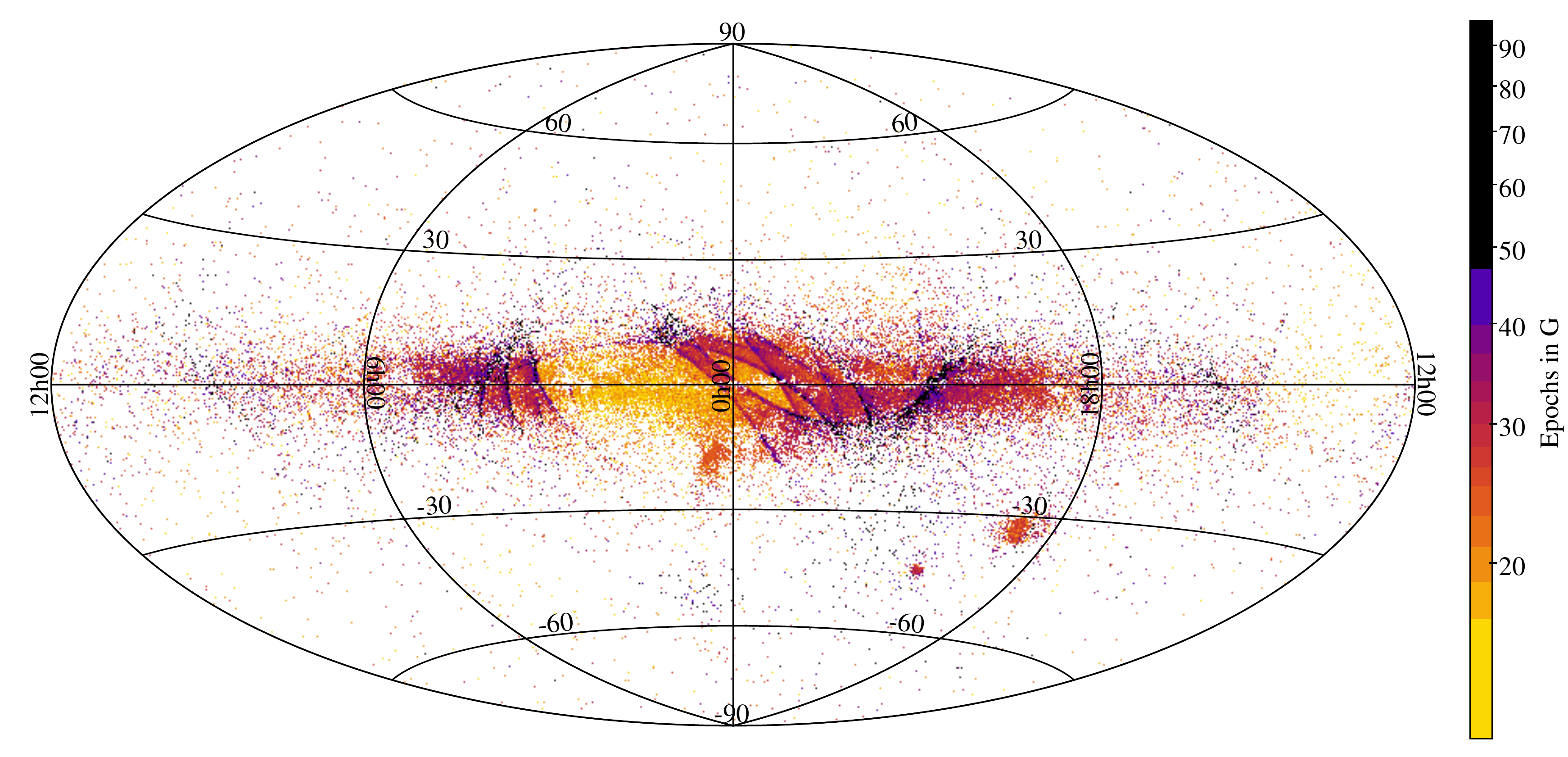}
   \caption{Distribution on sky, in galactic 
   coordinates of 70\,611 DR3 RR Lyrae stars that are new discoveries by {\it Gaia}. The sources are colour-coded according to the number of epochs in the $G$-band light curves, as
encoded in the colour scale on the right.}
              \label{newRRLs}%
    \end{figure*}

   The DR3 RR Lyrae clean catalogue contains 
   70\,611 
   new RR Lyrae stars discovered by {\it Gaia}. They are mainly faint sources (see Fig.~\ref{G-range}) and, according to Fig.~\ref{newRRLs}, they  
   are mainly concentrated in high reddening regions of the Galactic disc and  bulge,  that were poorly sampled with {\it Gaia} DR2 scanning law. 

    
\begin{figure*}
   \centering
   \includegraphics[scale=0.4]{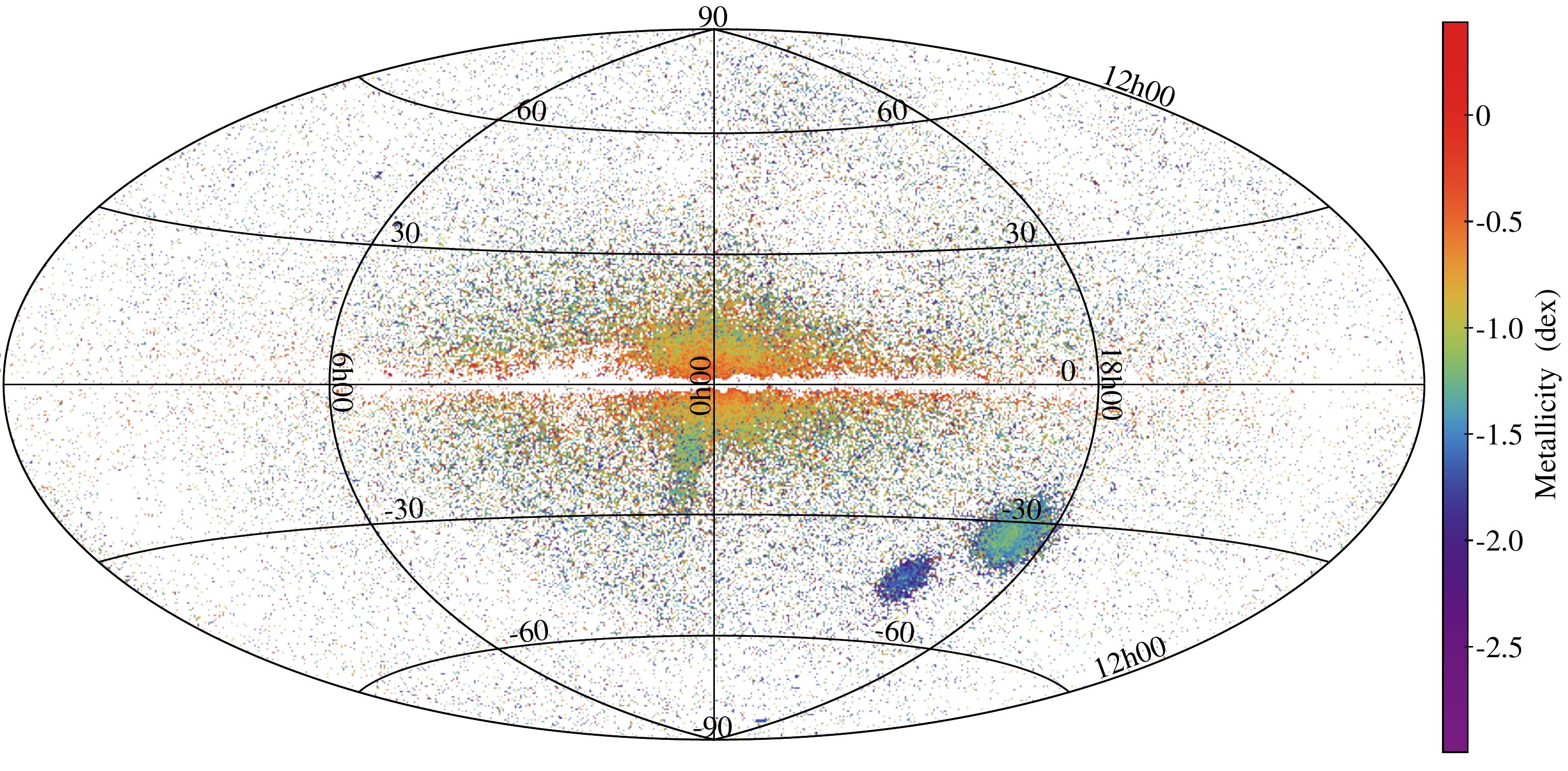}
   \caption{Map in galactic coordinates of 133\,559 
   RR Lyrae stars in the DR3 clean catalogue for which individual photometric  metallicity  estimates obtained by the SOS pipeline are published in the {\it Gaia} DR3  {\tt vari\_rrlyrae} table. The sources are colour-coded according to their
metallicity ([Fe/H]) as encoded in the colour scale on the right.}
              \label{MET-all-map}%
    \end{figure*}

\begin{figure*}
   \centering
      \includegraphics[scale=0.42]{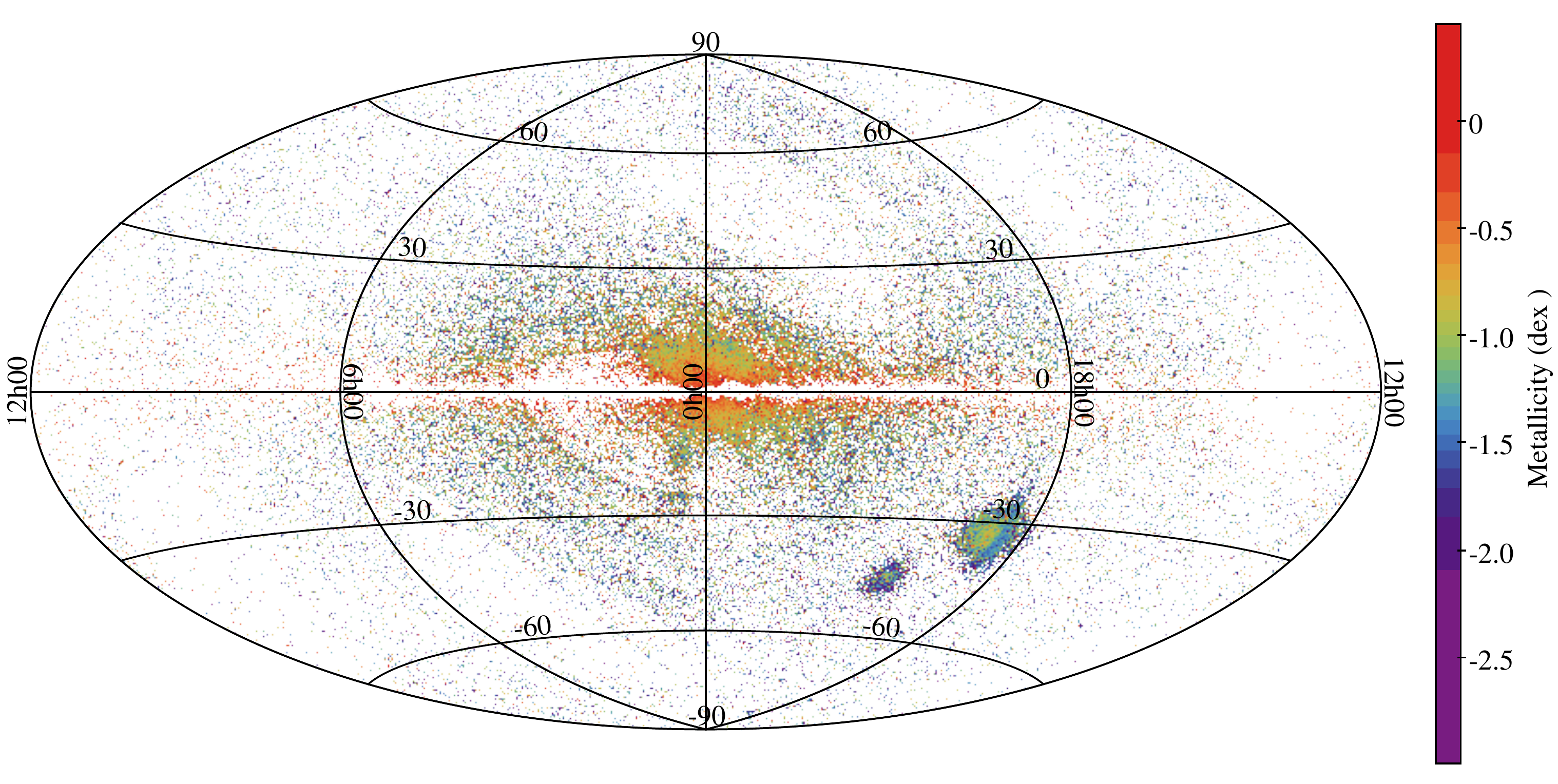}

   \caption{Same as in  Fig.~\ref{MET-all-map} but showing only 88\,440 RR Lyrae stars 
    with error in  metallicity  $\sigma_{\rm [Fe/H]}\leq$ 
   0.46 dex.}
              \label{MET-taglio-map}%
    \end{figure*}

\begin{figure*}
   \centering
   \includegraphics[scale=0.42]{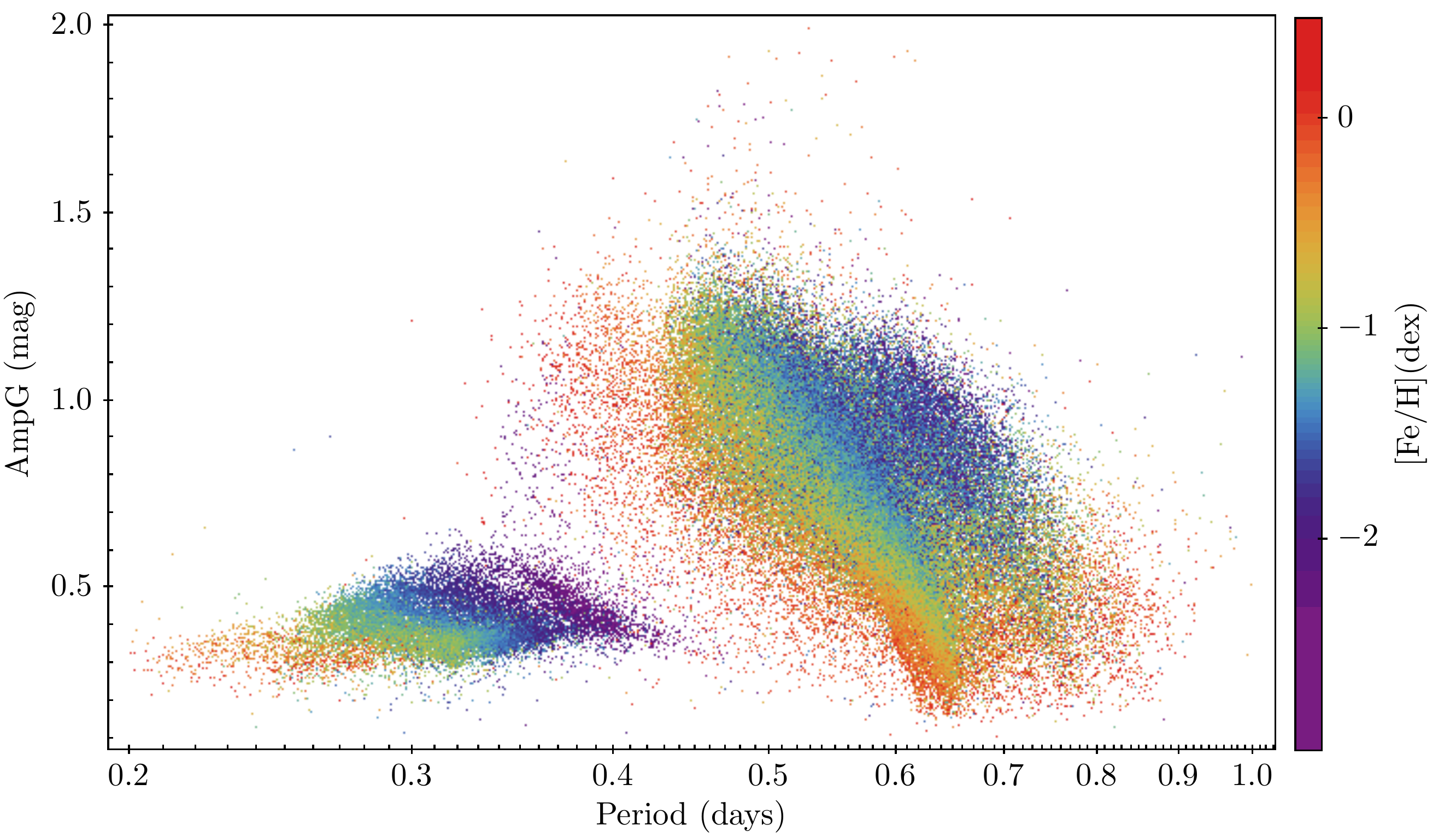}
   \caption{$PA$ diagram of the 133\,559 
   RR Lyrae stars with a metallicity estimate. 
   The sources are colour-coded according to their metal abundance ([Fe/H]) as encoded in the colour scale on the right.}
              \label{fig:pamv1}%
    \end{figure*}

Fig.~\ref{MET-all-map} presents a map, in galactic coordinates, drawn from 133\,559 RR Lyrae stars in the DR3 clean catalogue  
for which individual photometric metallicities ([Fe/H)] obtained by the SOS Cep\&RRL pipeline
(see Sect.~\ref{sec:sos-met}) are published in the 
{\it Gaia} DR3  {\tt vari\_rrlyrae} table. We have colour-coded the sources in the map according to their  
metallicity.
The higher metallicity of the RR Lyrae stars in the MW bulge and disc compared to variables in the
Galactic halo can easily be appreciated, as it is also clear the lower metal abundance of the RR
Lyrae stars in the SMC compared to the variables in the LMC. 
To make sure that the features we see in 
Fig.~\ref{MET-all-map} are not an artifact due to measurement errors, or to the adopted metallicity calibration \citep{Nemec-et-al-2013},  in Fig.~\ref{MET-taglio-map} we show only RR Lyrae stars with metallicity errors   $\sigma_{\rm [Fe/H]}\leq$ 0.46 dex (the mean error of the SOS Cep\&RRL metallicity estimates according to Fig.~\ref{fig:histo_met_err} in  Sect.~\ref{sec:sos-met}). 
The relative difference in metallicity of the RR Lyrae stars in the LMC, SMC,  Galactic halo, and in the bulge and thin disc of our Galaxy is still rather cleanly visible, and it would not change significantly were we to adopt, for instance, \citet{Crestani-et-al-2021} metallicity calibration. 
Fig.~\ref{fig:pamv1} shows the $PA$ diagram of the 133\,559 RR Lyrae stars with a metallicity estimate. 
We have again colour-coded the sources according to their metal abundance. The colour-coding highlights very nicely the difference in metallicity of the various sequences (Oosterhoff loci, \citealt{Oosterhoff-1939}) seen in the $PA$ diagram, in particular the higher metallicity of short period RRab and RRc stars, compared to the lower metal abundance of long period fundamental mode and first overtone RR Lyrae.

\begin{figure*}
   \centering
   \includegraphics[scale=0.43]{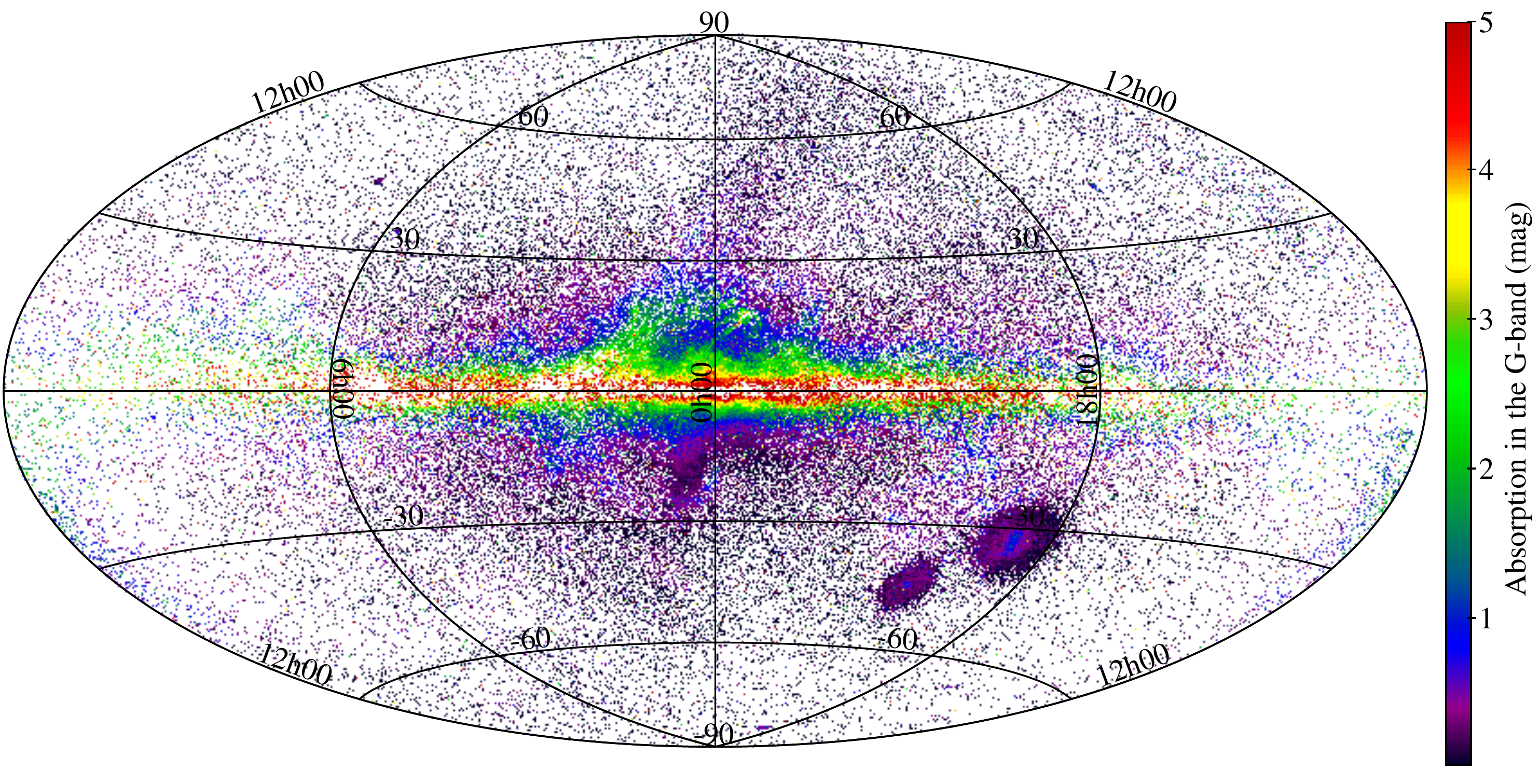}
   \caption{Map in galactic coordinates of 142\,660 
   RRab stars  with individual $G$ absorption values estimated by the SOS pipeline. The sources are colour-coded according to their $A(G)$ values as encoded in the colour scale on the right. 
   }
              \label{AGmap}%
    \end{figure*}

Finally, we show in Fig.~\ref{AGmap} a map in galactic coordinates  of 142\,660 fundamental-mode RR Lyrae stars with absorption in the $G$-band
computed  by the SOS pipeline (see Sect.~\ref{sec:sos-A(G)}). The sources 
are colour-coded according to their individual $A(G)$ values. 
They neatly trace the high extinction regions along  the Galactic plane.


\section{Conclusions and future developments}
\label{sec:conclusions}

\normalsize

An factor two  
in statistic 
with respect to DR2, a better characterization of the RR Lyrae stars and their pulsational and astrophysical parameters, 
along with the improved astrometry published with {\it Gaia} EDR3 \citep{EDR3-DPACP-130}, make the SOS Cep\&RRL DR3 sample, the largest, most homogeneous and parameter-rich catalogue of All-Sky RR Lyrae  stars published so far, in the magnitude range from 
$\langle G \rangle =$ 7.64 mag  
(the magnitude of RR Lyr itsef, sourceid: 2125982599343482624), and 
$\langle G \rangle =$ 21.14 mag (the magnitude of star 658193189673442944, 
the faintest RR Lyrae in the catalogue).

The DR3 RR Lyrae catalogue  
along with known RR Lyrae stars confirmed by the SOS pipeline, but for which the SOS periods differ from the literature periods (see Sect.~\ref{gold-sample}), lead to about 370\,000 the census of RR Lyrae stars, within {\it Gaia}'s limiting magnitude,  in   our Galaxy and its close companions.

{\it Gaia} Data Release 4 (DR4), 
will be based on 66 months of data, 
hence, it will allow the SOS Cep\&RRL pipeline to better pin down the pulsation period(s) and more efficiently characterise RR Lyrae stars and Cepheids. 
The  
pipeline is currently being re-designed and simplified, and a number of changes will be  implemented to improve the 
processing in view of \textit{Gaia} DR4.
Some of these changes are briefly outlined below.
\begin{enumerate}
\item Tests are in progress about how to improve  the 
algorithm for the period determination (currently the Lomb-Scargle, see Sect.~\ref{sec:sos-pip_genover}), in order to  reduce the number of spurious period detections, particularly when several peaks of similar power are present in the power spectrum of the sources; 
\item The number of harmonics for modeling the light curves of RRc stars with well sampled light/RV curves will be increased to more than 2, in order to increase the number of RR Lyrae stars for which the $\phi_{31}$ Fourier parameter is available;
\item The skewness of the light curve and the time of rise to maximum light ($\phi$-rise time) will be adopted, besides the position on the $PA$ diagram,  to  distinguish RRab from RRc stars, particularly in the short period regime 
($P<$ 0.4 d);
\item Luminosity-metallicity ($LZ$), $PW$ and  $PW$-metallicity ($PWZ$) relations in the {\it Gaia} bands, calibrated on  {\it Gaia} parallaxes \citep[such as those published by][]{MNRAS-in-press} will be adopted in the pipeline to classify RR Lyrae stars with reliable parallax values; 
\item CMDs in absolute magnitude, colour-colour diagrams  and the comparison with  theoretical instability strips will be used to improve the source classification, the derivation of intrinsic parameters (e.g. effective temperatures) and the removal of contaminants;
\item We will test different calibrations of the relations used to estimate photometric  metallicities 
from the $\phi_{31}$ Fourier parameter and the pulsation period of RR Lyrae stars;
\item We will revise how errors in metallicity are computed as currently they may be  overestimated;
\item Test will be made to derive   astrophysical parameters (metallicities, gravities and $T_\textrm{eff}$  values) to  
best select the template spectra for the cross-correlation to measure radial velocities of RR Lyrae stars, using also the time series of $BP/RP$ spectra.
\end{enumerate}
To conclude, the SOS Cep\&RRL clean catalogue of
270\,905 RR Lyrae stars published with {\it Gaia} DR3 already represents a significant step forward in the construction of a complete, homogeneous and parameter-rich catalogue of  All-Sky RR Lyrae
stars, within the limiting magnitude of {\it Gaia}, that was lacking so far. We anticipate this RR Lyrae catalogue to be consolidated and  further improved with  
{\it Gaia} DR4. 

\begin{table*}
\tiny
\setlength\tabcolsep{5pt}
\caption{
Links to the {\it Gaia} archive table to retrieve the 
pulsation characteristics: period(s), epochs of maximum light and minimum radial velocity (E), peak-to-peak amplitudes, intensity-averaged mean magnitudes, mean radial velocity, $\phi_{21}$, $R_{21}$, $\phi_{31}$, $R_{31}$ Fourier parameters with related uncertainties,  metallicity and absorption in the $G$ band computed by the SOS Cep\&RRL pipeline 
for  270\,891  confirmed  RR Lyrae stars 
(as well as for 888 rejected sources listed in Table~\ref{table:888_rejected}) that are  released in {\textit{Gaia}} DR3.}           
\label{table:3aRRL}     
\centering                         
\begin{tabular}{ll}       
\hline\hline                 
\noalign{\smallskip}
Table URL & \texttt{http://archives.esac.esa.int/gaia/}\\
\noalign{\smallskip}
\hline
\noalign{\smallskip}
\multicolumn{2}{c}{RR Lyrae 
main parameters computed by the SOS Cep\&RRL pipeline}\\
\hline
\noalign{\smallskip}
Table Name&  \texttt{gaiadr3.vari\_rrlyrae}\\
Source ID & \texttt{source\_id}\\
Type& \texttt{best\_classification} (one of \texttt{RRc}, \texttt{RRab} or \texttt{RRd})\\
$Pf, P1O$
&\texttt{p\_f, p1\_o}
\\
$\sigma (Pf, P1O)$
&\texttt{pf\_error, p1\_o\_error}
\\
E\tablefootmark{\rm (a)}($G$,  $G_{\rm BP}$, $G_{\rm RP}$,  RV)&\texttt{epoch\_g, epoch\_bp, epoch\_rp, epoch\_rv}\\
$\sigma {\rm E}(G, G_{\rm BP}, G_{\rm RP}, {\rm RV})$
&\texttt{epoch\_g\_error, epoch\_bp\_error, epoch\_rp\_error,  epoch\_rv\_error}\\
$\langle G \rangle, \langle G_{\rm BP} \rangle, \langle G_{\rm RP} \rangle$, $\langle {\rm RV} \rangle$ &\texttt{int\_average\_g, int\_average\_bp, int\_average\_rp, average\_rv}\\
$\sigma \langle G \rangle$ , $\sigma \langle G_{\rm BP} \rangle$ , $\sigma \langle G_{\rm RP} \rangle$, $\sigma \langle {\rm RV} \rangle$ &\texttt{int\_average\_g\_error, int\_average\_bp\_error, int\_average\_rp\_error, 
average\_rv\_error}\\
Amp$(G, G_{\rm BP}, G_{\rm RP}, RV)$&\texttt{peak\_to\_peak\_g, peak\_to\_peak\_bp, peak\_to\_peak\_rp, peak\_to\_peak\_rv}\\
$\sigma [{\rm Amp}(G)], \sigma [{\rm Amp}(G_{\rm BP}], \sigma [{\rm Amp}(G_{\rm RP}], \sigma [{\rm Amp}({\rm RV})]$ &\texttt{peak\_to\_peak\_g\_error, peak\_to\_peak\_bp\_error, peak\_to\_peak\_rp\_error, peak\_to\_peak\_rv\_error}\\
$\phi_{21}(G)$ &\texttt{phi21\_g}\\
$\sigma [\phi_{21}(G)]$ &\texttt{phi21\_\linebreak g\_error} \\
$R_{21}(G)$ &\texttt{r21\_g}\\
$\sigma [R_{21}(G)]$&\texttt{r21\_g\_error}\\ 
$\phi_{31}(G)$ &\texttt{phi31\_g}\\
$\sigma [\phi_{31}(G)]$ &\texttt{phi31\_\linebreak g\_error} \\
$R_{31}$ &\texttt{r31\_g}\\
$\sigma [R_{31}(G)]$&\texttt{r31\_g\_error} \\ 
${\rm [Fe/H]}\tablefootmark{\rm (b)}$&\texttt{metallicity}\\
$\sigma$ ([Fe/H]) &\texttt{metallicity\_error}\\
A($G$)\tablefootmark{\rm (c)} & \texttt{g\_absorption}\\
$\sigma$ A($G$) &\texttt{g\_absorption\_error}\\
$N_{\rm obs}$($G$ band) & \texttt{num\_clean\_epochs\_g}\\
$N_{\rm obs}$($G_{\rm BP}$ band) & \texttt{num\_clean\_epochs\_bp}\\
$N_{\rm obs}$($G_{\rm RP}$ band) & \texttt{num\_clean\_epochs\_rp}\\
$N_{\rm obs}$(RV)  & \texttt{num\_clean\_epochs\_rv}\\
\noalign{\smallskip}
\hline                                  
\end{tabular}
\tablefoot{To ease table access, we also provide the correspondence between parameter [period(s), E, etc.] and the name of the parameter in the \textit{Gaia}  archive table. 
$^{\rm (a)}$E corresponds to the time of   maximum in the light curve and the time of minimum in the RV curve. The BJD of all epochs 
is offset by JD 2455197.5 d (= J2010.0).  
$^{\rm (b)}$Photometric metal abundance derived from the period and the $\phi_{31}$ Fourier parameter of the $G$ light curve of  
RR Lyrae stars (see Sects.~\ref{sec:sos-met} and ~\ref{sec:results}).  $^{\rm (c)}$Absorption in the $G$ band computed from a relation that links the star intrinsic colour to the period and the amplitude of the $G$-band light variation 
of fundamental-mode 
RR Lyrae stars (see Sects.~\ref{sec:sos-A(G)} and ~\ref{sec:results}).}\\
\end{table*}

\begin{acknowledgements}
This work has made use of data from the ESA space mission {\it Gaia}, processed by the {\it Gaia} Data
Processing and Analysis Consortium (DPAC). Funding for the DPAC has been
provided by national institutions participating in
the {\it Gaia} Multilateral Agreement. In particular, the Italian participation in DPAC has been supported by Istituto Nazionale di
Astrofisica (INAF) and the Agenzia Spaziale Italiana (ASI) through grants
I/037/08/0,  I/058/10/0,  2014-025-R.0, 2014-025-R.1.2015 and  2018-24-HH.0 to INAF (PI M.G.
Lattanzi), the Swiss participation by the Swiss State Secretariat for Education, Research
and Innovation  through the ``Activit\'{e}s Nationales Compl\'{e}mentaires''. 
The work was supported in part by" the French Centre National de la Recherche Scientifique (CNRS), the Centre National d'Etudes Spatiales (CNES), the Institut des Sciences de l' Univers (INSU) through the Service National d'Observation (SNO) Gaia.
UK community participation in this work has been supported by funding
from the UK Space Agency, and from the UK Science and Technology
Research Council. RA has been funded by the DLR (German space agency) via grants 50QG0602, 50QG1001, 50QG1403, and 50QG201.
The {\it Gaia} mission website is:  \texttt{http://www.cosmos.esa.int/gaia}.
This research has made use of the 
SIMBAD database,
operated at CDS, Strasbourg, France. 
We warmly thank O. Creevey, A. Recio-Blanco and M. Fouesneau for advises on the use of products from the  GSP\_Spec and GSP\_Phot modules of the Apsis pipeline, D. Harrison for running the 
SEAPipe pipeline on the catalogue of RR Lyrae stars produced by the SOS Cep\&RRL pipeline and P. Montegriffo for the development and
maintenance of the GRATIS software.  
In this study we have largely made use of TOPCAT,  Taylor, M. B. (2005), ``TOPCAT \& STIL: Starlink Table/VOTable Processing Software", in Astronomical Data Analysis Software and Systems XIV, eds. P. Shopbell et al., ASP Conf. Ser. 347, p. 29.
 
\end{acknowledgements}

%
  \bibliographystyle{aa} 
  \bibliography{p.bib} 
%



\begin{appendix} 

\section{RVS radial velocities} 
\subsection{Tables with the   source-level parameters of 51  Cepheids and 45 RR Lyrae stars for which this information is lacking in the {\tt gaia\_source} table of the {\it Gaia} DR3 archive.}  
\label{app:var-nopubb}

\begin{table*}[h!]
\tiny
\setlength\tabcolsep{3pt}
\arrayrulecolor{black}
\caption{Source-level information for 51 Cepheids for which the all-epoch-combined RV was deemed of insufficient quality. 
{\tt radVelInvalidReason} provides the reason why the source-level parameters are excluded from publication: D: Double-lined detected in more than 10\% of the epochs; P: variations in the spectral-sample flux from epoch to epoch; H: hot star; CF: bad cross-correlation-function; T: potential template mismatch; E=the error of the all-epochs combined radial velocity is larger than 40 km/s
\citep[see][Chap. 6, for details]{CU6-DR3-documentation}.
This table is available in its entirety in the electronic version of the journal.
}
\label{table:51Cep-RVnopub}      
\centering 
\begin{tiny}
\begin{tabular}{lccccccc}     
\hline\hline             
\noalign{\smallskip} 
  \gaia sourceid &rv\_template\_teff&rv\_template\_logg& rv\_template\_feh&rv\_atm\_param\_origin&grvs\_mag& grvs\_mag\_error&radvelinvalidreason\\
\\
  \noalign{\smallskip} \hline \noalign{\smallskip}
5351161399793209984 & 5\,000 &1.5 &0.25&222&6.281&0.043&D\\
5329352998951582976&6\,000&1&$-$0.25&222&6.492&0.056&D\\
5355057622307185280&5\,500&3.5&0&333&6.607&0.038&D\\
5338036117182452096&6\,250&2&0.25&222&6.815&0.070&D\\
473043922712140928&5\,500&4.5&0&333&6.834&0.047&D\\
\noalign{\smallskip} \hline \noalign{\smallskip}
\end{tabular}
\end{tiny}
\end{table*} 

\begin{table*}[h!]
\arrayrulecolor{black}
\caption{Same as in table~\ref{table:51Cep-RVnopub} but for 45 RR Lyrae stars.
This table is available in its entirety in the electronic version of the journal.
}
\label{table:45RRL-RVnopub}      
\centering 
\begin{tiny}
\setlength\tabcolsep{3pt}
\begin{tabular}{lccccccc}     
\hline\hline             
\noalign{\smallskip} 
  \gaia sourceid & rv\_template\_teff & rv\_template\_logg & rv\_template\_feh& rv\_atm\_param\_origin& grvs\_mag & grvs\_mag\_error &radvelinvalidreason\\
\\
  \noalign{\smallskip} \hline \noalign{\smallskip}
2857456211775108480&6\,500&3&0&222&9.076&0.127&D\\
2142052889490819328&6\,250&3&$-$0.5&444&9.324&0.083&T\\
234108363683247616&6\,500&3&$-$0.25&555&9.411&0.091&D\\
6686289122295812736&6\,250&3.5&$-$0.5&444&9.534&0.051&D\\
1858568795812429056&6\,000&3&$-$0.5&444&10.242&0.059&D\\
\noalign{\smallskip} \hline \noalign{\smallskip}

\end{tabular}
\end{tiny}
\end{table*}

\arrayrulecolor{black}
\begin{sidewaystable*}\label{RRL-TN}
\begin{tiny}
  \setlength\tabcolsep{3pt}
\caption{
Information for 19 
RR Lyrae stars with high accuracy RV curves available in the literature.} 
\medskip
\begin{tabular}{clclrccrccccccccc}
\hline\hline             
\noalign{\smallskip} 
      Source\_id  &   ~~Name & Mode & ~~Period  & $\langle G \rangle$~~& N$_{\rm  RVS}$ & Amp(RV) & ${\bf \gamma_{\rm (RVS+Lit.)}}$~~
      & $\gamma_{\rm Lit.}$~~& 
${\rm {[Fe/H]}_{\rm Lit.}}$ & $\langle T_{\rm eff}\rangle_{\rm Lit.}$  & $\langle log g \rangle _{\rm Lit.}$  &  Ref1 & Ref2 & Notes  &Ins. \\ 
 & & & ~~~days & mag& & km/s~~&km/s~~~~~~& km/s~~& dex& K & km/s& & & &\\ 
(1) &~~~~(2) &(3) &~~~~~(4) &(5)~~&(6) &(7)&(8)~~~~~~~ &(9)~~ &(10) &(11) &(12) &(13) &(14) &(15)&(16)\\
\noalign{\smallskip} \hline \noalign{\smallskip}
1286188056265485952 & RS Boo & RRab& 0.3773523
&10.34&28&67.38
& $-3.96$~(2.57)
&$-$3.7&$-$0.50&6\,795 &3.261$^{(a,b)}$&8&8&Blaz.&b\\
1492230556717187456 & TV Boo & RRc  & 0.312551
&10.91&34& 27.07
&$-102.44$~(5.00)
&$-$104.0&$-$2.20$\pm$0.15&7\,057&2.90&1&1&-&b \\
6884361748289023488 & YZ Cap & RRc  &0.2734508&11.22&27&26.19
&$-106.91$~(2.29)
&$-106.5\pm1.6$&$-$1.25&7\,280$\pm$100&3.00$\pm$0.10&2&10&-&a\\
2558296724402139392 & RR Cet & RRab& 0.5530296&9.61&30&62.79
&$-74.42$~(2.50)
&$-$74.6/$-$75.1&$-$1.25$\pm$0.10&6\,434&2.71&1,6&1&-&b\\
2414817603803476864 & UU Cet & RRab&0.6060755&11.94&21&55.71
&$-113.88$~(3.10)
&$-114.4$&$-$0.87&6\,290&2.75$^{(c,b)}$&3&13&-&a\\
3546458301374134528 & W Crt    & RRab& 0.4120288
&11.45&31&75.76
&59.87~(3.64)
&58.8&$-$0.70&6\,715&2.70$^{(c)}$&4&4&-&c\\ 
1760981190300823808 & DX Del  & RRab&0.4726188&9.81&26&55.48
&$-55.15$~(2.80)
&$-$55.1&$-$0.25&6\,512&2.7$\pm$0.2$^{(d,b)}$& 5,6 &12&-&d,a\\   
1058066262817534336 & SU Dra & RRab& 0.6604232
&9.68&13&60.63
&$-167.08$~(1.87)
&$-166.9$&$-$1.60$\pm$0.20&6\,433&2.72&1&1&-&b \\ 
1683444631038133248  & SW Dra &RRab& 0.5696720
&10.38&20&62.04
&$-29.13$~(2.57)
&$-29.8$&$-$0.80/$-$1.40&6\,380$\pm$50$^{e}$ &2.78$\pm$0.10&2,7&11,8&-&a,b\\
2981136563934324224 & RX Eri   & RRab& 0.5872470
&9.56&26&67.15
&67.72~(3.89)
&66.4&$-$1.40$\pm$0.20&6\,326&2.67&1&1&-&b \\ 
5117708899055276416 & SS For  & RRab& 0.49544999
&10.10&28&62.81
&$-$109.93~(4.92)
&$-$111.7&$-$1.50&6\,710$\pm$50&2.85$\pm$0.10&2&11&Blaz.&a\\    
4596935593202765184 & TW Her & RRab& 0.3996157
&11.22&27&72.90
&$-$3.31~(2.66)
&$-$4.5&$-$0.50& 6\,770&3.424$^{(a,b)}$&8&8&-&b\\
6483680332235888896 & V Ind     & RRab& 0.4796017
&9.86&12&52.83
&202.71~(1.74)
&202.5&$-$1.51&-&-&3&-&-&a\\
1793460115244988800 & AV Peg &  RRab& 0.3903809
&10.39&16
&65.68
&$-$58.22~(1.70)
&$-$58.7&0.00$\pm$0.10&6\,603&2.99&1&1&-&b\\
6526559499016401408 & RV Phe & RRab&0.5963778
&11.81&37
&59.72
& 99.72~(6.65)
&99.1&$-$1.50
&6\,370$\pm$50/6\,290&2.76$\pm$0.10&2&10,13&Blaz.?&a\\
6771307454464848768 & V440 Sgr$^{(f)}$ & RRab& 0.4774854&10.23&10 &76.61&
$-$60.92~ (2.95)&$-$61.6&$-$1.40&6\,810$\pm$50&2.87$\pm$0.10&2&10&-&a\\ 
4709830423483623808 & W Tuc  &  RRab&0.6422386
&11.34&18
&69.22
&64.69~(1.90)
&64.60&$-$1.35&6\,440&2.75$^{(c,b)}$&3&13&-&a\\
3698725337376560512 & UU Vir  &  RRab&0.4756085&10.52&14
&72.32
&$-7.80$~(2.07)
&$-8.2$/$-7.1$&$-$0.40$\pm$0.10/$-$0.7&6\,520/6\,570&2.84/3.279$^{(a,b)}$&1,5,8 &1,8&Bin.?&b\\   
6045485228725626752 & M4-V32$^{(g)}$ & RRab& 0.5791070
&12.90& 34&63.50
&67.93~(5.31)
&66.4& $-$1.3 $\pm$ 0.2& 6\,324&2.67&9&9&-&b\\
\noalign{\smallskip} \hline \noalign{\smallskip}
\end{tabular}
\label{Tab:Table-rr}
\\
\tablefoot{Meaning
  of the different columns is as follows: (1) Gaia DR3 source id;
  (2) Literature Name; (3) Pulsation mode  (RRab=Fundamental; RRc= First
Overtone); (4) Pulsation period (P), as re-evaluated in the present analysis; (5) Intensity-averaged $G$-band  mean magnitude, as derived from the SOS Cep\&RRL pipeline; (6)
number of valid RVS RV measurements; (7) Peak-to-peak amplitude of the combined RVS and literature RV data. The RVS measurements of TV Boo and M4-V32 were shifted by +5km/s, those of TW Her and UU Vir by $-$4 km/s, and those of AV Peg by $-$3 km/s to better match the corresponding literature data}; (8) Centre of mass RV ($\gamma$) of the curve obtained by combining RVS and literature RV data, with in parentheses the rms scatter of the modeled RV curve; (9) Literature values for $\gamma$; 
(10) Iron abundance; (11) Mean effective temperature; (12)
Mean gravity; (13) References for the
literature RV; (14) References for the stellar parameters; (15) Notes: Binary (Bin) and/or Blazhko (Blaz.).; (16) Instrument type: a=CORAVEL; b=spectra; c=image-tube spectrograph; 
d=photoelectric RV meter. 
The  meaning of the numbers in columns (13) and (14) is as follows: 1=\citet{1989ApJS...69..593L}; 
2=\citet{1987A&AS...69..135C};
3=\citet{1990A&AS...85..865C};
4=\citet{1993MNRAS.265..301S};
5=\citet{1988ApJS...67..403B};
6=\citet{1986A&AS...64...25M};
7=\citet{1987ApJ...314..605J};
8=\citet{1988ApJ...332..206J};
9=\citet{1990ApJ...360..561L};
10=\citet{1989A&A...209..154C};
11=\citet{1989A&A...209..141C};
12=\citet{1989MNRAS.241..281S};
13=\citet{1992ApJ...396..219C};
The binarity information is taken from \citet{2019A&A...623A.116K}.\\ 
$^{a}$Gravity value at $\phi$=0.0.\\
$^{b}$\citet{1992ApJ...396..219C} 
derived a mean gravity of 2.83$\pm$0.11 dex by averaging the gravity values inferred from the radii and masses
published by 
\citet{1992ApJ...386..646J}, 
which were derived from the B-W analysis of 18 RRab and 2 RRc stars. This mean gravity value can be used in place of $^{(a)}$, $^{(c)}$ and $^{(d)}$ or when a  $\langle log g \rangle $ value is missing.\\
$^{c}$Constant gravity value adopted in \citet{1992ApJ...396..219C} 
B-W analysis of UU Cet, RV Phe and W Tuc, and in \citet{1993MNRAS.265..301S} 
analysis of W Crt.\\
$^{d}$2.7$\pm$0.2 is the `static' gravity, that is the gravity of an RR Lyrae star were it not pulsating \citep{1989MNRAS.236..447F}.\\
$^{e}$$\langle T_{\rm eff}\rangle$=6460 K estimated from the $V-K$ colour by \citep{1988ApJ...332..206J} 
.\\
$^{f}$The period derived by the SOS pipeline for this star ($P_{\rm SOS}$=0.428933547 d) is incorrect, we provide in the table the correct period ($P$=0.4774854 d). The $\langle G \rangle$, Amp(RV) and  $\gamma_{\rm RVS}$ values of V440~Sgr  listed in the table were computed using the correct period.\\
$^{g}$ Two RV measurements from $Gaia$ RVS around the curve maximum are outliers with large errors. They were discarded when computing the Amp(RV) and $\gamma$ values in columns (7) and (8).
\label{Tab:Table-rr}
\end{tiny}
\end{sidewaystable*}

\subsection{Comparison of literature and {\it Gaia} RV curves for 13 of the 19 RR Lyrae stars with high accuracy RVs available in the literature (see Sect.~\ref{sec:RV-validation}).}  
\label{app:var-19}

\begin{figure*}[h!]
\begin{center}
      \includegraphics[width=6.5cm]{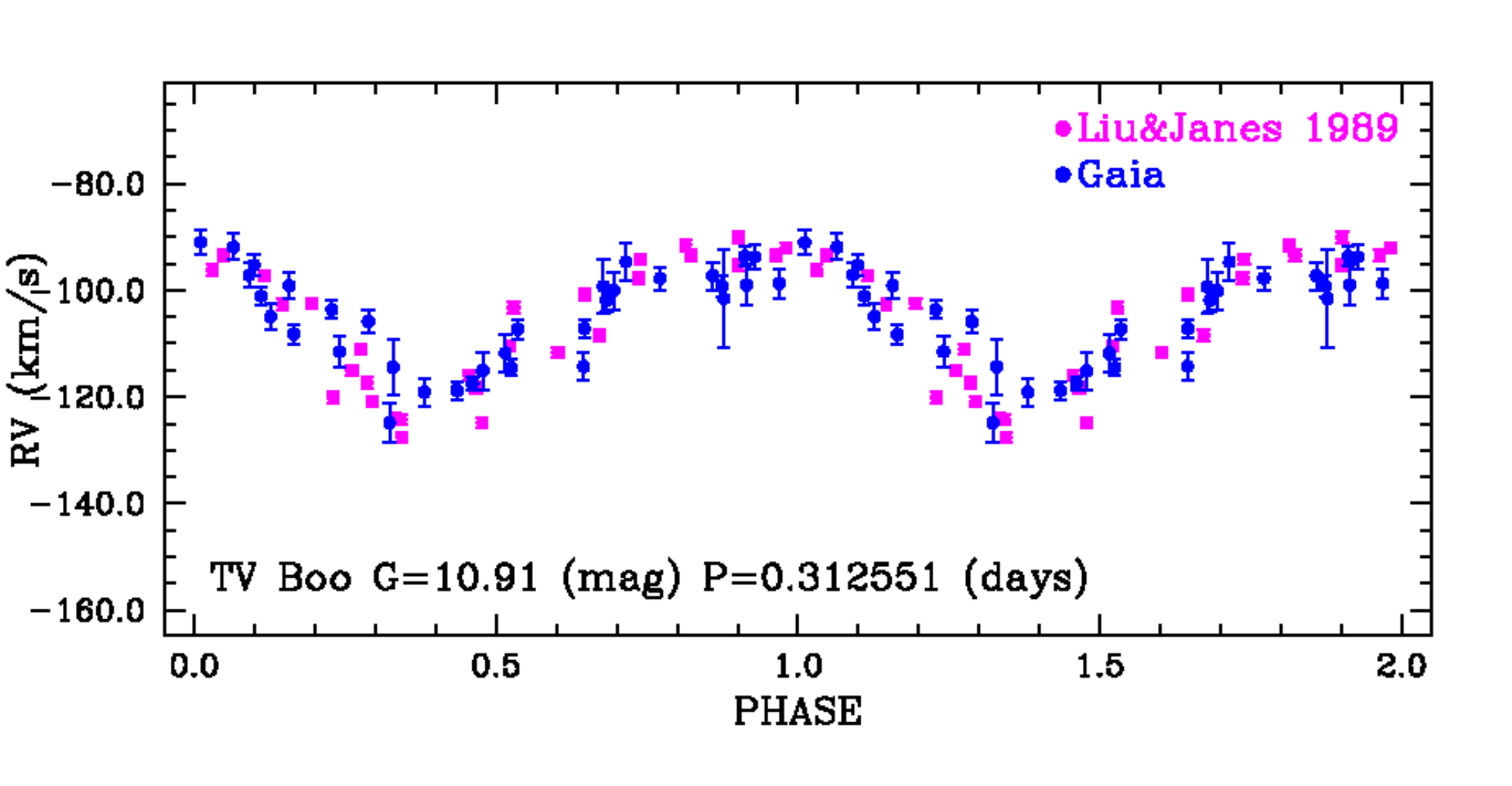}
      ~\includegraphics[width=6.5cm]{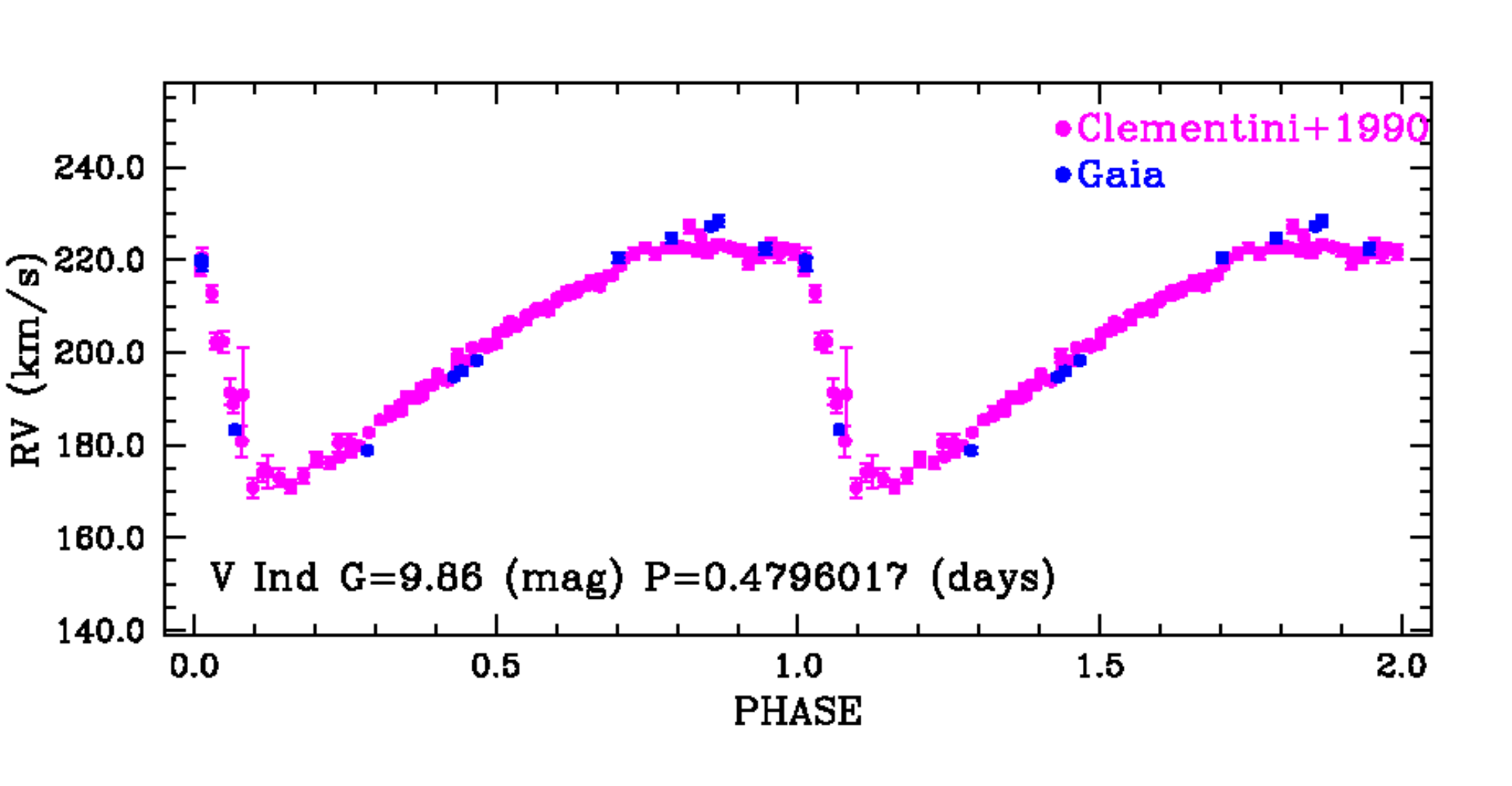}
      \includegraphics[width=6.5cm]{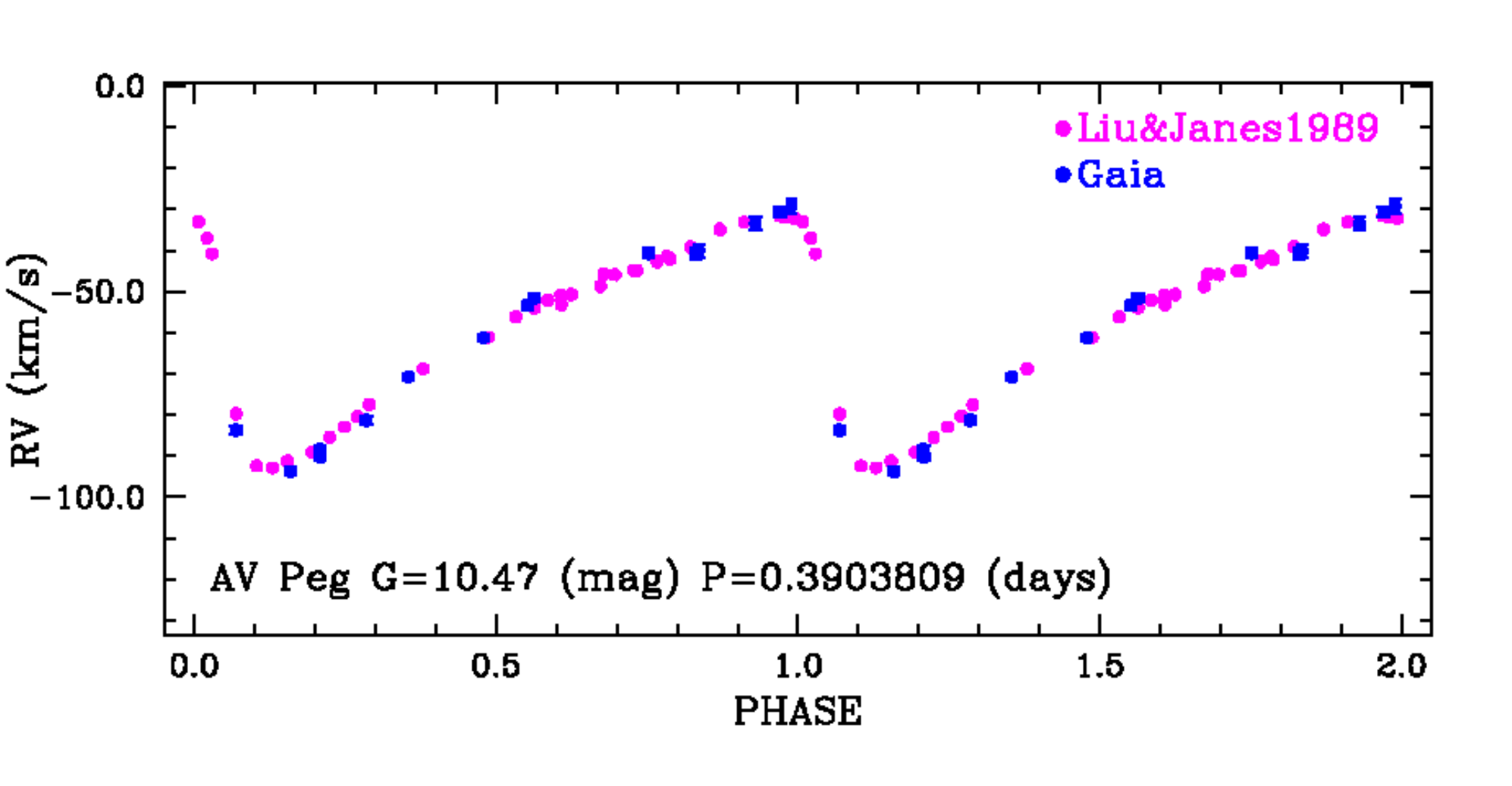}
      ~\includegraphics[width=6.5cm]{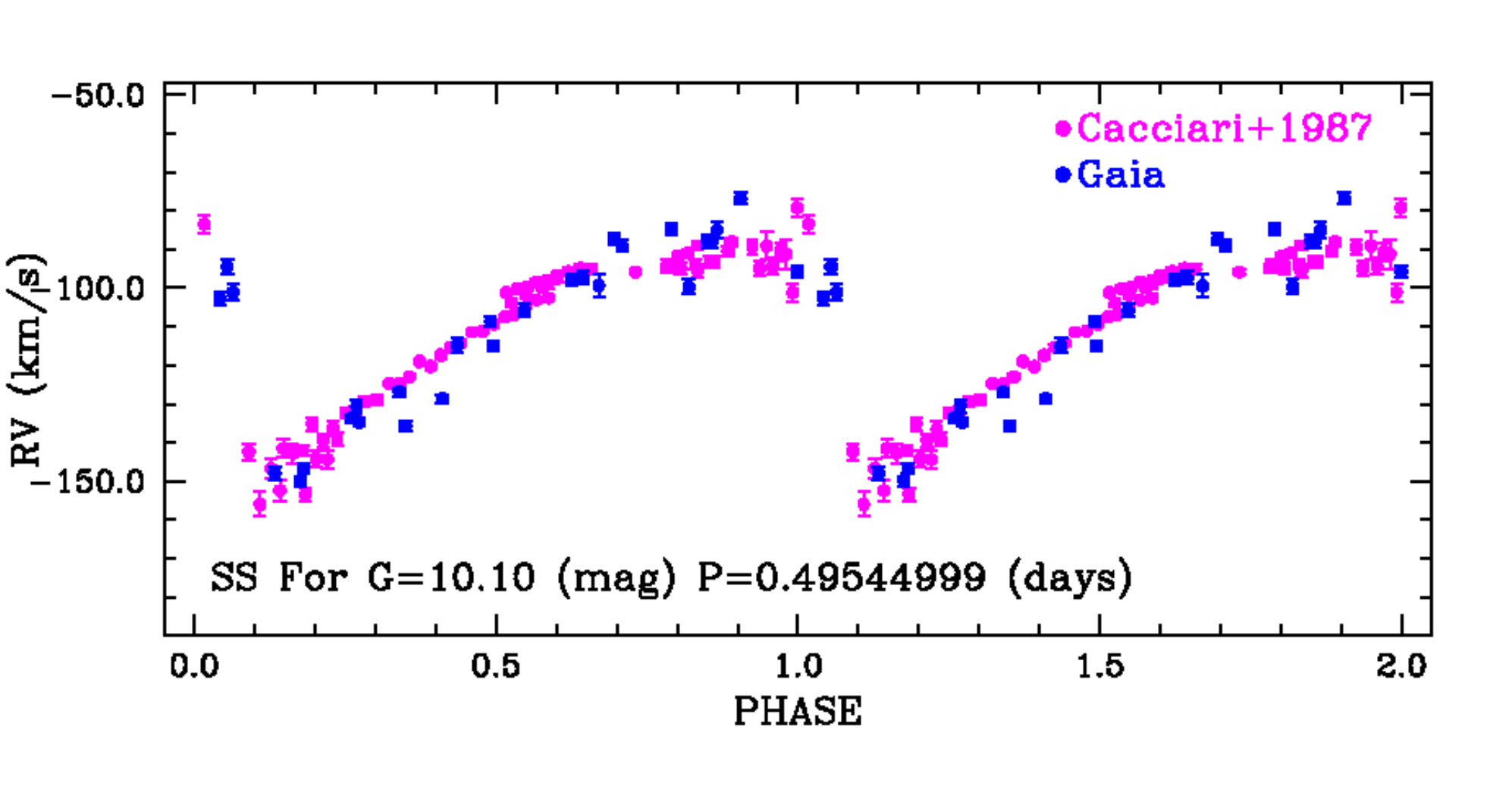}
      \includegraphics[width=6.5cm]{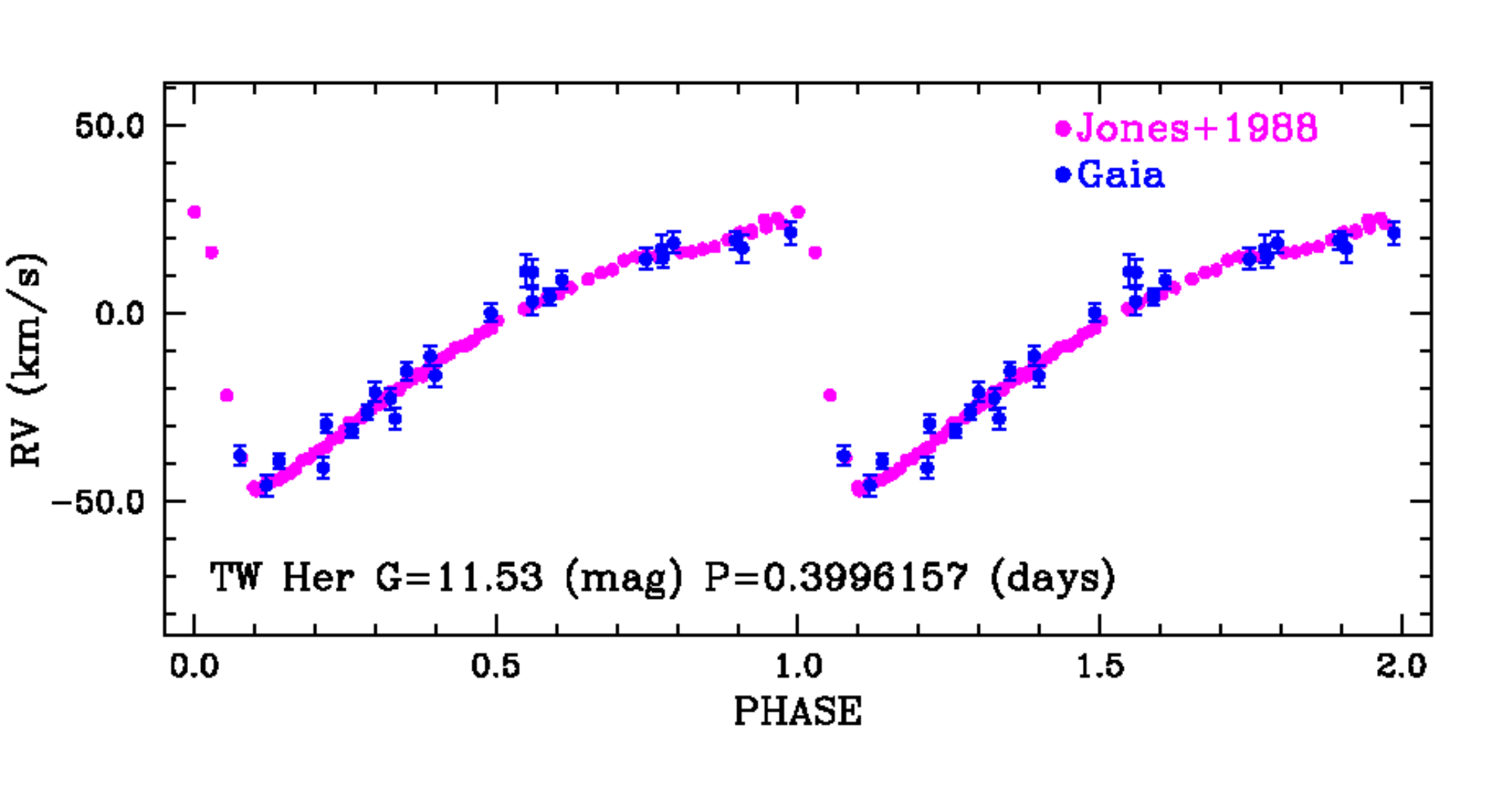}
      ~\includegraphics[width=6.5cm]{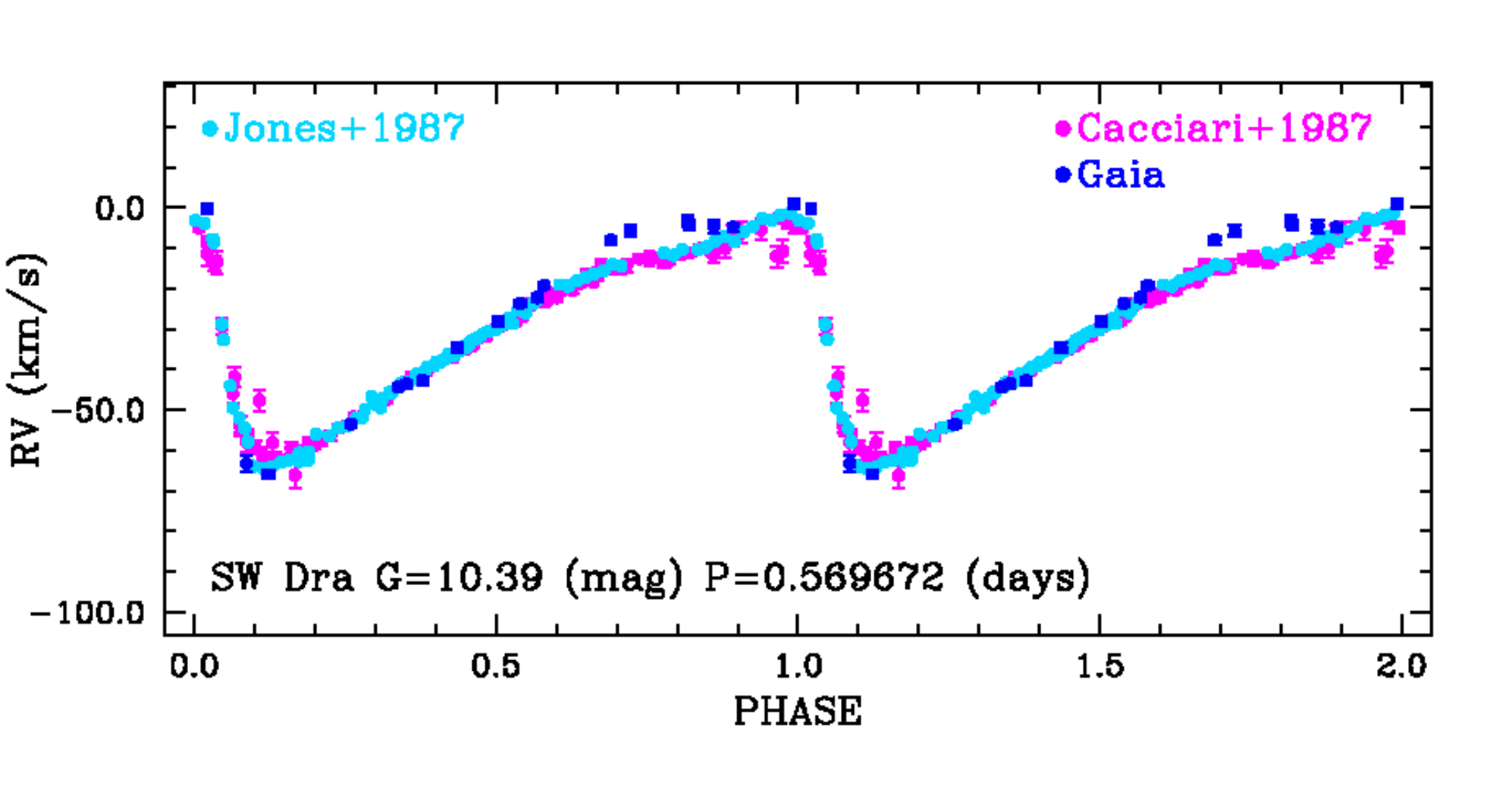}
      \includegraphics[width=6.5cm]{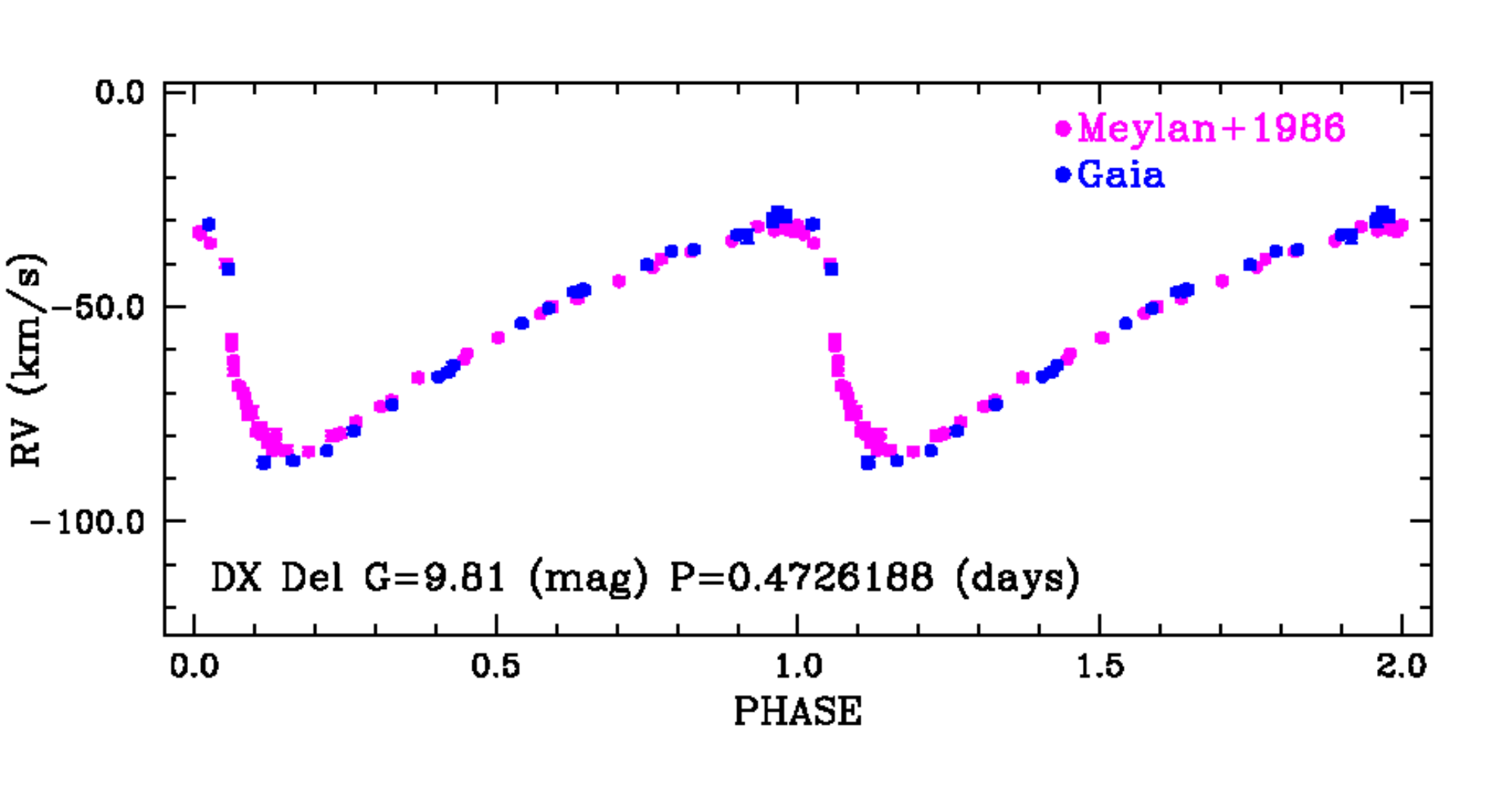}
      ~\includegraphics[width=6.5cm]{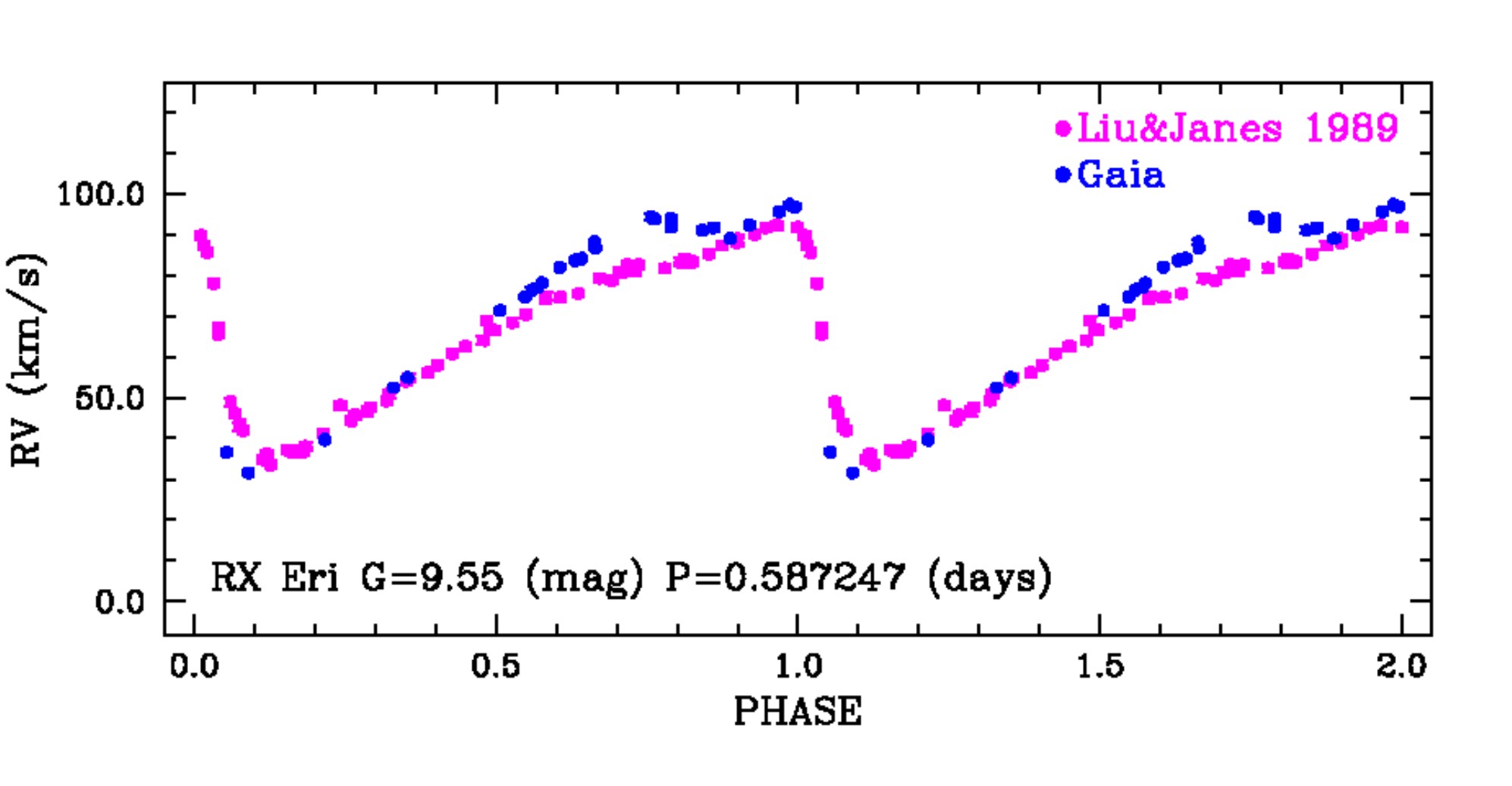}
      \includegraphics[width=6.5cm]{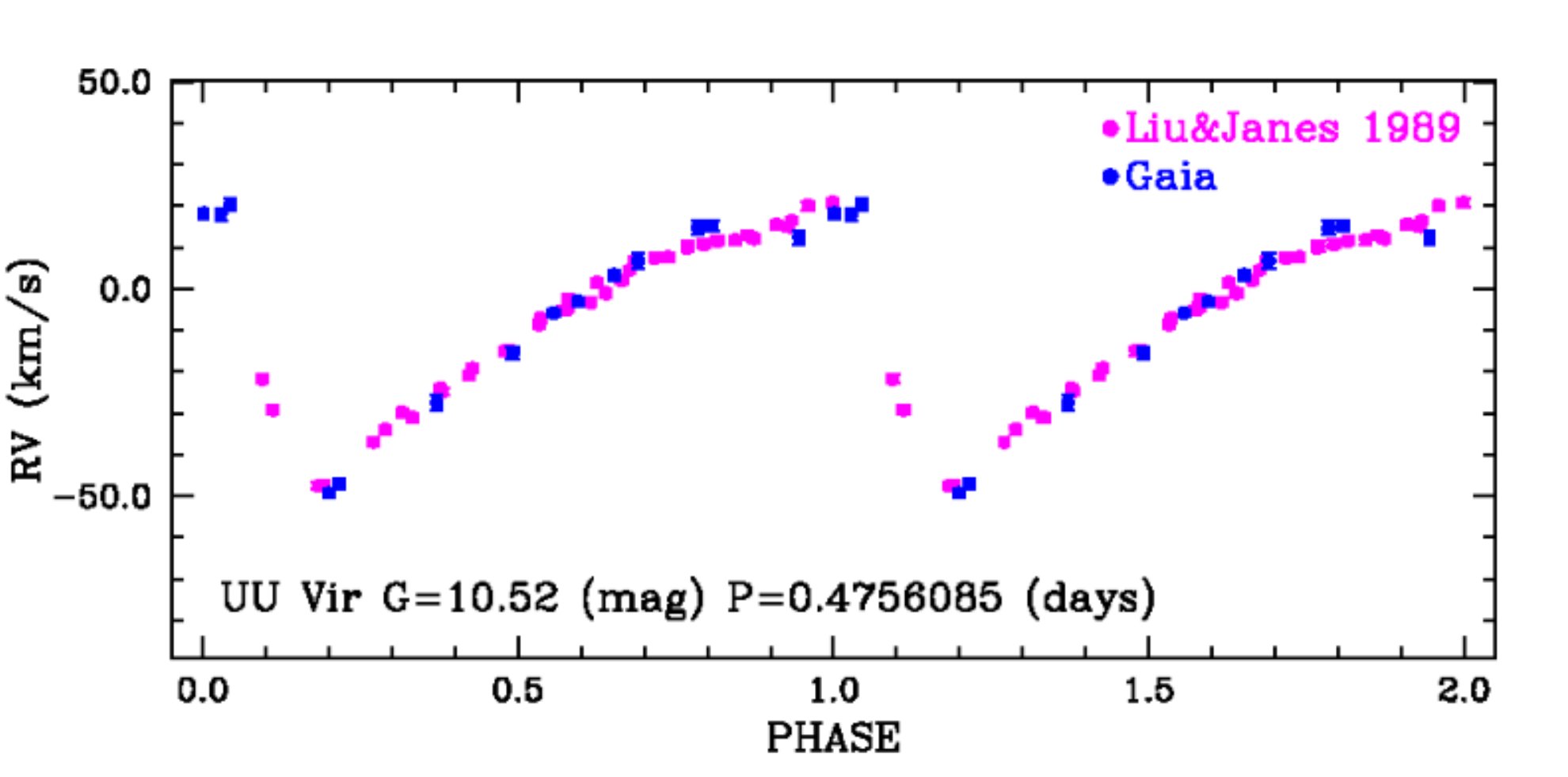}
      ~\includegraphics[width=6.5cm]{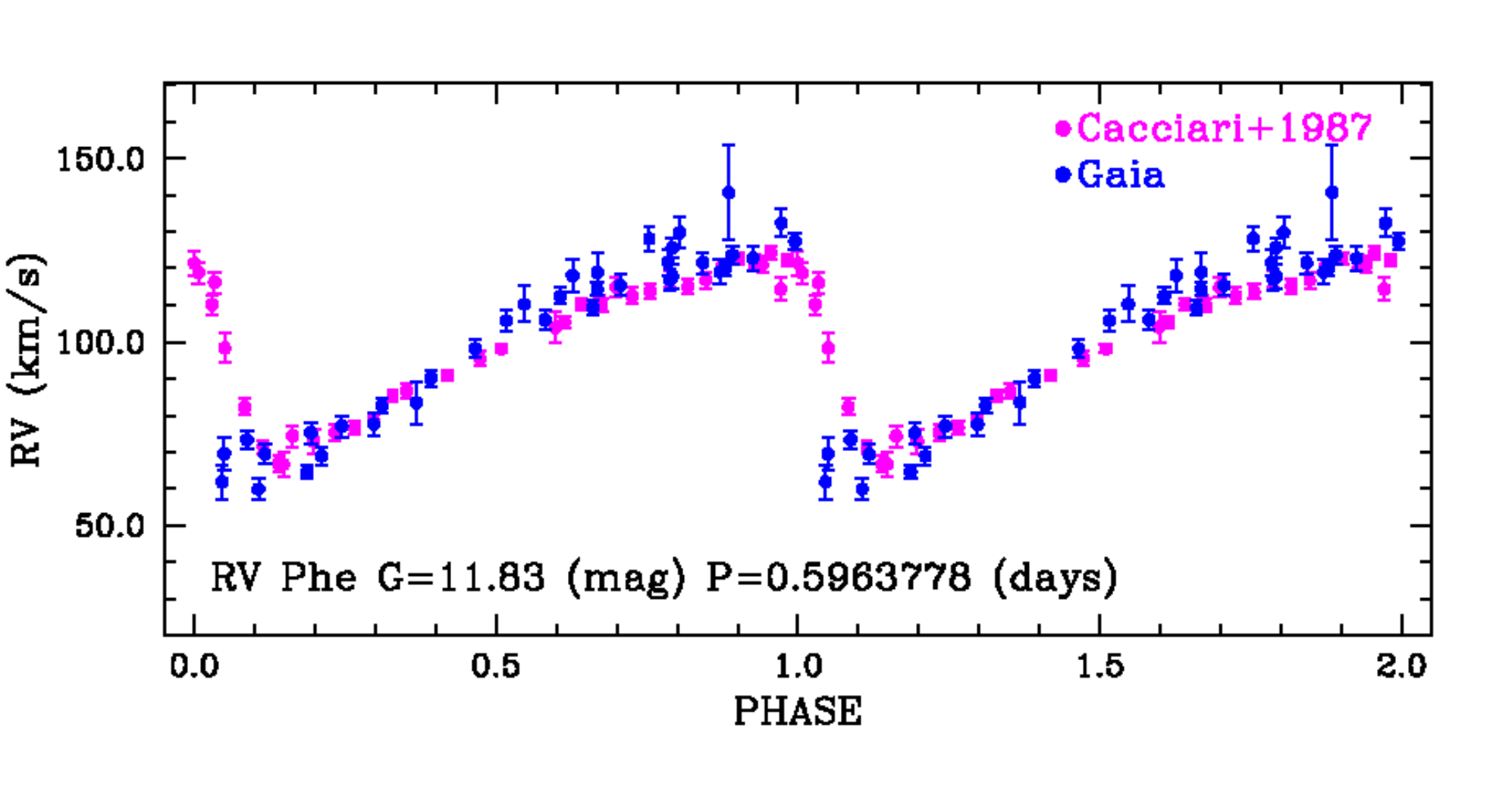}
      \includegraphics[width=6.5cm]{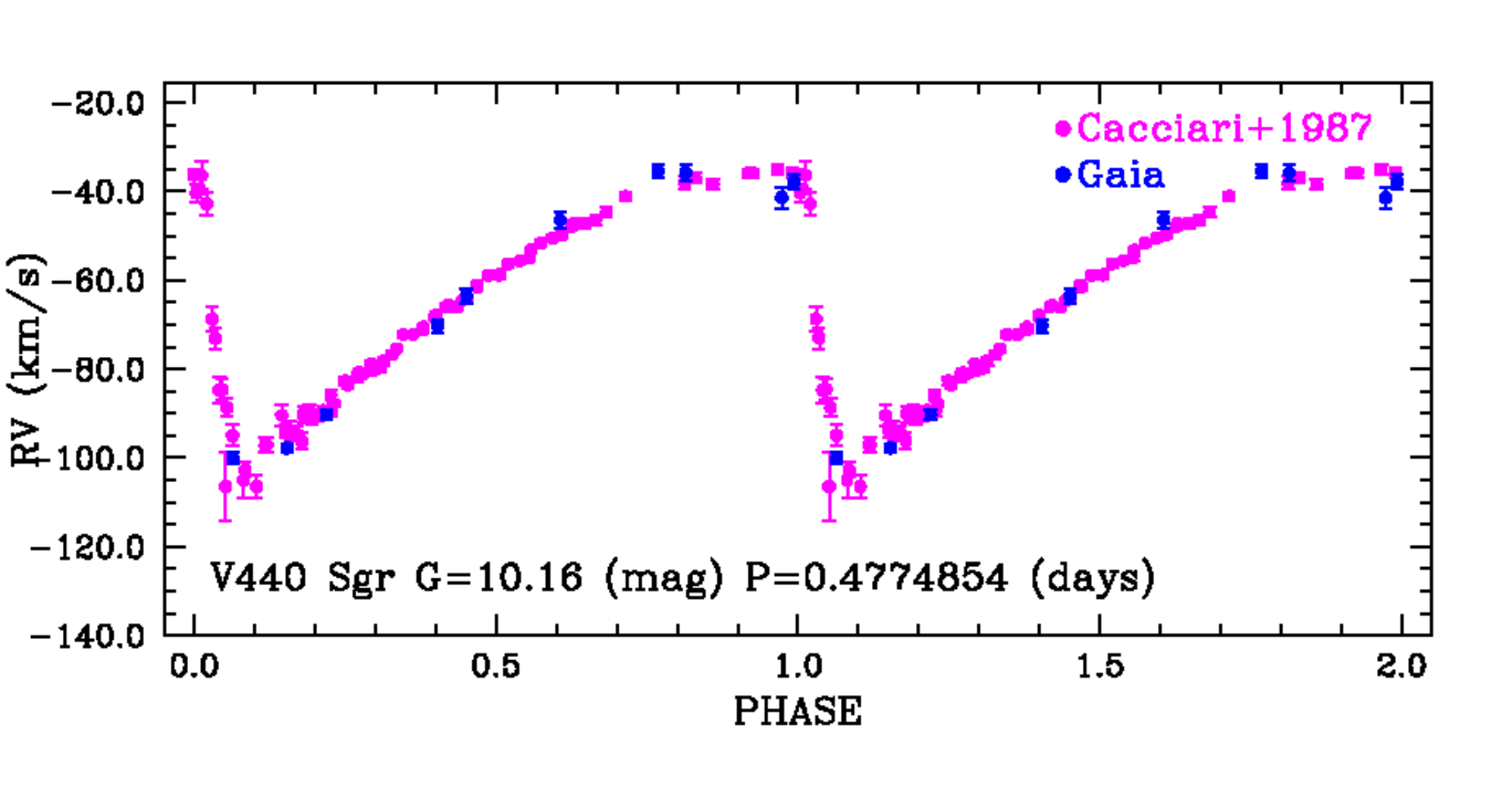}
      ~\includegraphics[width=6.5cm]{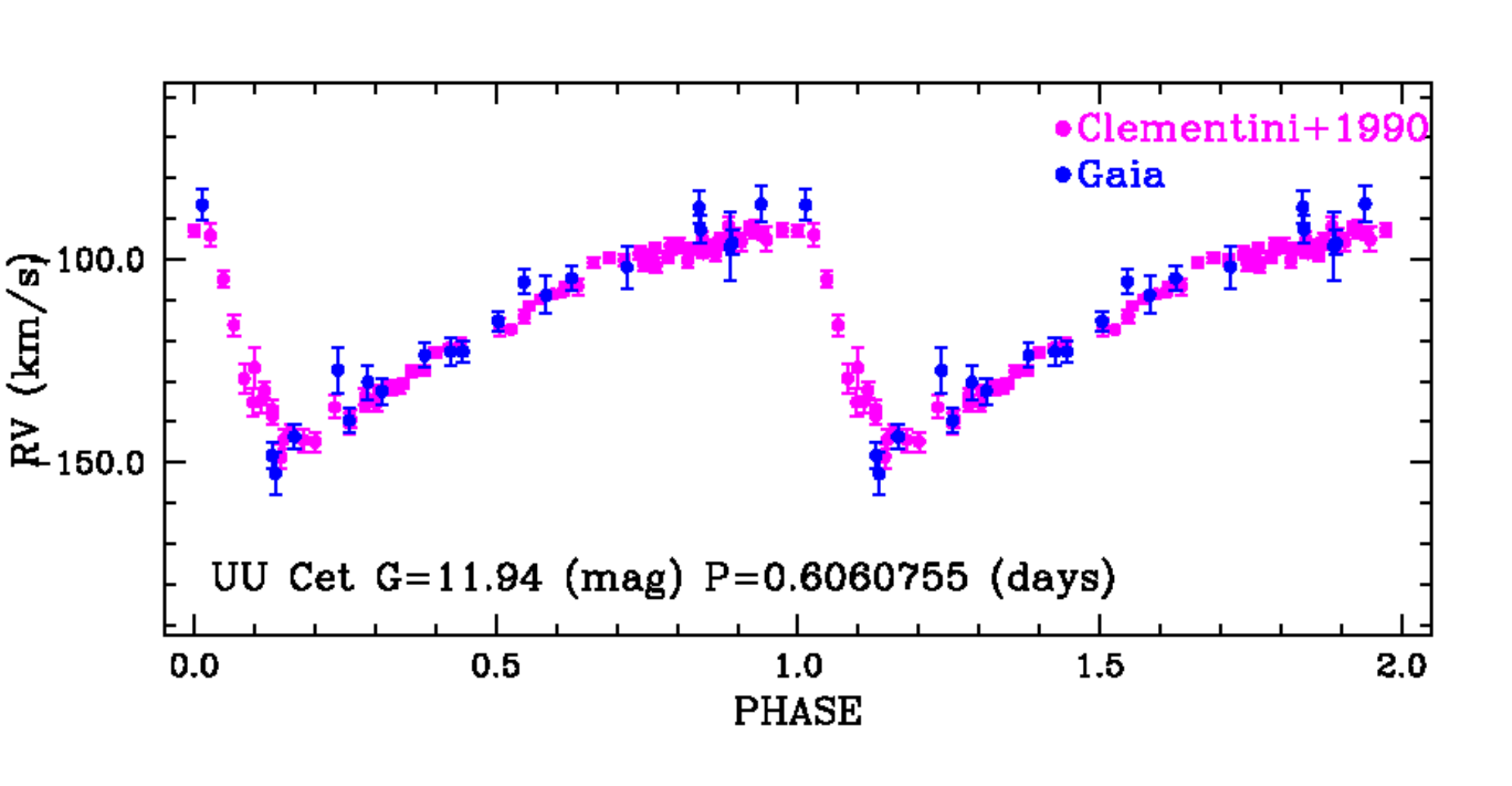}
      \includegraphics[width=6.5cm]{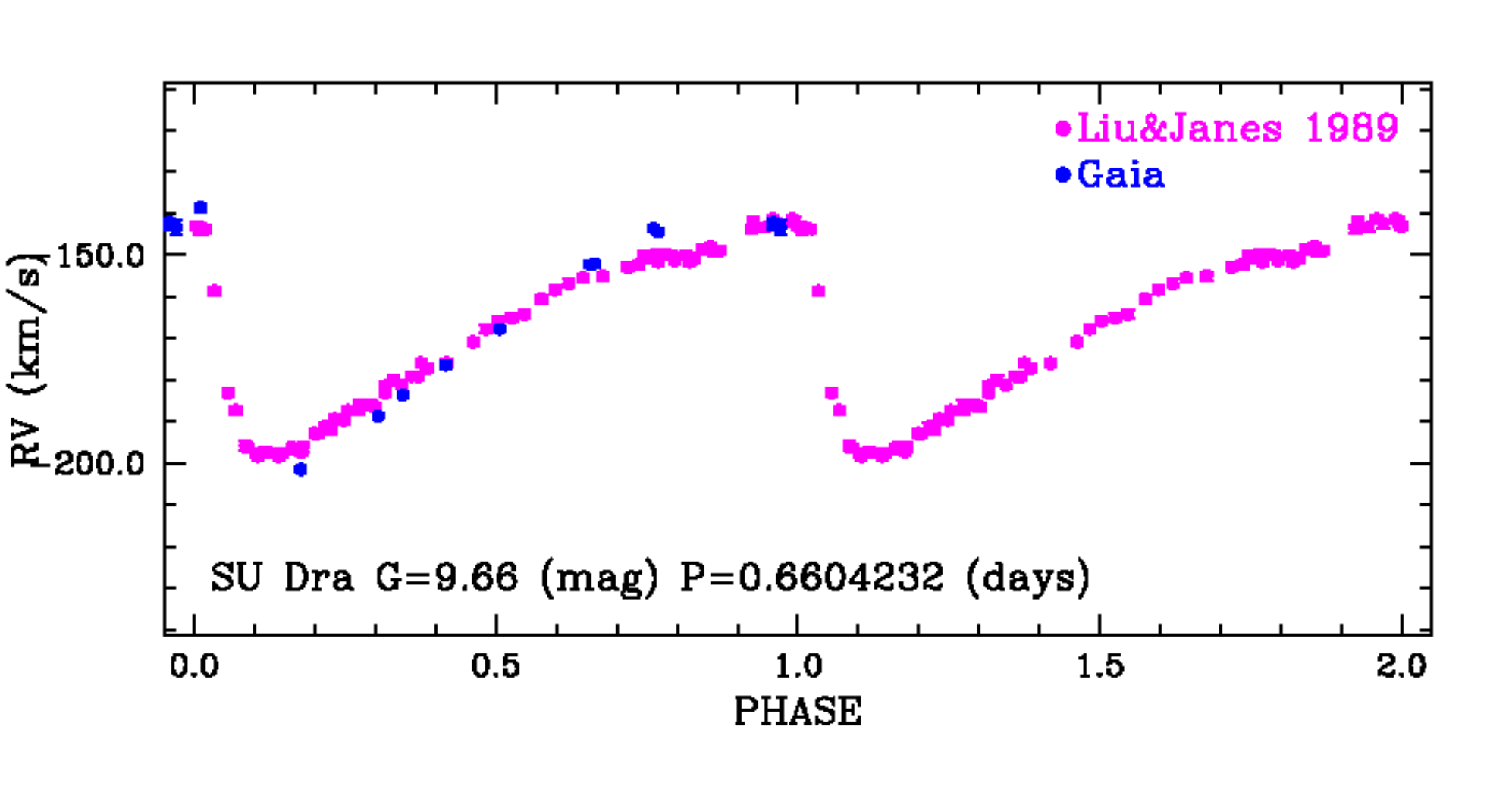}~
      \caption{ Same as in  Fig.~\ref{Fig:RR+CC} for the remaining 13 RR Lyrae stars in the sample of 19 sources discussed in Sect.~\ref{sec:RV-validation}. The Gaia RV curves of TW~Her and UU~Vir were shifted by $-$4 km/s, that of TV~Boo by +5 km/s and that of AV~Peg by $-$3 km/s,  to better match the literature
RV curves.}
     \label{Fig:13di19}
      
  \end{center}
\end{figure*}

\subsection{Atlas of the light and radial velocity curves for the RR Lyrae stars with RVS time series data published in DR3.}\label{appendix-atlas}
We show $G$, $G_{BP}$, $G_{RP}$ light and RV curves for a total sample of 1\,100 sources for which RVS time series RVs are released in DR3. The sample includes 1\,096 bona fide RR Lyrae stars and 4 sources that during validation were re-classified as different variable types. Specifically, source  53376268583351180 was re-classified as DCEP\_1O (see Sect.~\ref{sec:reclassifications}), and the sources: 4130380472726484608, 3062985235999231104, 
573418300979785817 were re-classified as ECLs 
(see Sect~\ref{sec:binaries-et-al}).
\begin{figure*}[h!]
\begin{center}
    \includegraphics[width=16cm]{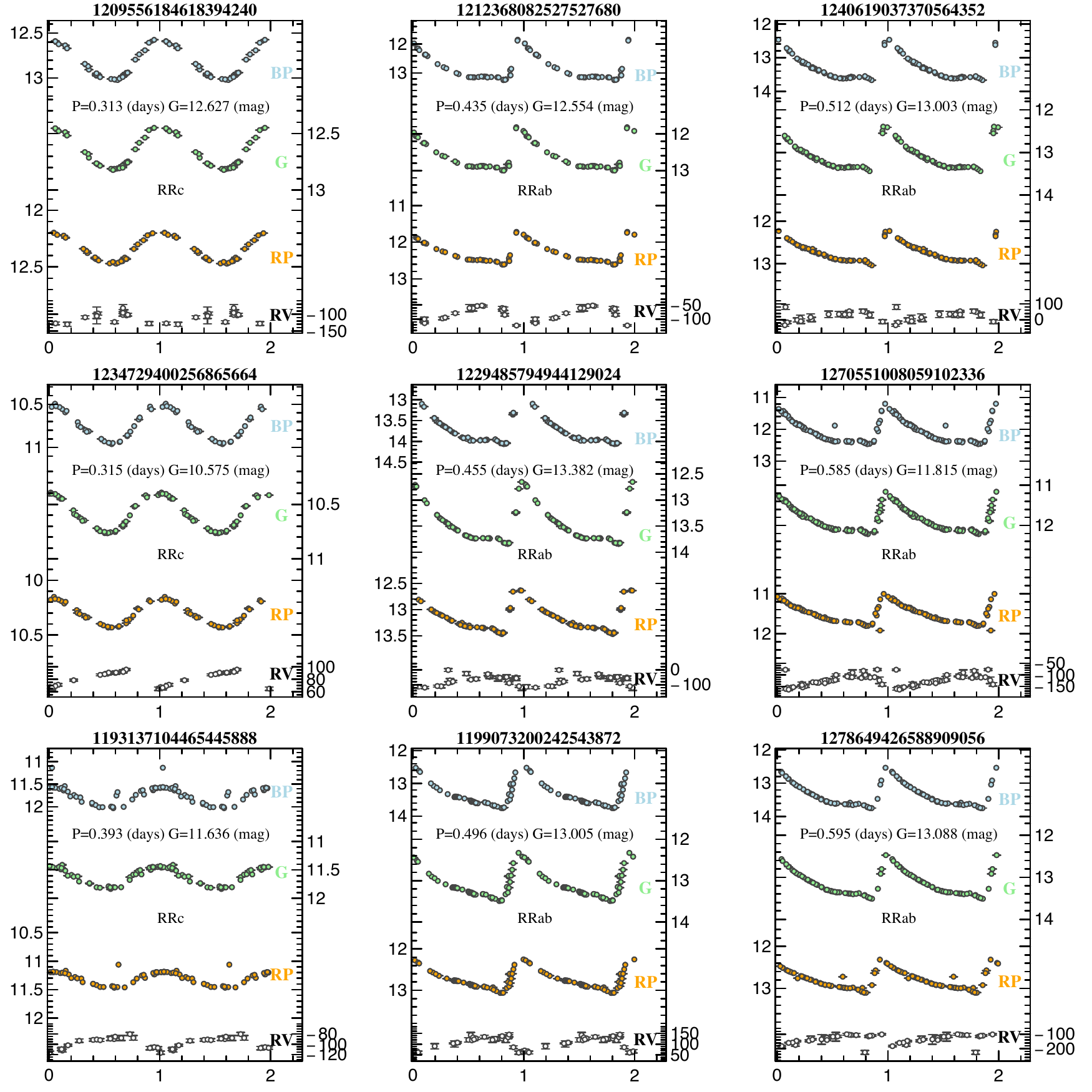}
   \caption{$G_{BP}$, $G$, $G_{RP}$ light and RV curves for RR Lyrae stars with RVS time series data published
in DR3. The atlas is published in its entirety only in the electronic version of the paper.}
   \label{Fig:atlas}
\end{center}
\end{figure*}

\section{Tests on the astrophysical parameters for RR Lyrae stars}
\subsection{Metallicities}\label{other-comparisons}
In this 
section we present 
comparisons 
of the SOS Cep\&RRL photometric metallicities with metal abundances from high resolution spectra of RR Lyrae stars
in  \citet{1995AJ....110.2319C}, \citet{1996ApJS..103..183L}, 
\citet{2015MNRAS.447.2404P},  and with the  metal abundances reported in column 10 of Table~\ref{RRL-TN} for the 19 RR Lyrae  stars used to validate the RV curves from the RVS.

\begin{figure*}
\includegraphics[scale=0.32]{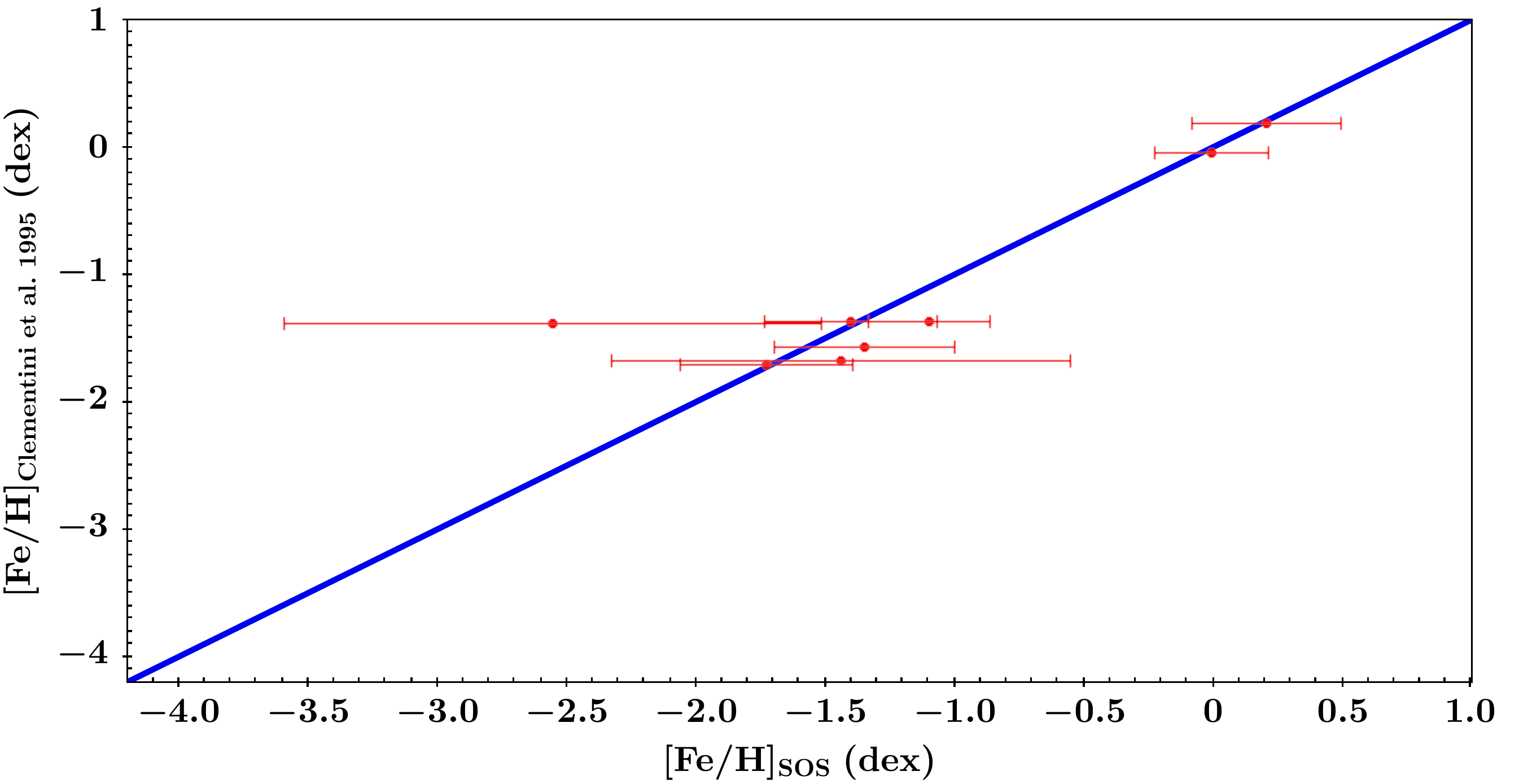}~\includegraphics[scale=0.32]{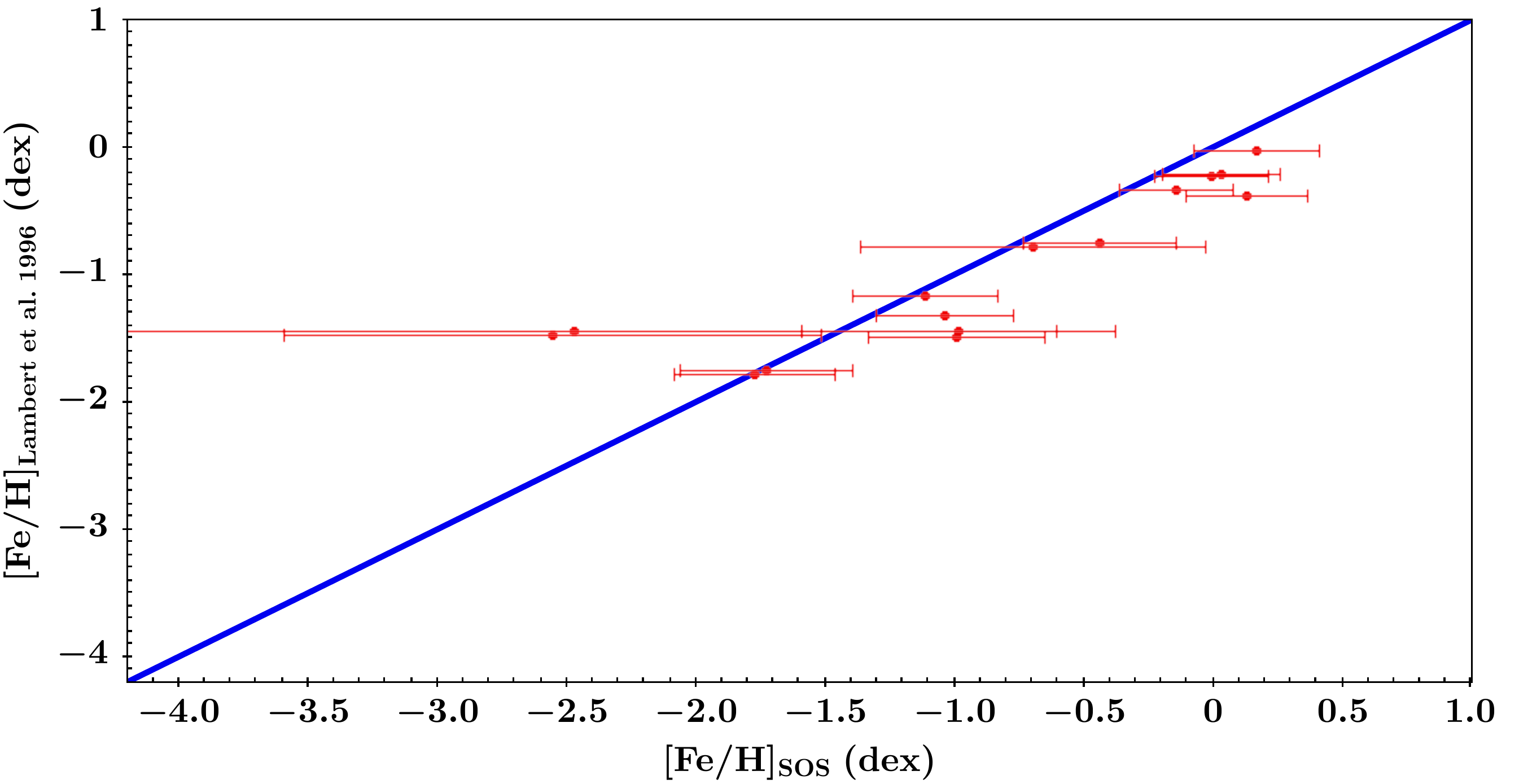}
\caption{{\it Left panel:} Comparison of the photometric metallicities derived by the SOS Cep\&RRL pipeline and metal abundances from high-resolution spectra of 8 RR Lyrae  stars  in \citet{1995AJ....110.2319C}; {\it Right panel:} Same as in the left panel but for metal abundances from high-resolution spectra of 15  
RR Lyrae  stars  in \citet{1996ApJS..103..183L}.
}
\label{fig:Clementini-Lambert}
\end{figure*}

\begin{figure*}
\includegraphics[scale=0.32]{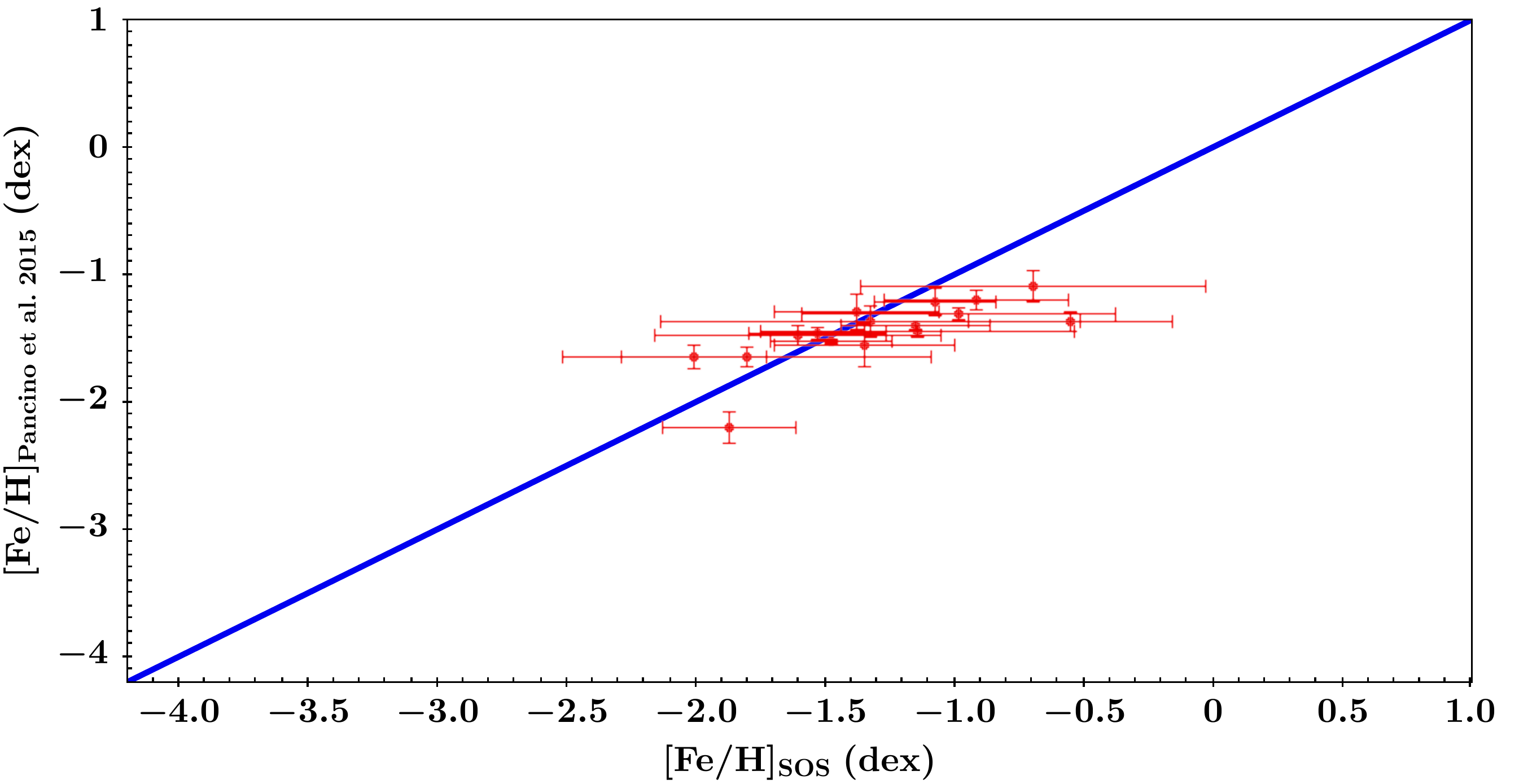}~\includegraphics[scale=0.32]{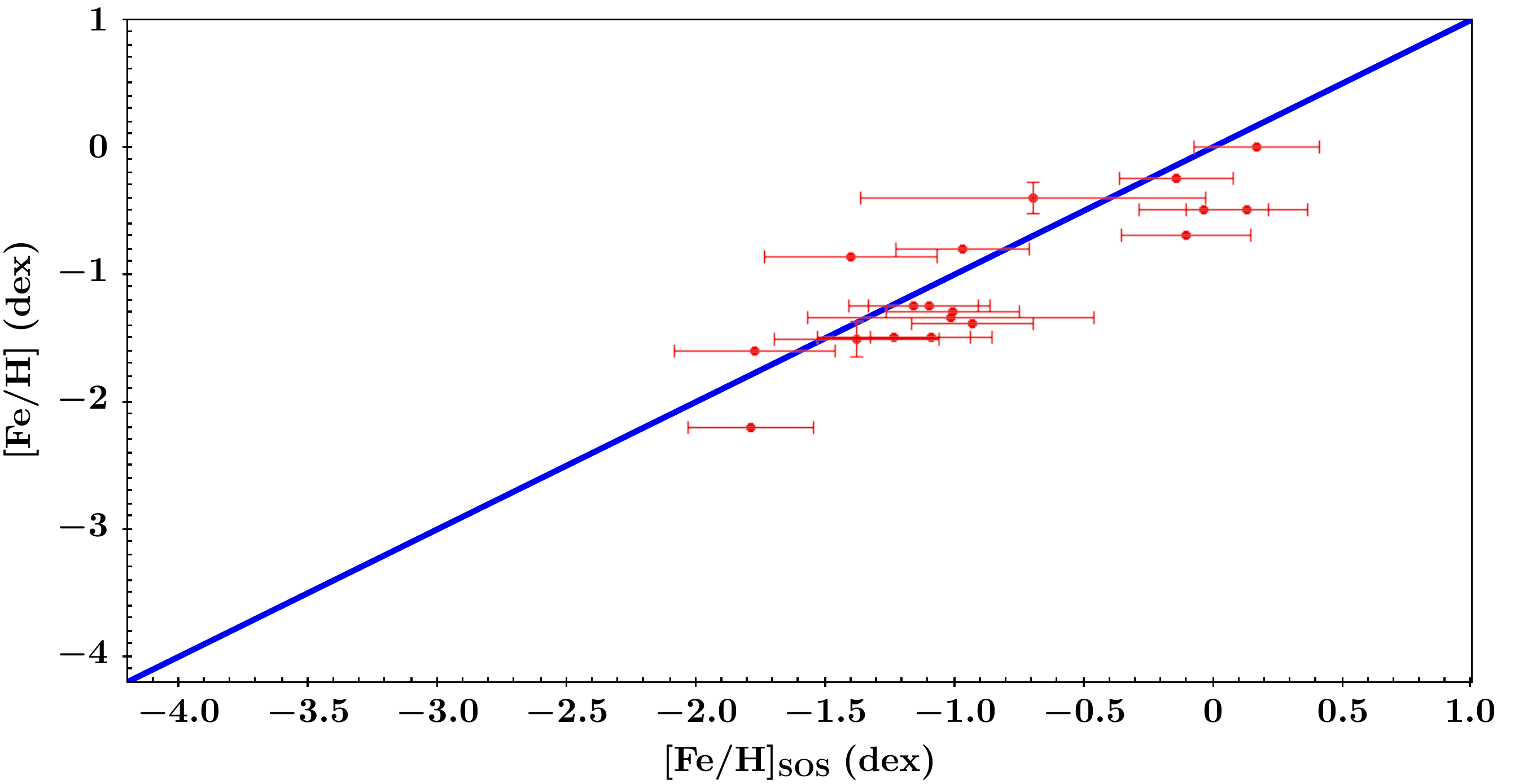}
\caption{{\it Left panel:} comparison of the photometric metallicities derived by the SOS Cep\&RRL pipeline and metal abundances from high-resolution spectra of 16 RR Lyrae  stars  in \citet{2015MNRAS.447.2404P};  {\it Right panel:} comparison of the SOS Cep\&RRL metallicities and the metal abundances listed in Column 10 of Table~\ref{Tab:Table-rr} of 19 RR Lyrae
stars used to validate the RV curves from the RVS.
}
\label{fig:For-Pancino}
\end{figure*}

\subsection{Absorption and metallicity  values from the GSP\_Phot module of Apsis}\label{andrae}

\begin{figure}
   \centering
\includegraphics[scale=0.35]{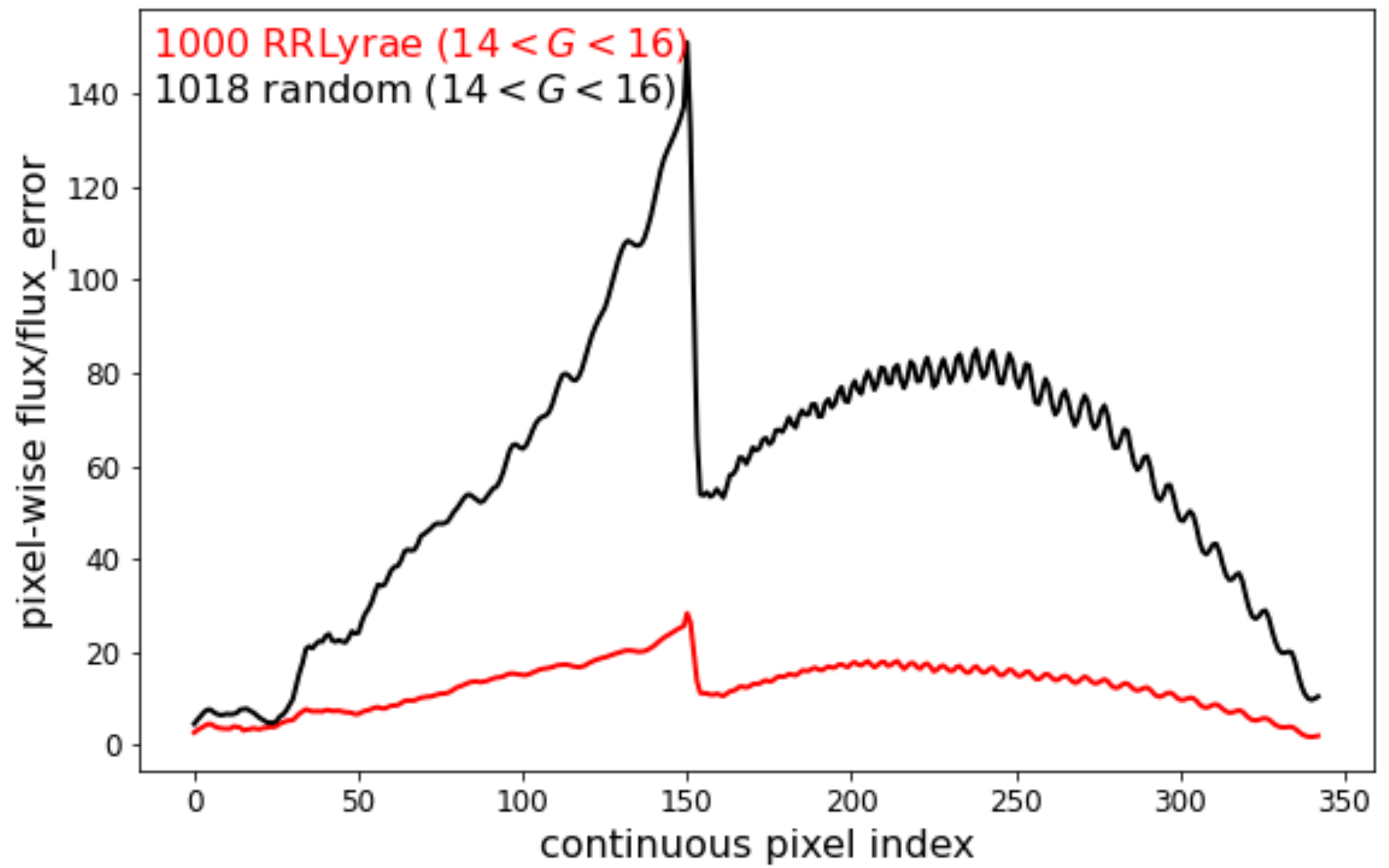}
 \caption{Comparison between the
signal-to-noise ratios (median of each sample in each pixel)
of the continuous 
$BP/RP$ spectra of a sample of 1\,000 RR Lyrae stars with $14<\langle G\rangle <16$ mag extracted from the {\tt vari\_rrlyrae} table (red line) and the continuous 
$BP/RP$ spectra of a 
control sample composed by  1\,000 random non-variable sources with
$14<G<16$ mag.
The random sources (black line) have much higher pixel-wise S/N
than the RR Lyrae stars (red line), even though both samples are drawn from the same apparent $G$ range. 
}
\label{cu8-XP-spectra-RRLs}%
\end{figure}

RR Lyrae stars have unusually large flux errors in
their time-averaged $BP/RP$  
spectra,  
compared to non-variable sources.
This is because 
when computing
the mean $BP/RP$ 
spectra,  
the amplitude of the light variation, that in the $G$ band can range from $\sim$0.2 mag for RRc  to more than 1 magnitude for RRab stars,  enters as an extra flux error in the computed 
time-averaged $BP/RP$ spectra. 
This is clearly shown by the comparison between the
signal-to-noise ratios (median of each sample in each pixel)
of the continuous 
$BP/RP$ spectra of a sample of 1\,000 RR Lyrae stars with $14<\langle G\rangle<16$ extracted from the {\tt vari\_rrlyrae} table and the continuous 
$BP/RP$ spectra of a comparison sample composed by  1\,000 random sources again within $14<G<16$ presented in  Fig.~\ref{cu8-XP-spectra-RRLs}.

The larger 
$BP/RP$ flux uncertainties
cause larger uncertainties in the temperatures derived for RR Lyrae stars, thus opening the room for temperature-extinction degeneracies in the GSP\_Phot absorption estimates for RR Lyrae stars as well as causing large uncertainties and systematic offsets in the GSP\_Phot metallicities  for these variable stars.

   \begin{figure}
   \centering
   \includegraphics[scale=0.35]{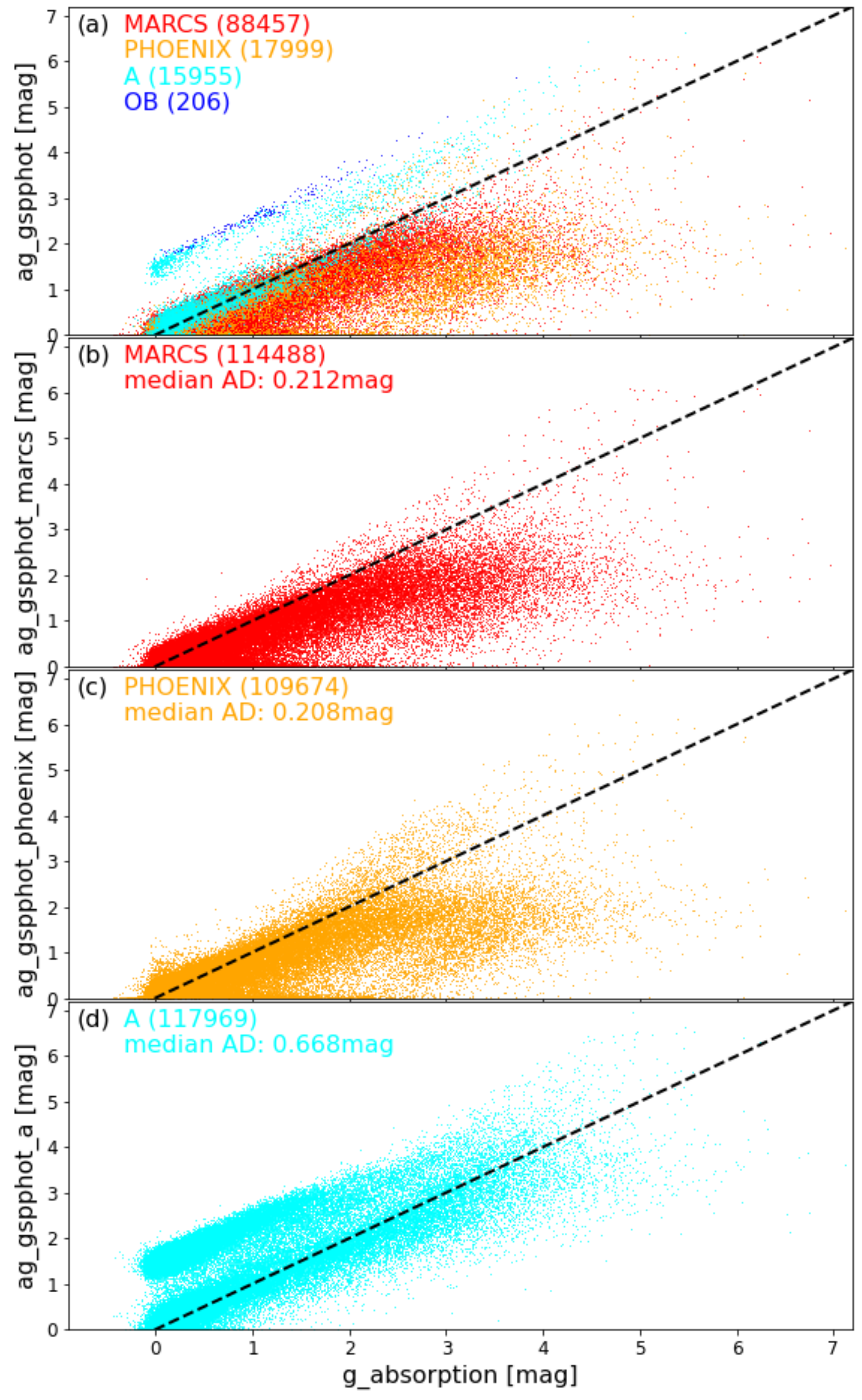}
   \caption{Panel (a), same as in Fig.~\ref{cu8-cu7-A(G)} but now with sources colour-coded according to the GSP\_Phot best library. The numbers behind each library name
indicates how often each library is ``best''. The outlier
group above the one-to-one relation are 
OB stars and some A stars
(though other A stars fall nicely onto the 1-1 relation). Panels (b) to (d) show the results from each
library individually (not best) as taken from the {\tt astrophysical\_parameters\_supp} table in the {\it Gaia} archive. Panels (b) and (c) show that the MARCS and
PHOENIX libraries do not have the outlier group above the identity relation, yet for large SOS Cep\&RRL $A(G)$ values the GSP\_Phot absorption appears to be
smaller. Also quoted in panels (b)-(d) 
are the median absolute differences (AD),
which are 0.212 mag for MARCS and 0.208 mag for PHOENIX. 
}
              \label{cu8-plot1}%
    \end{figure}

   \begin{figure}
   \centering
\includegraphics[scale=0.35]{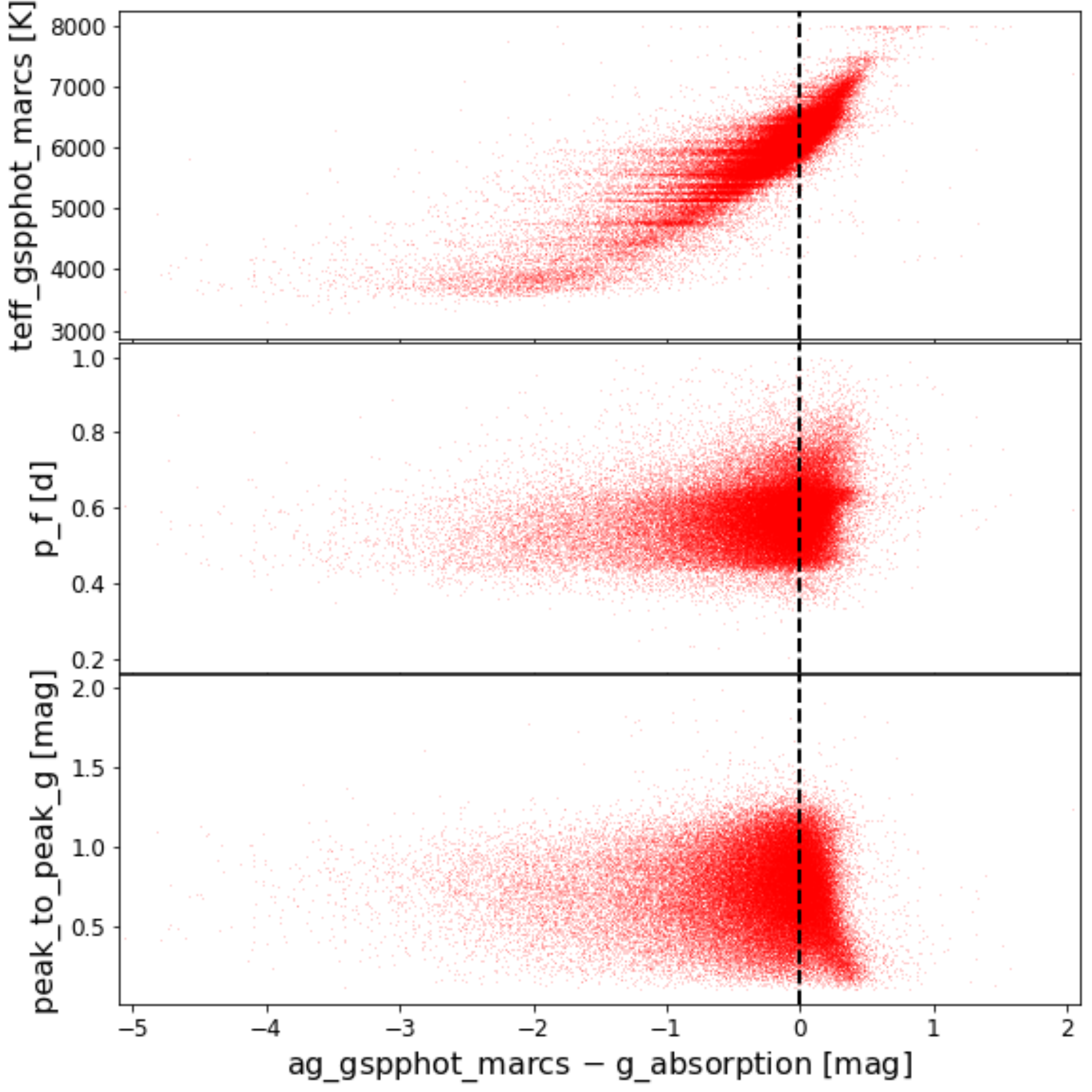}
   \caption{
   Run of the GSP\_Phot  $T_{\rm eff}$, pulsation period ($Pf$) and peak-to-peak amplitude in the $G$ band [Amp$(G)$] of RR Lyrae stars for which both the  GSP\_Phot and the SOS Cep\&RRL $G$ absorption [$A(G)$]  values are available,  as a function of the difference between the two absorption estimates, using for GSP\_Phot only the MARCS library results 
   in the {\tt astrophysical\_parameters\_supp} table of the DR3 archive. There is clearly no relation of the absorption differences to the pulsation period or the $G$ peak-to-peak amplitude.
   The top panel clearly shows that large absorption differences relate to GSP\_Phot MARCS getting the temperature wrong. 
   In fact, the absorption estimates agree for
   $6\,300 \lesssim T_{\rm eff} \lesssim 6\,800$ K,  
   which is the typical mean temperature range of RRab stars (see also Column 11 in Table~\ref{RRL-TN}).
   The variability (0.2-1.0 mag peak-to-peak amplitude in the $G$ band) of the RR Lyrae stars is "absorbed" as increased flux uncertainties in the time-averaged $BP/RP$ 
   spectra. Given the unusually large 
   $BP/RP$ flux uncertainties, the $\chi^2$ is "flattened" and thus the room for degeneracies is opened. In particular, a  temperature-extinction degeneracy takes place that would explain the difference between the SOS Cep\&RRL and the GSP\_Phot absorption estimates for RR Lyrae stars.   
}
              \label{cu8-plot2}%
    \end{figure}


\section{Parameters for 14 RR Lyrae missing in the {\tt vari\_rrlyrae} table.}

We provide in Table~\ref{table:SOS-params-15-RRLs}
parameters computed by the SOS Cep\&RRL pipeline for 14 bona fide RR Lyrae stars that were not published in the DR3 {\tt vari\_rrlyrae} table (see Sect.~\ref{sec:15-overlaps}).  
\begin{sidewaystable*}
\tiny
\setlength\tabcolsep{3pt}
\caption{
Main parameters computed by the SOS Cep\&RRL pipeline for 14 RR Lyrae stars that do not appear in the DR3 {\tt vari\_rrlyrae} table (see Sect.~\ref{sec:15-overlaps}).
The $G$, $G_{\rm BP}$, $G_{\rm RP}$ time series photometry of these sources can be retrieved from the {\tt gaiadr3.vari\_classifier\_results} table. 
}
\label{table:SOS-params-15-RRLs}     
\begin{tabular}{cccccccccccccc}       
\hline\hline                 
\noalign{\smallskip}
\multicolumn{13}{c}{Main parameters computed by the SOS Cep\&RRL pipeline}\\
\noalign{\smallskip}
\hline
\noalign{\smallskip}
Source ID &Type&$Pf$&$P1O$&E\tablefootmark{\rm(a)}($G$
)&$\langle G \rangle$/$\langle G_{\rm BP} \rangle$/$\langle G_{\rm RP} \rangle$&Amp$(G)$/$G_{\rm BP})$/$G_{\rm RP})$&$\phi_{21}(G)$&$R_{21}(G)$&$\phi_{31}(G)$&$R_{31}$&${\rm [Fe/H]}\tablefootmark{\rm (b)}$&A($G$)\tablefootmark{\rm (c)}&
$N_{G/G_{BP}/G_{RP}}$\\ 
&&$\sigma (Pf)$&$\sigma (P1O)$&&&$\sigma {\rm Amp}(G/G_{\rm BP}/G_{\rm RP})$&$\sigma [\phi_{21}(G)]$&$\sigma [R_{21}(G))]$&$\sigma [\phi_{31}(G)]$&$\sigma [R_{31}(G)]$&$\sigma$[Fe/H]&$\sigma$A($G$)&\\
\noalign{\smallskip}
\hline
\noalign{\smallskip}
4092009204924599040&RRc&-&0.2581538&1716.555440&15.4824/15.9069/14.866& 0.126/0.156/0.085&4.806&0.070&-&-&-&-&24/22/23\\
&&-&$\pm5$x10$^{-7}$& &$\pm9.7$x10$^{-4}$/$\pm$0.0023/$\pm$0.0033&$\pm$0.002/$\pm$0.007/$\pm$0.006&$\pm$0.232&$\pm$0.012&-&-&-&-&\\
4120414435009794048&RRc&-&0.256715&1748.953931&17.317/18.136/16.401&0.380/0.385/0.303&4.687&0.174&-&-&-&-&24/12/13\\
&&-&$\pm1$x10$^{-6}$&&$\pm$0.006/$\pm$0.018/0.012&$\pm$0.019/$\pm$0.040/$\pm$0.019&$\pm$0.173&$\pm$0.042&-&-&-&-&\\
4144246349481643392&RRab&0.81001&-&1713.903329&17.216/18.149/16.119&0.225/0.251/0.200&4.859&0.281&-&-&-&2.827&25/24/23\\
&&$\pm1$x10$^{-5}$&-&&$\pm$0.003/$\pm$0.026/$\pm$0.018&$\pm$0.013/$\pm$0.076/$\pm$0.064&$\pm$0.130&$\pm$0.045&-&-&-&$\pm$0.100&\\
5797652730842515968&RRc&-&0.2167765&1700.798893&16.3430/16.5870/15.8778&0.158/0.197/0.107&4.511&0.067&-&-&-&-&56/53/54\\
&&-&$\pm4$x10$^{-7}$&&$\pm7$x10$^{-4}$/$\pm$0.0056/$\pm$0.0030&$\pm$0.002/$\pm$0.017/$\pm$0.008&$\pm$0.182&$\pm$0.012&-&-&-&-&\\
5797917193442176640&RRc&-&0.2637340&1701.161373&15.6366/15.8718/15.2366&0.147/0.170/0.101&4.707&0.083&-&-&-&-&57/48/47\\
&&-&$\pm4$x10$^{-7}$&&$\pm$0.0005/$\pm$0.00196/$\pm$0.00245&$\pm$0.001/$\pm$0.009/$\pm$0.010&$\pm$0.130&$\pm$0.010&-&-&-&-&\\
5846086424210395520&RRc&-&0.2933843&1700.684493&16.6451/16.8922/16.2396&0.156/0.186/0.118&5.00&0.080&-&-&-&-&52/47/46\\
&&-&$\pm5$x10$^{-7}$&&$\pm$0.0008/$\pm$0.0036/$\pm$0.0035&$\pm$0.002/$\pm$0.011/$\pm$0.010&$\pm$0.178&$\pm$0.015&-&-&-&-&\\
5917239841741208576&RRab&0.638600&-&1705.946735&16.3224/16.6628/15.8031&0.147/0.188/0.116&4.422&0.189&2.937&0.040&$-$0.25&0.415&81/80/76\\
&&$\pm$2x10$^{-6}$&-&&$\pm$0.0005/$\pm$0.0026/$\pm$0.0024&$\pm$0.001/$\pm$0.009/$\pm$0.006&$\pm$0.050&$\pm$0.007&$\pm$0.229&$\pm$0.011&$\pm$0.344&$\pm$0.051\\
5991733644318583424&RRc&-&0.258867&1707.080357&16.444/17.10244/15.636&0.135/0.178/0.123&4.378&0.074&-&-&-&-&53/53/50\\
&&-&$\pm$1x10$^{-6}$&&$\pm$0.013/$\pm$0.004/$\pm$0.004&$\pm$0.003/$\pm$0.014/$\pm$0.012&$\pm$0.483&$\pm$0.015&-&-&-&-&\\
6017924835910361344&RRc&-&0.268742&1708.808652&17.762/18.319/17.037&0.214/0.294/0.183&4.847&0.178&-&-&-&-&32/32/30\\
&&-&$\pm$2x10$^{-6}$&&$\pm$0.002/$\pm$0.011/$\pm$0.006&$\pm$0.007/$\pm$0.023/$\pm$0.013&$\pm$0.184&$\pm$0.046&-&-&-&-&\\
6069336998880602240&RRab&0.73337&-&1699.822497&16.3183/16.6855/15.7082&0.181/0.265/0.152&4.516&0.242&-&-&-&0.712&37/35/36\\
&&$\pm$5x10$^{-6}$&-&&$\pm$0.0008/$\pm$0.0157/$\pm$0.0057&$\pm$0.004/$\pm$0.052/$\pm$0.019&$\pm$0.087&$\pm$0.020&-&-&-&-&\\
6707009423228603904&RRc&-&0.306965&1709.373150&16.16385/16.3367/15.7239&0.154/0.138/0.245&4.778&0.048&-&-&-&-&27/25/26\\
&&-&$\pm$1x10$^{-6}$&&$\pm$0.0009/$\pm$0.0155/$\pm$0.0261&$\pm$0.022/$\pm$0.043/$\pm$0.086&$\pm$0.320&$\pm$0.012&-&-&-&-&\\
5935214760885709440&RRc&-&0.340823&1706.070130&19.0373/20.624/17.757&0.158/0.819/0.113&5.373&0.193&-&-&-&-&40/24/27\\
&&-&$\pm$4x10$^{-6}$&&$\pm$0.004/$\pm$0.092/$\pm$0.011&$\pm$0.018/$\pm$0.301/$\pm$0.044&$\pm$0.495&$\pm$0.065&-&-&-&-&\\
4362766825101261952&RRc&-&0.2631889&1716.317264 &16.3937/16.9315/15.6715&0.140/0.153/0.115&4.849&0.0518&-&-&-&-&43/40/41\\
&&-&$\pm$5x10$^{-7}$&&$\pm$0.0008/$\pm$0.0071/$\pm$0.0036 &$\pm$0.002/$\pm$0.027/$\pm$0.011&$\pm$0.380&$\pm$0.019&-&-&-&-&\\
5967334102579505664&RRab&0.62513&-&1707.700232&18.426/19.999/17.229&0.130/0.372/0.124&-&-&-&-&-&3.429&37/38/37\\
&&$\pm$2x10$^{-5}$&-&&$\pm$0.005/$\pm$0.040/$\pm$0.012&$\pm$0.013/$\pm$0.310/$\pm$0.059&-&-&-&-&-&$\pm$0.090&\\
\noalign{\smallskip} \hline 
\end{tabular}
\tablefoot{
$^{\rm (a)}$The BJD of the 
epoch of maximum light 
is offset by JD 2455197.5 d (= J2010.0).  
$^{\rm (b)}$Photometric metal abundance derived from the $\phi_{31}$ Fourier parameter of the light curve of 
fundamental-mode RR Lyrae stars (see Sects.~\ref{sec:sos-met} and ~\ref{sec:results}).  $^{\rm (c)}$Absorption in the $G$ band computed from a relation that links the star intrinsic colour to the period and the amplitude of the $G$-band light variation 
of 
RR Lyrae stars (see Sects.~\ref{sec:sos-A(G)} and ~\ref{sec:results}).}\\
\end{sidewaystable*}
 

\section{OGLE-IV, ASAS-SN and  CATALINA control samples}
\begin{figure*}
   \centering
   \includegraphics[scale=0.49]{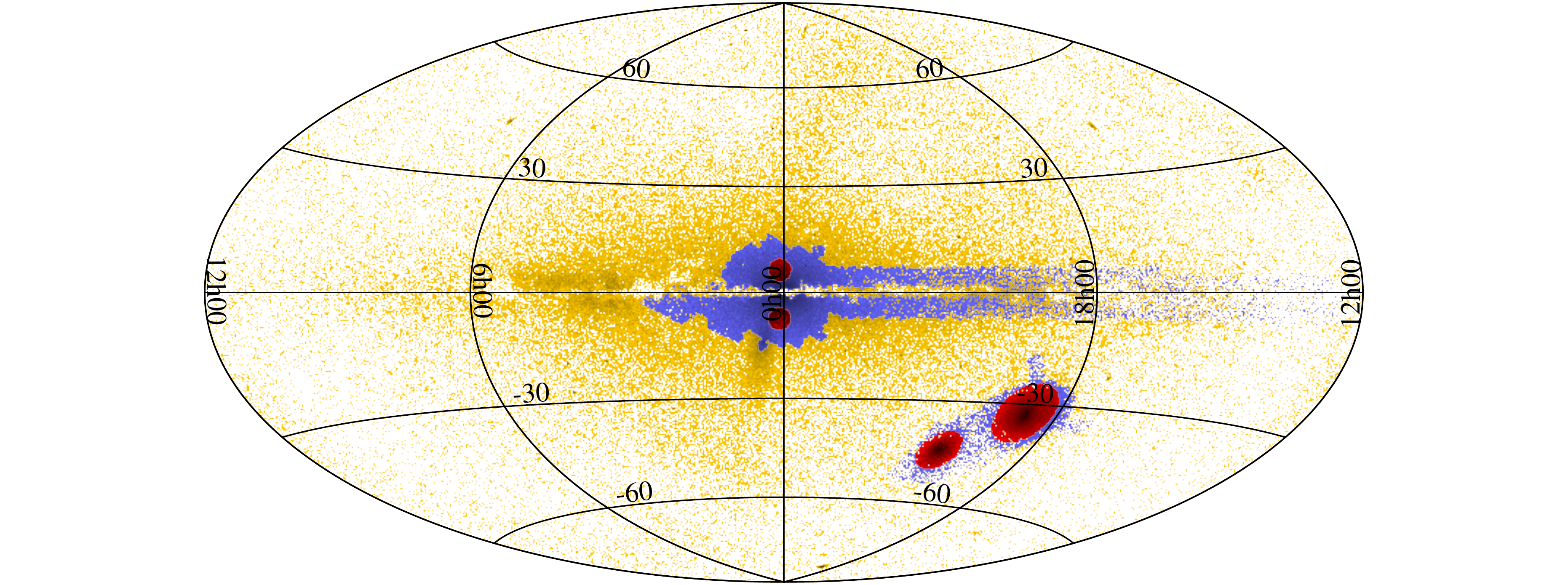}
   \caption{Regions (red areas) in the footprint of the OGLE-IV RR Lyrae stars (violet symbols) that we have used to estimate the  completeness, contamination and percentage of new discoveries of the  final, clean catalogue of DR3 RR Lyrae stars  (orange symbols), in the LMC,  SMC and in the Galactic Bulge.
   For the LMC we used a circular region with  $8.3^{\circ}$ in radius centred at (RA=$81.5^{\circ}$ Dec=$-70.1^{\circ}$), for the SMC a 
   region with $5.6^{\circ}$ in radius centred at (RA=$13.2^{\circ}$, Dec=$-72.9^{\circ}$) and for the MW Bulge two regions with $3^{\circ}$ degree in radius centred  
   at (RA=$274.7^{\circ}$, Dec=$-31.8^{\circ}$;  Bulge-up) and 
   (RA=$261.2^{\circ}$, Dec=$-24.9^{\circ}$;  Bulge-down), respectively.
}
               \label{OGLE-IV-footprint}%
    \end{figure*}

\begin{figure*}
   \centering
\includegraphics[scale=0.7]{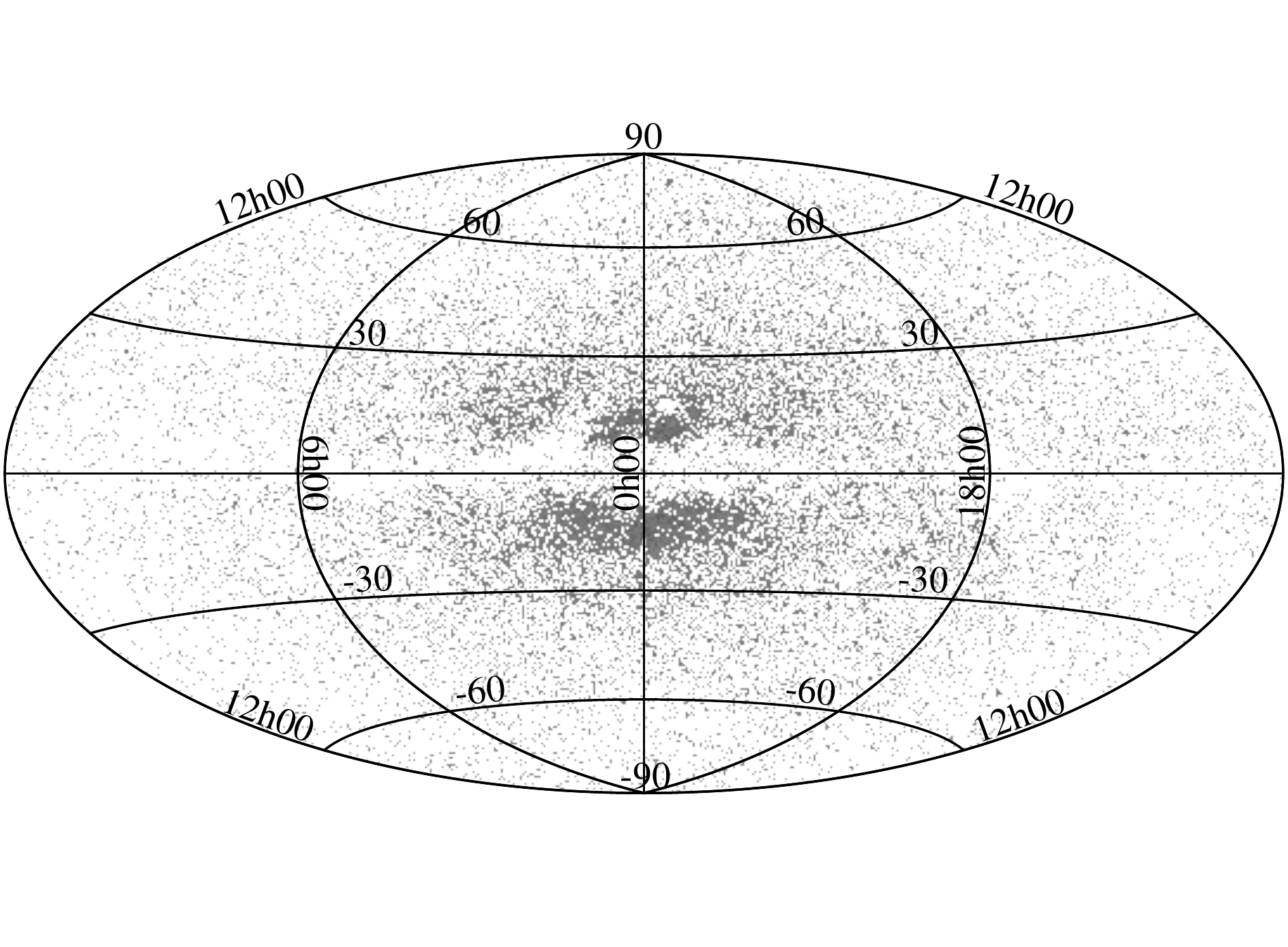}
   \caption{Footprint of the RR Lyrae stars in the ASAS-SN survey that we used to estimate the completeness of the {\it Gaia} DR3 RR Lyrae
clean catalogue. 
    In the region shown in the map there are 28\,377 all-sky RR Lyrae stars observed by ASAS-SN with magnitudes in the range 10.4 $<V<17.4$ mag. 
   In the RR Lyrae clean catalogue we recover 20\,921 of them, using a cross-match radius of 2.5 arcsec.}            \label{ASAS-SN-footprint}%
    \end{figure*}

\begin{figure*}
   \centering
   \includegraphics[scale=0.7]{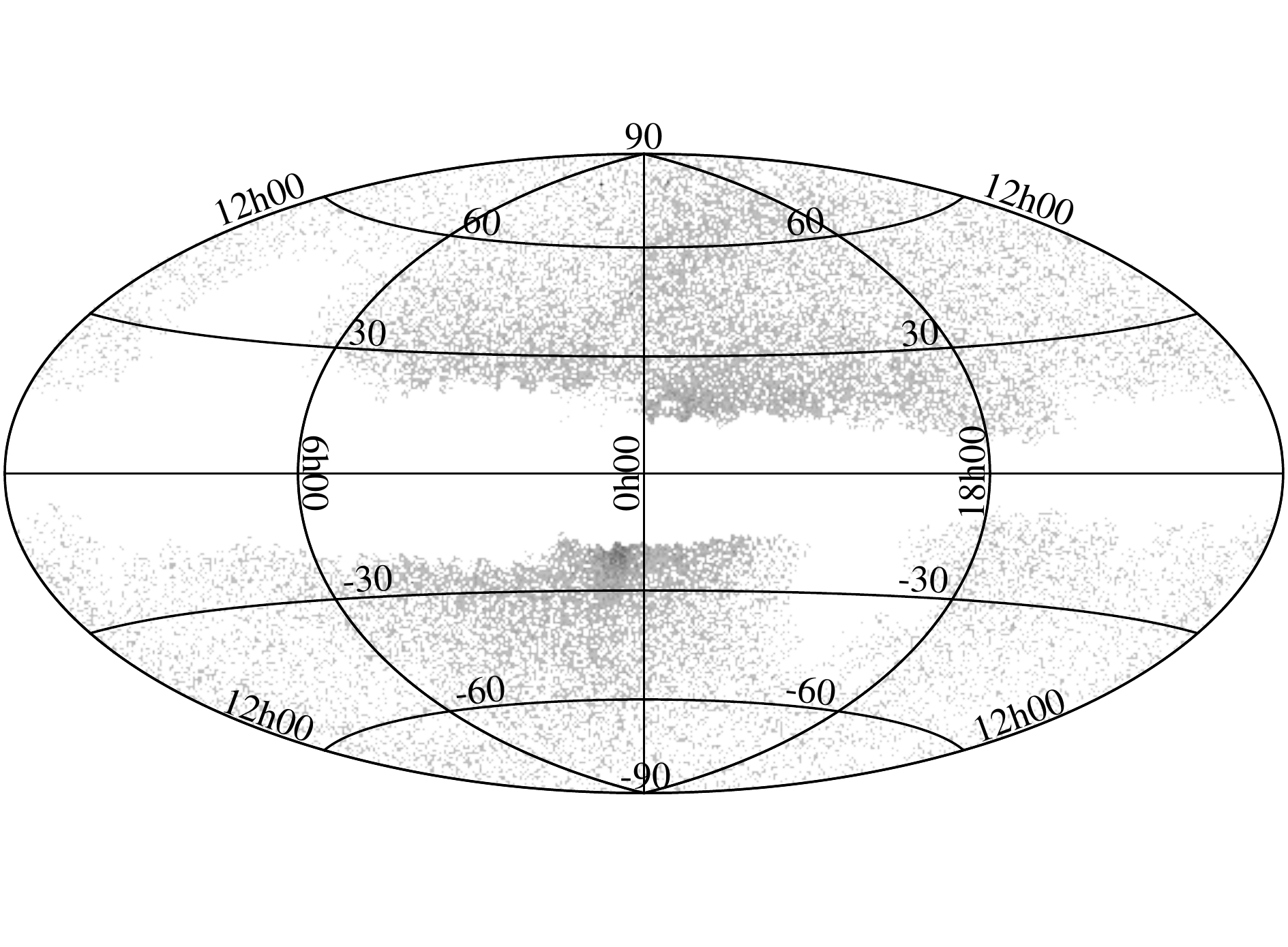}
   \caption{Footprint of the RR Lyrae stars of the CATALINA survey that we used to estimate the completeness  of the {\it Gaia} DR3 RR Lyrae clean catalogue. 
   In the region shown in the map there are 42\,775 RR Lyrae observed by CATALINA, with magnitudes in the range 11 $<V<21.1$ mag. 
   In the RR Lyrae clean catalogue we recover 36\,576 of them, using a cross-match radius of 2.5 arcsec.}           
   \label{CATALINA-footprint}%
   \end{figure*}

Figures~\ref{OGLE-IV-footprint}, ~\ref{ASAS-SN-footprint} and ~\ref{CATALINA-footprint} show the footprints of RR Lyrae control samples drawn from the OGLE-IV, ASAS-SN and CATALINA   catalogues, respectively, that were used to estimate the completeness (and contamination, only OGLE) of the final clean catalogue of 270\,905 
RR Lyrae stars 
that are released in DR3.

\section{Examples of \textit{\textbf{Gaia}} archive queries}
\label{app:queries}
 The names of the tables in the queries will change, this will be updated at the final submission of the paper. Also, functions might be added to the time series query to group results 

  \begin{table*}[!htbp]
    \caption{Queries to retrieve DR3 information on the 
    RR Lyrae stars from the {\it Gaia} archive in the Astronomical Data Query Language (\citealt{2008ivoa.spec.1030O}).}
    \label{tab:queries}
    \centering
    \begin{tabular}{p{\textwidth}}
    \footnotesize
      Query to retrieve time series of all RR Lyrae stars confirmed by the SOS Cep\&RRL pipeline in the {\it Gaia} DR3.
      \begin{verbatim}
 select gaia.source_id, epoch_photometry_url from gaiadr3.gaia_source as gaia
      inner join gaiadr3.vari_rrlyrae as rrl on gaia.source_id=rrl.source_id
      \end{verbatim} \\
      Query to retrieve the number of processed observations and SOS Cep\&RRL-computed parameters of all RR Lyrae in the {\it Gaia} DR3.
      \begin{verbatim}
      select rrl.*,tsr.num_selected_g_fov,tsr.num_selected_bp,tsr.num_selected_rp 
           from gaiadr3.vari_rrlyrae rrl 
              inner join gaiadr3.vari_time_series_statistics tsr on rrl.source_id=tsr.source_id
     \end{verbatim}
    \end{tabular}
  \end{table*}

\section{Acronyms}
\label{app:acro}

  \begin{table*}[!htbp]
      \caption{
      List of acronyms used in this paper.}
\begin{tabular}{ll}
\hline
\hline
\noalign{\smallskip}
{\bf Acronym} & {\bf Description}  \\
\hline 
\noalign{\smallskip}
ACEP & Anomalous Cepheid \\ 
$A(G)$& Absorption in the $G$ band\\
AGN&Active Galactic Nuclei\\
All\_Sky&The celestial region excluding  the LMC, SMC, M31 and M33 regions \\
Amp($G$)& Amplitude of the light variation in the $G$ band\\
Amp($G_{BP}$)& Amplitude of the light variation in the $G_{BP}$ band\\
Amp($G_{RP}$)& Amplitude of the light variation in the $G_{RP}$ band\\
Amp($RV$)& Amplitude of the RVS radial velocity variation\\
B-W&Baade-Wesselink\\
CMD & Colour Magnitude Diagram \\
DCEP &  Classical Cepheid (Population I) \\
DPAC& Data Processing and Analysis Consortium\\
dSph& Dwarf spheroidal galaxy \\ 
DR&Data Release \\
EDR3& Early Data Release 3\\
ECL&Eclipsing binary\\
ELL&Ellipsoidal variable\\ 
F&Fundamental mode of pulsation \\ 
$[$Fe/H$]$& Iron abundance\\
FO or 1O&First overtone mode of pulsation \\
$G$ & Gaia photometric $G$-band \\ 
$G_{BP}$ &Gaia photometric $G_{BP}$ band \\
$G_{RP}$&Gaia photometric $G_{RP}$ band \\
$G_{RVS}$&Gaia photometric $G_{RVS}$ band\\
GC& Globular cluster \\ 
LMC&Large Magellanic Cloud \\
LZ&Luminosity--Metallicity\\
M31& Andromeda galaxy\\ 
M33& Triangulum galaxy\\
MAD&Median Absolute Deviation\\
M$_{G_{0}}$&Absolute G magnitude\\
$[$M/H$]$&Total metal abundance\\
MORO&MultibandOutlierRemovalOperator\\
MW&Milky Way \\
PA&Period--Amplitude \\
Paper I& \citet{Clementini-et-al-2016}\\
Paper II& \citet{Clementini-et-al-2019}\\
PL&Period--Luminosity \\
PLZ&Period--Luminosity--Metallicity\\
PW&Period--Wesenheit \\
QSO& Quasi Stellar Object \\
ROFABO&RemoveOutliersFaintAndBrightOperator\\
RRab&RR Lyrae star of ab type \\
RRc&RR Lyrae star of c type \\
RRd& double-mode RR Lyrae star \\
RV& Radial Velocity \\
RVS& Radial Velocity Spectrometer \\
SEAPipe&Source Environment Analysis Pipeline\\
SMC&Small Magellanic Cloud \\
SOS& Specific Object Study \\
T2CEP & Type II Cepheid (Population II)\\
UFD &  Ultra-faint dwarf galaxy \\
 \hline  
\noalign{\smallskip}
\end{tabular} 
  \end{table*}

\end{appendix}

\end{document}